\documentstyle[amsfonts,amssymb,epsfig,twoside]{article}
\newcommand{\half}{{\textstyle {1 \over 2}}}
\newcommand{\be}{\begin{equation}}
\newcommand{\ee}{\end{equation}}
\newcommand{\bea}{\begin{eqnarray}}
\newcommand{\eea}{\end{eqnarray}}
\newcommand{\tpol}{{\textstyle {2 \pi \over L}}}
\newcommand{\gol}{\stackrel{\scriptscriptstyle >}{\scriptscriptstyle <}}

\newcommand{\Eq}[1]{Eq.~(\ref{#1})}
\newcommand{\Eqs}[1]{Eqs.~(\ref{#1})}

\newcommand{\SH}{{\scriptscriptstyle Sha}}
\newcommand{\HAL}{{\scriptscriptstyle Hal}}
\newcommand{\ZZ}{{\mathbb Z}}
\newcommand{\RR}{{\mathbb R}}
\newcommand{\bk}{{\bar k}}

\renewcommand{\thesection}{\arabic{section}}
\renewcommand{\thesubsection}{\thesection.\Alph{subsection}}
\renewcommand{\thesubsubsection}{\thesection.\Alph{subsection}.\arabic{subsubsection}}

\begin{document}

\vspace*{27mm}

\noindent
{\large {\bf Bosonization for Beginners --- Refermionization for Experts}}

\bigskip

\noindent
 {\bf Jan von Delft and Herbert Schoeller}

\bigskip

\noindent
{\small 
Institut f\"ur Theoretische Festk\"orperphysik, 
Universit\"at   Karlsruhe, D-76128 Karlsruhe, Germany 

\bigskip

\noindent
Submitted 25 May 1998, revised 27 September 1998. 

\bigskip

\noindent
{\bf Abstract.} 
This tutorial review gives an elementary and self-contained derivation
of the standard identities $(\psi_\eta (x) \sim F_\eta e^{-i \phi_\eta
  (x)}$, etc.)  for abelian bosonization in 1 dimension in a system of
finite size $L$, following and simplifying Haldane's constructive
approach.  As a non-trivial application, we rigorously resolve
(following Furusaki) a recent controversy regarding the tunneling
density of states, $\rho_{dos}(\omega)$, at the site of an impurity in
a Tomonaga-Luttinger liquid: we use finite-size refermionization to
show exactly that for $g= \half$ its asymptotic low-energy behavior is
$\rho_{dos}(\omega) \sim \omega$.  This agrees with the results of
Fabrizio \& Gogolin and of Furusaki, but not with those of Oreg and
Finkel'stein (probably because we capture effects not included in
their mean-field treatment of the Coulomb gas that they obtained by an
exact mapping; their treatment of anti-commutation relations in this
mapping is correct, however, contrary to recent suggestions in the
literature).  --- The tutorial is addressed to readers with little or
no prior knowledge of bosonization, who are interested in seeing ``all
the details'' explicitly; it is written at the level of beginning
graduate students, requiring only knowledge of second quantization,
but not of field theory (which is not needed here).  At the same time,
we hope that experts too might find useful our explicit treatment of
certain subtleties that can often be swept under the rug, but are
crucial for some applications, such as the calculation of $\rho_{dos}
(\omega)$ -- these include the proper treatment of the so-called Klein
factors that act as fermion-number ladder operators (and also ensure
the anti-commutation of different species of fermion fields), the
retention of terms of order $1/L$, and a novel, rigorous formulation
of finite-size refermionization of both $F e^{-i \Phi(x)}$ and the
boson field $\Phi (x)$ itself. \vspace*{3mm}

Changes relative to first version of cond-mat/9805275: We have
substantially revised our discussion of the controversy regarding the
tunneling density of states $\rho_{dos}$ at the site of an impurity in
a Luttinger liquid, with regard to the following points: (1) In a new
Appendix K, we confirm explicitly that Oreg and Finkel'stein's
treatment of fermionic anti-commutation relations is {\em correct}\/,
contrary to recent suggestions (including our own).  (2) 
To try to understand why their result for $\rho_{dos}$ differs from
that of Fabrizio \& Gogolin, Furusaki and (for g=1/2) ourselves, we
make a new suggestion in Sections 1.B and 10.D: this is probably because of
effects not captured by their {\em mean-field}\/ treatment of their
Coulomb gas. (3) In Sections 10.C and 10.D we have replaced the first
version of our calculation of $\rho_{dos}$ by a more explicit one (the
result is unchanged), in which we refermionize not only the
exponential $e^{i \Phi}$ but, for the first time, also the field
$\Phi$ itself (Section 10.C.4); this allows us to calculate various
correlation functions involving $\Phi$ explicitly in terms of {\em
  fermion}\/ operators (a new Appendix J contains several detailed
examples, and a new Figure 4 showing the corresponding Feynman
diagrams).  \bigskip

\\cond-mat/9805275;  Annalen der Physik, Vol. {\bf 4}, 225-305 (1998).
}

\bigskip


\tableofcontents

\section{Introduction}
\label{introduction}


1-Dimensional abelian\footnote{We discuss only abelian bosonization. 
For a review of non-abelian
bosonization, see \protect\cite{Lud94a} or 
\protect\cite{vondelftthesis}.}
 bosonization is a technique for representing
1-D fermion fields $\psi_\eta (x)$, where $\eta$ is a
species (e.g. spin) index,   in terms of  bosonic fields $\phi_\eta (x)$
through a relation of the form\footnote{Alternative
notations are discussed in Sections~\ref{Haldanesnotation},
\ref{sec:KFcompare}, \ref{Shankarsnotation} and \ref{thetaF}.}
$\psi_\eta \sim F_\eta \, e^{-i \phi_\eta}$,
where  $F_\eta$ is a so-called  Klein factor which lowers the number
of $\eta$-fermions by one. 
Over the years, it has become a
rather  popular tool for treating certain strongly-correlated electron
systems in 1 dimension. 
The reason for its popularity is that some problems which
appear intractable when formulated in terms
of fermions turn out to become easy, even trivial,
when formulated in terms of boson fields --
successful applications include
Tomonaga-Luttinger 
liquid theory (dealing with a quantum wire of interacting 1-D electrons), 
quantum Hall edge states and  quantum impurity
problems such as the Kondo problem.

Once one has learnt the ``bosonization rules'',
the formalism is very user-friendly -- 
one seldom needs to know more than how to work with free boson fields.
However, as often happens, actually {\em proving}\/
the  validity of the bosonization formalism 
 in explicit detail and ironing out all the subtleties is
substantially harder than simply applying it. 
The successive  efforts of quite
a number of pioneers was required to
piece together the puzzle, of which we mention just
a few milestones (brief overviews of historical developments
are also given in  Refs.~\cite{Emery79} and \cite{Haldane81}).
Tomonaga \cite{Tomonaga} was the first to identify
boson-like behavior of certain elementary excitations in a 1-D
theory of interacting fermions. A precise definition
of these bosonic excitations in terms of
bare fermions was given by Mattis and Lieb \cite{MattisLieb},
who took the first step towards a correct solution of a model
of interacting 1-D fermions 
earlier proposed by Luttinger
\cite{Luttinger}. A bosonic representation of a fermion
field at a single point, essentially of the form 
$\psi_\eta (x=0) \sim  e^{-i \phi_\eta (x=0)}$,
was first introduced by Schotte and Schotte \cite{SchotteSchotte}
to calculate x-ray edge transition rates.
The extension of their relation to arbitrary $x$, 
$\psi_\eta (x) \sim  e^{-i \phi_\eta (x)}$,
was discovered simultaneously 
by Mattis \cite{Mattis74}
and by Luther and Peschel \cite{LutherPeschel},
which made the systematic
calculation of general 
correlation functions very simple. 
However, their expressions for $\psi_\eta (x)$  were not
operator identities in Fock space, since they did not discuss
the number-lowering  Klein factors $F_\eta$. The first completely
precise bosonization relation in the solid-state literature
(though from a field-theoretical viewpoint) 
was given by Heidenreich \cite{Heidenreich75}
when discussing the model of \cite{LutherPeschel}
(there was an entirely parallel development in the
field-theoretical
literature on the the related ``massless Thirring model''
\cite{Coleman78,Mandelstam75}, 
a review of which lies beyound our scope).
The first explicit construction of the Klein factors $F_\eta$
in terms of bare fermionic operators was given by Haldane
\cite{Haldane79}, whose detailed 
discussion in  \cite{Haldane81} essentially completed
the development of the bosonization formalism.

These advances resulted in two somewhat different and not
entirely equivalent approaches,
which we shall call ``field-theoretical'' and ``constructive'',
respectively. 

\subsection{Field-theoretical versus constructive bosonization}
\label{intro:constr}

\noindent
{\em We  recapitulate the differences between 
field-theoretical and constructive bosonization,
and  explain why we strongly prefer the latter,
which is more rigorous and, we believe, more user-friendly.\vspace*{2mm}}

The  field-theoretical approach, examples
of which are summarized in 
Appendix~\ref{fieldconstruct}
(following Shankar \cite{Shankar}) 
or in Section~\ref{sec:KFcompare} (following Kane and Fisher 
\cite{KF}),
has a somewhat formal character: typically one starts by defining
bosonic fields  $\phi_\eta (x)$ with a set of prescribed 
properties, usually in a system of infinite size, 
and then uses field-theoretical machinery to 
calculate the commutation relations and Green's functions of
$e^{-i  \phi_\eta (x)}$ and $e^{i  \phi_\eta (x)}$. 
These turn out to be the
same as those of fermion fields $\psi_\eta (x)$ and
$\psi_\eta^\dagger (x)$, suggesting their formal correspondence,
so that one writes $\psi_\eta (x) \sim F_\eta e^{-i \phi_\eta (x)}$.
The so-called {\em Klein factors}\/  $F_\eta$ (often
denoted by $e^{-i \theta_\eta}$, or viewed as  Majorana fermions)
are included formally to guarantee
appropriate anti-commutation relations
($\{ \psi_\eta, \psi_{\eta'}^\dagger  \} = 0$
if $\eta \neq \eta'$). Some treatments also get by completely without
Klein factors, using instead 
appropriately contrived definitions of the $\phi_\eta
(x)$ fields (see  Appendix~\ref{fieldconstruct} for an example). 

Such field-theoretical approaches are completely adequate to
prove {\em that}\/ bosonization works; however, 
they do not really clarify {\em why}\/ it works.
Moreover, $\phi_\eta(x)$ and $F_\eta$ do not appear naturally
from first principles, but instead seem to have the status of mere
auxiliary quantities that are introduced somewhat 
artificially, with properties that must be fine-tuned to make things work.
Newcomers may be left with the sense that  
 $\psi_\eta (x) \sim F_\eta e^{-i \phi_\eta (x)}$,
though demonstrably true,  is a somewhat arbitrary 
coincidence, without clearly understanding
its origin and how it could possibly have been discovered. 

These issues are clarified in 
the more rigorous 
``constructive'' approach,  used e.g.\ by 
Mattis and Lieb \cite{MattisLieb}, Luther and Peschel
\cite{LutherPeschel} and  Emery \cite{Emery79},
 and nurtured to maturity  by Haldane,
whose 1981 paper \cite{Haldane81} is the standard reference.
This approach refrains  from using any formal
field-theoretical  machinery. Instead, 
it takes as starting point a fermion field
$\psi_\eta (x) = \left( \tpol \right)^{1/2} \sum_k
 e^{-i k x} c_{k \eta} $ in a system of finite size $L$ (this quantizes
the momenta, yielding a Hilbert space with a countable set of states,
which is crucial), and constructs (hence ``constructive'')
{\em all}\/ further  operators
and fields explicitly and naturally in terms of the initially given 
$c_{k \eta}$-operators (these constructions
are summarized in Table 1 on p.~\pageref{p:summary}). 
The entire bosonization formalism can then be derived 
deductively at a most elementary level
as a set of operator identities in Fock space, simply by judiciously
employing  standard operator
identities  like the Baker-Hausdorff lemma to manipulate functions of the
electron operators $c_{k \eta}$.

At present, the field-theoretical approach seems to be in much wider use
than the constructive one (perhaps because 
Haldane's discussion of his construction of Klein factors \cite{Haldane81} 
can appear complicated and hard to follow, though unneccessarily so). 
 For example, it is used extensively in the path-breaking work of 
Kane and Fisher on impurities in Tomonaga-Luttinger liquids
\cite{KF} (we review their notation in Section~\ref{sec:KFcompare}).
 Nevertheless, this
tutorial reviews and strongly advocates the constructive approach,
since in our opinion it is significantly superior, for a number of
reasons (admittedly the last three are subjective):
\begin{enumerate}
\item 
Constructive bosonization is more rigorous:
$\psi_\eta \sim F_\eta e^{-i \phi_\eta}$
has the status of an {\em operator identity in Fock space,}\/ and 
since its ingredients 
are constructed explicitly from the $c_{k \eta}$'s, their physical
meaning becomes explicit:
$\partial_x \phi_\eta  (x)$
represents local density fluctuations (at fixed total fermion
number) of the Fermi sea, and the Klein factor $F_\eta$  lowers the
total number of $\eta$-fermions  by one. 
In contrast, in the field-theoretical  approach  the
Fock space of states is not explicitly defined, and 
 $\psi_\eta \sim F_\eta e^{-i \phi_\eta}$
merely has the status of a formal correspondence.
Though $\partial \phi_\eta(x)$ can also be related to density fluctuations,
the Klein factors $F_\eta$ usually are viewed merely as formal tools 
ensuring proper anti-commutation relations, and the fact
that they lower the number of $\eta$-electrons is ignored.
This is particularly
evident in papers in which $F_\eta$ are viewed as a Majorana
fermions, which is imprecise: in fact $F_\eta^2 \neq 1$, since
removing two $\eta$-electrons is not equivalent to unity. 
\item 
In the constructive approach, refermionization is more rigorous too:
the refermionization identity, in which the 
bosonization identity is ``read backwards'', so to speak,
also has the status of an operator identity in Fock space.
\item 
The constructive approach is more ``user-friendly''
(because of, not in spite of, its higher rigor):
since all needed operators arise naturally and have a physical
interpretation, the formalism is easier to learn and to
work with. In contrast, field-theoretical bosonization is,
due to its formal nature, littered with formal pitfalls.
Of course it also yields correct results when used
with sufficient care -- 
however, regarding Klein factors
it is quite easy to make mistakes 
(Ref.~\cite{KotliarSi96} discusses an example). 
\item
Constructive bosonization offers a very simple answer to the
question: ``{\em Why}\/ is it possible at all to represent a fermionic
field as the exponential of a bosonic field?'' The essence of the answer is
that the state $\psi_\eta (x) | \vec 0 \rangle_0$,
where $|\vec 0\rangle_0$ is the Fermi ground state, turns out to be
an {\em eigenstate}\/ of
 the bosonic operators $b_{q \eta}$ from which
the boson field $\phi_\eta  (x) = -  \sum_{q>0} \left( {2 \pi \over q L}
\right)^{1/2}  (   e^{-  q (i x + a/2)}  b_{q \eta} + \mbox{h.c.})$. is
constructed. Therefore it must have a {\em coherent
state representation}\/ in terms of the $b_{q \eta}^\dagger$'s, which turns
out to be precisely $\sim F_\eta  e^{-i \phi_\eta  (x)} |\vec
0\rangle_0$. Thus,
one discovers that the
trick that makes bosonization work, its {\em raison d'etre,}\/ so
to speak, is that it cleverly exploits
some very convenient properties of bosonic   coherent states!
\item
In constructive bosonization, regularizing infinities
is easier: one simply 
consistently normal-orders all $c_k$'s and $b_q$'s.
Instead, field-theoretical treatments 
customarily employ a point-splitting prescription
that becomes rather cumbersome 
when terms of order $1/L$ are to be  retained, as here 
(or \cite{vDZF,ZvD}). 
Of course, normal-ordering and point-splitting are equivalent regularization 
schemes, see Appendix~\ref{point-splitting}. 
\end{enumerate}

In this tutorial we give a (at times very) detailed account of the
constructive approach to bosonization, at a level accessible
to beginning graduate students with a knowledge of second
quantization, but not of field theory.  Our development
of the formalism is a simplified version of 
that given by Haldane \cite{Haldane81} (the relation 
between his and our notation is given in Section~\ref{Haldanesnotation}).
 Ours differs from Haldane's
mainly in that we use only left-moving fields (it is easy to
rewrite some of them as right-movers, if required, see
Section~\ref{LRderivation}), and in
that we exploit to the full the above-mentioned
connection to the properties of boson \mbox{coherent states.}
(Another discussion of constructive
bosonization similar in spirit to ours 
was recently written by Sch\"onhammer and Meden
\cite{Schoenhammer1,Schoenhammer2}.)
 
\subsection{Application to Tomonaga-Luttinger liquid with impurity}
\label{intr:TLL}

\noindent
{\em To give a non-trivial example of how the formalism is used, 
we discuss  impurity scattering in a Tomonaga-Luttinger liquid;
this  requires  not only bosonization but also refermionization,
a rigorous, novel  treatment of which 
allows us to resolve a recent controversy.
\vspace*{2mm}}

In Section~\ref{TLL}  we consider the tunneling density of
states, $\rho_{dos} (\omega)$, at the site of an impurity in a
Tomonaga-Luttinger liquid \cite{Tomonaga,Luttinger}, i.e.\ a quantum
wire of interacting 1-D electrons
characterized by the dimensionless electron-electron
interaction parameter $g>0$ (for free electrons
$g=1$).  The  exponent $\nu$
governing the low-energy behavior of 
$\rho_{dos} (\omega) \sim \omega^{\nu - 1}$ as $\omega \to 0$, 
was the subject of a recent controversy: 
Without impurities, it is known
that $\nu_{free} = (g + 1/g)/2 $. 
In the presence of an impurity, 
Oreg and Finkel'stein (OF) \cite{OF}
found $\nu = {1/(2g)}$, using an exact mapping to a Coulomb gas problem,
which they treated in a mean-field approximation;
this would  imply that $\nu < \nu_{free}$, i.e.\ 
$\rho_{dos} (\omega)$ is  {\em enhanced,}\/ and actually
diverges for $\omega \to 0$ if $g > 1/2$. 
In contrast, Fabrizio and Gogolin (FG)
\cite{FG} and  Furusaki \cite{Furusaki}
found $\nu = 1/g$,  which would imply that for repulsive 
interactions ($g<1$) one has $\nu > \nu_{free}$,  i.e.\ 
  $\rho_{dos} (\omega) $ is {\em suppressed,}\/
with $\rho_{dos} (0 ) = 0$ (reminiscent
of the classical RG conclusion of Kane and Fisher
\cite{KF} that the  conductance accross a backscattering
impurity vanishes at zero temperature if $g < 1$). 
Furusaki \cite{Furusaki} checked
his result for 
the exactly solvable case $g= 1/2$ 
(using refermionization, the inverse, so to
speak, of bosonization), and indeed found $\nu = 2$. 
To explain why OF had obtained a different result, 
FG \cite{FG} suggested that OF 
had neglected the effects of Klein factors. 
OF disputed this \cite{OF97} and in turn alledged that FG
had incorrectly replaced
the impurity  by ``open boundary conditions'',
although the two are in general {\em not}\/ equivalent:
 a ``cut wire''  suppresses both current {\em and}\/
density  fluctuations, whereas an impurity suppresses {\em only}\/
current  fluctuations.

Our opinion of these matters is explained in 
detail in Section~\ref{sec:OF} and 
Appendix~\ref{sec:NNcor}. In brief, we believe
(i) that FG's analysis of the role Klein factors is correct, but not 
their criticism of OF; 
(ii) that  OF did treat Klein factors
 correctly  (see our Appendix~\ref{app:OF}), showing that 
they produce a Coulomb gas with a certain ``sign problem'',
but  that OF's mean-field treatment of the latter 
is not sufficiently accurate; 
(iii) that  OF's assertion 
about the inequivalence of  an 
impurity and a ``cut wire'' is correct, but that 
their criticism of FG is misguided nevertheless,
since FG do incorporate
density fluctuations and 
use the cut wire  only to find  the effects of
{\em current}\/ fluctuations;
(iv) that the relevant issues  become {\em much}\/
 clearer when reformulated
using {\em constructive}\/ instead of the field-theoretic bosonization
used by FG, OF and Furusaki. Using the former, 
we give an appealingly simple yet  more 
rigorous version of Furusaki's $g=1/2$ calculation
(he nonrigorously  treats Klein factors as
Majorana fermions), 
%
which  resolves the controversy in favor of 
FG and Furusaki. 

To set the stage  for this calculation, 
we discuss {\em refermionization}\/  in pedagogical detail in 
Section~\ref{fsrefermionization}. 
  We refermionize {\em at finite $L$}, 
 since then 
 the requisite Klein factors can be introduced as naturally and
rigorously as during bosonization. 
Moreover, we refermionize not only the usual combination $F e^{- i
  \Phi(x)}$, but, for the
first time,  also the bosonic field $\Phi
(x)$ itself; this enables us to calculate general
bosonic correlation functions in terms of fermionic 
 ones. Since refermionization is usually implemented less rigorously, 
our treatment of this 
topic might be  of
interest to experts too, hence the second part of the review's title.

\subsection{Outline and bosonization dictionary} 

The outline of the review is as follows:
The main ingredients of the constructive approach to bosonization
are summarized in Table~1 below, 
for ease of reference and to survey  what is to be
proven in subsequent sections.
In Section~\ref{prerequisites} we state
the properties required
to make a 1-D fermion theory amenable to bosonization,
and in Section~\ref{firstdeffermions} define 
the standard fermion fields $\psi_\eta (x)$
as Fourier sums over a given set of $c_{k \eta}$'s.
In Section~\ref{Fockbosons} we show that
the fermionic Fock space spanned by the $c_{k \eta}$'s can also be reorganized
in terms of the electron number operators
$\widehat N_\eta$, their raising and lowering operators
$F_\eta$, $F_\eta^\dagger$ and
 bosonic particle-hole operators $b_{q \eta}$,
 $b_{q \eta}^\dagger$, and construct from the latter 
the boson fields $\phi_\eta (x)$ in  Section~\ref{sec:bose}. 
The heart of this review is Section~\ref{schoeller},
where we give a very simple yet rigorous and detailed derivation
of the bosonization identity.  
In Section~\ref{ham} we consider fermions with
linear dispersion and bosonize the Hamiltonian,
in Section~\ref{bosfer}  derive a remarkable relation between free fermion
and boson Green's functions, and in Section~\ref{Vertex} derive some general
properties of the so-called vertex operators $V^{(\eta)}_\lambda \sim e^{i
  \lambda \phi_\eta }$.  In Section~\ref{TLL} we illustrate the formalism by
calculating the tunneling density of states $\rho_{dos} (\omega)$ at an
impurity site in a $g = \half$ Tomonaga-Luttinger liquid.

In Appendix~\ref{fieldconstruct} we make
explicit the connection between the
constructive and field-theoretical approaches to bosonization
(as described by Shankar \cite{Shankar}),
by showing how the operators used in the latter can be
constructed in terms of the former. 
The remaining appendices contain details somewhat too arduous
to appear in the main text. 

\newpage

\noindent \phantom{.} \vspace*{-6mm}
\begin{eqnarray*}
\nonumber
  \begin{array}{llr}
\mbox{starting point:}   &
 \{ c_{k \eta}, c^\dagger_{k' \eta'} \} =
        \delta_{\eta \eta'} \delta_{k k'} \qquad (\eta = 1, \dots, M) 
& \hfill (\ref{cancom})
\\
\mbox{$k$-quantization:} \rule{0mm}{6.5001mm}
&
 k = \tpol (n_k - \half \delta_b )    \qquad \qquad
        (\delta_b \in [0, 2), \quad n_k \in \ZZ)        
&  (\ref{momentumquant})
\\
\mbox{vacuum state:} \rule{0mm}{6.5001mm}
&
        c_{k \eta} |\vec 0\rangle_0 \equiv 0
        \quad \mbox{for} \quad  k > 0,  \qquad 
        c^\dagger_{k \eta} |\vec 0\rangle_0 \equiv 0
        \quad \mbox{for} \quad  k \leq  0
& 
 (\ref{vacuum1})
\\
\mbox{number operator:} \rule{0mm}{6.5001mm}
&
 \widehat N_\eta \equiv 
          \sum_k
        {}^\ast_\ast c^\dagger_{k \eta} c_{k \eta} {}^\ast_\ast
        = \sum_k \biggl[
        c^\dagger_{k \eta} c_{k \eta} -
        {}_0\langle \vec 0 | c^\dagger_{k \eta} c_{k \eta}
        | \vec 0 \rangle_0  \biggr] 
& 
 (\ref{eq:numberoperator})
\\
\mbox{boson creator:} \rule{0mm}{6.5001mm}
&
 b^\dagger_{q \eta} \equiv {\textstyle  { i \over \sqrt n_q}} 
 \sum_k 
c^\dagger_{k+q \; \eta} c_{k \eta}   
        \quad (q = \tpol n_q > 0, \; n_q \in \ZZ^+)
& 
 (\ref{defbq})
\\
\mbox{boson commutator:} \rule{0mm}{6.5001mm}
&
 [ b_{q \eta}  , b^\dagger_{q' \eta'} ] = \delta_{\eta \eta'} \delta_{q q'}
& 
 (\ref{rhorhoc})
\\
\mbox{$\vec N$-part.\ ground st.:} \rule{0mm}{6.5001mm}
&
\widehat N_\eta |\vec N \rangle_0 = N_\eta | \vec N \rangle_0 \; , 
\qquad  b_{q \eta} | \vec N \rangle_0 = 0 
& 
 (\ref{N00},\ref{a-vacuum})
\\
\mbox{def. of Klein factor:} \rule{0mm}{6.5001mm}
&
        F^\dagger_\eta f(b^\dagger)
 | \vec N \rangle_0 
\equiv    f(b^\dagger)
        c^\dagger_{N_\eta +1}
        | \vec N \rangle_0 
& 
\hspace{-12mm} (\ref{UNa}) \, \mbox{[or} \, (\ref{Kleinexplicit})\mbox{]}
\\
\mbox{$F$ commutators:} \rule{0mm}{6.5001mm}
& \mbox{[}F, b \mbox{]} = 0, \quad 
 \{ F^\dagger_\eta, F_{\eta'} \} =  2 \delta_{\eta \eta'} ,\quad
 [\widehat N_\eta, F_{\eta'}] = -  \delta_{\eta \eta'}
        F_\eta 
&
\hspace{-12mm} (\ref{Ucommutes},\ref{FFcommute},\ref{NF})
\\
\mbox{fermion field:} \rule{0mm}{6.5001mm}
&
        \psi_\eta (x)  \equiv  \left( \tpol \right)^{1/2}
        \sum_k e^{-i k x} c_{k \eta}
& 
 (\ref{deffermions})
\\
\mbox{$\psi_\eta$ commutator:} \rule{0mm}{6.5001mm}
&
 \{ \psi_\eta (x), \psi^\dagger_\eta (x') \} = 
\delta_{\eta \eta'} \,  2 \pi  \, \delta( x - x')
               \quad (\mbox{for} \;  |x-x'| < L) 
& 
 (\ref{antic})
\\
\mbox{boson field:} \rule{0mm}{6.5001mm}
&
        \phi_\eta (x) \equiv
        - \sum_{q > 0}
         {\textstyle {1  \over \sqrt n_q}}
        \left( e^{-i q x} b_{q \eta} +  e^{i q x} b^\dagger_{q \eta} \right)
          e^{-a q/2}      
& 
 (\ref{defphi})
\\
\mbox{$\phi_\eta, \partial_x \phi_\eta$ commutator:} \rule{0mm}{6.5001mm}
&
 \mbox{[} \phi_\eta (x), \partial_{x'} \phi_{\eta'} (x') \mbox{]} 
= 
        \delta_{\eta \eta'} \, 2 \pi i 
        \left( \delta (x -x') - {1 \over L} \right) 
& 
 (\ref{phiphicom})
\\
\mbox{\underline{bosonization identity:}} \rule{0mm}{6.5001mm}
\;
&
 \psi_\eta (x) = 
        F_\eta \,
        a^{- 1/2}
        e^{-i {2 \pi \over L} (\widehat N_\eta - {1 \over 2}
        \delta_b)x }
        e^{- i \phi_\eta (x) } \;         
& 
 (\ref{bf4})
\\
\mbox{$(2 \pi)$ density:} \rule{0mm}{6.5001mm}
&
  \rho_\eta (x) \equiv  {\,}^\ast_\ast
       \psi_\eta^\dagger (x) \psi_\eta (x)
       {\,}^\ast_\ast
    = \partial_x \phi_\eta (x)
        + \tpol \widehat N_\eta  
& 
 (\ref{defnormdens})
\\
\mbox{free ferm. Hamilton:} \rule{0mm}{6.5001mm}
&
H_{0\eta} \equiv  \sum_k
        k  {\,}^\ast_\ast  c^\dagger_{k \eta} c_{k \eta}  {\,}^\ast_\ast
        =  \int_{-L/2}^{L/2} \! 
        {\textstyle {dx \over 2 \pi}}
         {\,}^\ast_\ast \psi_\eta^\dagger (x) i \partial_x \psi_\eta  (x) 
        {\,}^\ast_\ast 
& 
 (\ref{H0},\ref{Hpsipoint})
\\
\mbox{bosonized Hamilton:} \rule{0mm}{6.5001mm}
&
\phantom{H_{0\eta} }
 = \sum_{q > 0} q \,
        b^\dagger_{q \eta} b_{q \eta}
         \; + \;
         \tpol \half \widehat N_\eta ( \widehat N_\eta + 1 -
\delta_b ) 
& 
 (\ref{Hboson})
\\
 \rule{0mm}{6.5001mm}
&
\phantom{H_{0\eta} } = 
\int_{-L/2}^{L/2} {\textstyle {dx \over 2 \pi}}
        {\textstyle {1 \over 2}}
         {\,}^\ast_\ast ( \partial_x \phi_\eta (x) )^2  {\,}^\ast_\ast
         \; + \;
         (\tpol   ) \half \widehat N_\eta ( \widehat N_\eta + 1 - \delta_b )
& 
 (\ref{bosonH})
\\
\mbox{Green's functions:} \rule{0mm}{6.5001mm}
&
\langle{\cal T} \psi_\eta (z) \psi_{\eta'}^\dagger (0) \rangle
= 
 \delta_{\eta \eta'} \:  a ^{-1}  \, \mbox{sign}(\tau) \, 
        e^{\langle  {\cal T}
          \phi_\eta (z) \phi_{\eta} (0) -
           \phi_\eta (0) \phi_{\eta} (0)
        \rangle} \hspace*{- 5mm}
& 
 (\ref{GF2a})
\\
(z \equiv \tau + ix )  \rule{0mm}{6.5001mm}
&
\phantom{\langle{\cal T}}
 =
        \delta_{\eta \eta'}  
     \left( {\textstyle {\beta \over \pi } 
       \sin [{\pi \over \beta}(z +  a \, \mbox{sign}(\tau))]}  \right)^{-1}
   \;\stackrel{T = 0}{\longrightarrow} \;  
{1 \over z + a \, \mbox{\tiny sign}(\tau)}
& 
 (\ref{ffG})
\\
\mbox{vertex operator:} \rule{0mm}{6.5001mm}
&
 V^{(\eta)}_{ \lambda} (z) \; \equiv  \;
        \left( {\textstyle {L \over 2 \pi}} \right)^{- \lambda^2 / 2} \:
        {}^\ast_\ast e^{i \lambda \phi_{ \eta} (z)} {}^\ast_\ast
        \;  = \;
        a^{- \lambda^2 /2} e^{i \lambda \phi_{ \eta} (z)}
& 
 (\ref{F.defvertex})
\\
\mbox{$V^{(\eta)}_{ \lambda}$ Green's funct.:} \rule{0mm}{6.5001mm}
&
 \langle  V^{(\eta)}_{  \lambda } (z)
         V^{(\eta')}_{  \lambda' } (0) \rangle =
 { \delta_{\eta \eta'} \:
        (L/ 2 \pi)^{- (\lambda + \lambda')^2/2}
        \over 
        \left(\textstyle {\beta \over \pi} \sin [{\pi \over
        \beta}(z +a)]
       \right)^{- \lambda \lambda'} } 
     \stackrel{L = \beta = \infty}{\longrightarrow}
     { \delta_{ \eta \eta'} \delta_{- \lambda, \lambda'} \over 
       (z + a)^{\lambda^2}} 
& 
 (\ref{VV2})
\end{array}
\end{eqnarray*}
{\small Table 1: \label{p:summary}
Bosonization Dictionary: 
A survey of the main ingredients and results of 
the constructive bosonization
formalism (with the equation numbers used below), listed here
for ease of reference.}
\newpage \noindent

\section{Bosonization prerequisites}
\label{prerequisites}


It is possible to constructively bosonize a
 theory involving $M$ species of fermions whenever the following
prerequisites are met: 
The theory can be formulated in terms of a set of fermion creation
and annihilation operators with canonical anti-commutation relations
\begin{equation}
\label{cancom}
        \{ c_{k \eta}, c^\dagger_{k' \eta'} \} =
        \delta_{\eta \eta'} \delta_{k k'} \; ,
        \qquad  k \in [ - \infty, \infty] , \qquad \eta = 1, \dots , M,
\ee
which are labelled by a
{\em species}\/ index $\eta = 1, \dots , M$ distinguishing the $M$ different
species from each other, and a discrete,  unbounded
momentum (or wave-number)
 index $k$  of the form 
\begin{equation}
\label{momentumquant}
        k = \tpol (n_k - \half \delta_b ) \; ,
        \qquad \mbox{with} \quad n_k \in \ZZ \quad \mbox{and} \quad
        \delta_b \in [0, 2) \; .
\ee
Here the $n_k$ are integers, $L$ is a length to be associated with
the system size, and $\delta_b$ is a parameter that will determine
the boundary conditions of the fermion fields defined 
below [see \Eq{boundarycond}].

For example, $\eta$ can denote electron spin:
$\eta = (\uparrow, \downarrow)$, $M=2$; or it can 
distinguish left-moving from right-moving spinless electrons, e.g.\ in
a one-dimensional wire: \mbox{$\eta =(L , R)$}, $M=2$,
see Section~\ref{TLL}; or both: 
\mbox{$\eta =(L \! \uparrow, R \! \uparrow, 
L \! \downarrow, R \! \downarrow)$,} $M=4$, etc.  
The momentum index  $k$ typically labels the
eigenergies $\varepsilon_k$ of the free, non-interacting system
(with $\varepsilon_0$ corresponding to the Fermi-energy $\varepsilon_F$),
and hence could equally well have been called an ``energy'' index.
That $k$ be  {\em discrete}\/ and {\em unbounded}
is an essential prerequisite  of a  detailed and  systematic derivation
of the  bosonization identities.
Its discreteness is needed to allow systematic book keeping
of states, its unboundedness
to allow the definition of proper bosonic operators, see \Eq{defbq} to
(\ref{rhorhoc}) below.

The manipulations required to cast  a given problem in a form
that meets the above prerequisites depends, of
course, on the details of the problem. However, they are
a prelude to bosonization, not part of the
technique itself. Therefore, we do not discuss them here,
but refer the reader to Section~\ref{LRmovers} 
for an example, 
a 1-D quantum wire containing spinless left- and right-moving
electrons
(for another example, the Kondo problem, see Refs.~\cite{vDZF,ZvD}).
Suffice it here to state the main ideas: To ensure that $k$ is discrete,
one considers a system of finite size $L$ and definite boundary
conditions,
thus quantizing  the momenta $k$ and energies
 $\varepsilon_k$. If the continuum limit $L \to \infty$ is required, 
it is taken  only
at the end, after the bosonization rules have been established.
If the dispersion relation does not automatically imply that $k$ is
unbounded, one  can make it so by adding
a set of  (unphysical) negative-energy ``positron'' states
(see  Section~\ref{LRderivation}).

For definiteness, the reader may think in terms 
of a linear dispersion relation,
$\varepsilon_k = \hbar v_F (k - k_F)$ (i.e.\ energies and momenta
are measured
relative to $\varepsilon_F$ and $k_F$), with an infinite bandwidth,
so that  $k \in [- \infty , \infty]$.
However, we emphasize that the bosonization identity
$\psi_\eta = F_\eta e^{-i \phi_\eta}$ can be derived {\em without}\/
specifying the dispersion
relation (though the nomenclature to be used, like ``ground state''
and ``particle-hole excitations'' only makes sense when the
dispersion is monotonic, i.e. $|k| > |k'|$ $\Rightarrow$  $\varepsilon_k
> \varepsilon_{k'}$). This is possible because
the bosonization identity is an {\em operator identity,}\/
i.e.\ valid when acting on any state in the full Fock space.
Hence it is independent of the Hamiltonian,  whose
detailed form only becomes relevant when one calculates
correlation functions. Therefore, we shall refrain from specifying a
Hamiltonian until Section~\ref{ham}, after all bosonization formalities
have been dealt with.

\section{Fermion fields -- definition and properties}
\label{firstdeffermions}


Starting from given a set of electron annihilation operators $c_{k \eta}$
with the properties (\ref{cancom}) and (\ref{momentumquant})
specified in Section~\ref{prerequisites},
a set of $M$ fermion fields $\psi_\eta (x)$ can be defined as 
follows:\footnote{There is 
no particular reason for the choice of phase in Eq.~(\ref{deffermions}),
namely $e^{-ikx}$ instead of $e^{ikx}$; the former
defines so-called left-moving fields, the latter right-moving fields
(this nomenclature is explained in footnote~\ref{f:LRmovers}), 
and the two are related simply by $x \leftrightarrow - x$.
 In Section~\ref{LRderivation} we shall use both kinds of fields.}
\bea
\label{deffermions}
        \psi_\eta (x) & \equiv & \left( \tpol \right)^{1/2}
        \sum_{k= - \infty}^{\infty} e^{-i k x} c_{k \eta} \;, 
\\
\mbox{with inverse}
\label{psiinverse}
                \qquad
                c_{k \eta} &=& (2 \pi L)^{-1/2} \int_{-L/2}^{L/2} \! dx \,
                e^{i k x} \, \psi_\eta (x) \; .
\eea
Though in applications
one usually takes $x \in [-L/2, L/2]$ (and often $L \to \infty$),
the formalism developed below holds for arbitrary $x \in [-\infty, \infty]$. 
The physical meaning of the   $\psi_\eta(x)$-fields and
the variable $x$ depends on the manipulations
required to formulate a given model in terms of the  $c_{k \eta}$'s.
For present purposes, the  $\psi_\eta(x)$'s
are to be regarded simply as mathematical constructs 
that have the useful  property, to be proven below, 
of being expressable in terms of bosonic fields. 

Given a set of discrete $k$'s 
of the form (\ref{momentumquant}), the $\psi_\eta$ obey the 
following periodicity condition
[the simplest cases are $\delta_b = 0$ (or 1) for complete periodicity 
(or anti-periodicity)]:
\be
\label{boundarycond}
        \psi_\eta (x + L/2) = e^{i \pi \delta_b} \, \psi_\eta (x- L/2) \; .
\ee
(Alternatively, one can view Eq.~(\ref{boundarycond}) as a boundary 
condition that is purposefully imposed on the $\psi_\eta$ in order
to obtain discrete $k$'s satisfying Eq.~(\ref{momentumquant}).) 
Furthermore, \Eqs{cancom} and (\ref{momentumquant})  and the identity
\cite{Gelfand}
\be
\label{peridelta}
     \sum_{n \in \ZZ}
     e^{i n y} = 2 \pi \sum_{\bar n \in \ZZ}
     \delta (y - 2 \pi \bar n)
\end{equation}
immediately imply the anti-commutation
relations 
\bea
        \{ \psi_\eta (x), \psi^\dagger_{\eta'} (x') \}
        &=& \delta_{\eta \eta'}  \, \tpol 
        \sum_{n \in \ZZ}  e^{- i (x-x')(n - \delta_b /2) 2 \pi / L} 
\\
\label{antic}
        &= & \delta_{\eta \eta'} \,  2 \pi   \sum_{\bar n \in Z}
        \delta( x - x' - \bar n L) e^{i \pi \bar n \delta_b} \; ; 
\\
        \{ \psi_\eta (x), \psi_{\eta'} (x') \} &=& 0 \; .
\eea
For  $x, x' \in [-L/2, L/2]$, 
these are  just the standard\footnote{Note though, that many 
authors use normalization $(1/L)^{1/2}$ instead of
our $(2 \pi / L)^{1/2}$ in Eq.~(\ref{deffermions}), 
so that  in Eq.~(\ref{antic}) their
fields are normalized to 1 instead
of our $2 \pi$.   The advantage of our normalization
(used e.g.\ in conformal field theory), is that
correlation functions are normalized to unity, 
$\langle \psi (x) \psi^\dagger (0) \rangle 
= 1 / (i x)$, see \Eq{ffG}.}
relations obeyed by fermion fields. 
For unrestricted values of
$x,x'$, one obtains a  more general  $\delta$-function with
appropriate periodicity.

\section{Bosonic reorganization of Fock space}
\label{Fockbosons}

 \noindent
{\em 
 ``Bosonizing''  a fermionic theory means rewriting it
in terms of bosonic degrees of freedom. The ``deep reason''
why this is possible for 1-D fermion theories is that the Fock space 
${\cal F}$ of
states spanned by the $c_{k \eta}$ operators can
be reorganized as a direct sum,
 ${\cal F} = \sum_{\oplus \vec N} {\cal H}_{\vec N}$ over
Hilbert spaces ${\cal H}_{\vec N}$ characterized by a {\bf fixed
particle number} $\vec N$,
 within each of which  all excitations 
are are particle-hole-like and hence 
have {\bf bosonic} character.  
In this section, we define the concepts and operators needed to accomplish
this reorganization. 
\vspace*{2mm}}

\subsection{Vacuum state $| \vec 0 \rangle_0$}
\label{sec:vacuum}

Let  $| \vec 0 \rangle_0$ be the state defined  by the properties 
\bea
\label{vacuum1}
        c_{k \eta} |\vec 0\rangle_0 \equiv 0
        & \quad \mbox{for} \quad & k > 0,  \quad (\mbox{i.e.} \; 
          n_k > 0 ) \; ,
\\
\label{vacuum2} 
        c^\dagger_{k \eta} |\vec 0\rangle_0 \equiv 0
        & \quad \mbox{for} \quad & k \leq  0, \quad (\mbox{i.e.} \; 
          n_k \leq 0 ) \; .
\eea
and illustrated for $M=1$ in
Fig.~\ref{fig:levels}(a). 
In other words,  for all $\eta$, the  highest filled level of
$| \vec 0 \rangle_0$ 
is by definition labeled by $n_k = 0$ and the lowest empty
level by $n_k = 1$
 (irrespective of $\delta_b \in [0,2)$). 
We shall call $|\vec 0 \rangle_0$   the {\em vacuum state}\/ ({\em Fermi
sea}\/ would be equally appropriate)
and  use it as reference state relative to which the
occupations of all other states in ${\cal F}$  are specified. 
In particular, we define the operation of
{\em fermion-normal-ordering}\/, to be
denoted by  ${}^\ast_\ast \quad {}^\ast_\ast$,
with respect to this vacuum state: 
to fermion-normal-order a function of $c$ and $c^\dagger$'s,
all $c_{k\eta}$ with $k>0$ 
and all $c_{k \eta}^\dagger$ with $k \le$ are to be moved 
to the right of all other operators (namely
 all $c_{k\eta}^\dagger$ with $k > 0$
and  $c_{k \eta}$ with $k \le 0 $) , so that 
\be
\label{fermionnormalordering}
{}^\ast_\ast ABC \dots {}^\ast_\ast 
= ABC \dots  \; - \; {}_0 \langle \vec 0 | ABC \dots  | \vec 0 \rangle_0
\qquad \mbox{for}\quad A, B, C , \dots
\in \{c_{k \eta}; c_{k \eta}^\dagger \}
\; .
\ee

\begin{figure}[htbp]
  \begin{center}
        \leavevmode 
    \epsfig{
width=0.8\linewidth,%
file=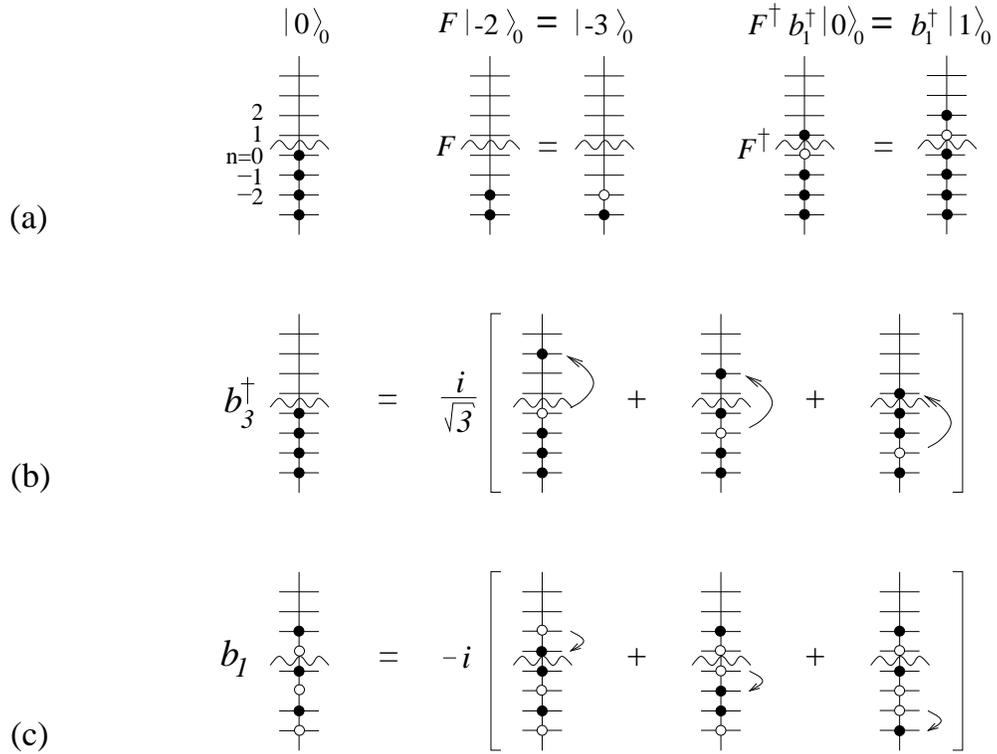}
\vspace*{5mm}
\caption{\label{fig:levels} For the case $M=1$ (i.e.\ the
index $\eta$ suppressed) we depict 
(a) the vacuum state $|0 \rangle_0$ (the wiggly line indicates
the ``Fermi surface''),
the
action of $F$ on the $-2$-particle ground state $| \! -\! 2 \rangle_0$, 
which yields $|\! -\! 3 \rangle_0$, 
and the action of $F^\dagger$ on  the $0$-particle excited state
$ i c_1^\dagger   c_{0}| 0 \rangle_0 = b_1^\dagger
|0 \rangle_0$, which yields $b_1^\dagger
|1 \rangle_0$ \mbox{[}see Eqs.~(\protect\ref{UNa}-\protect\ref{UNb})\mbox{]};
(b) the action of $b_3^\dagger$ on $|0 \rangle_0$ 
[see Eq.~(\protect\ref{defbq})]; 
(c) the action of $b_1$ on 
$ c^\dagger_{2} c^\dagger_{0}   c_{-2}^\dagger |\! - \! 3 \rangle_0$
[see Eq.~(\protect\ref{defbq})].}
\end{center}
\end{figure}

\subsection{$\vec N$-particle ground states $| \vec N\rangle_0$}
\label{Noperator}

Let $ \widehat N_\eta$ be the operator that counts the
number of $\eta$-electrons relative to $|\vec 0 \rangle_0$:
\be
\label{eq:numberoperator}
        \widehat N_\eta \equiv 
          \sum_{k = - \infty}^\infty
        {}^\ast_\ast c^\dagger_{k \eta} c_{k \eta} {}^\ast_\ast
        = \sum_{k = - \infty}^\infty \biggl[
        c^\dagger_{k \eta} c_{k \eta} -
        {}_0\langle \vec 0 | c^\dagger_{k \eta} c_{k \eta} 
        | \vec 0 \rangle_0  \biggr] \; .
\ee
The set of all states with the same 
$ \widehat N_\eta$-eigenvalues 
$\vec N = (N_1, \dots, N_M) \in \ZZ^M $ will 
be called the  $\vec N$-particle Hilbert space $H_{\vec N}$. 
It contains infinitely many states, corresponding to 
different configurations of particle-hole
excitations, all of which will  generically be denoted by 
$| \vec N \rangle$. 

Furthermore, for given  $\vec N$, let $|\vec N\rangle_0$ be
that particular $\vec N$-particle state
which has {\em no}\/ particle-hole excitations;
since it is the lowest-energy state in  $H_{\vec N}$,
we shall call it the  $\vec N$-{\em particle ground state}\/.
To resolve ambiguities in its phase, we define it by
specifying a particular ordering of operators
(for ease of notation, we here use $n_k$ instead of $k$ as index):
\bea
        | \vec N_\eta \rangle_0
        &\equiv &
 (C_1)^{N_1} (C_2)^{N_2} \dots (C_M)^{N_M} | \vec 0 \rangle_0 \; , 
\\
\label{N00}
         (C_\eta)^{N_\eta} & \equiv & 
        \left\{ \begin{array}{ll}
        c^\dagger_{N_\eta \, \eta} c^\dagger_{(N_\eta -1)\, \eta}
        \dots c^\dagger_{1 \, \eta} 
        & \mbox{for}\quad
        N_\eta > 0 \; ,  \\ 
        1         & \mbox{for}\quad
        N_\eta = 0 \; ,  \\ 
        c_{(N_\eta +1 ) \, \eta} c_{(N_\eta +2)\, \eta}
        \dots c_{0\, \eta} 
        & \mbox{for}\quad  N_\eta < 0 \; .
        \end{array} \right.
\eea

\subsection{Bosonic particle-hole operators $ b^\dagger_{q \eta}$
and  $b_{q \eta}$}

Since all excited states within a given  $\vec N$-particle Hilbert
space have the same $\vec N$, they can all be regarded as
particle-hole excitations built on the ground state $|\vec N \rangle_0$.
We shall see below that for a systematic treatment of all
such operators it suffices to consider only the following
{\em bosonic creation and annihilation}\/ operators
(both defined {\em only}\/ for $q  > 0$):
\be
\label{defbq} 
       b^\dagger_{q \eta} \equiv {\textstyle  { i \over \sqrt n_q}}
          \sum_{k = -\infty}^\infty
          c^\dagger_{k+q \; \eta} c_{k \eta}   \; , 
         \qquad \qquad
        b_{q \eta} \equiv  {\textstyle  {- i \over \sqrt n_q}}
         \sum_{k = -\infty}^\infty
         c^\dagger_{k-q \; \eta} c_{k \eta}   \; , 
 \ee
with $q \equiv \tpol n_q > 0 $, where  
$n_q \in         \ZZ^+$ is a positive integer.
For any $|\vec N\rangle$, the state
$b^\dagger_{q \eta}|\vec N\rangle $ (or $b_{q \eta}|\vec N\rangle $)
consists of  a linear combination of
particle-hole excitations relative to $|\vec N\rangle$,
each term having  $q$ units of momentum more (or less) 
than $|\vec N\rangle$, as illustrated in Fig.~\ref{fig:levels}(b-c).
 In this sense, $b^\dagger_{q \eta}$
and $b_{q \eta}$ can be viewed as
momentum-raising or -lowering operators
[they can also be identified with the Fourier-components
of the electron density, see \Eq{densityfourier}].
Their normalization is purposefully chosen to produce harmonic oscillator
commutation relations:
\be
\label{bobo}
        \qquad [b_{q \eta},b_{q' \eta'}] = 
        [b_{q \eta}^\dagger, b_{q' \eta'}^\dagger] = 0 \; ,
        \qquad [N_{q \eta}, b_{q' \eta'}] = 
        [N_{q \eta}, b_{q' \eta'}^\dagger] = 0 \; , 
        \quad \mbox{for all} \quad q, q', \eta, \eta' \; ;
\ee
\bea
\nonumber 
\lefteqn{        [ b_{q \eta}  , b^\dagger_{q' \eta'} ]
        = \delta_{\eta        \eta'} 
        \sum_{k = - \infty}^\infty 
        {\textstyle \frac{1}{n_q} }
        \left(c^\dagger_{k+q - q' \eta} c_{k \eta}
        -  c^\dagger_{k+q \eta} c_{k+q' \eta}  \right)}
\\
\nonumber 
        &=& \delta_{\eta \eta'}  \delta_{q q'} 
        \sum_k 
        {\textstyle \frac{1}{n_q} }
        \left\{
        \left[ {\,}^\ast_\ast c^\dagger_{k \eta} c_{k \eta}
        {\,}^\ast_\ast - 
        {\,}^\ast_\ast c^\dagger_{k+q \eta} c_{k+q \eta} {\,}^\ast_\ast
        \right] +
        \left( {}_0\langle \vec 0 | 
          c^\dagger_{k \eta} c_{k \eta} | \vec 0 \rangle_0
        -  {}_0\langle \vec 0 | 
        c^\dagger_{k+q \eta} c_{k+q \eta} | \vec 0 \rangle_0
          \right) \right\}
\\
\label{rhorhoc} 
\label{rhorho}
        &=& \delta_{\eta \eta'} \delta_{q q'} \; .
\eea
\Eqs{bobo} are easily checked, but the derivation
of (\ref{rhorhoc}) requires some care, as first
pointed out by Mattis and Lieb \cite{MattisLieb}:
For $q \neq q'$
the two terms in  the first line are both normal-ordered,
and hence no subtleties can arise when subtracting them from
each other to get zero (by shifting $k \to k - q'$ in the second term).
However, for  $q = q'$,  before subtracting we first have to  construct
in  the second line
normal-ordered expressions (else we would be subtracting
infinite expressions in an uncontrolled way).
The normal-ordered terms in the second line cancel,
as can be seen by writing $k+q = k'$ in the second term (this
is allowed, since they are normal ordered, hence relabellings
cannot produce problems). The definition
of $| \vec 0 \rangle_0$ in Eqs.~(\ref{vacuum1}) and (\ref{vacuum2})
 implies that the remaining difference
in expectation values in the second line  gives  
\be
        {1 \over n_q}
         \left( \sum_{n_k =  -\infty}^0
           - \sum_{n_k = - \infty}^{-n_q}\right) =
                  {1 \over n_q} \, n_q = 1 \; . 
\ee
Note that the construction (\ref{defbq}) of $b_{q \eta}$ and the derivation 
of its commutators (\ref{rhorhoc})  heavily rely on 
the set of $k$'s being  unbounded from below, which is
why this property was stipulated in  Section~\ref{prerequisites} to be
a prerequisite for bosonization.

\subsection{Bosonic ground states $| \vec N \rangle_0$}
Using \Eq{N00}, it is easy to verify that in each
$\vec N$-particle Hilbert space $H_{\vec N}$, 
 $|\vec N\rangle_0$ serves as vacuum state for the bosonic excitations:
\be
\label{a-vacuum}
        b_{q \eta} | \vec N \rangle_0 = 0 \; , \qquad
        \mbox{for all }\quad q, \eta \; .
\ee
Intuitively, the reason is clear:
Since $|\vec N\rangle_0$ is the $\vec N$-particle {\em ground}\/
 state, it does not contain any particle-hole excitations.

With respect to these boson vacuum states, one can define
the operation of {\em boson-normal-ordering}\/,
as follows: by
definition, to boson-normal-order a function of $b$ and $b^\dagger$'s,
all $b_{q\eta}$'s are to be  moved  
to the right of all $b_{q\eta}^\dagger$'s, so that 
\be
\label{bosonnormalordering}
{}^\ast_\ast ABC \dots {}^\ast_\ast 
= ABC \dots  \; - \; {}_0\langle \vec N | ABC \dots  |  \vec N \rangle_0\ ;,
\qquad \mbox{for}\quad A, B, C, \dots \in \{b_{q \eta};  b_{q \eta}^\dagger \}
\; .
\ee
We use the same notation ${}^\ast_\ast \quad {}^\ast_\ast$ for boson as for
fermion normal ordering, because a boson normal-ordered
expression is automatically fermion normal ordered, as
follows by taking $\vec N = \vec 0$ in 
\Eq{bosonnormalordering}. Conversely, if a product purely of
boson operators is fermion normal-ordered, i.e.\ if
${}_0 \langle \vec 0 | ABC\dots | \vec 0 \rangle_0 = 0$, 
then automatically 
${}_0 \langle \vec N | ABC\dots | \vec N \rangle_0 = 0$ will hold
for any $\vec N$, i.e.\ the product is also boson normal-ordered.

\subsection{Completeness of states in bosonic representation}
\label{bosoncompleteness}

It is obvious that every state $| \vec N\rangle$ 
in the  $\vec N$-particle Hilbert space $H_{\vec N}$ can 
be obtained by acting on the corresponding ground  state  $| \vec
N\rangle_0$ with  some function 
of bilinear combinations of fermion operators:
$| \vec N\rangle = \overline f(c^\dagger_{k \eta}
 c_{k' \eta})| \vec N\rangle_0 $.
Remarkably, a much less obvious representation 
in terms of $b^\dagger_{q \eta}$'s also exists, namely:
\bea
\label{NFN0}
 &&  \mbox{For every $| \vec N\rangle$, 
there exists a function
$f(b^\dagger)$ of $b^\dagger$s such that
$| \vec N\rangle = f (b^\dagger) | \vec N\rangle_0 $.} \qquad 
\\ &&
\nonumber
\mbox{\em i.e.\ the $b^\dagger$'s, 
acting on $|\vec N \rangle_0$,
span the complete $\vec N$-particle Hilbert space $H_{\vec N}$.}\/
\eea
(It is clearly not necessary to consider functions
$f(b^\dagger, b)$ of $b$ too, since $b | \vec N\rangle_0 = 0$.)
This is a highly non-trivial statement, since 
the $b^\dagger$'s, being infinite sums,
create complicated linear combinations of particle-hole
excitations. For example, it is not at all obvious that
even  a state as simple as
 $c^\dagger_k c_{k'}| \vec N\rangle_0$ can be written 
in the form of \Eq{NFN0}. 

To make plausible the validity of assertion (\ref{NFN0}),
we offer here a (logically non-rigorous)  
``circular argument'': By \Eq{psiinverse}, we have
\be
\label{cc}
c^\dagger_{k \eta} c_{k' \eta} = 
\frac{1}{2 \pi L} \int_{-L/2}^{L/2}  \! dx \int_{-L/2}^{L/2} \!  dx'
               \,  e^{i( k' x' - k x)} \, \psi^\dagger_\eta (x)
                \psi_\eta (x') \; .
\end{equation}
Now, if one assumes that assertion (\ref{NFN0}) holds,
the validity of the bosonization rules can readily
be established (as shown below).  As we shall see, they imply that
$\psi^\dagger_\eta (x)  \psi_\eta (x')$ 
can be expressed purely in terms of the 
$b^\dagger_{q \eta}$ and $b_{q \eta}$s.
Therefore  
$\overline f(c^\dagger_{k \eta} c_{k' \eta}) |\vec
  N\rangle_0$ has the form (\ref{NFN0})
[using \Eq{cc}, rearranging its right-hand-side into boson-normal-ordered
form and exploiting \Eq{a-vacuum}], so that we have
``proven our starting assumption''. 

Readers that find circular arguments unconvincing are referred to 
Appendix~\ref{app:partition} for a rigorous proof of assertion (\ref{NFN0}).

\subsection{Klein factors $F_\eta^\dagger$ and $F_\eta$}
\label{ladder}

As final bosonization ingredient, one has  to define
``ladder operators'' that connect the various
$\vec N$-particle Hilbert spaces, i.e.\  
{\em  raise or lower the total fermion number
by one}\/, which no combination of bosonic operators can ever do.
As a bonus, they also ensure that fermion fields of different species
anticommute. 
Following the notation of Kotliar and Si
\cite{KotliarSi96}, we shall call these ladder operators
{\em Klein factors} 
and denote them by $F^\dagger_\eta$ and $F_\eta$.
(In Haldane's paper, they are denoted by 
$U$ and $U^{-1}$, see p.~2593 of Ref.~\cite{Haldane81};
for earlier discussions of such operators, see also 
\cite{Heidenreich,Neuberger,Kleinhistory}.) 

We define the Klein factors $F_\eta^\dagger$ and $F_\eta$
to be operators with the following properties: Firstly,
they commute with all bosonic operators:
\be
\label{Ucommutes}
        [b_{q \eta}, F^\dagger_{\eta'} ] =
        [b^\dagger_{q \eta}, F^\dagger_{\eta'} ] =
         [b_{q \eta}, F_{\eta'} ] =
        [b^\dagger_{q \eta}, F_{\eta'} ]
        = 0 \qquad \mbox{for all}\quad q, \eta, \eta' \; .
\ee
Secondly, their action on a general $\vec N$-particle state 
$| \vec N \rangle$, which can always  be ``factorized'' 
as in ({\ref{NFN0})
into a set of particle-hole excitations
$f(b^\dagger)$ acting on the  corresponding $\vec N$-particle ground state,
$| \vec N \rangle = f(b^\dagger) |\vec N\rangle_0$, 
is thus defined:
\bea
\label{UNa}
        F^\dagger_\eta  | \vec N \rangle
        & \equiv    f(b^\dagger)
        c^\dagger_{N_\eta +1}
        | N_1, \dots , N_\eta, \dots N_M \rangle_0 
        & \equiv  f(b^\dagger) 
        \widehat T_\eta  | N_1, \dots , N_\eta +1 , \dots N_M \rangle_0   
        \; ; \quad
\\
\label{UNb}
        F_\eta  |  \vec N \rangle
        & \equiv 
        f(b^\dagger)
        c_{N_\eta} | N_1, \dots , N_\eta , \dots N_M \rangle_0 
        & \equiv   f(b^\dagger)
        \widehat T_\eta | N_1, \dots , N_\eta-1 , \dots N_M \rangle_0  \;
        . \quad
\eea
This is illustrated in Fig.~\ref{fig:levels}(a) and
can be visualized as follows:
 $F_\eta^\dagger$ (or  $F_\eta$) commutes past $ f(b^\dagger)$
and then adds (or removes)
an $\eta$-electron to the
lowest empty (from the highest occupied)
$\eta$-level of  $| \vec N \rangle_0$; 
this results in a new ground state, 
namely $ c^\dagger_{N_\eta +1} | \vec N \rangle_0 $
(or  $c_{N_\eta}  | \vec N \rangle_0 $), on which
the set of particle-hole excitations $f(b^\dagger)$ is then
recreated. Thus the state $ F^\dagger_\eta  | \vec N \rangle$ 
(or  $ F_\eta  | \vec N \rangle$) has the
same set of bosonic excitations as the state $|\vec N\rangle$, but
created on a ground state with one more (or less) $\eta$-electron. 

The operator $\widehat T_\eta$ occuring in the last
equalities in \Eqs{UNa} and
(\ref{UNb}), to be called the  {\em phase-counting operator}\/,  is 
defined by 
\be
\label{phasecounting}
        \widehat T_\eta \equiv (-)^{\sum_{\bar \eta = 1}^{\eta -1}
        \widehat N_{\bar \eta}} \; .
\ee
$\widehat T_\eta$ keeps track of the number of signs picked up when 
acting with a fermion
operator $c_{k \eta}$ on $|\vec N\rangle_0$ to obtain a different
$|\vec N' \rangle_0$:
\be
        c_{k \eta}  
        (C_1)^{N_1} \dots (C_\eta)^{N_\eta} \dots (C_M)^{N_M} 
        | \vec 0 \rangle_0 
        = \widehat T_\eta 
        (C_1)^{N_1} \dots (C_{\eta - 1})^{N_{\eta - 1}}
        c_{k \eta} (C_\eta)^{N_\eta} 
        \dots (C_M)^{N_M} | \vec 0  \rangle_0
        \; .
\ee

The properties (\ref{Ucommutes}) to (\ref{UNb}) completely
specify $F^\dagger_\eta$ and $F_\eta$, and it is in principle not
necessary to give a ``more explicit'' construction of  operators
with these properties. Nevertheless,
such a construction is in fact easy to achieve:
in Appendix~\ref{explicitF} we verify that, roughly
speaking, the inverse of the bosonization identity,
$F^\dagger_\eta \simeq a^{1/2} e^{- i \phi_\eta (0)} \psi^\dagger_\eta (0)$,
does the job [see \Eq{UUdx}]. The most straightforward
explicit representation of $F$, however, is in the bosonic
representation [as first emphasized by
Sch\"onhammer\cite{Schoenhammer2},
see his Eq.(B17)]: since the Fock space of all states 
is spanned by a set of orthornormal basis states of the form 
$
          |N; \{ m_q \} \rangle = 
          \prod_{q > 0}^\infty 
          \frac{b_{q}^{\dagger m_q} }{(m_q !)^{1/2}}
          | N \rangle_0 
$
 [compare Eq.~(\ref{bosonexcited})], 
the  properties (\ref{Ucommutes}) to (\ref{UNb})
immediately imply  
\begin{equation}
\nonumber 
  F_\eta = \sum_{\vec N} \sum_{\{ m_q \} }
   | N_1, \dots, N_{\eta} - 1 ,  \dots N_M; \{ m_q \} \rangle
 \langle N_1, \dots, N_{\eta}, \dots N_M; \{ m_q \} | \;  \widehat T_\eta \; .
\end{equation}
In fact, this equation can be viewed as a
self-sufficient definition of $F_\eta$, 
alternative but equivalent to  (\ref{Ucommutes}) to (\ref{UNb}).

The defining properties (\ref{Ucommutes}) to (\ref{UNb})
have the following consequences\footnote{
Above, we took Eqs.~(\protect\ref{Ucommutes}) to (\protect\ref{UNb})
as the defining relations for  $F_\eta$ and $F^\dagger_\eta$,
and derived (\protect\ref{FFcommute}) to  (\protect\ref{NF}) from them.
Note that if instead Eqs.~(\protect\ref{Ucommutes})
and (\protect\ref{FFcommute}) to  (\protect\ref{NF}) were
used as definitions, 
the action of  $F_\eta$ and $F^\dagger_\eta$ would
be defined only up to a phase, since Eq.~(\protect\ref{NF})
implies that $F^\dagger_\eta  | \vec N \rangle_0$
is equal to 
$ |\dots , N_\eta + 1, \dots \rangle_0  $ modulo a phase. 
To fix this phase, additional definitions such as 
$F^\dagger_\eta  | \vec N \rangle_0
        \equiv   \hat T_\eta |\dots , N_\eta + 1, \dots \rangle_0 
$
are needed. Therefore, the approach chosen above is not only
more explicit, but also more economical.}: 
Firstly, since the spectrum of $\widehat N_\eta$ is unbounded 
from above or below,
$F_\eta$ is unitary: $F_\eta^{-1} = F_\eta^\dagger$.
For this reason, it is often written as $F_\eta^\dagger \equiv 
e^{i \theta_\eta}$, with $\theta = \theta^\dagger$.
We prefer not to use this notation, customary in
the ``field-theoretic'' approach to bosonization, 
since it involves some (insufficiently well-known)
subtleties and  can lead to mistakes if used incorrectly, as 
 discussed in Appendix~\ref{thetaF}.

Secondly, the Klein
factors can be checked to obey the following commution relations:
\bea
\label{FFcommute}
        \{ F^\dagger_\eta, F_{\eta'} \} &= & 2 \delta_{\eta \eta'}
         \qquad \mbox{for all}\quad      \eta, \eta' 
\qquad (\mbox{with}\; F_\eta F_\eta^\dagger = F_\eta^\dagger F_\eta = 1 ) \; ;
\qquad \phantom{.}
\\
\label{FdFdcommute}
        \{ F^\dagger_\eta, F^\dagger_{\eta'} \} &=&
        \{ F_\eta, F_{\eta'} \} = 0  \; ,
        \qquad \mbox{for}\quad      \eta \neq \eta' \; ; \\
\label{NF}
        [\widehat N_\eta, F^\dagger_{\eta'}] &=&  \delta_{\eta \eta'}
        F^\dagger_\eta \; , \qquad \qquad
        [\widehat N_\eta, F_{\eta'}] \, = \,  - \delta_{\eta \eta'}
        F_\eta \; . 
\eea

\section{Boson fields -- definition and properties}
\label{sec:bose}

\noindent
{\em 
We define the boson  fields $\phi_\eta (x)$
as Fourier sums over the  $b_{q \eta}$ and $b^\dagger_{q \eta}$'s
 and derive some
of their commonly-used properties, treating factors of $1/L$ with 
uncommon care. 
\vspace*{2mm}}

When bosonizing $\psi_\eta (x)$ below, we shall find it useful
to introduce the boson fields
\bea
\label{defpsia} \label{defbosonsb}
        \varphi_\eta (x) &\equiv & - \sum_{q > 0}
        {\textstyle {1  \over \sqrt n_q}}
        e^{-i q x} b_{q \eta} e^{-a q/2} \; , \qquad
\label{defpsib} \label{defbosonsa}
        \varphi^\dagger_\eta (x) \equiv  - \sum_{q > 0}
        {\textstyle {1  \over \sqrt n_q}}
        e^{i q x} b^\dagger_{q \eta} e^{-a q/2}, 
\eea
and their Hermitian combination
\be
\label{defphi}
        \phi_\eta (x) \equiv \varphi_\eta (x) + \varphi^\dagger_\eta
        (x) \; = 
        - \sum_{q > 0}
         {\textstyle {1  \over \sqrt n_q}}
        \left( e^{-i q x} b_{q \eta} +  e^{i q x} b^\dagger_{q \eta} \right)
          e^{-a q/2}      \; .   
\ee
Here $a > 0$ is an infinitessimal 
 mathematical regularization parameter needed to regularize
 ultraviolet  ($q \to \infty$) divergent momentum sums
that arise in certain non-normal-ordered expressions and commutators.
Although $a$ is often taken to be on the order
of a lattice spacing, i.e. $a \simeq 1/k_F$, it was emphasized
by Haldane \cite{Haldane81} that it ``in no way plays the role of a 
`cut-off' length''. Nevertheless, $1/a$ can be interpreted
as a kind of  ``effective band-width''
(cf.\ the end of Section~\ref{fsrefermionization}), in the sense that it 
represents the ``maximum 
momentum difference'' for the $c_{k \pm q}^\dagger
        c_k$-combinations occuring in $\phi$.
By construction, $\varphi_\eta (x)$ and $\phi_\eta (x)$ are
periodic in $x$ with period $L$. 
All properties of these fields follow directly from those
[\Eqs{defbq} to (\ref{rhorhoc})]
 of the $b_{q \eta}$, $b_{q \eta}^\dagger$ operators. Below
we list some  useful ones. 

The normal-ordered electron density can be expressed in terms of the 
derivative field $\partial_x \phi_\eta (x)$, as follows: 
\bea
\label{defnormdens}
      \rho_\eta (x) &\equiv & {\,}^\ast_\ast
       \psi_\eta^\dagger (x) \psi_\eta (x)
       {\,}^\ast_\ast
        =  \tpol  \sum_q e^{-i q x} \sum_{k}
        {\,}^\ast_\ast
        c^\dagger_{k-q, \eta} c_{k \eta}
         {\,}^\ast_\ast \\
         & = & 
\label{densityfourier}
        \tpol  \sum_{q > 0} 
         i \sqrt n_q  \left(  e^{-i q x} b_{q \eta} - 
            e^{i q x} b^\dagger_{q \eta} \right) 
          + \tpol 
          \sum_k  {\,}^\ast_\ast
        c^\dagger_{k \eta} c_{k \eta}
         {\,}^\ast_\ast 
\\
\label{normorddens}
      & = &   \partial_x \phi_\eta (x)
        + \tpol \widehat N_\eta  \;  \qquad (\mbox{for} \quad a \to 0)
        \; . 
\eea
\Eq{densityfourier} shows that 
the $ b_{q \eta}$ and $b^\dagger_{q \eta} $ are simply
proportional to the Fourier-components of the electron density.
(Note that the {\em actual}\/ electron density
is $\rho_\eta / (2 \pi)$, 
since we normalized our $\psi_\eta$-fields to
$2  \pi$ instead of 1 in \Eq{antic}.)

The following commutators are often needed:
\bea
\label{varphis}
      [ \varphi_\eta (x), \varphi_{\eta'} (x') ] &=&
       [ \varphi^\dagger_\eta (x), \varphi^\dagger_{\eta'} (x') ] 
       = 0 \; ,
\\
\label{rhorhocom}
        [ \varphi_\eta (x), \varphi^\dagger_{\eta'} (x') ]
        &=&  \delta_{\eta \eta'} \, \sum_{q > 0}
        {\textstyle  {1 \over n_q}}
        e^{-  q [i (x-x') + a]}
\\
\label{comln}
        &=& - \, \delta_{\eta \eta'} \,  \ln \left[
        1 - e^{- i {2 \pi \over L}(x - x' - ia)} \right]
\\
\label{comL}
        &       \stackrel{L \to \infty}{\longrightarrow} &
            -  \, \delta_{\eta \eta'}\, 
        \ln \left[ i \tpol (x - x' - ia) \right] \, .
\eea
\Eq{comln} was obtained using $\ln(1-y) = - \sum_{ n = 1}^\infty
y^{n}/  n$.
Note that $a$ cuts off the ultraviolet divergence
at $x=x'$ that is typical of 1-D boson fields. 
The commutator (\ref{rhorhocom}) occurs, 
for example, when combining exponentials  of boson fields
as follows, using  identity (\ref{lemma2}) of Appendix~\ref{identities}:
\bea
\label{unnormal}
        e^{i \varphi_\eta^\dagger (x)} e^{i \varphi_\eta (x)}
        &=  e^{i (\varphi_\eta^\dagger + \varphi_\eta)(x)}
        e^{[i \varphi_\eta^\dagger (x), i \varphi_\eta (x)]/2} \, 
        & = \, 
        \left( {\textstyle { L \over 2 \pi a }} \right)^{1/2}
        e^{i \phi_\eta (x)} \; , 
\\
\label{unnormal2}
        e^{-i \varphi_\eta (x)} e^{-i \varphi_\eta^\dagger (x)}
        &=  e^{- i (\varphi_\eta + \varphi^\dagger_\eta)(x)}
        e^{[-i \varphi_\eta (x), -i \varphi^\dagger_\eta (x)]/2} \, 
        & = \, 
        \left( {\textstyle { 2 \pi a \over L }} \right)^{1/2}
        e^{-i \phi_\eta (x)} \; .
\eea
Note that the left-hand side of \Eq{unnormal} is boson-normal-ordered,
whereas the right-hand side is not. This is reflected in its
prefactor  factor $a^{-1/2}$, which would diverge in the limit $a\to 0$.

The commutator of $\phi_\eta (x)$ with its derivative\footnote{
\label{Pi} In field-theoretical treatments,
one often encounters the canonically conjugate field to $ \phi_\eta
(t,x)$,  defined (in the Heisenberg picture) by
$ \Pi_\eta (t,x) \equiv \partial_t \phi_\eta (t,x)$. 
If (as is usual) a linear dispersion is assumed, 
with $v_F = 1$, then $ \phi_\eta (t,x) =  \phi_\eta (t+ x)$ and 
$
        \Pi_\eta (t+x) 
        = \partial_x \phi_\eta (t+x) ,
$
so that Eq.~(\ref{phidphiaa}) or
(\protect\ref{1/L}) is the usual canonical commutation
relation for boson fields.
}
can be evaluated in two  ways, depending on the order
of limits for $L\to \infty$, $a \to 0$. 
If one takes the limit $L\to \infty$ first
(but keeps terms of order $1/L$), 
one obtains, using  \Eq{comln}:
\bea
\lefteqn{  \phantom{.} \hspace{-1cm}
       [ \phi_\eta (x), \partial_{x'} \phi_{\eta'} (x') ]
         =   \delta_{\eta \eta'} \;  i \tpol
        \left[ {1 \over e^{i{ 2 \pi \over L} (x-x' - ia)} -1 }
          +  {1 \over e^{-i { 2 \pi \over L} (x-x' +  ia)} -1 } \right] }
\\
\label{phidphiaa}      
      &\stackrel{L \to \infty }{\longrightarrow} & 
            \, \delta_{\eta \eta'} 2 \pi i \left[
        { a/\pi   \over (x-x')^2 + a^2 } -  {\textstyle {1 \over L}} \right]
        \quad \stackrel{a \to 0}{\longrightarrow} \quad
        2 \pi i \, \left[ \delta (x-x')  -  {\textstyle {1 \over L}} \right]
        \; .
\eea
Alternatively, if for some reason one wants
to recover the periodic $\delta$-function, 
the limit $L\to \infty$ can not be taken and  instead one has to
the limit $a\to 0$ first; using 
\Eqs{varphis}, (\ref{rhorhocom}) and (\ref{peridelta}),
one obtains
\bea
\label{phidphi}      
       [ \phi_\eta (x), \partial_{x'} \phi_{\eta'} (x') ]
        & = &
         \delta_{\eta \eta'} \; 
         i \tpol \sum_{q >0}  e^{-qa} \left(     e^{- i q  (x-x') }
         + e^{i q( x-x')} \right)
\\
\label{1/L}
      & \stackrel{a \to 0}{\longrightarrow} & 
        \delta_{\eta \eta'} \; 
        2 \pi i \left( \sum_{\bar n \in \ZZ} \delta(x-x' - \bar n L) -
        {\textstyle {1 \over L}} \right) \; ,
\eea
where the $1/L$ term in \Eq{1/L} is caused by 
the absence of a $q=0$ term in \Eq{phidphi}. 
Note that retaining the 
leading $1/L$ term in 
\Eqs{phidphiaa} and (\ref{1/L})  ensures their consistency with 
\be
\label{intphidphi}
       \int_{-L/2}^{L/2} dx'
        [ \phi_\eta (x), \partial_{x'} \phi_{\eta'} (x') ] = 0 \; , 
\ee
which must hold since $\int_{-L/2}^{L/2} dx'  \partial_{x'} \phi_{\eta'} (x')
= \phi_{\eta'} (L/2) - \phi_{\eta'} (-L/2) = 0$.

Finally,  the commutator of $\phi_\eta (x)$ with itself
can be found by integrating \Eq{phidphiaa}    
over $x'$ over a region near $x$, and
fixing the integration constant by requiring
that the commutator be zero for $x=x'$:
\bea
\label{phiphicomA}
\lefteqn{ \phantom{.} \hspace{-1cm}
        \mbox{[} \phi_\eta (x), \phi_{\eta'} (x') \mbox{]}
        \stackrel{L \to \infty }{\longrightarrow} 
        -  \delta_{\eta \eta'} \, 2  i 
        \biggl[ \arctan \mbox{[}(x-x')/a\mbox{]} - \pi  (x-x')/L 
        \vspace{1mm} \biggr]}
\\
\label{phiphicom}
       &\stackrel{ L= \infty, a \to 0 }{\longrightarrow} &      
        -  \delta_{\eta \eta'} \, i  \pi  \, \epsilon (x -x') 
        \qquad \mbox{where}\quad \; \epsilon (x) \equiv
        \left\{ \begin{array}{cc}
        \pm 1 & \quad \mbox{for} \quad x \gol 0 \; ,  \\
        0  & \quad \mbox{for} \quad x =  0 \; .
        \end{array} \right.
\eea
\Eq{phiphicom} is the form cited most often, but \Eq{phiphicomA}
shows that the step-function is actually smeared over
a range $a$, and that there is a term of order $1/L$.

\begin{figure}[htbp]
  \begin{center}
        \leavevmode 
    \epsfig{
width=0.8\linewidth,%
file=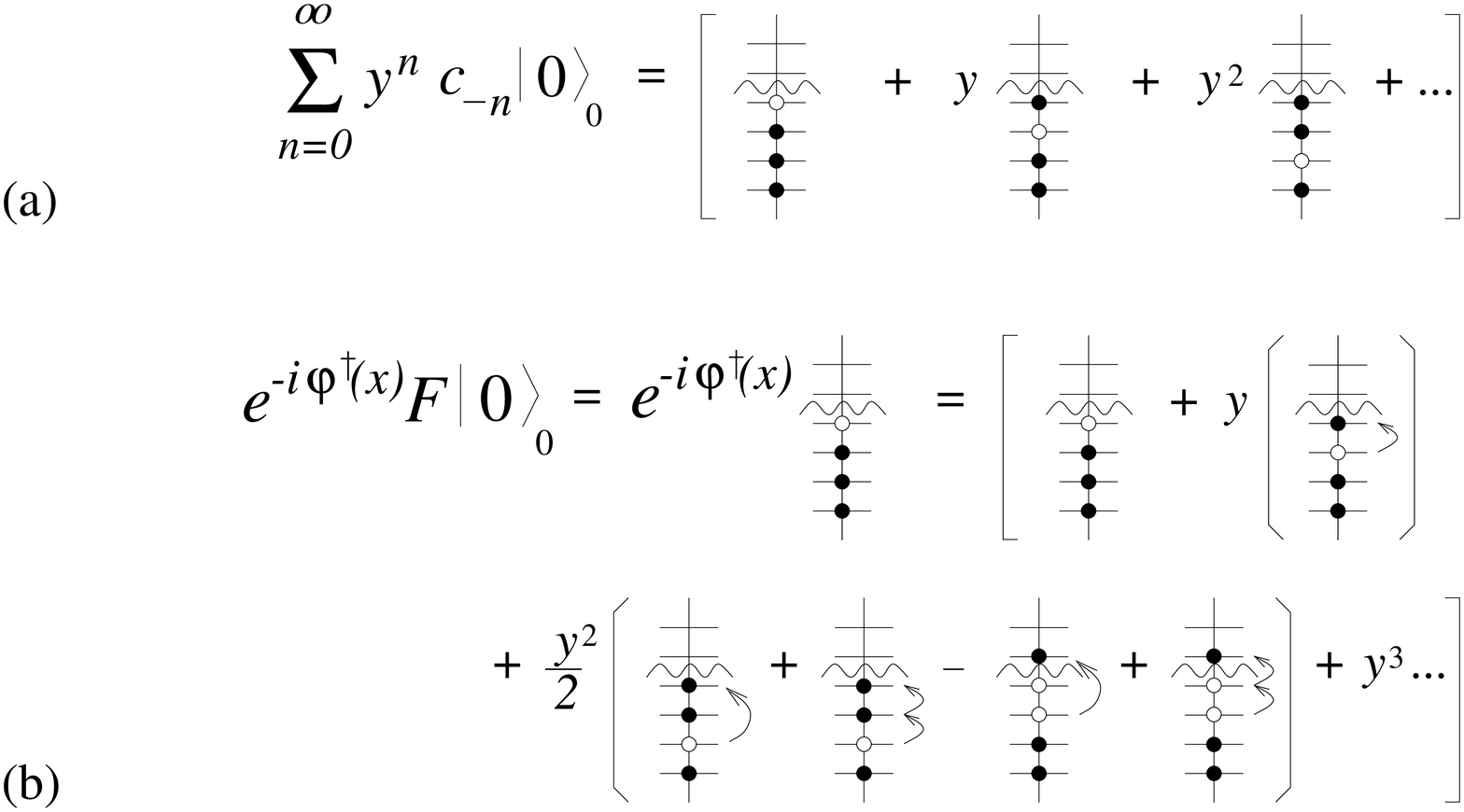}
\vspace*{5mm}
\caption{\label{fig:psiaction}  For the case $M=1$ (i.e.\ the
index $\eta$ suppressed) we depict the action of $\psi (x)$
on $|0 \rangle_0$ in two ways, using either (a)
the Fourier expansion of \Eq{deffermions}, $\psi (x) |0 \rangle_0 \sim
\sum_{n = 0}^\infty y^n c_{-n}  |0 \rangle_0$, where $y = e^{i 2 \pi x / L}$;
or (b) the coherent state representation of
Eq.~(\protect\ref{coherent}),
$ \psi_{\eta} (x) | 0 \rangle_0  \sim  e^{-i \varphi^\dagger (x)} F 
|0 \rangle_0$.
Although the second expansion appears to contain many more
excited states than the first, remarkable cancellations
of prefactors (some of which are
investigated explicitly in Appendix~\protect\ref{explicitholes})
 guarantee that they are identical.}
\end{center}
\end{figure}

\section{Derivation of the bosonization identity}
\label{schoeller}

\noindent
{\em 
We use the definitions of the preceding sections 
to give a novel derivation of
 the famous bosonization identities [(\ref{bf1}) to
(\ref{bf5})]. 
It is  really quite straightforward,
since we exploit some elementary 
properties of boson coherent states, which follow
from standard operator identities stated
and derived in Appendix~\ref{identities}.}

\subsection{$\psi_\eta | \vec N\rangle_0$ is a boson coherent state}

\noindent
{\em 
We show that $\psi_\eta | \vec N \rangle_0$ is an
eigenstate of $b_{q \eta}$ and hence has a coherent-state
representation.
\vspace*{2mm}}

The definitions (\ref{deffermions}) 
of $\psi_\eta (x)$ and (\ref{defbq}) of $b_{q \eta}$ 
imply that 
\bea
\label{bpsi} 
       \mbox{[}b_{q \eta'}, \psi_{\eta} (x)  \mbox{]}& =&
        \delta_{\eta \eta'}  \alpha_q (x)  \psi_{\eta} (x)
       \; , 
\\
\label{bpsibb} 
         \mbox{[} b^\dagger_{q \eta'}, \psi_{\eta} (x)  \mbox{]} &=&
        \delta_{\eta \eta'}  \alpha_q^\ast (x)  \psi_{\eta} (x) \; ,
\eea
where $\alpha_q (x) =
        {i \over \sqrt n_q} e^{i q x}$.
Now, since $ b_{q \eta} | \vec N \rangle_0 = 0$ [see \Eq{a-vacuum}], 
\Eq{bpsi} shows that 
 $\psi_{\eta} (x) | \vec N \rangle_0$
is an {\em eigenstate of the boson annihilation operator}\/
$b_{q \eta}$, with eigenvalue $ \alpha_q (x)$:
\be
\label{eigen}
         b_{q \eta'} \psi_{\eta} (x) | \vec N \rangle_0
        = \delta_{\eta \eta'} \alpha_q (x)  \psi_{\eta} (x)
         | \vec N \rangle_0 \; .
\ee
Hence this state  must have a coherent-state
representation of the form (see e.g.\ \cite{Negele})
\be
\label{coherent}
         \psi_{\eta} (x) | \vec N \rangle_0 \; = \;
         \exp \left[ \sum_{q > 0}  \alpha_q (x) b^\dagger_{q \eta}  \right]
         F_{\eta}  \hat \lambda_\eta (x) | \vec N \rangle_0 
         \, = \,
        e^{- i \varphi^\dagger_\eta (x) }
        F_{\eta}  \hat \lambda_\eta (x) | \vec N \rangle_0
        \; ,
\ee
where \Eq{defpsib} was used for the second equality.\footnote{
\label{f:a=0}
In \Eq{coherent}, strictly speaking
$ \sum_{q > 0}  \alpha_q (x) b^\dagger_{q \eta}
= - i \varphi_\eta^\dagger (x)$ holds only if 
the regularization parameter in definition (\ref{defpsib}) of  
$\varphi^\dagger_\eta (x)$ equals zero, $a=0$.
Thus, all the manipulations
below, up to and including \Eq{bf1}, hold even if one strictly
takes $a=0$ (and regards $ - i \varphi_\eta^\dagger (x)$ as 
shorthand for $ \sum_{q > 0}  \alpha_q (x) b^\dagger_{q \eta}$).
However, a $a\neq 0$  regularization parameter is required 
if one wants to un-normal-order the final result  (\ref{bf1})
to obtain (\ref{bf4}) --- but un-normal-ordering is usually done merely
for notational convenience, and is never essential.}
Here $\hat \lambda_\eta$ is a phase operator to be derived below,
and $F_{\eta} $ is needed because $\psi_\eta$ removes exactly
one $\eta$-particle, which the boson field $\varphi_\eta^\dagger (x)$
 of course {\em cannot}\/ accomplish. 
The representation (\ref{coherent})
guarantees that \Eq{eigen} is satisfied, as can be seen by using
identity (\ref{lemma1a}), with $A = b_{q \eta'}$,
$B = - i  \varphi^\dagger_\eta (x)$, $C =   \delta_{\eta \eta'} \alpha_q
 (x) $.

\Eq{coherent} is a rather remarkable relation, 
since it shows that the action of  $\psi_\eta (x)$ 
on $|N_\eta \rangle$ can be visualized in two different ways,
illustrated in Fig.~\ref{fig:psiaction} (and discussed in
more detail in Appendix~\ref{explicitholes}):
when  $\psi_\eta (x)$ is represented by its standard 
Fourier expansion (\ref{deffermions}),  it creates an
infinite linear combination of single-hole states,
$ \left( \tpol \right)^{1/2}    \sum_k e^{-i k x} c_{n_k \eta} 
| \vec  N \rangle_0$.
The right-hand side of  \Eq{coherent} states that
the same  result can be obtained  in a different way:
first $F_\eta$ removes the highest $\eta$-electron 
in the $\vec N$-particle ground state
$| \vec N \rangle_0$ to obtain a different
ground state $c_{N_\eta \eta} |\vec N \rangle_0$, 
and then $e^{- i \varphi^\dagger_\eta (x)}$ creates
 on this a linear combination of hole states
through the action of the raising operators $b^\dagger_{q \eta}$ which
it contains.
\Eq{coherent} states that both ways produce the {\em same} combination
of single-hole states. This is a highly non-trivial
statement, since intuitively one might have expected 
that $e^{- i \varphi^\dagger_\eta (x)}$, which is after all an exponential
of an infinite sum of particle-hole operators of arbitrarly large
 momenta $q$,
could produce much more complicated particle-hole excitations than the simple
linear combination of {\em single-hole} states produced by $\psi_\eta (x)$. 
However, exploiting the remarkable properties of coherent states,
\Eq{coherent} guarantees that of all the multitude of combinations
of particle-hole excitations contained in 
 $e^{- i \varphi^\dagger_\eta (x)}$, only those terms contribute, 
when acting on  $c_{N_\eta \eta} |\vec N \rangle_0$, that fill its 
empty $N_\eta$-level by moving to the latter a single $\eta$-electron 
 from some lower filled state, leaving behind a single lower-lying hole;
remarkably, {\em all}\/
 other particle-hole contributions (all of which would produce
$\eta$-electrons {\em above}\/ $N_\eta$)  cancel out to zero. 
[In Appendix~\ref{explicitholes} 
we perform the instructive but cumbersome exercise 
of verifying  this explicitly, for the highest few hole states,
by expanding the exponential  $e^{- i \varphi^\dagger_\eta (x)}$.]

To evaluate the operator  $\hat \lambda_\eta (x)$, 
we calculate  the following expectation value in two different ways:
On the one hand, 
\be
\label{exNFPex}
        {}_0\langle \vec N |
        F^\dagger_\eta \psi_\eta (x) | \vec N \rangle_0
        = \lambda_\eta (x) \; , 
\ee
where we used  \Eq{coherent} for $\psi_\eta (x) $,
commuted $e^{- i \varphi^\dagger_\eta (x) }$ to the left past
$F^\dagger_\eta$
[using \Eq{Ucommutes}], and used $ {}_0\langle \vec N |
e^{- i \varphi^\dagger_\eta (x) } = {}_0\langle \vec N | $
[by \Eq{a-vacuum}]. On the other hand, inserting the Fourier
series (\ref{deffermions}) 
for $\psi_\eta (x) $ into \Eq{exNFPex}, we note that since
$ | \vec N \rangle_0 $ and
${}_0 \langle \vec N |  F^\dagger_\eta$ don't contain any
particle-hole pairs, 
only the term in the sum with $n_k = N_\eta$ [i.e.\
$k = \tpol (N_\eta - \half \delta_b)$] can contribute:
\be
\label{TTT}
       {}_0 \langle \vec N |
        F^\dagger_\eta \psi_\eta (x) |  \vec N \rangle_0
        =
         \left( \tpol \right)^{1/2}
         e^{- i {2 \pi \over L} (N_\eta - \half
        \delta_b) x} \; .
\ee

Thus, we conclude that the operator $\hat \lambda_\eta (x)$ is
given by:
\label{lambdasec}
\be
\label{lambdahat}
        \hat \lambda_\eta (x) =
         \left( \tpol \right)^{1/2} e^{- i {2 \pi \over L} (\widehat N_\eta
- \half
        \delta_b) x} \; .
\ee

\subsection{Action of $\psi_\eta (x)$ on an arbitrary state 
$|\vec N\rangle$}

\noindent
{\em 
We derive the bosonization identity
by studying the action of $\psi_\eta (x)$ on an arbitrary state 
$|\vec N\rangle$.
\vspace*{2mm}}

Next we examine how $\psi_\eta (x)$ acts on an arbitrary state
$|\vec N \rangle$ in Fock space, which by  \Eq{NFN0} we
write as 
 $|\vec N \rangle = f( \{ b_{q \eta'}^\dagger \} )
         |  \vec N \rangle_0$.
To this end, two identities are extremely useful:
\bea
\label{id1}
        \psi_{\eta} (x) f( \{ b_{q \eta'}^\dagger \} )
        &=& f( \{ b_{q \eta'}^\dagger - \delta_{\eta \eta'}
        \alpha^\ast_q (x) \} ) \, \psi_{\eta} (x) \; ,
\\
\label{id2}
         f( \{ b_{q \eta'}^\dagger - \delta_{\eta \eta'}
        \alpha^\ast_q (x) \} )
        &=& e^{- i \varphi_{\eta} (x) } \, 
         f( \{ b_{q \eta'}^\dagger \} )  \, e^{ i \varphi_{\eta} (x) } .
\eea
The first follows from \Eqs{deffermions} and (\ref{lemma3}),
 with $A =  b_{q \eta'}^\dagger - \delta_{\eta \eta'}  \alpha^\ast_q (x)$,
$B = \psi_{\eta} (x) $ and $D =  \delta_{\eta \eta'} \alpha^\ast_q (x)$;
the second from \Eqs{defpsia} and (\ref{lemma1}), with
$A = b^\dagger_{q \eta'}$, $B = i \varphi_{\eta} (x)$ and 
$C =  - \delta_{\eta \eta'} \alpha^\ast_q (x)$.

Now evaluate $ \psi_{\eta} | \vec N \rangle$ by
commuting $\psi_\eta$ past $ f( \{ b_{q \eta'}^\dagger \} )$,
then inserting \Eq{coherent} for $\psi_\eta |\vec N\rangle_0$,
and finally rearranging:
\be
\begin{array}{rll}
         \psi_{\eta} (x) | \vec N \rangle 
         & = 
         \psi_{\eta} (x) \,        f( \{ b_{q \eta'}^\dagger \} )
          |  \vec N \rangle_0 & 
\vspace*{1mm}
\\ &=
         f( \{ b_{q \eta'}^\dagger - \delta_{\eta \eta'} 
        \alpha^\ast_q (x) \} ) \, 
        \psi_{\eta} (x) 
         |  \vec N \rangle_0
        & \qquad \mbox{[by \Eq{id1}]}
\vspace*{1mm}
\\ &=
         f( \{ b_{q \eta'}^\dagger - \delta_{\eta \eta'}
        \alpha^\ast_q (x) \} ) \, 
        e^{- i \varphi^\dagger_{\eta} (x) } \, 
        F_{\eta}  \hat \lambda_{\eta} (x) 
         |  \vec N \rangle_0
        & \qquad  \mbox{[by \Eq{coherent}]}
\vspace*{1mm}
\\ &=
         F_{\eta}  \hat \lambda_{\eta} (x) \, 
        e^{- i \varphi^\dagger_{\eta} (x) } \, 
         f( \{ b_{q \eta'}^\dagger - \delta_{\eta \eta'}
        \alpha^\ast_q (x) \} )  
         |  \vec N \rangle_0
        & \qquad  \mbox{[by \Eq{Ucommutes}]}
\vspace*{1mm}
\\ &=
         F_{\eta}  \hat \lambda_{\eta} (x) \, 
        e^{- i \varphi^\dagger_{\eta} (x) } \, 
        \biggl[
        e^{- i \varphi_{\eta} (x) }
         f( \{ b_{q \eta'}^\dagger \} )  e^{i \varphi_{\eta} (x) }
        \biggr]
        |  \vec N \rangle_0
        & \qquad  \mbox{[by \Eq{id2}]}
\vspace*{1mm}
\\ &=
         F_{\eta}  \hat \lambda_{\eta} (x)
        e^{- i \varphi^\dagger_{\eta} (x) } \, 
        e^{- i \varphi_{\eta} (x) }  \,
        f( \{ b_{q \eta'}^\dagger \} )
         |  \vec N \rangle_0
        & \qquad  \mbox{[by \Eq{a-vacuum}]}
\vspace*{1mm}
\\ &=
         F_{\eta}  \hat \lambda_{\eta} (x) \, 
        e^{- i \varphi^\dagger_{\eta} (x) }
        e^{- i \varphi_{\eta} (x)} | \vec N \rangle \; .
        & \qquad  \mbox{[by \Eq{NFN0}]} \; 
\end{array}
\nonumber
\ee
Since $| \vec N \rangle$ is an arbitrary state in the Fock space
${\cal F}$, (and {\em all}\/ states in ${\cal F}$ have
the form (\ref{NFN0}), see Section~\ref{bosoncompleteness}
and appendix~\ref{app:partition}), 
we conclude that the following so-called {\em bosonization formulas}\/ 
for $ \psi_{\eta } (x)$ hold as {\em operator identities} in Fock
space,\footnote{
It can readily be checked
that Eq.~(\protect\ref{bf1}) satisfies Eqs.~(\ref{bpsi}) [and
(\protect\ref{bpsibb})], using identity 
(\ref{lemma1a}), with $A = b_{q \eta'}$,
$B = - i \varphi^\dagger_\eta (x)$, $C =  \delta_{\eta \eta'} \alpha_q  (x) $ 
[or with $A = b^\dagger_{q \eta'}$,
$B = - i  \varphi_\eta (x)$, $C =    \delta_{\eta \eta'} \alpha_q^\ast
 (x) $].
In fact, Emery uses this observation \cite[Eq.~(52)]{Emery79}
to infer directly from Eqs.~(\ref{bpsi}) and
(\protect\ref{bpsibb})  that $\psi_\eta (x) \sim 
e^{-i \phi_\eta}$ (see also \cite{LutherPeschel}). 
However, in order to also obtain
the Klein factor $F_\eta$ in Eq.~(\protect\ref{bf0}) (which
Emery did not), the more elaborate derivation given above is needed.}
 valid for all $L$ (i.e.\ all orders in an 
$1/L$ expansion):
\bea
\label{bf0}
        \psi_\eta (x)
        &=&  F_{\eta}  \hat \lambda_{\eta} (x)
        e^{-i \varphi^\dagger_{\eta} (x) }
        e^{-i \varphi_{\eta} (x)}
\\
\label{bf1}
         &=&  F_{\eta} \,
         \left( \tpol \right)^{1/2}
        e^{-i {2 \pi \over L} (\widehat N_\eta - {1 \over 2}
        \delta_b)x }
        e^{-i \varphi_\eta^\dagger (x)}
        e^{-i \varphi_\eta (x)}
        \qquad \qquad \mbox{[by Eq.~(\ref{lambdahat})]}
\\
\label{bf4}
        &=&
        F_\eta \,
        a^{- 1/2}
        e^{-i {2 \pi \over L} (\widehat N_\eta - {1 \over 2}
        \delta_b)x }
        e^{- i \phi_\eta (x) } \;         
        \qquad \qquad \qquad \qquad \; \mbox{[by Eq.~(\ref{unnormal})]} \\
\label{bf5}
        &=&
        F_\eta \,
        a^{- 1/2} \,
        e^{- i \Phi_\eta (x) } \;         
        \qquad  \; \mbox{with} \qquad
        \Phi_\eta (x) \equiv  \phi_\eta (x) + \tpol 
        (\widehat N_\eta - \half \delta_b)x  \; .
\eea 
These forms are all equivalent. 
(Alternative
notations used by Haldane \cite{Haldane81}, Kane \& Fisher \cite{KF},
Shankar \cite{Shankar} and others 
are discussed in Sections~\ref{Haldanesnotation},
\ref{sec:KFcompare}, \ref{Shankarsnotation} and \ref{thetaF}, respectively.)
\Eq{bf1} is the ``most rigorous'', since it is normal ordered and hence
valid even for $a=0$ (compare footnote~\ref{f:a=0}).
  \Eq{bf4} is the un-normal-ordered
version of \Eq{bf1}, obtained using (\ref{unnormal2}),
and evidently requires $a \neq 0$ [which is needed when unnormalordering
to evaluate $[ \varphi, \varphi^\dagger]$ in  (\ref{unnormal2})].
The most common form is \Eq{bf5}, which absorbs the factor $e^{-i
  {2 \pi \over L} (\widehat N_\eta - {1 \over 2} \delta_b)x}$ into the
definition of a new Boson field $\Phi_\eta (x)$, following Haldane
\cite{Haldane81}. However, we prefer not to use this notation, for two
reasons: firstly, $\phi_\eta(x)$ conveniently commutes with all Klein factors,
whereas $[\Phi_\eta (x), F^\dagger_{\eta'}] = \delta_{\eta \eta'} \tpol x
F^\dagger_\eta$ (by \Eq{NF}); and secondly, the Klein factor $F_\eta$ usually
has a time-dependence $e^{- {2 \pi \over L} (\widehat N_\eta -
  \delta_b/2)\tau}$ (see \Eq{Kleintime}), so it is natural to view $e^{-i {2
    \pi \over L} (\widehat N_\eta - {1 \over 2} \delta_b)x}$ as its
$x$-dependence.   If one is only interested in the limit $L\to \infty$, the
factor $e^{-i {2 \pi \over L} (\widehat N_\eta - {1 \over 2} \delta_b)x }$ can
be neglected.

This completes our derivation of the  bosonization formulas.
Since we deduced them  explicitly and step by step 
from first principles, using only elementary operator identities, 
there is no need to  ``check'' their validity by using them to calculate,
for example, the anti-commutator  $\{ \psi_\eta, \psi^\dagger_{\eta'} \}$
or the correlator $\langle \psi_\eta \psi^\dagger_{\eta'}  \rangle$,
although these are instructive excercises, performed in 
Appendix~\ref{checkanticom} and Section~\ref{bosfer}, respectively
[further such checks are performed in Appendix~\ref{Bosepointsplit},
where the density $\rho_\eta$ and free Hamilton of
(\ref{Hpsipoint}) below are bosonized using (\ref{bf1})].
This is one of the differences
between the constructive approach to bosonization and
the more formal field-theoretical one. In the latter, after 
fields $\psi_\eta (x)$ and $\phi_\eta (x)$ have been defined,
the bosonization formula (\ref{bf4}) is simply written down
as a gift from the gods, whereupon its validity has to be established
by calculating the anti-commutators and Green's functions of
$e^{-i \phi_\eta (x)}$. 

\section{Hamiltonian with linear dispersion}
\label{ham}

\noindent
{\em 
We consider fermions with
linear dispersion and bosonize the Hamiltonian
in both the position and the momentum representation.
\vspace*{2mm}}

So far, no assumptions have been made about the Hamiltonian.
Now assume a linear dispersion, $\varepsilon (k) = v_F \hbar k$,
and measure all energies in units of $v_F \hbar$,
i.e.\ set $v_F \hbar = 1$:
\bea
\label{H0}
        H_0 \equiv \sum_\eta H_{0\eta}, \qquad \mbox{with} \qquad 
        H_{0\eta} &\equiv&   \sum_{k = - \infty}^\infty
        k  {\,}^\ast_\ast  c^\dagger_{k \eta} c_{k \eta}  {\,}^\ast_\ast
\\
\label{Hpsipoint}
       & = &   \int_{-L/2}^{L/2} \! 
        {\textstyle {dx \over 2 \pi}}
         {\,}^\ast_\ast \psi_\eta^\dagger (x) i \partial_x \psi_\eta  (x) 
        {\,}^\ast_\ast \; .
\eea
The second form is equivalent to the first,
since inserting the Fourier expansions (\ref{deffermions}) for
        $\psi_\eta (x)$ into \Eq{Hpsipoint}
reproduces (\ref{H0}). 

Since $[H_{0 \eta}, \widehat N_{\eta'} ] = 0$ for all $\eta, \eta'$, 
any $\vec N$-particle ground state is an
eigenstate of $H_{0 \eta}$, i.e.\ $H_{0 \eta} |\vec N \rangle_0
= E^{\vec N}_{0 \eta} |\vec N \rangle_0$. By inspection, its
eigenvalue is
\bea
\nonumber
       E^{\vec N}_{0 \eta} =
       {}_0 \langle \vec N | H_{0 \eta} |\vec N \rangle_0
       &=& \tpol   \left \{
         \begin{array}{ll}
        \! {\displaystyle   \sum_{n=1}^{N_\eta}}
        (n - \delta_b / 2) &=
           \half N^2_\eta + \half N_\eta (1 - \delta_b)
           \quad \; \; \mbox{if} \; N_\eta \ge 0  \; ,
\\
          \! {\displaystyle  \sum^{0}_{n= N_\eta + 1}}
          - (n - \delta_b / 2) &=
           \half N^2_\eta + \half | N_\eta | (1 - \delta_b)
           \quad \mbox{if} \; N_\eta < 0  \; ,
           \end{array}
           \right.
\\        
\label{EvecNeigenvalues} 
           &=&
           \tpol  \half N_\eta ( N_\eta + 1 - \delta_b ) .
\eea
Furthermore, $b^\dagger_{q \eta}$ raises the energy
of any eigenstate $|E\rangle$ by $q$ units,
as expected intuitively from
 $b^\dagger_{q \eta}$'s
 {\em fermionic}\/ definition (\ref{defbq}), which yields
\be
\label{bqraises}
      [H_{0 \eta},  b^\dagger_{q \eta'} ] =
         \, q \,  b^\dagger_{q \eta} \, \delta_{\eta \eta'}
         \; , \qquad
\mbox{implying} \quad
 H_{0 } b^\dagger_{q \eta} |E \rangle =
 (E + q) b^\dagger_{q \eta} |E \rangle \; .
\ee
Now, the fact that 
the $b^\dagger$'s, 
acting on $|\vec N \rangle_0$,
span the complete $\vec N$-particle Hilbert space $H_{\vec N}$
[recall \Eq{NFN0}] implies that
$H_{0 \eta}$  must also
have a representation {\em purely}\/ in terms of bosonic
variables. The only form that reproduces 
 \Eqs{EvecNeigenvalues} and (\ref{bqraises}) is:
\bea
\label{Hboson}
        H_{0 \eta} &= &  \sum_{q > 0} q \,  
        b^\dagger_{q \eta} b_{q \eta}
         \; + \;
         \tpol \half \widehat N_\eta ( \widehat N_\eta + 1 -
\delta_b )
\\
\label{bosonH}
       & = &  
         \int_{-L/2}^{L/2} {\textstyle {dx \over 2 \pi}}
        {\textstyle {1 \over 2}}
         {\,}^\ast_\ast ( \partial_x \phi_\eta (x) )^2  {\,}^\ast_\ast
         \; + \;
         (\tpol   ) \half \widehat N_\eta ( \widehat N_\eta + 1 - \delta_b ) .
\eea
The second form is equivalent to the first,
since inserting (\ref{defphi}) for 
$\phi_\eta (x)$ into (\ref{bosonH}) 
reproduces (\ref{Hboson}).  Both 
 contain no $F_\eta$'s, since $H_{0 \eta}$ conserves
particle number. 

In field-theoretical treatments \Eqs{Hpsipoint} and (\ref{bosonH}) are
often written using a so-called {\em point-splitting}\/ prescription
(denoted by $: \quad :$) instead of the above normal-ordering
prescription.  Point-splitting, discussed in some detail in
Appendix~\ref{point-splitting}, is another method of regularizing the
product of fields at the same point in position space. It is in most
cases equivalent to normal-ordering, in that it subtracts off
diverging constants.  In Appendix~\ref{point-splitting} we show that
the point-split version of (\ref{bosonH}) can be obtained from the
point-split version of (\ref{Hpsipoint}) using the bosonization
formula (\ref{bf4})
[provided that regularization parameter 
used for point-splitting is the {\em same}\/ $a$ as 
that of the bosonic momentum cut-off $e^{-a q/2}$ 
in \Eq{defpsia}].

For future reference, note that $H_{0\eta}$ transforms
as follows under the unitary transformation 
$U \equiv e^{i c \phi_\eta (x)}$ [use \Eq{lemma1}, with
$[b_{q \eta} , \phi_\eta (x) ] =  
- {\textstyle {1 \over \sqrt n_q}}      
 e^{ i q x -a q/2}$, in \Eq{Hboson}]:
\begin{eqnarray}
\nonumber
  U H_{0\eta} U^{-1} &=&  \sum_{q > 0} q \!
        \left[b^\dagger_{q \eta} - i c   
{\textstyle {1 \over \sqrt n_q}}      
        e^{- i q x -a q/2} \right]\!\! 
        \left[b_{q \eta} + i c  
{\textstyle {1 \over \sqrt n_q}}      
        e^{i q x -a q/2} \right] \\
& & \nonumber
        +  \tpol \half \widehat N_\eta ( \widehat N_\eta + 1 -
\delta_b ) 
\\  \label{eq:UHU-1}
 &=& H_{0 \eta} - c \partial_x \phi_\eta (x) + 
 c^2 (1/a -  \pi / L)  + O(a/L^2) \; .
\end{eqnarray}
The first two terms of (\ref{eq:UHU-1}) can also easily be derived
from the position representation (\ref{bosonH})
for $H_{0 \eta}$, using   \Eqs{lemma1} and 
 (\ref{phidphiaa}), which (for $L \to \infty$) imply 
$U \partial_{x'}  \phi_\eta 
(x') U^{-1} = \partial_{x'} \phi_\eta (x') - 2 \pi c  \, 
\delta (x-x')$; however, this 
method is  too crude to 
correctly reproduce the constants in \Eq{eq:UHU-1}, which
is why we used the momentum representation here.

Also for future reference,
note that  the  $\widehat N_\eta$-dependent terms in $H_{0 \eta}$
imply that the Klein factors pick up an explicit time-dependence
in the imaginary-time Heisenberg picture (with 
 $\tau \in (- \beta, \beta]$ as time parameter):
\be
\label{Kleintime}
       F_\eta (\tau) \equiv e^{H_{0\eta} \tau} F_\eta  e^{-H_{0\eta} \tau} 
         = e^{- {2 \pi \over L} (\widehat N_\eta - \delta_b/2)\tau} F_\eta \; , 
\qquad
              F^\dagger_\eta (\tau) 
         = e^{{2 \pi \over L} 
           (\widehat N_\eta - \delta_b/2)\tau} F^\dagger_\eta \; .
\ee
Of course, ${2 \pi \over L} (\widehat N_\eta - \delta_b/2)$ is just  the 
energy of the particle removed by $F_\eta$ from 
the topmost occupied level of $|\vec N\rangle_0$. 
In the limit $L\to \infty$ in which this energy can be neglected,
the time-ordered expectation value of Klein
factors is simply  ${}_0\langle \vec N | {\cal T}
F_\eta (\tau) F_{\eta'}^\dagger (0) | \vec N \rangle_{0}
=  \delta_{\eta \eta'} \, \mbox{sgn} (\tau)$.

\section{Relation between fermion and boson Green's functions}
\label{bosfer}

\noindent
{\em 
We show that the two-point  Green's function
for free fermions can be expressed in terms of that
of free bosons [Eq.~(\ref{GF2a})], a fact 
which is sometimes used as the starting point
for field-theoretical bosonization.
\vspace*{2mm}}

When discussing Green's functions, we shall always work
in the imaginary-time Heisenberg picture (except in Section~\ref{TLL}).
Given the above Hamiltonian with linear dispersion,
the free fields $\psi_\eta (\tau,x)$ and $\phi_\eta (\tau, x)$
only depend on the combination
$z \equiv \tau + ix$ (and not on $\bar z \equiv \tau - ix$)
(since $c_{k \eta} (\tau) = e^{-k \tau} c_{k \eta} $,
etc., see Appendix~\ref{freefermbos}). Such fields
are often called ``chiral fields''
(or ``chiral left-movers'', since after rotating back
to real time, they depend only $t+ x$).
Therefore, we shall henceforth adopt the notation
$\psi_\eta (z) \equiv \psi_\eta (\tau, x)$ 
and $\phi_\eta (z) \equiv \phi_\eta (\tau, x)$,
$i \partial_z \phi_\eta (z) = \partial_x \phi_\eta (\tau, x)$.
[This use of notation is somewhat sloppy: 
up to now we had used 
$\psi_\eta (0,x) = \psi_\eta (x)$, whereas in the new
notation $\psi_\eta (0,x) = \psi_\eta (ix)$,  but this should
not cause confusion --- if it does, the reader should imagine
changing the old notation by 
replacing all preceding $\psi_\eta (x) $ by $\psi_\eta (ix)$, etc.]
 Real-time functions 
can obtained from imaginary-time ones
 by simply analytically continuing $\tau \to i t$.
It should remembered, though, that this works only
for {\em free}\/ fields: in the presence of interactions,
the operators' time-development is more complicated,
i.e. $c_{k \eta} (\tau) \neq e^{-k \tau} c_{k \eta} $.

\subsection{The limit $L \to \infty$ for $T \neq 0$}

In the limit $L \to \infty$, 
the free imaginary-time ordered fermion and boson Green's
functions (derived explicitly
  in Appendix~\ref{L-to-infty-T=0}) have the following 
forms for $T\neq0$\vspace{-2mm}:
\bea
\label{ffG}
        \langle{\cal T} \psi_\eta (z) \psi_{\eta'}^\dagger (0) \rangle
    &=&  {\delta_{\eta \eta'}  \over
     {\textstyle {\beta \over \pi} \sin [{\pi \over \beta}(z+\sigma a)]}}
\\
\label{bbG}
    \langle {\cal T} \phi_\eta (z) \phi_{\eta'} (0) \rangle
&=&
\label{thermalbosonsexp}
  - \: \delta_{\eta \eta'}
     \: \ln\left( {\textstyle {2 \beta \over L} 
         \sin [{\pi \over \beta}
         (\sigma z + a )]} \right) \; .
\eea
$\langle \; \rangle$ is a thermal expectation value,
${\cal T}$ is the time-ordering operator for $\tau$ and  $\sigma =
 \mbox{sign}(\tau)$.

By observing that (up to a constant)
 \Eq{bbG} is the logarithm of \Eq{ffG}, one
might, even withouth prior knowledge of the bosonization formula,
be led to conjecture that $\psi_\eta$ must 
somehow be the exponential of $\phi_\eta$;
indeed, this is a common  starting point for the field-theoretical
 treatment of bosonization. In the constructive approach
to bosonization, however, the bosonization formula has
already been established as an 
operator identity; hence, showing that 
fermion Green's functions can be calculated
in terms of  boson Green's functions  merely has the status of
a consistency check.

To perform this check, one exploits a 
remarkably identity \cite{Emery79} [see (\ref{expboson}), or
more simply, (\ref{eq:provemagiceasy})], 
valid for any function $\hat B = \sum_{q>0}
(\lambda_q b_q^\dagger + \tilde \lambda_q b_q)$ that is a linear  combination
of  free boson operators
governed by  the boson Hamiltonian \Eq{Hboson}: 
\be
\label{Vexp}
        \langle e^{ \lambda \hat B} \rangle =
        e^{ \langle \hat B^2 \rangle \lambda^2/2 } \; .
\ee
When combined with \Eq{lemma2}, this implies:
\be
\label{magicid}
        \langle e^{\lambda_1 \hat B_1} e^{\lambda_2 \hat B_2} \rangle
        = e^{\langle \lambda_1 \hat B_1 \lambda_2 \hat B_2 + {1 \over 2}
        ( \lambda_1^2 \hat B_1^2 + \lambda_2^2 \hat B_2^2 ) \rangle }
\ee
Using the bosonization formula (\ref{bf4}) 
[and (\ref{Kleintime}) for the time-dependence
of $F_\eta (\tau)$] we therefore have
\bea
\nonumber
\langle{\cal T} \psi_\eta (z) 
        \psi^\dagger_{\eta'} (0) \rangle  &=& 
          a ^{-1} 
      \left[ \theta(\tau)
        \langle 
        F_\eta 
 e^{- {2 \pi \over L} (\widehat N_\eta - {1 \over 2}
        \delta_b)z}
e^{- i \phi_\eta (z) }  e^{ i \phi_{\eta'} (0) }
        F_{\eta'}^\dagger \rangle \right.
   \\
\label{GF2b}
& &   \qquad \left. - \theta(- \tau)
       \langle  e^{ i \phi_{\eta'} (0) } F_{\eta'}^\dagger 
         F_\eta   e^{-  {2 \pi \over L} (\widehat N_\eta - {1 \over 2}
        \delta_b) z}
      e^{- i \phi_\eta (z) } \rangle \right]
\\
\label{GF2a}
        &=&  \delta_{\eta \eta'} \: \sigma a ^{-1} 
              e^{\langle  {\cal T}
          \phi_\eta (z) \phi_{\eta} (0) -
           \phi_\eta (0) \phi_{\eta} (0)
        \rangle}    \; .
\eea
Eq.~(\ref{GF2a}) was obtained using (\ref{magicid}) 
(and taking $  e^{-  {2 \pi \over L} (\widehat N_\eta - {1 \over 2}
        \delta_b)z} \simeq 1$
for $L \to \infty$); 
inserting \Eq{bbG} into \Eq{GF2a} then readily reproduces
\Eq{ffG}.

\subsection{The limit $T=0$ for $L \neq \infty$}

For the case $T=0$ but $L \neq \infty$, 
the fermion and boson Green's
functions (derived explicitly
  in Appendix~\ref{app:freefermbosT=0}) have the following 
forms:
\bea
\label{ffGT=0}
        \langle{\cal T} \psi_\eta (z) \psi_{\eta'}^\dagger (0) \rangle_{T=0}
    &=&  {\delta_{\eta \eta'}  e^{{\pi \over L} (\delta_b + \sigma) z} 
      \over
     {\textstyle {L \over \pi} \sinh [{\pi \over L}(z+ \sigma a)]}}
\\
\label{bbGT=0}
    \langle {\cal T} \phi_\eta (z) \phi_{\eta'} (0) \rangle_{T=0}
&=&
\label{thermalbosonsexpT=0}
  - \: \delta_{\eta \eta'}
     \: \ln\left( 1 - e^{-{2 \pi \over L} (\sigma z + a)} \right)
\eea
Here, too, \Eq{ffGT=0} can be recovered from (\ref{bbGT=0}),
by inserting the latter into the $T = 0$, $L \neq \infty $ version
of (\ref{GF2b}-\ref{GF2a}) [with 
$\langle {\cal T} \phi_\eta (0) \phi_{\eta'} (0) \rangle_{T=0}
=  \delta_{\eta \eta'} \ln (L / 2 \pi a)$], namely
\begin{eqnarray}
  \label{eq:GreensT=0}
         \langle{\cal T} \psi_\eta (z) 
        \psi^\dagger_{\eta'} (0) \rangle_{T=0} &=&
      \delta_{\eta \eta'} \: \sigma a ^{-1} e^{{\pi \over L} \delta_b z}
              e^{\langle  {\cal T}
          \phi_\eta (z) \phi_{\eta} (0) -
           \phi_\eta (0) \phi_{\eta} (0)
        \rangle_{T=0}} \; .
 \end{eqnarray}

\section{Vertex operators -- some general properties}
\label{Vertex}


Exponentials of boson fields, $V^{(\eta)}_\lambda (\tau,x)
\sim e^{i \lambda \phi_\eta (\tau,x)}$, are 
called {\em vertex operators}\/ in the field theory literature and 
are the natural generalizations to $\lambda \neq \pm 1$
of the combinations  $e^{\pm i \phi_\eta (\tau,x)}$ encountered so far.
They occur in many applications of bosonization, e.g.\
Luttinger liquids (see Section~\ref{TLL}) or the
Kondo problem~\cite{vDZF,ZvD}. Here we derive some of their general
properties.

Since along the imaginary time axis all fields only
depend on the combination $z \equiv \tau + i x$
(as is evident from Section~\ref{bosfer}), we henceforth
use $z$ as argument for all fields, writing e.g.\  $\phi_\eta (z)$.
Moreover, all non-equal-time products of field-operators below
will implicitly be assumed to be time-ordered
(i.e.\ when we write $\hat O_1 (z_1) \hat O_2 (z_2) \dots $,
it is to be understood that $\tau_1 > \tau_2 \dots$).
This is important, since non-time-ordered products
are ill-defined along the imaginary-time axis \cite[p.~245]{Negele}.

\subsection{Definition of vertex operator}

The boson normal-ordered form of the 
exponential $ e^{i \lambda \phi_\eta (z)}$ is
defined as follows:
\be
\label{defexpnorm}
        {}^\ast_\ast  e^{i \lambda \phi_\eta (z)}  {}^\ast_\ast
        \, \equiv \,
         e^{i \lambda \varphi^\dagger_\eta (z)}
        e^{i \lambda \varphi_\eta (z)}
        = \left( {\textstyle {L \over 2 \pi a}} \right)^{\lambda^2 /2}
         e^{i \lambda  \phi_\eta (z)}
        = {e^{i \lambda \phi_\eta (z)} \over
        \langle e^{i \lambda \phi_\eta (x)} \rangle}
\ee
The second equality is analogous to (\ref{unnormal});  the third follows
from (\ref{Vexp}) and (\ref{bbG}), and implies that
normal-ordered exponentials indeed satisfy
$\langle {}^\ast_\ast  e^{i \lambda \phi_\eta (z)}  {}^\ast_\ast
\rangle = 1$, as they should.

To normal order the  product of two normal-ordered
exponentials like (\ref{defexpnorm}), proceed as follows: 
\bea
\label{F.opidentities}
        {}^\ast_\ast   e^{i \lambda \phi_\eta (z)} {}^\ast_\ast  \: 
        {}^\ast_\ast e^{i \lambda'  \phi_{\eta'} (z')}
        {}^\ast_\ast 
        &=&
         e^{i (\lambda \varphi_\eta^\dagger (z) 
           +  \lambda'  \varphi^\dagger_{\eta'} (z'))} \, 
         e^{i (\lambda \varphi_\eta (z) +  
           \lambda'  \varphi_{\eta'} (z'))} \,
        e^{-\lambda \lambda' [ \varphi_\eta(z), \varphi_{\eta'} (z') ]} 
\\
\label{almostvOPE}
        & = &
       {}^\ast_\ast e^{i(\lambda \phi_\eta (z) + 
           \lambda'  \phi_{\eta'} (z'))} {}^\ast_\ast 
         \left[\tpol (z - z' + a)\right]^{\lambda \lambda'}
\eea
We used  \Eq{lemma1b} to commute
$         e^{i \lambda' \varphi^\dagger_{\eta'} (z')}$
to the left past 
$         e^{i \lambda \varphi_\eta (z)}$, 
and used \Eq{comL} to evaluate $[\varphi_\eta,  \varphi^\dagger_{\eta'}] $ 
to leading order in $L^{-1}$. In this section we
neglect finite-size corrections and hence
always drop subleading order $L^{-2}$ terms relative
to  $L^{-1}$ terms. However, the former have to be kept
if one is interested in finite-size corrections, see 
Appendix~\ref{Bosepointsplit}.

A {\em vertex operator} is a 
normal-ordered exponential, characterized by a real number 
$\lambda$, 
\be
\label{F.defvertex}
        V^{(\eta)}_{ \lambda} (z) \; \equiv  \;
        \left( {\textstyle {L \over 2 \pi}} \right)^{- \lambda^2 / 2} \:
        {}^\ast_\ast e^{i \lambda \phi_{ \eta} (z)} {}^\ast_\ast
        \;  = \;
        a^{- \lambda^2 /2} e^{i \lambda \phi_{ \eta} (z)} \; .
\ee
Its  normalization (motivated below) 
is a generalization to the case $\lambda \neq 1$ of
that of \Eqs{bf1} and~(\ref{bf4}).
Evidently $\langle V^{(\eta)}_{ \lambda} (z) \rangle =
\delta_{\lambda, 0}$ in the limit $L \to \infty$. 

\subsection{Two-point correlator $ \langle  V^{(\eta)}_{  \lambda } 
         V^{(\eta')}_{  \lambda' }  \rangle$}

The correlation function of two vertex operators can be
derived precisely as in Section~\ref{bosfer}, using \Eqs{magicid}
and  (\ref{thermalbosonsexp}) 
with the result [here $\sigma \equiv \mbox{sgn}(\tau - \tau')$]:
\bea
\label{F.vertexGreen}
         \langle {\cal T}  V^{(\eta)}_{  \lambda } (z)
         V^{(\eta')}_{  \lambda' } (z') \rangle
        & =  &
        \delta_{\eta \eta'} \:
        \left(  a^{- {1 \over 2} (\lambda_1^2 + \lambda_2^2)} \right) \!
         \left( e^{\lambda \lambda' \ln
        \left[{2 \beta \over L}   
          \sin [{\pi \over \beta}(\sigma z - \sigma z' +a)]
        \right] } \right)  \!
         \left( e^{{1 \over 2}  (\lambda_1^2 + \lambda_2^2)
          \ln (       {2 \pi a \over L}) } \right) 
\\
\label{VV2} 
       &=&
        { \delta_{\eta \eta'} \:
        (L/ 2 \pi)^{- (\lambda + \lambda')^2/2}
        \over 
        \left(\textstyle {\beta \over \pi} \sin [{\pi \over
        \beta}(\sigma z-\sigma z' +a)]
       \right)^{- \lambda \lambda'} }    
\; \stackrel{L \to \infty, T \to 0}{\longrightarrow} 
{\delta_{- \lambda,\lambda' } \over (\sigma z - \sigma z' + a)^{\lambda^2}}
\; .
\eea
The reason for including the factor
 $\left( {L\over 2 \pi } \right)^{- \lambda^2 / 2}$
  in definition (\ref{F.defvertex}) now becomes apparent:
by producing the numerator in \Eq{VV2},
it ensures that the above correlator is non-zero in the limit  $L \to \infty$
only if $\lambda + \lambda' = 0$. This latter property is required
on general grounds: since the boson Hamiltonian (\ref{bosonH})
is invariant under a shift $\phi_\eta (z)  \to \phi_\eta (z)+ const$, the 
same is expected for correlators of two properly normalized
normal-ordered exponentials of boson fields. But  for a correlator
containing $\langle e^{i \lambda \phi (z)}
e^{i \lambda' \phi(z')}\rangle$ this can clearly be
true only  if $\lambda + \lambda' = 0$,  
implying that such correlators   must vanish otherwise.

The $T\to 0$ limit of \Eq{VV2} gives 
 $ \langle {\cal T}  V^{(\eta)}_{  \lambda } (z)
         V^{(\eta)}_{ - \lambda } (0) \rangle_{T=0} = (\sigma z)^{-\lambda^2}$.
Thus, the scaling dimension (as defined in Appendix~\ref{OPEss})
of   $ V^{(\eta)}_{  \lambda } $ and  $ V^{(\eta) \dagger}_{  \lambda } = 
 V^{(\eta)}_{ - \lambda } $ is $\lambda^2/2$.

\subsection{OPEs involving vertex operators}

The short-distance behavior of  a product $ V^{(\eta)}_{ \lambda}
V^{(\eta')}_{ \lambda'}$ of vertex
operators is summarized in its {\em operator product expansion}\/
(OPE) (a concept reviewed in Appendix~\ref{OPEss}). 
To derive its OPE, we simply have to normal order
$ V^{(\eta)}_{ \lambda} V^{(\eta')}_{ \lambda'}$. 
Since it is already normal-ordered if $\eta \neq \eta'$
(since then $[ V^{(\eta)}_{ \lambda},  V^{(\eta')}_{ \lambda'} ] = 0 $),
it suffices to consider $\eta = \eta'$: 
\bea
\label{F.vertexOPEa}
  V^{(\eta)}_{  \lambda } (z) \:
                V^{(\eta)}_{  \lambda' } (z')
       &=& 
         \left( {\textstyle {L \over 2 \pi}} \right)^{
           -( \lambda^2 + {\lambda'}^2)/ 2} 
          \left[\tpol (z \!-\! z' \!+\! a)\right]^{\lambda \lambda'} 
          \left(1 + \lambda (z-z') i \partial_{z'} 
            \varphi^\dagger_{\eta} (z')\right) 
\\
\nonumber
    & & \times  \; {}^\ast_\ast e^{i(\lambda \phi_\eta (z) + 
           \lambda'  \phi_{\eta'} (z'))} {}^\ast_\ast 
          \left(1 + \lambda (z-z') i \partial_{z'} 
            \varphi_{\eta} (z') \vphantom{\varphi^\dagger} \right)  
          \quad + \dots
 \\
\label{F.vertexOPE}   & = &
         { V^{(\eta)}_{ \lambda + \lambda'} (z')
           \over (z\! - \! z' \! + a)^{- \lambda \lambda'} } 
         \; + \; 
         {\lambda \: {}^\ast_\ast   V^{(\eta)}_{ \lambda + \lambda'} (z')
           \, i \partial_{z'} \phi_\eta (z') {}^\ast_\ast  
           \over
           (z\! - \! z' \! + a)^{- \lambda \lambda' - 1} }
           \quad  + \dots
\eea
We used \Eq{almostvOPE}  to bring the left-hand side of
                \Eq{F.vertexOPEa} into normal-ordered form, and
then took the limit  $z \to z'$.
Since all expressions within a normal-ordering symbol are well-defined
(i.e. non-diverging), we  Taylor-expanded inside the exponential,
$\varphi_\eta (z) = \varphi_\eta (z') + 
(z\!-\!z') \partial_{z'} \varphi_\eta (z')$, but took care
to maintain the normal order (which is why $\partial_{z'} \varphi^\dagger_\eta$ 
and $\partial_{z'} \varphi_\eta$  are right-most and 
left-most in \Eq{F.vertexOPEa}, and why  the second term of 
\Eq{F.vertexOPE} explicitly needs the normal-ordering symbol).

Note that taking the expectation value
of the OPE (\ref{F.vertexOPE}), namely 
$ \langle V^{(\eta)}_{  \lambda } (z) \:
                V^{(\eta)}_{ \lambda' } (z') \rangle = 
                 (L/ 2 \pi)^{- (\lambda + \lambda')^2/2}
                   (z\! - \! z' \! + a)^{- \lambda^2},
$
reproduces the $z/\beta \to 0$ limit of \Eq{F.vertexGreen}, 
i.e.\ its $T=0$ limit.
This illustrates a rule of thumb, which can be proven quite
generally:\footnote{The general proof of
this rule exploits conformal invariance 
of correlation functions, by using a conformal
mapping of the complex plane onto a cylinder with radius $\beta$
(see e.g.\ Eq.~(3.9) of Ref.~\cite{Affleck}).}
 $T\neq 0 $ correlators  of free (massless) fields
can be obtained from $T=0$ ones
by replacing $(z\! - \! z')$ by ${\beta \over \pi}
\sin[ {\pi \over \beta} (z \! - \! z')]$.

Another important OPE, 
\bea
\label{F.phivertexOPE}
        i \partial_z \phi_{ \eta} (z)
        \; V^{(\eta')}_{  \lambda' } (z')
        &\!=\!& \! {  \delta_{\eta \eta'} \: \lambda' \over
        z \!-\! z' \! + \! a}
        \;  V^{(\eta')}_{  \lambda' } (z')
        \; + \; {}^\ast_\ast   V^{(\eta')}_{\lambda'} (z')
           \, i \partial_{z'} \phi_\eta (z') {}^\ast_\ast \; ,
\eea
is obtained by commuting the $\partial_z  \varphi_{ \eta} (z)$ 
part of $\partial_z \phi_\eta$ past
$ V^{(\eta')}_{  \lambda' } $ using \Eq{lemma1a},  
and evaluating $[\partial_z \varphi_\eta,  
\varphi^\dagger_{\eta'}] $ using \Eq{comL}.

\subsection{Fermions as vertex operators}
\label{fvertex}

Comparing expressions (\ref{bf4}) and (\ref{F.defvertex}),
we see that {\em fermion}\/ operators can
be expressed in terms of vertex operators with $\lambda = \pm 1/2$:
\be
\label{fermionvertex}
        \psi_\eta (z) =
        F_\eta
        e^{- {2 \pi \over L} (\widehat N_\eta - {1 \over 2}
        \delta_b)z  }
        V_{-1/2}^\eta (z)
\ee
The factor  
 $e^{- {2 \pi \over L} (\widehat N_\eta - {1 \over 2} \delta_b)z  }$  
is a combination of the phase factor  in \Eq{bf1} for  $\psi_\eta (x)$ and the 
time dependence (\ref{Kleintime}) of $F_\eta (\tau)$. 
Using Eq.~(\ref{F.vertexOPE}), we thus find the following
OPE for two fermion fields of the same species:
\bea
\label{fermionOPE}
        \psi^\dagger_\eta (z) \psi_{\eta} (z')
        &\stackrel{z \to z'}{\longrightarrow}&  
        {1 \over (z \!-\!  z' \! + \! a)}
       \;\; + \;\; 
        i \partial_{z'} \phi(z') \; + \; 
        \mbox{Order}({\textstyle {1 \over L}},a)
\eea

\subsection{General expectation values of vertex operators}

It is possible to give the expectation value of a general
time-ordered product of vertex operators in closed form [\Eq{vertexproduct}].
 To derive this result, we need some identities:
Let $\hat B_i$ be linear in free boson variables, so that
$[\hat B_i, \hat B_j] = c$-number. Then repeated application of
\Eq{lemma2} gives:
\be
        e^{\hat B_1} e^{\hat B_2} \dots e^{\hat B_n}
        = e^{ \sum_{j = 1}^n \hat B_j }
        e^{{1 \over 2} \sum_{i <  j} [ \hat B_i, \hat B_{ j} ]} \; . 
\ee
By \Eq{Vexp}, we thus have:
\be
\label{EEEepx}
        \langle e^{\hat B_1} e^{\hat B_2} \dots e^{\hat B_n} \rangle
        = e^{{1 \over 2}
        \langle \left(  \sum_{j = 1}^n \hat B_j \right)^2 \rangle }
        e^{{1 \over 2} \sum_{i < j} [ \hat B_i, \hat B_{ j} ]}
        = e^{{1 \over 2}  \sum_{j = 1}^n  \langle \hat B_j^2 \rangle }
        e^{ \sum_{i <  j}
        \langle \hat B_i \hat B_{ j} \rangle } \; . 
\ee
Now apply this identity  to a product of
vertex operators
$V^{(\eta)}_{\lambda_j} (z_j)  \equiv
 a^{- \lambda^2 /2} e^{i \lambda_j \phi_{ \eta} (z_j)}$,
with $j = 1, \dots, n$. 
Using  \Eqs{EEEepx} and (\ref{bbG}) we readily obtain
the following generalization of \Eq{VV2}:
\bea
\label{VVVexp}
        \langle {\cal T} V^{(\eta)}_{\lambda_1} (z_1) \dots
                 V^{(\eta)}_{\lambda_n} (z_n) \rangle 
        &=& \hspace{-3mm}
        \left[ a^{- {1 \over 2} \sum_{j } \lambda_j^2} \right] \! \! 
        \left[ e^{\sum_{i <  j}  \lambda_i \lambda_{j} \ln
        \left[ {2 \pi \over L}  s(z_i, z_j) 
        \right] }  \right] \! \!
        \left[ e^{{1 \over 2} 
         \sum_{j} \lambda_j^2
          \ln (       {2 \pi a \over L}) }  \right]
\\
\label{vertexproduct}
        &=&
        \left( {\textstyle {2 \pi \over L}
        }\right)^{ {1 \over 2} \left( \sum_{j = 1}^n \lambda_j  \right)^2 }
        \prod_{i <  j}
        [s (z_i,z_j)]^{\lambda_i \lambda_{ j}} \; ,
\eea
where $s (z_i,z_j) \equiv  { \beta \over \pi} 
        \sin ({\pi \over \beta}[(z_i - z_{j})  \mbox{sgn}(\tau_i - \tau_j)
+ a])$. 
Taking the limit $L \to \infty$, we see that this expectation
value is non-zero only if 
$
        \sum_{j = 1}^n \lambda_j = 0 \; .
$
This is an important generalization of the corresponding result
for two-point functions, discussed after (\ref{F.vertexGreen}),
and again reflects invariance of the correlator under 
 $\phi_\eta (z)  \to \phi_\eta (z)+ const$.

Suppose that all the $\lambda_j$ are equal to $\pm \lambda$, with $
        \sum_{j = 1}^n \lambda_j = 0, 
$
and denote the corresponding arguments by $z^{(\pm)}$,
i.e. use vertex operators
$V_{+ \lambda}^{(\eta)} ( z^{(+)}_i) $ and
$V_{- \lambda}^{(\eta)} ( z^{(-)}_{ j}) $.
Then it can be shown by simple combinatorical algebra 
that the product  in \Eq{vertexproduct} can be rewritten as a sum:
\bea
\label{Wick}
\lefteqn{       \langle V^{(\eta)}_{+ \lambda} (z^{(+)}_1) \, 
                 V^{(\eta)}_{- \lambda} (z^{(-)}_1) \,
        V^{(\eta)}_{+ \lambda} (z^{(+)}_2) \, 
                 V^{(\eta)}_{- \lambda} (z^{(-)}_2) \, 
        \dots \rangle }
\\
\label{Wick2}
        &= & \left[ \sum_{\{z_i^{(+)},z_j^{(-)} \} } \;
        {1 \over s (z^{(+)}_1, z^{(-)}_1 ) \, s (z^{(+)}_2 ,z^{(-)}_2 ) \dots }
        +
        {1 \over s(z^{(+)}_1, z^{(-)}_2 )  \, s(z^{(-)}_1, z^{(+)}_2  ) \dots }
        + \dots \right]^{\lambda^2} 
\eea
where the sum goes over all  combination of pairs 
$\{z_i^{(+)},z_j^{(-)} \}$, with $i< j$.
Actually, the algebra can be sidestepped via the following observation:
Set $\lambda = 1$, so that 
(\ref{Wick}) is an expectation value of free fermion fields.
Then it has, on the one hand, a Wick-expansion in terms of
two-point correlators 
$\langle \psi^\dagger (z_i^{(+)})  \psi (z_i^{(-)}) \rangle =
 1/ s(z^{(+)}_i, z^{(-)}_j) $,
which is simply (\ref{Wick2}) with $\lambda^2 = 1$; but on the
other hand (\ref{Wick})
is also equal to (\ref{vertexproduct}), with $\lambda_i \lambda_j = \pm 1$. 
Thus the product $\prod$ in (\ref{vertexproduct}) must equal the
Wick-sum $\sum$ in (\ref{Wick2}), implying that also for $\lambda \neq 1$
one must have $(\prod)^{\lambda^2} = (\sum)^{\lambda^2}$.
Hence (\ref{Wick2}) is, remarkably, a 
generalization of Wick's theorem to $\lambda \neq 1$!

\section{Impurity in a Tomonaga-Luttinger liquid}
\label{TLL}

\noindent
{\em To illustrate bosonization ``in practice'', we
calculate the 
tunneling density of states, $\rho_{dos} (\omega)$,
at the site of an
impurity in a Tomonaga-Luttinger liquid.
We resolve the recent controversy 
regarding $\rho_{dos} (\omega)$ by using a  rigorous treatment of 
finite-size refermionization.
\vspace*{2mm}}

Though treating some elements of the
theory  of Tomonaga-Luttinger liquids \cite{Tomonaga,Luttinger}
in great detail, this section
is by no means intended as a complete review of this subject
(for such a review, see \cite{Schoenhammer2});
instead, it aims merely to 
illustrate at a detailed introductory level
(and with more attention to subtleties than usual) 
the application of bosonization to a specific, non-trivial 
problem, namely the calculation of $\rho_{dos} (\omega)$. 
So non-trivial, in fact, that the exponent $\nu$
governing the low-energy behavior of the tunneling
density of states, namely  
$\rho_{dos} (\omega) \sim \omega^{\nu - 1}$ as $\omega \to 0$, 
has   recently been subject to quite some controversy (summarized in 
Section~\ref{intr:TLL} of the introduction). 

We begin in Section~\ref{LRmovers} with 
a quantum wire of {\em free,}\/
 spinless $L$- and $R$-moving 1-D electrons and discuss the
manipulations required to make the problem amenable to 
bosonization. In Section~\ref{Coulomb} we switch on 
an electron-electron interaction 
(of dimensionless strength $g$) and diagonalize it
in the boson basis. In Section~\ref{scatterer}
we add a single impurity to the wire and,
 at a special value
of the coupling constant ($g= 1/2$), refermionize
and diagonalize the Hamiltonian and calculate
a number of useful correlation functions.
Section~\ref{controversy} contains the culmination of
the preceding developments: following 
a strategy due to Furusaki but implementing it more rigorously,
we calculate $\rho_{dos} (\omega)$ at $g=1/2$ and
show that $\nu = 2$, confirming Fabrizio and Gogolin's  \cite{FG} result
and contradicting that of Oreg and Finkel'stein \cite{OF}. 
The calculation is appealingly straightforward ---
if it appears lengthy, this is only because for
pedagogical reasons we show
all details in full\vspace*{5mm}. 

\begin{figure}[htbp]
  \begin{center}
        \leavevmode 
   \epsfig{
width=0.7\linewidth,%
file=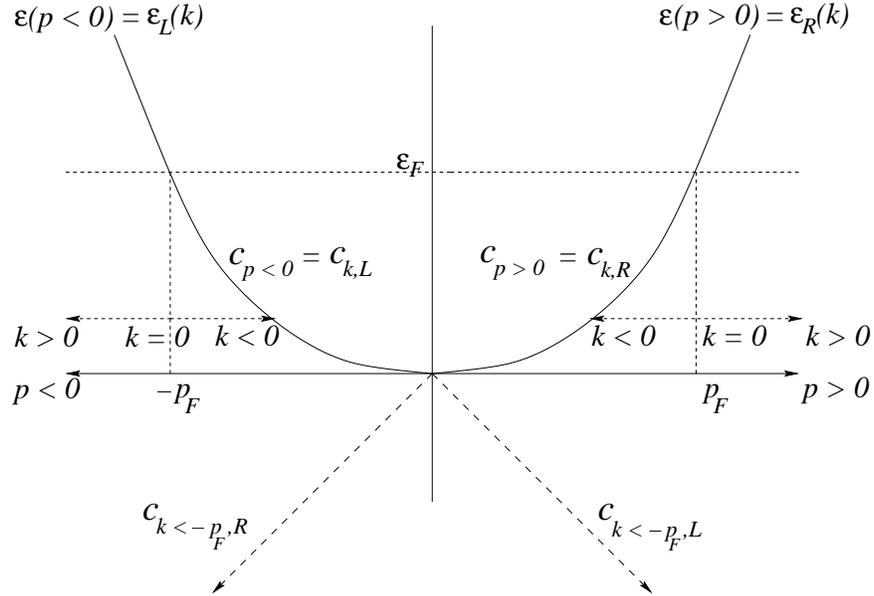}
\vspace*{5mm}
\caption{\label{fig:positron} \label{fig:LR-movers}
Schematic depiction of the dispersion relation
$\varepsilon (p)$ of a 1-D wire containing
$R$- and $L$-moving electrons with $p >0$ and $p<0$,
respectively. From the original electron creation 
operators $c_p$, we construct $c_{k,L/R} \equiv
c_{\mp(k + k_F)}$  for $k \in [-k_F , \infty]$, with corresponding dispersion
$\varepsilon_{k, L/R} \equiv \varepsilon ( 
\mp (k+ k_F))$, see Eq.~(\protect\ref{cLR}).
Then we extend the Hilbert space by hand, by 
taking $k \in [-\infty, \infty]$, i.e.\ by adding
``positron states''  with $k <  - k_F$, whose dispersion we took as
$ \varepsilon_{k, \nu} \equiv 
 \varepsilon (0)  + v_F (k+ k_F)$. 
}
\end{center}
\end{figure}

\subsection{1-Dimensional wire with free left- and right-moving electrons:}
\label{LRmovers}
\label{LRderivation}

\noindent
{\em We introduce a quantum wire of {\bf free,}
 spinless $L$- and $R$-moving 1-D electrons, discuss the
manipulations required to make the problem amenable to 
bosonization, and bosonize.}

\subsubsection{Definition of $c_{k \eta}$ operators}

Consider a 1-D conductor of length $L$ containing free spinless left- and
right-moving electrons, labelled by a momentum index $p \in ( -\infty,
\infty)$, with dispersion $\varepsilon (p) $ that is bounded from below, e.g.\ 
$\varepsilon (p) =(p^2 - p^2_F) /2m$.  The standard definition for the 
physical fermion
field is
\be
\label{LRdeffermions}
        \Psi_{phys} (x) \equiv \left( \tpol \right)^{1/2}
        \sum_{p=- \infty}^{\infty}
        e^{i p x} c_{p}  \; = \;
        \left( \tpol \right)^{1/2} \sum_{k = - k_F}^\infty
        \left( 
        e^{- i (k_F + k)  x}     c_{- k_F - k}  + 
           e^{i( k_F+ k) x}    c_{ k_F + k} 
           \right) \; .
\end{equation}
Here we wrote $p = \mp (k + k_F)$ with $k \in [-k_F, \infty)$, where 
$p \, {\stackrel{\scriptscriptstyle <}{\scriptscriptstyle >}}\,  0$
corresponds to $L$ and $R$-moving electrons.
They can be viewed as two separate, independent ``species'', which we shall
distinguish by an index $\nu = (L,R)$ (analogous to the $\eta$ used hitherto),
writing 
\be
\label{cLR}
c_{k\nu}  \equiv c_{k, L/R} \equiv c_{\mp (k+ k_F)} \qquad
\mbox{with} \quad \varepsilon_{k, L/R} \equiv \varepsilon ( 
\mp (k+ k_F)) \; .  
\end{equation}
Our definition of $k$ purposefully ensures  that 
$\varepsilon_{k, \nu} \gol 0$ if  $k \gol 0$  for both $L$- and $R$-movers.

\subsubsection{Defining $L$- and $R$-moving fermion fields $\tilde \psi_{L/R}$}
\label{sec:defLR}

We now have to cast the problem in a form that meets
the prerequisites for bosonization specified in 
Section~\ref{prerequisites}. This is not yet the case,
since $k \in [-k_F, \infty)$   is {\em bounded from below},
and {\em not discrete}.\/ To remedy this, we proceed below in
three steps: firstly, we extend the range of $k$ to be unbounded,
secondly introduce $L$- and $R$-moving fermion fields
$\tilde \psi_{L/R}$ and thirdly impose boundary conditions
on these to quantize $k$. 

We begin by extending  the  single-particle Hilbert space by 
introducing
(following Haldane \cite{Haldane81}) additional unphysical ``positron states''
at the bottom of the Fermi sea:
we simply extend the range of $k$ to be {\em unbounded}\/
by taking $k \in (- \infty, \infty)$, and
define the corresponding energies in such a way that they
all lie below $ \varepsilon (p=0)$,  e.g.\  $ \varepsilon_{k, \nu} \equiv 
 \varepsilon (0)  + v_F (k+ k_F)$ for $k < - k_F$. 
The introduction of extra ``unphysical'' states does 
not change the low-energy physics of the system, since
by construction they require very high energies ($> \varepsilon_F)$ 
for their excitation. (However,  they would be
excited if a perturbation 
such as an electric field or impurity potential were  sufficiently
strong, so that strong perturbations  cannot be dealt with using 
bosonization.) 

Next, we factor out the rapidly fluctuating 
$e^{\mp i k_F x}$ phase factors  and express
$\Psi_{phys} (x)$ in terms of two fields 
$\tilde \psi_{\nu}(x)$ that vary {\em slowly}\/ on the scale of $1/k_F$:
\bea
\label{oldpsi}
     \Psi_{phys} (x) & \; \mbox{``=''} \; & 
        e^{- i k_F x}  \tilde \psi_L ( x) 
        + e^{+ i k_F x}  \tilde \psi_R ( x) , 
\\
\label{LReta} \label{eq:LRpsifields} \label{psiLR}
 \tilde \psi_\nu (x) \equiv
\tilde \psi_{L/R} (x) &\equiv& \left( \tpol \right)^{1/2}
\sum_{k = - \infty}^\infty e^{\mp i k x} c_{k, L/R}  \; .  \eea The ``=''
indicates that the r.h.s.\ of \Eq{psiLR} differs from that of
\Eq{LRdeffermions} by the inclusion of positron states.  But since these do
not change the low-energy physics, this difference does not matter and ``for
low-energy purposes'' the first of \Eq{psiLR} can effectively be regarded as a
true equality.  Our notation of using the  index $\nu$ (instead of $\eta$) and
putting a $\tilde{\,}$ on $\tilde \psi_\nu$ serves as a reminder that 
$\tilde \psi_L$ and $\tilde \psi_R$ are ``mathematical $L$- and
$R$-movers'',\footnote{\label{f:LRmovers} A field is called a ``mathematical
  $L$- or $R$- mover'' if in the Heisenberg picture
it depends on (real) time $t$ and position $x$ only
  via the combination $(t +x)$ or $(t-x)$, respectively.  For example, the
  fields $\tilde \psi_{L/R} (t,x)$
 defined in Eq.~(\protect\ref{eq:LRpsifields}) are
  mathematical $L$- or $R$-movers if $ c_{k, \pm} (t) = e^{-i k t} c_{k, \pm}
  $, which holds if (i) $\varepsilon_{k \pm} = v_F \hbar k$ (with $v_F \hbar =
  1$) and (ii) no interactions are present. However, even if these two
  conditions do not hold, it is customary to refer to fields constructed as in
  Eq.~(\protect\ref{eq:LRpsifields}) as $L$- or $R$-movers.} respectively, in
  contrast to the $\psi_\eta$'s of earlier sections, which were {\em
  all} mathematical $L$-movers. (If one prefers to
 work purely with the latter, as is sometimes convenient, 
one can simply define purely $L$-moving fields by
$\psi_{1,2} (x) \equiv \tilde\psi_{L,R} ( \pm x)$,
and similarly $\phi_{1,2} (x) \equiv \tilde\phi_{L,R} ( \pm x)$ 
for the boson fields of \Eq{LRbosonizewe} below.)

Finally, to quantize the allowed 
electron momenta $k$  in units of $\tpol$, we
impose boundary conditions on the fermion
fields, choosing (for definiteness) anti-periodic ones:
 $\tilde \psi_\nu ( L/2) = - \tilde \psi_\nu ( -L/2)$
(i.e. $\delta_b = 1$ in \Eqs{momentumquant} and (\ref{boundarycond})
--- the specific choice of boundary condition
becomes unimportant in the continuum
limit $L \to  \infty$).

\subsubsection{Defining $L$- and $R$-moving boson fields $\tilde \phi_{L/R}$}

The inclusion of
``positron states'' in the single-particle Hilbert space 
and the imposition of definite boundary conditions 
in the previous subsection 
should be viewed merely as formal tricks that
make the problem amenable to bosonization. Now that the
prerequisites of Section~\ref{prerequisites} are met, 
we can  rigorously define number
operators $\widehat N_{L/R}$, Klein factors $F_{L/R}$,  and 
boson operators $b_{q L/R}$ in terms of the $c_{k L/R}$'s
as in  Sections~\ref{Noperator}, \ref{ladder} and~\ref{sec:bose}, 
and procede to bosonize. 
Since the fields  $\tilde\psi_{L}$ and $\tilde\psi_{R}$
formally differ from each other only by the factor $e^{\mp i kx}$
 in \Eq{eq:LRpsifields}, the only change needed for $\tilde\psi_R$ relative
to $\tilde\psi_L$ is to replace 
$x$ by $-x$ [and $\partial_x$ by $-\partial_x$
and $\epsilon(x)$ by $- \epsilon(x)$),
cf.\ Eqs.~(\ref{defbosonsb}-\ref{defphi}), 
(\ref{bf4}), (\ref{normorddens}) and (\ref{phiphicom})]:
\bea
\label{LRbosonizewe}
   \tilde \phi_{L/R} (x) &\equiv & %
        - \sum_{ n_q \in \ZZ^+}
        {\textstyle  {1 \over \sqrt{ n_q}}}
       e^{-a q/2}\left[
         e^{\mp i q x} b_{q L/R }  +
          e^{\pm i q x} b^\dagger_{q L/R} \right]  \quad
        (q = \tpol n_q > 0) \; , \qquad \phantom{.}
\\
\label{nubosonization}
      \tilde  \psi_{{L/R}} (x) &=& a^{-1/2} F_{{L/R}} \, 
        e^{\mp  i {2 \pi \over L} (\widehat N_{L/R} - {1 \over 2}
        \delta_b) x }
        e^{- i \tilde \phi_{L/R} ( x) } \;  ,
\\
\label{LRrho}
\tilde \rho_{L/R} (x) &\equiv&        
{\,}^\ast_\ast \tilde \psi^\dagger_{{L/R}} (x) \tilde \psi_{{L/R}} (x)
      {\,}^\ast_\ast       = 
        \pm \partial_x \tilde \phi_{{L/R}} ( x) +  \tpol \widehat
        N_{L/R} \; 
\; .
\eea
Note that $q = \tpol n_q$ implies that the boson fields
and densities are periodic:
 $\tilde \phi_\nu ( L/2) =  \tilde \phi_\nu ( -L/2)$ and
 $\tilde \rho_\nu ( L/2) =  \tilde \rho_\nu ( -L/2)$.

The case of linear dispersion, $\varepsilon( p) = 
v_F (|p| -p_F)$, implying $\varepsilon_{k L/R} = {v_F \hbar k}$ for all
        $k$, is particularly simple. 
Then, in the imaginary-time Heisenberg picture used in
Section~\ref{bosfer}  (with $v_F \hbar = 1$), 
the free $L$- (and $R$-) fields depend only
on $z = \tau + ix$ (and $\bar z = \tau - ix$).
Thus, we have  (the comments just before \Eq{ffG} regarding
the notation $\psi (z)$ apply here, too): 
\begin{eqnarray}
\nonumber
 \tilde  \psi_L (z) \equiv  \tilde \psi_L (\tau, x) 
\; \; \mbox{[} \!  = \! \psi_1 (\tau,x)  \mbox{]} , &\qquad &
 \tilde   \psi_R(\bar z)  \equiv   \tilde  \psi_R (\tau, x)
\; \;  \mbox{[} \! = \!  \psi_2 (\tau, - x) \mbox{]},   
\\
  \label{eq:LRzfields}
\tilde  \phi_L (z) \equiv \tilde \phi_L (\tau, x) 
\; \; \mbox{[} \! = \! \phi_1 (\tau,x)  \mbox{]},  
 &\qquad &
 \tilde    \phi_R(\bar z) \equiv  \tilde \phi_R (\tau, x) 
\; \; \mbox{[} \! = \! \phi_2 (\tau, - x)  \mbox{]},   
\\
\nonumber
 i \partial_z \tilde \phi_L (z) \equiv
\partial_x \tilde \phi_L (\tau,x) 
\; \; \mbox{[} \! = \! \partial_x \phi_1 (\tau,x)  \mbox{]},  
 &\qquad &
  i \partial_{\bar z} \tilde \phi_R(\bar z) \equiv
  - \partial_x  \tilde \phi_L (\tau,   x)
\; \; \mbox{[} \! = \! -  \partial_x \phi_2 (\tau, - x)  \mbox{]}.
\end{eqnarray}
These relations, given here for the sake of completeness,
show that formulas involving relations between free
$R$-movers can be obtained from ones involving relations
between free $L$-movers
by simply replacing $L$ by $R$ and 
$z$ by $\bar z$ (and $\partial_z$ by $\partial_{\bar z}$).
This notation is used extensively, for example,
 by Affleck and Ludwig \cite{Affleck}
in their confromal field theory solution of the Kondo problem.

\subsubsection{Relation between our notation and that of Haldane}
\label{Haldanesnotation}

Readers interested in comparing our definitions with those
used by Haldane in Ref.~\cite{Haldane81} (denoted by the
subscript $\HAL$ below) should note that the
normalization and particularly the phase factor in his  definition 
(3.3), namely $\psi_\HAL (x) 
\equiv L^{-1/2} \sum_{p= - \infty}^{\infty} e^{-i p x} c_{p \HAL}$, 
differ from ours in \Eq{LRdeffermions}. 
Therefore he calls $R/L$-movers what we
call $L/R$-movers. By 
identifying his index $p= (+,-) = (R,L) $
 with our $ \nu = (L,R)$ and
making the identification
$  c_{\pm (k+ k_F) \HAL}$ $\mbox{=}  
c_{k , L/R}$, 
one finds that his and our definitions are related as follows:
for the fermion fields $\psi_{\pm \HAL} (x)
= e^{\mp i k_F x} (2 \pi)^{-1/2}
\tilde \psi_{L/R} (x)$, for the Klein factors
$U_{\pm}= F^\dagger_{L/R}$ and $U_{\pm}^{-1}= F_{L/R}$, 
for the boson operators (which he defines for both $q >0$ and $<0$) 
one has 
$a^\dagger_{q \HAL} = -i 
 [\theta(q) b^\dagger_{q,L} + \theta (-q) b^\dagger_{|q|,R}]$, 
and for the boson fields
$\phi_{\pm \HAL } (x) = 
{\pm \pi x \over L} \widehat N_{L/R} + 
\tilde \varphi_{L/R} ( x)  $. 

\subsection{Diagonalizing an electron-electron interaction by bosonizing}
\label{Coulomb}

\noindent
{\em We consider a simple model
with linear free-electron dispersion and a local electron-electron
interaction. Diagonalizing it
explicitly in the boson basis, we arrive at the ``standard''
bosonic form of the Tomonaga-Luttinger model, with dimensionless
coupling $g$. 
\vspace*{2mm}}

\subsubsection{Turning on an electron-electron interaction}
\label{turnonint}

To illustrate the basic physics of a Tomonaga-Luttinger liquid,
i.e.\ a system of interacting 1-D fermions, we shall consider
the following simple model Hamiltonian:
\begin{eqnarray}
  \label{eq:HL001}
       H_{kin} & = &   \int_{-L/2}^{L/2} \! 
        {\textstyle {dx \over 2 \pi}} 
         {\,}^\ast_\ast \! \left[ \psi_L^\dagger (x) i \partial_x \psi_L  (x) 
         \; + \;
          \psi_R^\dagger (x) (- i \partial_x ) \psi_R  (x) \right] \!
        {\,}^\ast_\ast  \; , 
\\
\label{eq:HLRVV}
H_{int} &=&  \int_{-L/2}^{L/2} \! 
        {\textstyle {dx \over 2 \pi}} {\,}^\ast_\ast \! \left[
        g_2 \,   \tilde \rho_L (x) \tilde \rho_R  (x) 
        \; + \;
        \half g_4 \left( \tilde \rho_L^2 (x) + 
          \tilde \rho^2_R  (x) \right) \right] \! 
        {\,}^\ast_\ast  \; .
\end{eqnarray}
The kinetic term assumes a linear dispersion, 
$\varepsilon (p) \equiv \hbar v_F | p - p_F|$, i.e.\
$\varepsilon_{k, L/R} = \hbar v_F k$, with $\hbar v_F = 1$, 
cf.\ \Eq{Hpsipoint}.
$H_{kin}$ describes a simple local (or point-like) electron-electron 
interaction,\footnote{
A more general local interaction such as 
$ {\,}^\ast_\ast [\Psi_{phys}^\dagger (x) \Psi_{phys} (x)]^2 {\,}^\ast_\ast$
also contains so-called {\em Umklapp} processes
that do not conserve the number of $L$- or $R$-movers,
e.g. $e^{i 2 k_F x} 
\tilde \psi_L^\dagger \tilde \psi_R  \tilde \rho_{\nu}$ or
 $e^{i 4 k_F x} 
\tilde \psi_L^\dagger \tilde \psi_L^\dagger 
\tilde \psi_R \tilde \psi_R$.
However, for the present spinless case they are zero,
since Fermi statistics ensures that $\psi_\nu (x) \psi_\nu (x) = 0$. 
Note also that using a point-like interaction in conjunction
with bosonization is strictly speaking somewhat sloppy:
as emphasized in Section~\ref{sec:defLR}, the extention of
Fock space to include infinitely many negative-energy
``positron states'' is strictly speaking justified only if these are not excited
by a perturbation; however, a point-like interaction
in position space is a constant in momentum space, i.e.\
it includes processes that couples 
states of arbitrarily large momentum differences,
including physical with unphysical states 
(though  their contribution to the low-energy physics is 
still negligible, because of the very large energy cost
involved in exciting a positron state). A cleaner approach
would thus be to explicitly use an interaction with
a finite range, say $\tilde a$, corresponding 
to a finite cut-off $1/\tilde a$ in momentum space.
For a more detailed discussion of such and other cut-off-related matters,
see Ref.~\cite{Schoenhammer2}.
}
parametrized by the dimensionless coupling strengths $g_2$ and $g_4$.

It is convenient to write the Hamiltonian
as follows in terms of the densities $\tilde \rho_\nu$:
\begin{eqnarray}
\label{rhoH00}
 H_{kin} &=&    
   \sum_{\nu = L,R}  \left[  \tpol \half \widehat N_\nu^2 + \!\!
 \int_{-L/2}^{L/2} \! 
        {\textstyle {dx \over 2 \pi}} {\,}^\ast_\ast 
           \half  \! \left( \partial_x \phi_\nu (x) \right)^2  \! 
            {\,}^\ast_\ast \right]\;  = \; 
  \int_{-L/2}^{L/2} \! 
        {\textstyle {dx \over 2 \pi}} {\,}^\ast_\ast 
           \half  \! \left[ \tilde \rho_L^2  + \tilde \rho^2_R   \right] (x)
            {\,}^\ast_\ast 
\\
  \label{eq:Hfull}
 H_0 &=& H_{kin} + H_{int} =   
 {\textstyle {v \over 4 }} \int_{-L/2}^{L/2} \! 
        {\textstyle {dx \over 2 \pi}} {\,}^\ast_\ast \! \left[
        {\textstyle {1 \over  g}} \!  
        \left( \tilde \rho_L + \tilde \rho_R   \right)^2
                 \; + \;          g \!  \left(
         \vphantom{\tilde \rho^2}\tilde \rho_L - \tilde \rho_R   \right)^2
        \right] (x) \!  {\,}^\ast_\ast  \;  ,
\\
& & \mbox{where} \qquad v  \equiv 
  \left[(1+g_4)^2 - g_2^2 \right]^{1/2} \; ,
\qquad g \equiv {\textstyle 
\left[{1 + g_4 - g_2 \over 1 + g_4 + g_2 }\right]^{1/2}} \; . 
\end{eqnarray}
\Eq{rhoH00} follows from (\ref{eq:HL001}) via 
(\ref{bosonH}) and (\ref{normorddens})
(the $\hat N_\nu \partial_x \phi_\nu $ cross-terms
in $\tilde \rho_\nu^2$ vanish when integrated, since 
$\tilde \phi_\nu$ is periodic);  simple algebra then produces 
(\ref{eq:Hfull}) for $H_{kin} + H_{int}$. 

\subsubsection{Diagonalizing $H_0$ in the boson basis}
\label{diagH0boson}

Now, the tremendous advantage of the bosonic representation
is that $\tilde \rho_\nu$, though quadratic in the fermion field
$\tilde \psi_\nu$, is linear in the boson field $\tilde\phi_\nu$
(see \Eq{LRrho}). Thus $H_0$ is {\em quadratic}\/ in
bosonic variables and can be diagonalized straightforwardly
by a Bogoljubov transformation of the $b_{q \nu}$'s:
\begin{eqnarray}
\nonumber 
H_0 \!\! &=& \!\!  \tpol {\textstyle {v \over 2}} \left\{ \!\!
  \left({\textstyle {1 \over g}} \! + \! g \right) \!\!
  \sum_{\nu = L,R} \! \! \left[ \half \widehat N_\nu^2 + \!\!
    \sum_q n_q b^\dagger_{q \nu} b_{q \nu} \right] +
  \left({\textstyle {1 \over g}} \!-\! g \right) \!\!
  \left[ \widehat N_L \widehat N_R - \!\!
    \sum_q n_q (b_{q R} b_{q L} + b^\dagger_{q R} b^\dagger_{q L} )
  \right] \!\!\right\} 
\\
\label{H0bbb}
&=& v \tpol    \sum_{\nu = \pm } \! \! 
\left[ g^{\nu } \widehat {\cal N}_\nu^2 + \!\!
    \sum_q n_q B^\dagger_{q \nu} B_{q \nu} \right] \\
&=& 
\label{fullHdiag}
v \sum_{\nu = \pm}  \left[  \tpol  g^{\nu} \widehat N_\nu^2 + \!\!
 \int_{-L/2}^{L/2} \! 
        {\textstyle {dx \over 2 \pi}} {\,}^\ast_\ast 
           \half  \! \left( \partial_x \Phi_\nu (x) \right)^2 \! 
            {\,}^\ast_\ast \right]  \; \equiv \;
            H_{0+} + H_{0-} \, ,
\end{eqnarray}
where we have defined the following quantities:
\begin{eqnarray}
\label{Bogol}
B_{q \pm} &=& {\textstyle {1 \over \sqrt 8}}
\left\{ 
 \left({\textstyle {1 \over \sqrt g}} \! + \! \sqrt g \right)
 \left(b_{q L} \mp b_{q R}\right) \, \pm \,  
  \left({\textstyle {1 \over \sqrt g}} \! - \! \sqrt g \right)
 \left(b^\dagger_{q L} \mp b^\dagger_{q R} \right) \right\}\; ,
\\
\label{newNNN}
\widehat { N}_+ &=& \half  (\widehat N_L - \widehat N_R) \; , \quad 
\widehat { N}_- = \half  (\widehat N_L
+ \widehat N_R) \; , 
\\
\label{pmnewbosonizewe}
   \Phi_{\pm} (x) &\equiv & %
        - \sum_{q > 0} {\textstyle  {1 \over \sqrt{n_q}}}
       e^{-a q/2}\left[
       e^{- i q x} B_{q \pm }  +
       e^{+ i q x} B^\dagger_{q \pm} \right] \; 
\\
\label{pmnewbosonizewe2}
&=& {\textstyle {1 \over \sqrt 8}}
\left\{ 
 \left({\textstyle {1 \over \sqrt g}} \! + \! \sqrt g \right) \!\!
 \left[\tilde \phi_L (x) \mp \tilde \phi_R (-x) \right] \, \pm \,  
  \left({\textstyle {1 \over \sqrt g}} \! - \! \sqrt g \right) \!\!
 \left[ \tilde \phi_L (- x) \mp  \tilde \phi_R ( x) \right]
\right\} , \qquad \phantom{.}
\\
\label{pmnewdensities}
 \rho_{\pm} (x) &\equiv&        
        \partial_x  \Phi_{{\pm}} ( x) +  \tpol \sqrt 2 g^{\pm 1/2} 
        \widehat   N_{\pm} \; 
\\
\label{pmnewdensities2}
&=& {\textstyle {1 \over \sqrt 8}}
\left\{ 
 \left({\textstyle {1 \over \sqrt g}} \! + \! \sqrt g \right)
 \left[\tilde \rho_L (x) \mp \tilde \rho_R (-x) \right] \, \mp \,  
  \left({\textstyle {1 \over \sqrt g}} \! - \! \sqrt g \right)
 \left[ \tilde \rho_L (- x) \mp  \tilde \rho_R ( x) \right] \right\}\; .
\end{eqnarray}
The first line for $H_0$ follows by inserting (\ref{LRbosonizewe}) into
(\ref{eq:Hfull}). The diagonal form (\ref{H0bbb}) was obtained by making the
Bogoljubov transformation \footnote{ A simple way to derive the Bogoljubov
  transformation is to insert the general Ansatz $B_{q \nu} \equiv \sum_{q'
    \nu'} ({\cal A}_{qq'}^{\nu \nu'} b_{q' \nu'} + \bar {\cal A}_{qq'}^{\nu
    \nu'} b^\dagger_{q' \nu'})$ into the equation of motion implied by
  (\protect\ref{H0bbb}), namely $[B_{q \nu}, H] = v q B_{q \nu}$, and
  solve this for the ${\cal A}$'s, under the condition that $[B_{q \nu} ,
  B_{q' \nu'}^\dagger ] = \delta_{\nu \nu'} \delta_{q q'}$.
Actually, the alternative combinations
$B_{q1}$ and $B_{q2} \equiv {1 \over \sqrt{2}} (B_{q+} \pm B_{q-})$
are more commonly used in the literature, since the corresponding
fields $\Phi_{1}$ and $\Phi_2$, constructed as in (\ref{pmnewbosonizewe}),
do not contain ``non-local'' combinations such 
as the $\tilde \phi_L (x) \mp \tilde \phi_R (-x)$ of 
(\ref{pmnewbosonizewe2}). However, 
for the purpose of discussing scattering from a point-like impurity, as we
do below, our ``non-local'' combinations cause no
problems, and in fact are more convenient, since
forward and backward scattering turn out to
depend only on $\Phi_-$ and $\Phi_+$, respectively, at $x=0$.
}
(\ref{Bogol}) to
a new set of orthonormal boson operators $B_{q \pm}$ (with $[B_{q \nu} , B_{q'
  \nu'}^\dagger ] = \delta_{\nu \nu'} \delta_{q q'}$) and number operators
${\cal N}_\pm$.  From these we constructed in \Eqs{pmnewbosonizewe} and
(\ref{pmnewdensities}) two new boson fields $\Phi_{\pm} (x)$ and densities
$\rho_\pm (x) $ [by analogy to \Eqs{LRbosonizewe} and (\ref{LRrho})]. By
construction they {\em both}\/ are manifestly $L$-moving (since $B_{q\pm} (t)
= e^{- i v q t} B_{q \pm} $) and hence obey all the boson-field identies
derived in Section~\ref{sec:bose} [we could equally well have defined
$\Phi_\pm$ to be $L/R$-moving by replacing $x$ by $\pm x$ in
(\ref{pmnewbosonizewe}-\ref{pmnewdensities2})]. Finally, (\ref{fullHdiag})
follows from (\ref{H0bbb}) just as (\ref{bosonH}) follows from (\ref{Hboson}).

%

\subsubsection{Relation between our notation and that of
Kane and Fisher}
\label{sec:KFcompare}

For the sake of completeness, we briefly explain
the notation used in the path-breaking papers of
Kane and Fisher \cite{KF},   denoting it by a subscript ${}_{kf}$.
They  use field-theoretic bosonization
(another example of which is summarized in Appendix~\ref{fieldconstruct})
and write
\begin{eqnarray}
  \label{eq:kfbosonize}
  \Psi_{phys} (x) \sim \!\! \sum_{n = odd} \!\! 
  e^{- i \sqrt \pi \phi_{kf} (x)}
      e^{-i n [\sqrt \pi \theta_{kf} (x) + k_F x]} \, ,
\end{eqnarray}
where $\theta_{kf}$ and $\phi_{kf} $ are  so-called
``dual fields'' that by definition satisfy
\begin{equation}
\label{eq:kfbosonize-2}
{[} \phi_{kf} (x), \theta_{kf} (x') \mbox{]} = - \half 
i    \epsilon (x - x') .
\end{equation}
Kane and Fisher state that  ``the sum on $n$ enforces the constraint
that the particle density be discrete'' \cite{KF}. Although
this $n$-sum is also used in an early paper by Haldane \cite{Haldane80},
the present authors see no need for it:
 using only $n = \pm 1$ and comparing \Eq{eq:kfbosonize}
with our (\ref{oldpsi}) and (\ref{nubosonization}-\ref{LRrho}),
we can make contact with the rigorous constructive approach
(in the continuum limit $L = \infty$) 
through the identifications
$\tilde \phi_{L/R,here} := \sqrt \pi ( \phi_{kf} \pm \theta_{kf} )$
i.e.\ 
\begin{eqnarray}
  \label{eq:KF-identify}
  \theta_{kf} (x) &:=& \! \! {\textstyle {1 \over 2\sqrt \pi}}
    \left[\tilde \phi_L (x) - \tilde \phi_R (x) 
  \right]_{here}  , \; \mbox{thus} \quad
  \partial_x \theta_{kf} (x) :=  {\textstyle {1 \over 2\sqrt \pi}}
   \left[\tilde  \rho_L (x) +  \tilde \rho_R (x) \right]_{here} , 
\quad \phantom{.}
\\
  \phi_{kf} (x) & :=& \!\! {\textstyle {1 \over 2\sqrt \pi}}
  \left[ \tilde \phi_L (x) + \tilde \phi_R (x) \right]_{here} ,
\; \mbox{thus} \quad
  \partial_x \phi_{kf} (x) :=  {\textstyle {1 \over 2\sqrt \pi}}
  \left[ \tilde \rho_L (x) - \tilde \rho_R (x) \right]_{here}  .
\quad \phantom{.}
\end{eqnarray}
\Eq{eq:kfbosonize-2} is then consistent with 
our (\ref{phiphicom}) (in which one must set $\epsilon (x) \to
- \epsilon(x)$ for $\tilde \phi_{R,here}$). 
Since  $\phi_{L/R,here} = \phi_{L/R,here} (x \pm t)$, it
follows that $\partial_x \phi_{kf} = \partial_t \theta_{kf}$,
thus $\partial_x \phi_{kf}$ is the canonically conjugate
field to $\theta_{kf}$. Translating our Hamiltonian $H_0$
as given by (\ref{eq:Hfull}) into Kane and Fisher's notation, it 
takes a form often encountered in the literature, namely 
\begin{equation}
  \label{eq:KFHamilton}
   H_0  = {v \over 2 } \int_{-L/2}^{L/2} \! 
        dx {\,}^\ast_\ast \! \left[
        {1 \over  g} \!  
        \left( \partial_x \theta_{kf} (x) \right)^2
                 \; + \;          g \!  \left(
         \partial_x \phi_{kf} (x)  \right)^2
        \right] \!  {\,}^\ast_\ast  \;  .
\end{equation}

\subsection{Adding an  impurity to a Tomonaga-Luttinger liquid}
\label{scatterer}

\noindent
{\em We add to the wire a single impurity at $x=0$ that causes
both forward and backward scattering  and bosonize 
the Hamiltonian $H_0 + H_F + H_B$. 
We  diagonalize $H_F$ exactly
for arbitrary $g$ by a unitary transformation.
We diagonalize $H_B$ exactly for $g=\half$ using
refermionization, which we introduce in pedagogical detail,
and calculate some useful correlation functions.}

\subsubsection{Adding an impurity}
\label{sec:addimpurity}

\noindent
{\em We turn on the impurity scattering
term $H_F + H_B$, bosonize,
show that $H_{F/B}$ depends only on $\Phi_\mp$,
and diagonalize $H_{0-} + H_F$ 
for arbitrary $g$ by a unitary transformation $U_- = e^{i c_- \Phi_-}$. 
\vspace*{2mm}}

Assuming an impurity at $x=0$ 
 acts like a point scatterer  causing both forward ($L$-$L$, $R$-$R$) 
and backward scattering ($L$-$R$, $R$-$L$), we 
consider the following perturbations
($\lambda_F$, $\lambda_B$ and the phase $\theta_B$
are real, dimensionless constants, with $\lambda_B > 0$):
\bea
\label{H-F}
 H_F &=& \sum_{\nu = L,R} 
 {\textstyle {v \lambda_F \over 2 \pi}}
 {\,}^\ast_\ast \tilde \psi^\dagger_{\nu} (0) \tilde \psi_{\nu} (0)
 {\,}^\ast_\ast =   {\textstyle {v \lambda_F \over 2 \pi}}
 \left( \tilde \rho_L (0)  + \tilde \rho_R (0)  \right) = 
 {\textstyle {v \lambda_F \over 2 \pi}} \sqrt{2 g} \, \rho_- \, ,
\\
H_B  &=&  {\textstyle {v \lambda_B \over 2 \pi}}
\left[ e^{i \theta_B} \tilde \psi_L^\dagger (0) \tilde\psi_R (0)
 + e^{- i \theta_B} \tilde\psi^\dagger_R (0)  \tilde \psi_L (0) \right]
\\
&=& {\textstyle {v \lambda_B \over 2 \pi a}}
\left[ F_L^\dagger F_R e^{i ( \tilde\phi_L(0) - \tilde\phi_R (0) + \theta_B)}
+ F_R^\dagger F_L e^{i ( \tilde\phi_R(0) - 
\tilde\phi_L (0) - \theta_B)} \right]
\\
&=& 
{\textstyle {v \lambda_B \over 2 \pi a}}
\left[ F_L^\dagger F_R e^{i ( \sqrt{2 g} \Phi_+ + \theta_B)}
+ F_R^\dagger F_L e^{- i ( \sqrt{2 g} \Phi_+ + \theta_B)} \right] \, ,
\label{H_B}
\eea
We  bosonized these using Eqs.~(\ref{nubosonization}-\ref{LRrho})
for the old and (\ref{pmnewbosonizewe2}), (\ref{pmnewdensities2})
for the new boson fields, with 
\begin{equation}
  \label{eq:phizero}
\Phi_\pm \equiv \Phi_\pm (0) = 
{\textstyle {1 \over \sqrt 2} } g^{\mp {1 \over 2}}
\left( \tilde \phi_L (0) \mp \tilde \phi_R (0) \right) , \qquad
\rho_\pm \equiv \rho_\pm (0) = 
{\textstyle {1 \over \sqrt 2} } g^{\pm {1 \over 2}}
\left( \tilde \rho_L (0) \mp \tilde \rho_R (0) \right) .   
\end{equation}
Note that $[\Phi_+, \Phi_-] = [\rho_+, \rho_-] = 0$, since 
 $[B_{q-}, B_{q'+}^\dagger] = 0$.
%
The full Hamiltonian,
\begin{equation}
  \label{eq:finalfullH}
  H \equiv H_0 + H_F + H_B = (H_{0+} + H_B) + (H_{0-} + H_F) \equiv
  H_+ ( \Phi_+) + H_- (\Phi_-)  \; ,
\end{equation}
falls apart into two commuting parts, depending only
on $\Phi_+(x)$ and $\Phi_-(x)$, respectively. 
The second of these can be written as 
\begin{equation}
\label{H_-}
  H_{-} = H_{0-} + H_F 
  =  v\left[
         \int_{-L/2}^{L/2} {\textstyle {dx \over 2 \pi}}
        {\textstyle {1 \over 2}}
         {\,}^\ast_\ast ( \partial_x \Phi_- (x) )^2  {\,}^\ast_\ast
         \; + \;
         (\tpol   ) {\textstyle {1 \over g}}  \widehat N_-^2 + 
         c_- \left( \partial_x \Phi_- + \tpol 
           {\textstyle {\sqrt {2 \over g}}} 
           \widehat N_- \right)
         \right]  , 
\end{equation}
with $ c_- = {\textstyle {\lambda_F \over 2 \pi}}  (2g)^{1\over2}$, 
see \Eqs{fullHdiag} and (\ref{H-F}). It 
 can be diagonalized  using
the unitary transformation 
 $U_- = e^{i c_- \Phi_-}$, which  maps it 
onto a Hamiltonian $H'_{-}$ which is essentially free
[we evaluate $U_- H_{0-} U_-^{-1}$
using  (\ref{eq:UHU-1}) and $U_- H_{F} U_-^{-1}$
using (\ref{phidphiaa})]:
\begin{equation}
\label{UUU-}
\label{eq:H-0prime}
    H'_{-} \equiv  U_- H_{-} U_-^{-1} =  v\left[
         \int_{-L/2}^{L/2} {\textstyle {dx \over 2 \pi}}
        {\textstyle {1 \over 2}}
         {\,}^\ast_\ast ( \partial_x \Phi_- (x) )^2  {\,}^\ast_\ast
         \; + \;
         \tpol  {\textstyle {1 \over g}} 
   \widehat N_-^2 + c_- \tpol  {\textstyle {\sqrt{ 2 \over g}}} \widehat N_- 
         - c_-^2 \left({\textstyle {1 \over a} - { \pi \over L}}\right) 
  \right]  .
\end{equation}

\subsubsection{Finite-size refermionization of $H_+$ at $g = \half$}
\label{fsrefermionization}

\noindent
{\em We give a rigorous introduction to the
technique of finite-size refermionization,
the ``inverse'' of bosonization. Then we 
refermionize $H_{+} = H_{0+} + H_B$
for $g = \half$ and make a unitary 
transformation $U_+ \sim e^{i {\pi \over 2} \widehat {\cal N}^2}$
such that $H'_+ = U_+ H_+ U_+^{-1}$ is quadratic
in refermionized operators. 
\vspace*{2mm}}

The problem posed by  the second term in \Eq{eq:finalfullH}, 
$H_+ = H_{0+} + H_B$, is not exactly solvable for general
values of the coupling constant $g$,
since $H_B$  also involves Klein factors, i.e.\
is not expressed purely in bosonic language. 
 However,
$H_+$ {\em can}\/ be diagonalized exactly for 
$g = \half$, to which we
henceforth restrict our attention. 

$g = \half$ is special, since then $\Phi_+$ occurs in the backscattering term
$H_B$ only in the combination $e^{\pm i\Phi_+}$, which is precisely what occurs
on the right-hand side of a bosonization identity!\footnote{ In the language
  of Section~\ref{fvertex}, $e^{-i\Phi_+}$ is a vertex operator with scaling
  dimension $\half$, which can be refermionized because a free fermion field
  has {\em also}\/ has scaling dimension $\half$.}  This can be exploited by
{\em refermionizing:}\/ we invert the line of reasoning of
Section~\ref{Fockbosons} to~\ref{schoeller}, where bosons and Klein factors
were constructed from fermions, and here construct new fermions from bosons
and Klein factors.  We shall refermionize {\em at finite $L$,}\/ since this
allows us to discuss refermionization at the same level of rigor as
bosonization, namely as an operator identity in Fock space. (Our treatment is
an adaption of that invented by Zar\'and for the 2-channel Kondo model
\cite{vDZF,ZvD}; our way of defining the requisite new Klein factor $F_+$ is
considerably more precise and natural than previous treatments in the
literature.)

Besides the boson field $\Phi_+$ and number operator
$\widehat N_+$ occuring in $H_+$, we need 
new Klein factors $F_+$, $F_+^\dagger$
as ladder operators for  $\widehat N_+$. Since 
\Eq{newNNN} gives $\widehat N_+ = \half (\widehat N_L - \widehat N_R)$, it 
is natural to simply define 
\begin{equation}
  \label{eq:newKlein}
\phantom{.} \!\!\!\!   F_+ \equiv F_R^\dagger F_L , 
\quad 
   \mbox{implying} \quad
 \{ F_+, F_+^\dagger\} = 2 , \quad [\widehat N_+, F_+^\dagger ] = F_+^\dagger,
\quad  [\widehat N_-, F_+^\dagger ] = [\Phi_\pm (x), F^\dagger_+] = 0 .
\end{equation}
Thus  $F_+$, $F_+^\dagger$  and $\widehat N_+$ satisfy the requisite standard
relations (\ref{FFcommute}) and (\ref{NF}).
Next we define a new fermion field $\Psi_+ (x)$ and its Fourier
coefficients $c_\bk$ via the ``refermionization identity''
\begin{equation}
  \label{eq:refermionize}  
        \sqrt{\textstyle {2 \pi \over L}}
        {\textstyle \sum_{\bar k}} e^{- i {\bar k} x} c_{{\bar k} } 
        \; \equiv \; \Psi_+(x) \; \equiv  \;  F_+
      {\textstyle { 1 \over \sqrt a}} 
        e^{-i (\widehat N_+ - \frac12){ 2 \pi x \over L}}
        e^{- i \Phi_+  (x) }   ,
\end{equation}
which  should be read as follows:  the combination 
$F_+{\textstyle { 1 \over \sqrt a}} 
        e^{-i (\widehat N_+ - \frac12){ 2 \pi x \over L}}
        e^{- i \Phi_+  (x) }  $ 
is denoted by the short-hand notation $\Psi_+ (x)$,
since it is known (from Sections~\ref{schoeller},~\ref{bosfer}
and Appendix~\ref{checkanticom})
{\em to behave precisely like a standard fermion field.}\/ 
When Fourier-expanded,
its Fourier coefficients $c_{\bk}$ will thus be
standard, easy-to-work-with fermion operators satisfying
$\{ c_\bk, c_{\bk '}^\dagger \} = \delta_{\bk \bk'}$.
Formally, they can be defined by 
inverting the Fourier sum of \Eq{eq:refermionize}
(cf. \Eq{psiinverse}):
\begin{equation}
  \label{eq:cinvert}
                c_{\bk} \equiv  
                \!\! \int_{-L/2}^{L/2} \!
                {\textstyle {dx \over (2 \pi L )^{1/2}}} \,
                  e^{i k x} \, \Psi_+ (x) \; 
                =  \!
                \int_{-L/2}^{L/2} \! 
                {\textstyle {dx \over (2 \pi L )^{1/2}}}
                 \,  e^{i k x} \, F_+
      {\textstyle { 1 \over \sqrt a}} 
        e^{-i (\widehat N_+ - \frac12){ 2 \pi x \over L}}
        e^{- i \Phi_+  (x) }  .                
  \end{equation}
Since {\em all}\/ operators on the right-hand side
{\em were explicitly definded in terms of the original fermionic
$c_{k \pm}$ operators,}\/ \Eq{eq:cinvert} constitutes an
extremely non-linear yet explicit and well-defined
{\em construction of the new $c_\bk$'s  in terms of the old
 $c_{k \pm}$'s.}\/ That such a direct, explicit construction  is possible
is one of the main advantages of constructive over
field-theoretical bosonization.

What is the nature of the Fock space in which $\widehat N_+$, $F_+$
and the $c_\bk$'s act? Since $N_L, N_R \in \ZZ$,
we have  $ N_+ \in \ZZ + P/2$, where $P = 0$ or 1 if  $(N_L - N_R)$ is
even or odd. Since  $H_B$ contains only the combinations
$F_+^\dagger$ and $F_+$, which leave
$P = (2 \widehat N_+ )\mbox{mod} \,2$ invariant,
the Fock space of states
separates into two decoupled subspaces, 
labelled by $P=0,1$, with $N_+$ integer 
or half-integer in the $P=0$ or
1 subspaces, respectively. The latter fact implies via
\Eq{eq:refermionize}  that the boundary
condition on $\Psi_+$ is $P$-dependent, 
$\Psi_+(L/2) = e^{i \pi (1-P)} \Psi_+(-L/2)$, 
so that the $\bk$-quantization must be too, with 
$ \bk = \tpol \left(n_\bk - {1 - P \over 2} \right)$, $n_\bk \in \ZZ$
(cf. \Eqs{boundarycond} and~(\ref{momentumquant}), now with
$\delta_b = 1-P$)

The definition of a number operator $\widehat N_\eta$
and electron-hole operators $b_{q \eta}$ in Section~\ref{Fockbosons}
of course have analogues in the
refermionized Fock space of $c_\bk$'s. 
Firstly, we note that the following relation holds:
\begin{equation}
{\widehat {{\cal N}}} \equiv \sum_{\bar k} 
{}_\ast^\ast  c^\dagger _{\bar k}
c_{\bar k} {}_\ast^\ast = {\widehat N}_+ -  P/ 2\;,
\qquad (\mbox{with eigenvalues}\; {\cal N} \in \ZZ). 
\label{N_k}
\end{equation}
The left-hand side defines the number operator $\widehat {{\cal N}}$
of the new fermions,
where now  ${}_\ast^\ast\phantom{n}{}_\ast^\ast$ 
denotes normal ordering of the $c_{\bar k}$'s with respect
to a reference state, say $|0_+ \rangle$, defined by
$        c_{\bar k } |0_+\rangle \equiv 0$
for $\bar k > 0$ and 
$        c^\dagger_{\bar k}  |0_+\rangle \equiv 0$
for $\bar k \le 0$
[cf.\ Eqs.~(\ref{vacuum1}-\ref{vacuum2})].
The right-hand side of (\ref{N_k})
is an identity which can be proven by verifying
that  $\lim_{a \to 0} \int_{-L/2}^{L/2} 
{dx \over 2 \pi}  \left(\Psi_+^\dagger(x +a )\Psi_+(x) - {1 \over a}\right)$
yields the left- or  right-hand sides of Eq.~(\ref{N_k})
when evaluated (to $ {\cal O} (a, 1/L)$) 
using either the left- or the right-hand
sides of \Eq{eq:refermionize} 
[see \Eqs{pointsplit} or (\ref{checkbosonpoint}) for details],
respectively. 
More intuitively,  since 
$\Psi_+ \sim F_+ \sim c_{\bar k}$ [by Eq.~(\ref{eq:refermionize})], 
the action of $\Psi_+$ (or $\Psi_+^\dagger$) on any state 
decreases (or increases)
{\em both}\/ $N_+$ and ${\cal N}$ by one. These can thus
differ only by a constant, which must 
be chosen such that  ${\cal N}$ always is integer.
Our definition of $|0_+ \rangle$ above 
sets this constant equal to $P/2$, 
by setting 
${\cal N} = 0$ for  $N_+ = P/2$. 

Secondly, the $B_{q+}$'s in terms of which
 $\Phi_+ (x)$ was defined in \Eq{pmnewbosonizewe}  in fact
can be expressed as  particle-hole operators built from $c_\bk$'s.
To see this, 
we exploit the analogy between the refermionization
identity  (\ref{eq:refermionize}) and the 
 original bosonization identity  (\ref{bf4}) combined with
 its Fourier expansion (\ref{deffermions}),
 to conclude that 
\begin{equation}
  \label{eq:BBreferemionize}
\label{defBBq} 
       B^\dagger_{q +} = {\textstyle  { i \over \sqrt n_q}}
          \sum_\bk c^\dagger_{\bk+q } c_{\bk}   \; , 
      \quad
        B_{q +} =  {\textstyle  {- i \over \sqrt n_q}}
         \sum_\bk c^\dagger_{\bk-q } c_{\bk}   \; , 
         \quad (q = \tpol n_q > 0 ) ,
\end{equation}
in analogy to Eq.~(\ref{defbq}) for the $b_{q \eta}$ in the
$\phi_\eta (x)$-fields of  Eqs.~(\ref{defpsia}-\ref{defphi}).

We are now ready to refermionize $H_+$ by
expressing it in terms of the $c_\bk$'s: 
\begin{eqnarray}
H_{0+} \!\! &=& \!\! v \left[
         \tpol \left( \half \widehat{\cal N} 
           ( \widehat {\cal N}
        \! + \!  P ) + {\textstyle {P\over 8}} \right) +          
         \int_{-L/2}^{L/2} {\textstyle {dx \over 2 \pi}}
        {\textstyle {1 \over 2}}
         {\,}^\ast_\ast ( \partial_x \Phi_+ (x) )^2  {\,}^\ast_\ast \right]
\label{oldHphi}        
\\
  \label{eq:H0cbk}
& = & \!\! \sum_\bk \varepsilon_\bk 
         {}_\ast^\ast  c^\dagger _{\bar k} c_{\bar k} {}_\ast^\ast
         + \Delta_L  {\textstyle {P\over 8}} \ , 
\quad \mbox{where} \quad
\Delta_L \equiv  v \tpol \; , \quad 
\varepsilon_\bk \equiv v \bk = \Delta_L 
\left(n_\bk - {\textstyle {1 - P \over 2}}\right) ; \qquad \phantom{.}
\\
  H_B \!\! & = & \!\!  {v \lambda_B \over 2 \pi a}
  \left( F^\dagger_+ e^{i (\Phi_+ + \theta_B )} + 
 F_+ e^{- i (\Phi_+ + \theta_B )}  \right) = 
 {v \lambda_B \over 2 \pi \sqrt a} \left[ \Psi_+^\dagger (0)  
       e^{i  \theta_B} +     \Psi_+ (0)  e^{-i \theta_B} \right] \\ 
  \label{eq:refermionizeHB}
& = & \!\!
 \sqrt{\Delta_L \Gamma} \sum_\bk \left( c_\bk^\dagger e^{i \theta_B}  + 
    c_\bk   e^{-i \theta_B} \right) , 
\quad \mbox{where} \quad 
 \Gamma \equiv 
  {\textstyle {v \lambda_B^2 \over a (4 \pi)^2}} \; .
\end{eqnarray}
For  $H_{0+}$ in (\ref{oldHphi}) we started from \Eq{fullHdiag} and
wrote  $\half \widehat N_+^2 = \half 
 \widehat{\cal  N}   ( \widehat {\cal N} 
 +  P ) + {\textstyle {P\over 8}}$ 
[using \Eq{N_k}]; to obtain (\ref{eq:H0cbk}) we 
inverted the line of reasoning that
lead from \Eq{H0} to~(\ref{bosonH}) [or from (\ref{H0point}) to
(\ref{H0bospoint})].
For $H_B$ we simply inserted (\ref{eq:refermionize})
into (\ref{H_B}). 
\Eq{eq:refermionizeHB} manifestly shows that 
$\widehat { N}_+$ is {\em not}\/ conserved,
as expected for a back-scattering term that converts
$L$- into $R$-movers and vice versa. (In contrast, the total number
of fermions, $2 \widehat N_- = \widehat N_L + \widehat N_R$, 
obviously {\em is}\/ conserved, since by \Eq{eq:newKlein}
$[F_+, \widehat N_-] = 0$.)

It is easiest to diagonalize $H_+$  if it is nominally
{\em quadratic}\/ in fermionic operators. 
It can be made so (and the phase $e^{i \theta_B}$ absorbed) by 
a trivial unitary phase transformation: 
\begin{equation}
  \label{eq:U+definition}
 U_+ \equiv e^{i ({ \pi \over 2} \widehat {{\cal N}}^2 
 -\theta_B \widehat {\cal N} )}  
\qquad \mbox{gives}\qquad
U_+ \left( F_+^\dagger  e^{i \theta_B} \right) U_+^{-1} = 
F_+^\dagger \, e^{ i \pi (\widehat {\cal N} + \half) } 
 = F_+^\dagger (i \sqrt 2 \,\alpha_d) \; .
\end{equation}
Here $\alpha_d \equiv {\textstyle {1 \over \sqrt 2}}
  e^{i \pi \widehat {\cal N}} $ is a 
``local Majorana fermion'', since  its definition implies the following
properties [the first two follow from the fact that  ${\cal N} \in
  \ZZ$, the last four from \Eq{eq:newKlein}]:
\begin{equation}
  \label{eq:alphad}
  \{ \alpha_d, \alpha_d \} = 1 \; , \qquad \alpha_d^\dagger = \alpha_d \;
  , \qquad    \{F_+, \alpha_d \}  = \{ F_+^\dagger , \alpha_d \} = 
  \{c_\bk, \alpha_d \}  = \{ c_\bk^\dagger , \alpha_d \} =   0 \; .
\end{equation}
Since \Eqs{eq:U+definition} and (\ref{eq:cinvert}) imply
$U_+ \left( c_\bk^\dagger  e^{i \theta_B} \right) U_+^{-1} = 
 c_\bk^\dagger (i \sqrt 2 \, \alpha_d ) $, the transformed version
of $H_+$ is quadratic in fermions, as desired:
\begin{equation}
  \label{eq:h+prime}
  H'_+ \equiv U_+ H_+ U_+^{-1} =
  \Delta_L {\textstyle {P\over 8}} + 
 \sum_\bk  \left[\varepsilon_\bk 
         {}_\ast^\ast  c^\dagger _{\bar k} c_{\bar k} {}_\ast^\ast
         + \sqrt{\Delta_L \Gamma} \left( c_\bk^\dagger  + 
    c_\bk \right) \left( i \sqrt{2} \, \alpha_d \right) \right] \; .
\end{equation}
The trick of converting a term linear in fermions to a quadratic form 
using a Majorana fermion was also used by Matveev \cite{Matveev}. In contrast
to his work, however, our use of constructive bosonization, in which
electron counting operators such as $\widehat {\cal N}$ play a fundamental
role, allows us to precisely formulate the unitary transformation that causes
this Majorana fermion to appear naturally. 

The fact that $\Gamma \propto 1/a$ is consistent with the
well-known fact that for $g<1$, an impurity in a Tomonoga-Luttinger
liquid ``scales into the strong-coupling regime''. By 
this statement one means that under a renormalization
group transformation designed to focus on the low-energy regime
of the model, the effective strength of the impurity
scattering {\em increases.}\/ To see this explicitly,
one can adopt for example
Anderson's poor man's scaling approach, in which
the RG is generated by reducing 
(at fixed $L$, usually $= \infty$) the bandwidth while
adjusting the couplings to keep the dynamical properties invariant.
Since the cut-off used when bosonizing is $1/a\, (\sim p_F)$,
reducing the bandwidth means changing  $a$ to a larger value $a'$,
which must be accompanied by a change in coupling constant
from its initial value 
$\lambda_B (a) $ to a new value $ \lambda'_B (a')$. Since $a$
occurs in $H'_+$ {\em only}\/ through $\Gamma$, one immediately concludes
that $\lambda'_B (a') = \lambda_B \sqrt{a'/a}$, 
which implies that $\lambda_B$ grows under rescaling.
(This is completely analogous to what happens for
the 2-channel Kondo model \cite{vDZF,ZvD}.)

\subsubsection{Finite-size diagonalization of $H'_+$ at $g = \half$}
\label{fsdiagonalization}

\noindent
{\em We give the linear transformation 
(derived in 
Appendix~\ref{app:fsdiagonalization})
that diagonalizes 
the refermionized   $H'_+$ (for finite $L$)
by expressing the  $c_\bk$'s in terms of 
new fermions $\tilde \alpha_\varepsilon$
and $\beta_\bk$. 
\vspace*{2mm}}

As pointed out by Oreg and Finkel'stein  \protect\cite{OFphysrevB}
and Furusaki \cite{Furusaki}, the form (\ref{eq:h+prime}) for
$H'_+$ is related to that arising after bosonizing and refermionizing
the 2-channel Kondo model. 
Following the latter's solution in Refs.~\cite{vDZF,ZvD},
this fact can be exploited to explicitly diagonalize $H'_+$
for finite $L$ in the
$P=0$ sector\footnote{The differences between
the $P=0$ and 1 sectors disappear in the continuum limit
$L \to 0$ that we are ultimately interested in.}
 [i.e.\ with $\bk = \tpol (n_\bk - \half)$].
 This elementary
excercise is performed in Appendix~\ref{app:fsdiagonalization},
with the following results:
\begin{eqnarray}
\label{oldck} \label{oldad}
  c_\bk (t)  \!\! &=& \!\! 
{\textstyle {1 \over \sqrt 2}}
\left( \alpha_\bk (t) + i \beta_\bk (t)  \right) \; ,
\\
\label{betak-betadagger}
\alpha_{- \bar k} \!\! &\equiv & \!\!  \alpha_{\bar k}^\dagger \; , \qquad
\beta_{- \bar k} \equiv \beta_{\bar k}^\dagger \; ,
\qquad
\{ \alpha_\bk , \alpha_{\bk'}^\dagger \} =
\{ \beta_\bk , \beta_{\bk'}^\dagger \} =  \delta_{ \bk \bk'} \; ,
\qquad \{ \alpha, \beta \} = 0 \; ;
\qquad \phantom{.}
\\
\label{alpha-tildealpha}
\alpha_{\bar k} (t) \!\! &=& \!\!
\sum_\varepsilon A_{\bk , \varepsilon}
 e^{-i \varepsilon t} \tilde \alpha_\varepsilon \;  , 
\qquad 
\alpha_d (t) = 
\sum_\varepsilon A_{d , \varepsilon}  e^{-i \varepsilon t}
 \tilde \alpha_\varepsilon \; ,
\qquad 
\beta_{\bar k} (t) = e^{-i \varepsilon_{\bar k} t} \beta_{\bar k} \; , 
\phantom{.}
\\
\label{e-enotindependent}
\tilde \alpha_{-\varepsilon} \!\! &=& \!\!
  \tilde \alpha^\dagger_{\varepsilon} \; , 
\qquad 
 \{ \tilde \alpha_\varepsilon, \tilde \alpha^\dagger_{ \varepsilon'} \} =
         \delta_{\varepsilon \varepsilon'} \; , \qquad
\{ \tilde \alpha,  \beta \} =  0 \; ; 
\\  \label{finalAen}
  A_{\bk, \varepsilon }  \!\! &=& \!\!  { i  2  \sqrt{\Delta_L \Gamma} 
A_{d, \varepsilon}  \over \varepsilon - \varepsilon_\bk}
\, , \quad
 A_{d, \varepsilon} = 
- i \, \mbox{sign} (\varepsilon) \! \left[{ 4 \Delta_L \Gamma \over
4 \Delta_L \Gamma + \varepsilon^2 + (4 \pi \Gamma)^2 } \right]^{1/ 2}  
\!\!\!\!\! ,
\quad \mbox{[sign} (0 )  \equiv i \mbox{]} \qquad \quad \phantom{.}
\\
  \label{eq:H+diagonalmain}
   H'_+    \!\! &=& \!\!   \sum_{\varepsilon > 0} \varepsilon  \, 
         \left( \tilde \alpha^\dagger_{\varepsilon} \tilde \alpha_{\varepsilon}
           - \half \right)  \; + \; 
          \sum_{\bk > 0}  \varepsilon_\bk   \,
          \left( \beta^\dagger_{\bar k} \beta_{\bar k} 
           + \half \right)  \; ; 
\\
  \label{eq:eigenvaluesmain}
  {\varepsilon \over 4 \Gamma}  \!\! &=& \!\! 
 \Delta_L \sum_{\bk = - \infty}^\infty 
 {1 \over \varepsilon -\varepsilon_\bk}    \; ;  \\
& & 
\label{newexpt} \hspace{-1.5cm}
\langle G'_B| \beta_{\bk} \beta^\dagger _{\bk'} |G'_B \rangle = 
\delta_{\bk \bk'} \theta (\varepsilon_\bk)  , \qquad
\langle G'_B| \tilde \alpha_\varepsilon
\tilde \alpha_{\varepsilon'}^\dagger |G'_B \rangle = 
\delta_{\varepsilon \varepsilon'} \theta (\varepsilon)  , 
\qquad (\mbox{with} \; \theta (0) \equiv \half) \; .
\qquad \quad \phantom{.}
\end{eqnarray}
\Eqs{oldck} and (\ref{alpha-tildealpha})
 express $c_\bk$ and $\alpha_d$ in terms of 
two sets of fermions, $\{ \beta_{\bar k} \}$ and 
$\{ \tilde \alpha_\varepsilon \}$, that diagonalize $H_+'$
[cf.\ (\ref{eq:H+diagonalmain})]. The sums
 $\sum_{\varepsilon}$ in  (\ref{alpha-tildealpha})
run over all real solutions $\varepsilon$ of
the eigenvalue equation (\ref{eq:eigenvaluesmain}).
Analyzing  (\ref{eq:eigenvaluesmain})
 graphically (cf. \cite{ZvD}) shows that 
each $\varepsilon$ lies within $\Delta_L/2$ of 
some $\varepsilon_{\bar k}$, to which it reduces as $\Gamma \to 0$,
with the exception
of one solution, namely $\varepsilon = 0$. The latter is
associated with the Majorana fermion 
$\tilde \alpha_0 = \tilde \alpha_0^\dagger$, which reduces
to $\alpha_d$ as $\Gamma \to 0$ (and whose contribution
to $c_{\bar k}$ is negligible for $L \to \infty$).
For each $\varepsilon > 0$ that solves (\ref{eq:eigenvaluesmain}), 
$-\varepsilon$  does too;  however, by  (\ref{betak-betadagger}) and 
(\ref{e-enotindependent}) the negative-energy operators
$\beta^\dagger_{- |\bar k|}$ and $\tilde \alpha^\dagger_{- |\varepsilon|}$
are not independent, but should be viewed as 
shorthand (making some equations more compact)
for $\beta_{ |\bar k|}$ 
and $\tilde \alpha_{ |\varepsilon|}$. The latter  annihilate
the ground state $|G_+' \rangle$ of $H_+'$ [cf.\ (\ref{newexpt})].

Correlation functions with respect to $H'_+$ and $|G'_B \rangle$,
which we denote by $\langle \quad \rangle'$, i.e.\ 
\begin{equation}
  \label{eq:HGprime}
  \langle O_1 (t) O_2 (0) \rangle' \equiv
  \langle G'_B | e^{i H'_+ t} O_1  e^{-i H'_+ t} O_2 | G'_B \rangle \; ,
\end{equation}
are straightforward to calculate using the above results, 
provided that $O_1$ and 
$O_2$ can be expressed in terms of the $c_\bk$'s and $\alpha_d$.
In the process it is often convenient to take 
the continuum limit $L \to \infty$, in which 
the spectrum of $\varepsilon$'s
and $\varepsilon_{\bar k}$'s becomes continuous:
\begin{equation}
  \label{eq:continuumlimit}
 \Delta_L  \sum_{\bk} 
\stackrel{L\to \infty}{\longrightarrow}
\int 
\! \! d \varepsilon_\bk , \qquad
 \Delta_L  \sum_{\varepsilon }
\stackrel{L\to \infty}{\longrightarrow}
\int 
\! \! d \varepsilon  , 
\qquad {1 \over \varepsilon - \varepsilon_\bk} 
\stackrel{L\to \infty}{\longrightarrow} P
{1 \over \varepsilon - \varepsilon_\bk}  + {\varepsilon \over 4
 \Gamma } \delta(
\varepsilon - \varepsilon_\bk) \; .
\qquad \phantom{.}
\end{equation}
In the third relation,  $P$ denotes principle value
and the $\delta$-function  is needed to ensure consistency
with  \Eq{eq:eigenvaluesmain}. 
Where necessary,
divergent integrals will be regularized  by 
inserting a factor $e^{-|\bar k| a} = e^{- |\varepsilon_{\bar k}| a/v}$
or $e^{-|\varepsilon | a/v}$ 
[just as one does 
for free Green's functions, cf.\ Appendix~\ref{sec:fermT=0Lneqinf}].
For example (see also Appendix~\ref{asymptotic})
\begin{eqnarray}
  \label{eq:betacorrelator}
  D_{\beta} (t) &\equiv &\Delta_L \sum_{\bk \bk'} \langle \beta_\bk (t)
  \beta_{\bk'}^\dag (0) \rangle' \, 
\stackrel{L\to \infty}{\longrightarrow} \, 
\int_0^\infty \! \!
d \varepsilon_\bk \, e^{-\bk (it +a/v)}  = {1 \over i t + a/v} \; ;
\\
  \label{eq:alphalphadref1}
D_{\alpha_d} (t) & \equiv & 
 \langle \alpha_d (t) \alpha_d (0) \rangle'  =
 \sum_{\varepsilon \varepsilon'} e^{-i \varepsilon  t}  
 A_{d, \varepsilon} A^\ast_{d,  \varepsilon'}
 \langle \alpha_\varepsilon \alpha^\dag_{\varepsilon'}  \rangle'  
 \\
  \label{eq:alphaalphadref4}
& \stackrel{L\to \infty}{\longrightarrow}  &
\!\! \int_0^\infty \! d \varepsilon \,
{ e^{-i \varepsilon  t} \, 4 \Gamma \over 
 \varepsilon^2 + (4 \pi
    \Gamma)^2 } \; 
= \; 
  \left\{ \begin{array}{ll}
 {1 \over 2} & 
\quad (t = 0)   \vspace{1mm} , 
\\
  \label{eq:alphaalphadref5}
 {1  \over 4 \pi^2  \Gamma i t} 
 \left[ 1 + 
{\cal O} \left( { 1 \over \Gamma t}  \right) \right] & 
\quad ( \Gamma t \gg 1)  .
\qquad \phantom{.}
\end{array} \right. 
\end{eqnarray}
We used  (\ref{alpha-tildealpha}) and (\ref{finalAen}) for $\alpha_d $
in (\ref{eq:alphalphadref1}), and (\ref{eq:continuumlimit})
when taking  the limit $L\to \infty$. 
The asymptotic  $\Gamma t \gg 1$ behavior of 
(\ref{eq:alphaalphadref5})   was obtained using 
the general result
\begin{eqnarray}
 \label{eq:asymtotics}
\int_0^\infty \!\!  d \varepsilon
\, {e^{- \varepsilon (i t + a)} \, 
\varepsilon^n \over (\varepsilon^2 + c^2)^{m}
(\varepsilon + \bar c)^{\bar m} } 
\;\sim \; {n! \over  c^{2m} \, \bar  c^{\bar m} \, (it)^{1 + n}} 
\;  \qquad \mbox{for} \quad  c t, \bar c  t \gg 1 ,
\phantom{.}
\end{eqnarray}
(with $n , m, \bar m \ge 0$ integer, 
$c, \bar c >0$ real), which follows 
by noting that for $c t, \bar c  t \gg 1 $
the integrals are dominated by the regime $\varepsilon \ll c, \bar c$,
in which $(\varepsilon^2 + c^2)^m (\varepsilon + \bar c)^{\bar m}
\simeq c^{2m} \bar c^{\bar m}$.

\subsubsection{Bosonic correlation functions at $g = 1/2$}
\label{bosonCorrelationfunctions}

\noindent
{\em We express $\Phi_+ (t,x)$ 
in terms of the fermions $ \beta_{\bar k}$ and $\tilde \alpha_\varepsilon$,
which enables us to express  {\bf arbitrary}\/ bosonic correlation functions 
in terms of fermionic ones. We then show explicitly that
for $\Gamma t \gg 1$,  
$\langle \Phi_+ (t,0) \Phi_+ (0,0) \rangle' \sim t^{-2}$ and 
$\langle e^{i \lambda \Phi_+ (t,0)}
e^{-i \lambda \Phi_+ (0,0)} \rangle' \sim \mbox{const}$.
}
\vspace{2mm}

The key to this endeavour  is that the Fourier coefficients
$B_{q+}$ of $\Phi_+(t,x)$ in (\ref{pmnewbosonizewe})
have a refermionized representation, namely (\ref{eq:BBreferemionize}),
a fact that has to our knowledge not been exploited before.
Using  (\ref{oldad}) and (\ref{betak-betadagger}), $B_{q+}$ 
 can be rewritten as 
\begin{equation}
  \label{eq:Bqbetaalpha}
  B_{q+} = {\textstyle {- i \over \sqrt n_q}} 
          \sum_{\bar k} ( \alpha_{-\bar k + q}  -i \beta_{- \bar k + q} ) 
                 (\alpha_{\bar k} + i \beta_{\bar k} )
          = - {\textstyle { 1 \over \sqrt n_q}} 
          \sum_{\bar k}\beta_{\bar k} \alpha_{-\bar k + q} \; .
\end{equation}
(Since $\sum_\bk = \sum_{\bk - q}$, we have
$
 \sum_{\bar k}  \alpha_{-\bar k + q} \alpha_{\bar k}
=  {1 \over 2} \sum_{\bar k}  \{ \alpha_{-\bar k + q} , \alpha_{\bar k} \}
= 0$, etc.) 
Inserting (\ref{eq:Bqbetaalpha}) into (\ref{pmnewbosonizewe})
for $\Phi_+ (t,x)$,  (\ref{alpha-tildealpha})
for $\alpha_{- \bar k + q}$,  and (\ref{finalAen})
 for $A_{- \bar k + q, \varepsilon}$
yields 
\begin{eqnarray}
  \label{eq:phi-t-x-alpha-beta}
  \Phi_+ (t,x) \!\! & =&  \!\!  \sum_{q \neq 0}
  {e^{- a |q|/2} e^{-i q x}  \over n_q} 
  \sum_{\bar k} \beta_{\bar k} (t)  \alpha_{ - \bar k + q} (t)
  = \sum_{\bar k \varepsilon} \Phi_{\bar k ,\varepsilon} (x)
    \, \beta_{\bar k} (t)  \tilde \alpha_{ \varepsilon} (t) \, ,
\qquad \phantom{.} 
\\
\label{Phikex}
 \Phi_{\bar k ,\varepsilon} (x) \!\! & \equiv & \!\! 
  \sum_{q \neq 0}
  {e^{- a |q|/2}e^{ -i q x}  \over n_q}
 A_{- \bar k + q, \varepsilon} \; 
= { \mbox{sgn} ( \varepsilon)\, 4 \Delta_L \Gamma 
\over [ 4 \Delta_L \Gamma + \varepsilon^2 + (4 \pi \Gamma)^2 ]^{1/2} }
\sum_{q \neq 0}
  { e ^{- a |q|/2} e^{ -i q x}  \over 
n_q ( \varepsilon + \varepsilon_{\bar k} - \varepsilon_q)} \;  .
\qquad \phantom{.}
\end{eqnarray}
Here  $\varepsilon_q = \Delta_L n_q$, and 
$\sum_{q \neq 0}$ means a sum over all $n_q \in \ZZ$ (positive
and negative) except $q = 0$. 
Using (\ref{eq:continuumlimit}) to perform this sum in
the continuum limit, we obtain for  $x=0$:  
\begin{eqnarray}
  \label{eq:doqsum}
 \sum_{q \neq 0} 
{e^{-a |q|/2} \over n_q( \varepsilon + \varepsilon_{\bar k} - \varepsilon_q)}
\! & = & \!   {\Delta_L \over \varepsilon + \varepsilon_{\bar k}} 
\sum_{q \neq 0}  \Bigl[   {e^{-a |\varepsilon_q|/2v}
 \over \varepsilon + \varepsilon_{\bar k} - 
\varepsilon_q}
- {e^{-a |\varepsilon_q|/2v} \over \varepsilon_q} \Bigr] 
\label{eq:doqsum-2}
\stackrel{L \to \infty}{\longrightarrow}
 { \varepsilon \, e^{- | \varepsilon + \varepsilon_{\bar k}|a/2v} 
 \over 4 \Gamma (\varepsilon + \varepsilon_{\bar k})} 
\qquad \phantom{.}
\end{eqnarray}
(we kept only terms that do not vanish when  $a \to 0$).
Thus (\ref{Phikex}) yields [again using (\ref{eq:continuumlimit})]:
\begin{equation}
  \label{eq:phiek(0)}
 \Phi_{\bar k ,\varepsilon} \equiv
 \Phi_{\bar k ,\varepsilon} (0)  \stackrel{L \to \infty}{\longrightarrow}
{ \Delta_L | \varepsilon| \over
[ \varepsilon^2 + ( 4 \pi \Gamma)^2 ]^{1/2} }
 \left[e^{- | \varepsilon + \varepsilon_{\bar k}|a/2 v} 
 {P \over \varepsilon + \varepsilon_{\bar k} } + 
{ \varepsilon \over 4 \Gamma} \delta (\varepsilon + \varepsilon_{\bar k}) 
\right] \; .
\end{equation}
Eqs.~(\ref{eq:phi-t-x-alpha-beta}) and (\ref{eq:phiek(0)}) for
$\Phi_+ (t,0)$ and 
$\Phi_{\bar k \varepsilon}(0)$ instructively show that the infrared
divergence inherent in a free boson field
(due to the $1/q$ in $\Phi_+ (0) =  \sum_{q \neq 0} {i \over q}
\sum_k c^\dagger_{\bar k -q } c_{\bar k}$)
is cut off by backscattering at a scale $\Gamma$. 
This has dramatic consequences for the 2-point correlator:
\begin{eqnarray}
  \label{eq:phi(t)phi(0)}
\lefteqn{
D_{\Phi_+} (t) \equiv  \langle \Phi_+ (t,0) \Phi_+ (0,0) \rangle' }
\\
&=&
\sum_{\bar k \bar k' \varepsilon \varepsilon'} 
\Phi_{\bar k ,\varepsilon} \Phi^\ast_{\bar k', \varepsilon'}
\langle \beta_{\bar k} (t) \tilde \alpha_\varepsilon (t)
\tilde \alpha_{\varepsilon'}^\dagger (0) \beta_{\bar k'}^\dagger (0) \rangle'
\, = \, 
\sum_{\bar k> 0} \sum_{ \varepsilon \ge 0} 
e^{-i (\varepsilon_{\bar k} + \varepsilon)t} \theta (\varepsilon)
| \Phi_{\bar k ,\varepsilon}|^2 
\\
&=& 
P \!\!\!\!\!\!
 \int_0^\infty \!\!\!  d      \varepsilon  \:
d \varepsilon_{\bar k}   \,
{ e^{- (\varepsilon_{\bar k} + \varepsilon)(it + a/v)}
  \, \varepsilon^2 \over
  [\varepsilon^2 + (4 \pi \Gamma)^2 ] 
(\varepsilon + \varepsilon_{\bar k})^2} 
\label{scatteredbosoncor}
 =    \left\{ \begin{array}{l}
 -   \ln\left( e^\gamma 4 \pi \Gamma a /v\right) 
\quad \mbox{for} \; t = 0 , \;  \Gamma a \ll 1 
\vspace{1mm} ; \qquad \phantom{.}
\\
   {1 \over   (4 \pi \Gamma i t)^2} \left[ 1 + {\cal O} 
\left( {1 \over \Gamma t } \right) \right] 
\quad \mbox{for} \; \Gamma t \gg 1 \, .
\end{array} \right. 
\end{eqnarray}
[The  $\Gamma t \gg 1$ result was obtained 
by doing first the $\varepsilon_{\bar k}$, then
the $\varepsilon$ integral, using Eq.~(\ref{eq:asymtotics}).]
Compare the results (\ref{scatteredbosoncor})
with (\ref{bbGT=0})  for a free boson correlator, 
namely $- \ln [{ 2 \pi \over L} (it + a)]$:
For $t = 0$, the infrared divergence is now cut off
by $\Gamma$  instead of $1/L$ ($\gamma = 0.577\dots$ is Euler's
constant); and for $\Gamma t \gg 1$ 
the $t^{-2}$ decay of the correlator 
is much faster than the free logarithmic behavior. 
This strong suppression of  the long-time fluctuations of $\Phi_+(t,x=0)$
has been paraphrased \cite{OF,FG} by saying that {\em at $x = 0$ the
backscattering impurity ``pins'' the field $\Phi_+ (t,0)$ to its
average value $\langle \Phi_+ (t, 0) \rangle'$, the fluctuations
around which are ``massive''\footnote{
By Eq.~(\ref{eq:phi-t-x-alpha-beta}),
this average
value is $ \langle \Phi_+ (t,0) \rangle' = 0$, but in general it 
depends on one's choice of gauge: if we had not 
``gauged away'' the phase factor $e^{i \theta_B}$ in
$H_+$ of (\ref{H_B}) by including the factor $e^{- i \theta_B {\cal N}_+}$
in the unitary transformation $U_+$ of (\ref{eq:U+definition}),
we would have obtained $\langle \Phi_+ \rangle' = \theta_B$ here \cite{FG}.
The fluctuations are called ``massive'' because 
similar behavior ($ \langle \Phi_+ (t) \Phi_+ (0) \rangle \sim t^{-2}$)
occurs in models involving only bosonic fields, but with a mass term,
e.g.\ with Hamiltonian $H_M = 
         \int_{-L/2}^{L/2} {\textstyle {dx \over 2 \pi}}
        {\textstyle {1 \over 2}}
         {\,}^\ast_\ast ( \partial_x \Phi_+ (x) )^2  {\,}^\ast_\ast
         \; + \; \half M [\Phi_+ (x=0)]^2 $.
In fact, such effective models can be used to approximately treat
the general case $g \neq \half$, which cannot be refermionized 
\protect\cite{FGGGG,Furusaki}.}}.\/
In more physical terms, the {\em current fluctuations}\/ at the impurity site,
which are governed by $\partial_x \Phi_+ (x)|_{x=0}$,
are suppressed by the backscattering impurity.
This  eminently plausible result
was first found by Kane and Fisher \cite{KF},
who showed via an RG analysis that the
conductance past such an impurity is 0 at $T=0$
whenever $g < 1$. Note, however, that
density fluctuations, governed by $\partial_x \Phi_- (x)|_{x=0}$,
are {\em not}\/ suppressed by backscattering (since $[H_B, \Phi_-] = 0$).

The pinning of $\Phi_+$ has important consequences; for example,
it immediately implies that the correlator
of two vertex functions is asymptotically constant:
\begin{eqnarray}
  \label{eq:vertexcorrel}
D_{V_\lambda} (t) & \equiv & a^{- \lambda^2}
  \langle e^{i \lambda \Phi_+(t,0) } e^{-i \lambda \Phi_+(0,0)} \rangle'
\\
  \label{eq:vertexcorrel-expand}
&=& a^{- \lambda^2}
\sum_{n = 0}^\infty \sum_{n' = 0}^\infty { (i \lambda)^n
(-i \lambda)^{n'} \over n! \, n'!}
\langle \Phi_+^n (t,0)\, \Phi_+^{n'} (0,0) \rangle' 
\\
 \label{eq:vertexcorrel-const} 
&=& a^{- \lambda^2} \langle e^{i \lambda \Phi_+(t,0) } \rangle' \langle
e^{-i \lambda \Phi_+(0,0)} \rangle' + {\cal O}  (\Gamma t)^{-2} \; 
\qquad \mbox{for} \quad \Gamma t \gg 1 \; . 
\end{eqnarray}
To obtain  (\ref{eq:vertexcorrel-const}),  
we invoked the fermionic expression
(\ref{eq:phi-t-x-alpha-beta})  for $\Phi_+$ 
and Wick's theorem for fermions to
envisage each  correlator in (\ref{eq:vertexcorrel-expand})
as a sum of products of contractions
of the kind $\langle \tilde \alpha \tilde \alpha^\dagger \rangle'$ and
$\langle \beta \beta^\dagger \rangle'$. All 
terms involving no contractions at all
between an operator at time $t$ and one at 0, to be called ``disconnected'',
can be reorganized to yield the first term of  (\ref{eq:vertexcorrel-const}),
a $t$-independent constant;
all other, ``connected'', terms  decay at least as $t^{-2}$,
since $D_{\Phi_+}(t)$ of (\ref{eq:phi(t)phi(0)}) is the leading
such term. Note that the value of the leading constant in
(\ref{eq:vertexcorrel-const}) is {\em not}\/ simply equal to 
$a^{- \lambda^2} e^{- \lambda^2 \langle \Phi_+(0,0)^2 \rangle' } = 
(e^\gamma 4 \pi \Gamma /v )^{\lambda^2}$
[by (\ref{scatteredbosoncor})], since
the identity (\ref{Vexp}) which would yield this result
holds only for {\em free}\/ boson fields
(cf.\ the end of Appendix \ref{sec:NNcor}).

\subsection{Tunneling density of states at the impurity site}
\label{dos}
\label{controversy}

\noindent
{\em For $g = \half$, we calculate  the low-energy ($\omega \to 0$) asymptotic 
behavior of the tunneling density of states at the impurity site,
$\rho_{dos} (\omega)  \sim \omega^{\nu - 1}$,
following Furusaki. Our result $\nu = 2$ 
 resolves the controversy
between Fabrizio \& Gogolin and Furusaki vs.\ and Oreg \& Finkel'stein
in favor of the former authors.
\vspace*{2mm}}

The tunneling electron density
of states at the impurity site,  $ \rho_{dos} (\omega)$,
 is defined by 
\begin{eqnarray}
\label{eq:dosDt}
  \rho_{dos} (\omega) &\equiv & \int_{- \infty}^\infty
  {\textstyle {d t \over 2 \pi }} \, e^{i \omega t} 
\langle G| \Psi_{phys} (t)  \Psi_{phys}^\dagger (0) +
 \Psi^\dagger_{phys} (0)  \Psi_{phys}(t) |G \rangle \; , 
\end{eqnarray}
where $\Psi_{phys} (t) \equiv e^{i H t} \Psi_{phys}(x=0) e^{-i H t}$
and $|G \rangle$ is the ground state of $H = H_0 + H_F + H_B$.
Both terms in (\ref{eq:dosDt}) are real, and 
$\rho_{dos} (\omega) = \rho_{dos} ( -\omega)$ by particle-hole
symmetry ($c_{k, L/R} \to c_{-k, L/R}^\dagger$
maps $H (\lambda_B)$ to $ H(- \lambda_B)$, both of which have 
the same spectrum). Since at $T=0$ the second term in (\ref{eq:dosDt}) does
not contribute to the $\omega>0$ part of $\rho_{dos} (\omega)$,
its asymptotic $\omega \to 0^+$ behavior is 
%
determined by the
asymptotic $t \to \infty$ behavior of 
\begin{eqnarray}
\label{eq:Dtt}
 D_{phys} (t) &  \equiv & \langle G|
 \Psi_{phys}(t) \Psi_{phys}^\dagger (0) |G \rangle 
\;  \sim \; (it)^{-\nu} \quad \mbox{for} \; t \to \infty \; , 
\end{eqnarray}
which implies  $\rho_{dos} (\omega) \sim 
\omega^{\nu - 1}$ for $ \omega \to 0^+$.
 We shall throughout take the continuum limit $L \to \infty$
and neglect all $1/L$ terms.  
The reason why $D_{phys} (t) $ asymptotically does not contain
a fluctuating factor $ e^{-i \Delta t}$ for $t \to \infty$ is
that $H$ has gapless excitations, implying that $\Delta$ must be zero
\cite{Furusaki}. Our goal is to calculate the exponent $\nu$, 
which has recently been subject to quite some controversy, as
mentioned in the introduction to Section~\ref{TLL}.

We start by bosonizing the physical fermion field $\Psi_{phys}$
occuring in (\ref{eq:Dtt}), using 
(\ref{oldpsi}),  (\ref{nubosonization}) and (\ref{eq:phizero}):
\begin{eqnarray}
\nonumber
  \Psi_{phys} (x=0) &=&  \tilde\psi_L(x=0) + \tilde\psi_R(x=0)  = 
\nonumber
a^{-1} \! \left(
 F_L e^{-i \tilde \phi_L} 
    \!+\!  F_R e^{-i \tilde\phi_R} \right) \\
&=&
a^{-1}  e^{- {i\over \sqrt{2g}} \Phi_-}
 \! \left(
 F_L e^{- i \sqrt {g \over 2} \Phi_+}
    \!+\! F_R e^{i  \sqrt {g \over 2} \Phi_+} \right) .
\end{eqnarray}
Since $H = H_+ + H_-$ 
and  $[H_-, H_+ ] = 0$ [see Eq.~(\ref{eq:finalfullH})],  $D_{phys}(t)$ can be 
factorized as $  D_{phys} (t)  \equiv D_F (t) D_B(t)$, where
\begin{eqnarray}
  \label{eq:factorDF}
 D_F(t) &\equiv& \langle G_F | 
 e^{i H_- t } \left(  e^{ -{i\over \sqrt{2g}} \Phi_-} \right)
 e^{-i H_- t} \left( e^{ {i\over \sqrt{2g}} \Phi_-} \right)
e^{i \widehat E t} |G_F \rangle
 \; , \\
\nonumber
 D_B(t) &\equiv& a^{-1} \langle G_B | 
 e^{i H_+ t }  \! \left(F_L e^{- i\sqrt {g \over 2} \Phi_+}
+ F_R  e^{  i\sqrt {g \over 2} \Phi_+} \right) \! 
 e^{-i H_+ t}   \left( F_L^\dagger e^{ i\sqrt {g \over 2} \Phi_+}
+ F_R^\dagger  e^{- i\sqrt {g \over 2} \Phi_+} \right) \! 
|G_B \rangle \; 
\\
\label{DLLLRRR}
 & \equiv & D_{LL} (t) + D_{RR} (t) + D_{LR} (t) + D_{RL} (t) \; .
\end{eqnarray}
Here  $|G_F \rangle$ and $|G_B \rangle$ are the ground states
of $H_-$ and $H_+$, respectively; $\widehat E$ is defined by 
$e^{i H_- t } F_{L/R}  e^{-i H_- t } \equiv
e^{i \widehat E t}$ and, being of order $\tpol$, will be neglected
henceforth.
 
\subsubsection{Free tunneling density of states}
\label{freeDOS}

In the absence of an impurity the calculation of $D_F (t)$ and
$D_B(t)$ is straightforward, since $H_{0\pm}$ are free boson Hamiltonians, so
that the free-boson relations (\ref{bbG}) and (\ref{magicid})
[or equivalently \Eq{VV2}] can be used:
\begin{eqnarray}
  \label{DFevalAfree}
D_F (t) &=& \langle G_F | e^{i H_{-}t} 
\! \left( e^{- {i \over \sqrt{2 g}} \Phi_- (0)} \right) \! 
e^{-i H_{-}t} \! \left( e^{{i \over \sqrt{2 g}} \Phi_- (0)} \right)
| G_F\rangle \\
&\simeq & e^{{1\over 2g} \langle G_F | \Phi_- (t) \Phi_- (0) \, - \, 
\Phi_- (0) \Phi_- (0) | G_F \rangle} 
  \label{DFevalAfree2}
\; = \; (1 + i v t/a)^{-{1 \over 2g}} 
\\
 \label{DGevalAfree}
D_B (t) &=& a^{-1} \langle G_B | e^{i H_{+}t} 
\! \left( e^{-{i  \sqrt{g \over 2 }} \Phi_+ (0)} \right) \! 
e^{-i H_{+}t} \! \left( e^{{i \sqrt{ g \over 2 }} \Phi_+ (0)} \right)
| G_B\rangle  + \mbox{h.c.} \\
&\simeq & a^{-1} e^{{g \over 2 } \langle G_B | \Phi_+ (t) \Phi_+ (0) \, - \, 
\Phi_+ (0) \Phi_+ (0) | G_B \rangle}  + \mbox{c.c.} 
  \label{DBBBeval}
\; = \; a^{-1} (1 + i v t/a)^{-{g \over 2}} + \mbox{c.c.} 
\qquad \phantom{.}
\end{eqnarray}
It follows that $\nu = \half (g + {1 \over g})$. For the free-fermion
case $g = 1$ we have $\nu = 1$, i.e.\ for $\omega \to 0$ we recover
the standard ``Fermi-liquid'' property $\rho_{dos} (\omega = 0) \neq 0$. 
However, for any $g \neq 1$ we have
$\nu > 1$, i.e.\ $\rho_{dos} (\omega) \to 0 $ for
$\omega \to 0$. Thus, the {\em interactions in 1-D
cause the density of states to  vanish at the
Fermi energy}\/. This property is one of the most spectacular differences
between a Tomonaga-Luttinger liquid (without impurities) and a Fermi
liquid. 

\subsubsection{Effect of an impurity on $\rho_{dos} (\omega)$}
\label{effectimpurity}

Let us now consider the effect of turning on $H_F$ and $H_B$. The calculation
of $D_F(t)$ is again simple, since we can use the 
unitary transformation $U_- = e^{-i c_- \Phi_-}$ of  
Section~\ref{sec:addimpurity} to map $H_-$ onto a free Hamiltonian
 $H'_- \equiv U_- H_- U_-^{-1}$, see \Eq{UUU-}. 
Denoting the corresponding ground state
by $|G'_F \rangle = U_- |G_F \rangle$, the function
$D_F(t)$ can  be evaluated by first making this transformation
 in \Eq{DFevalAfree}, and proceeding as before:
\begin{eqnarray}
  \label{DFevalA}
D_F (t) &=& \langle G_F' | e^{i H'_{-}t} 
\! \left( e^{-{i \over \sqrt{2 g}} \Phi_- (0)} \right) \! 
e^{-i H'_{-}t} \! \left( e^{{i \over \sqrt{2 g}} \Phi_- (0)} \right)
| G_F'\rangle \\
&\simeq & e^{{1\over 2g} \langle G_F' | \Phi_- (t) \Phi_- (0) \, - \, 
\Phi_- (0) \Phi_- (0) | G_F' \rangle} 
  \label{DFeval}
\; = \; (1 + i v t/a)^{-{1 \over 2g}} 
\end{eqnarray}
Now  $\Phi_- (t) \equiv \Phi_- (t, x=0)$ denotes time-development
w.r.t.\ $H'_{-}$.

To calculate $D_B(t)$, we restrict ourselves to the case $g=\half$
for which we have refermionized  $H_+$ above. 
First, using a trick due to 
Furusaki\footnote{\label{f:Furu}
Since Furusaki uses {\em field-theoretical}\/ bosonization, 
he somewhat nonrigorously treats
Klein factors (which he denotes by $\eta$) as though they were Majorana
fermions (which they are not, since $F^2 \neq 1$), 
and hence uses the same
$\eta$ for what here are three distinct operators, $F_+$, $F_+^\dagger$ and
$\alpha_d$.  Moreover, he uses a
 different argument [when discussing {\em his}\/ Eqs.~(25-28)] to
evaluate $D_{LL/RR} (t)$ than our
Eqs.~(\ref{eq:LL-RRcorrelation1}-\ref{alphavertex-result}): He points out that
in \Eq{eq:LL-RRcorrelation2} one can use (\ref{eq:UHU-1}) to write $e^{\pm {i
    \over 2} \Phi_+} e^{- i \bar H'_+ t} e^{\mp {i \over 2} \Phi_+} = e^{- i (
  \bar H'_+ \mp {v \over 2} \partial_x \Phi_+ + {v \over 4 a})t}$, then argues
that $\partial_x \Phi_+ \equiv \partial_x \Phi_+ (x=0)$ is an irrelevant
operator (since, as can be readily confirmed using our methods,
 $\langle \partial_x \Phi_+ (t) \partial_x \Phi_+(0) \rangle
\sim t^{-4}$), which hence 
does not affect the asymptotic behavior
of $D_{LL/RR}$. This argument leads to $D_{LL/RR} (t) \sim \langle G^\prime_B
| e^{i H'_+ t} e^{-i ( \bar H'_+ + {v \over 4a} )t} | G^\prime_+ \rangle \sim
t^{-1} e^{-i \Delta t}$, which produces the desired asymptotic $t^{-1}$ decay.
However, it also produces an oscillatory
factor $\Delta = {v \over 4 a}$, which 
Furusaki seems to have overlooked but which cannot be correct:
since $H$ is gapless, we know that $\Delta$ must be zero, as Furusaki points
out himself earlier in his paper [after his Eq.~(11)].  Thus, his
argumentation  subtly contradicts itself --- the resolution is
probably \cite{Fabrizio:priv} that infinite-order 
perturbation theory in
the irrelevant operator $\partial_x \Phi_+ $ will produce an oscillatory
factor $e^{i \Delta t}$ that exactly cancels the above $e^{-i \Delta t}$
(without affecting the asymptotic $t^{-1}$ decay), but showing this explicitly
seems like a rather non-trivial task.}
 \cite{Furusaki} and Fabrizio and Gogolin
\cite{FG}, we note that the following relations hold: 
\begin{eqnarray}
  \label{eq:aHa}
  2 \, \alpha_d H'_+ (\lambda_B) \alpha_d &= & H'_+ (- \lambda_B) \equiv \bar
  H'_+ (\lambda_B)  \; , \\
  \label{eq:FHF}
  F_+ H'_+ (\lambda_B) F_+^\dagger &=& \bar
  H'_+  (\lambda_B) +  v \tpol ( \widehat {\cal N} + \half P + \half )\; ,
\\
   \label{eq:FLRHFLR}
  F_{L/R} H_+ (\lambda_B) F_{L/R}^\dagger
 &=& H_+ (- \lambda_B) +  O(\tpol)
  \equiv \bar   H_+ (\lambda) , 
\end{eqnarray}
The first two follow from (\ref{eq:h+prime})
for $H_B'$ and (\ref{eq:alphad}), and 
the third from using $\{F_L, F_R \} = 0$, etc., in (\ref{H_B}) for $H_B$.
Thus, commuting
$\alpha_d$ or  $F_+$  past $H'_+ (\lambda_B)$
yields a similar Hamiltonian
 with backscattering term of 
opposite sign, $\bar   H'_+ (\lambda_B) \equiv  H'_+ (- \lambda_B) $,
and similarly for commuting $F_{L/R}$ past $H_+ (\lambda_B)$.
The extra  ${\cal O}(\tpol)$ 
term in (\ref{eq:FHF}) results from the $\half \hat {\cal N} 
(\hat {\cal N} + P)$ in $H'_{0+} = H_{0+}$
[see (\ref{oldHphi})] and can be neglected in the continuum limit,
and similarly for that in (\ref{eq:FLRHFLR}).

To calculate the $LL$ and $RR$ contributions 
in \Eq{DLLLRRR}, we now proceed as follows:
\begin{eqnarray}
  \label{eq:LL-RRcorrelation1}
  D_{LL/RR} (t) &\equiv &
a^{-1} \langle G_B | e^{i H_+ t} 
\left( F_{L/R} e^{\mp {i\over 2} \Phi_+} \right)
 e^{-i H_+ t} 
\left(F_{L/R}^\dagger e^{ \pm {i \over 2} \Phi_+ } \right)
| G_B \rangle \\
  \label{eq:LL-RRcorrelation2}
 &=  &  a^{-1} \langle G'_B | e^{i H'_+ t} e^{\mp {i \over 2 }\Phi_+ } 
 e^{-i \bar H'_+ t}  e^{ \pm {i \over 2} \Phi_+ } | G'_B \rangle \\
  \label{eq:LL-RRcorrelation2a}
& \sim & 
 2 a^{-1} \langle  e^{i H'_+ t} \alpha_d 
 e^{\mp {i \over 2 }\Phi_+ } 
 e^{-i  H'_+ t}  e^{ \pm {i \over 2} \Phi_+ }  \alpha_d  \rangle' 
\equiv 2 a^{-3/4}  D_{\alpha_d V_{\mp 1/2}} (t) \; .
\end{eqnarray}
To obtain (\ref{eq:LL-RRcorrelation2}), we first
commuted $F^\dagger_{L/R}$ to the front (changing $H_+$ into $\bar H_+$),
where it drops out via $F_{L/R} F_{L/R}^\dagger = 1$,
and then performed the unitary transformation $U_+ = 
e^{i ({ \pi \over 2} \widehat {{\cal N}}^2 
 -\theta_B \widehat {\cal N} )}  $ of \Eq{eq:U+definition} to change
$H_+ (\pm \lambda_B) $ to $H'_+ (\pm \lambda_B)$ 
and $|G_B \rangle$ to $|G'_B \rangle$. To obtain
(\ref{eq:LL-RRcorrelation2a}), we used (\ref{eq:aHa}) to change
$\bar H_+'$ back to $H_+'$. The resulting correlator, or
its generalization to arbitrary $\lambda$, can be 
asymptotically evaluated in a way analogous to $D_{V_\lambda}$
of (\ref{eq:vertexcorrel}):
\begin{eqnarray}
  \label{eq:alphaVlambda}
  D_{\alpha_d V_{\lambda}} (t) & \equiv &
  a^{-\lambda^2} \langle \alpha_d (t) 
 e^{i \lambda \Phi_+(t) } e^{-i \lambda \Phi_+(0)} 
\alpha_d (0) \rangle'
\\
  \label{eq:alphavertexcorrel-expand}
&=& a^{-\lambda^2} \sum_{n = 0}^\infty \sum_{n' = 0}^\infty { (i \lambda)^n
(-i \lambda)^{n'} \over n! \, n'!}
\langle \alpha_d (t) \, \Phi_+^n (t) \, \Phi_+^{n'} (0) \, 
 \alpha_d (0) \rangle' 
\\
 \label{eq:alphavertexcorrel-const} 
&=& a^{-\lambda^2} \langle \alpha_d (t) \alpha_d (0) 
\rangle' \langle  e^{i \lambda \Phi_+(t) } 
e^{-i \lambda \Phi_+(0)} \rangle' + 
\mbox{connected terms} 
\\
\label{alphavertex-result}
& = & {C_\lambda \over ( i t)} \left[ 1 + {\cal O} \left( 
{1 \over \Gamma t } \right) \right] 
\qquad \mbox{for} \quad \Gamma t \gg 1 \; . 
\end{eqnarray}
First note that the factors of $\alpha_d$ in  $D_{\alpha_d V_\lambda}
(t)$ guarantee that it can not 
approach a non-zero constant for $\Gamma t \to \infty$;
if it did, then  $\langle \alpha_d e^{i \lambda \Phi_+}\rangle'$ would 
itself have to be non-zero, which it trivially is not
(since $\langle \alpha_d \rangle' = 0$ and $[\alpha_d, \Phi_+] = 0$). 
That $D_{\alpha_d V_\lambda}(t) \sim t^{-1}$ follows
  from the observation that each
correlator in (\ref{eq:alphavertexcorrel-expand}), when expressed [via
(\ref{eq:phi-t-x-alpha-beta})] in terms of $\tilde \alpha$'s and $\beta$'s and
evaluated using Wick's theorem, contains {\em at least one}\/ contraction of
the kind $\Delta_L \sum_\bk C_\bk \langle \beta_{\bar k} (t)
\beta^\dagger_{\bar k} (0) \rangle'$ or $\Delta_L \sum_{\varepsilon}
C_\varepsilon \langle \tilde \alpha_\varepsilon (t) \tilde
\alpha^\dagger_\varepsilon (0)\rangle'$, i.e.\ between fermions at times $t$
and 0.  This necessarily gives rise to a factor of at least $t^{-1}$, 
as illustrated by (\ref{eq:betacorrelator}) for $D_{\beta} (t) \sim t^{-1}$ 
[where $C_\bk = 1$] or
(\ref{eq:alphaalphadref4}) for $D_{\alpha_d} (t) \sim t^{-1}$ 
[where $C_\varepsilon = 4 \Gamma / (\varepsilon^2 + (4 \pi
\Gamma)^2)$], or (\ref{eq:phi(t)phi(0)}) for 
$D_{\Phi_+} (t) \sim t^{-2}$ [which features a product of 
two such contractions, with $C_\bk C_\varepsilon =
\varepsilon^2 (
  [\varepsilon^2 + (4 \pi \Gamma)^2 ]
[\varepsilon + \varepsilon_{\bar k}]^2)^{-1} $].
(In general, the coefficients
$C_\bk$ and $C_\varepsilon$ that occur in the Wick expansion are
of the form occuring under the integral in (\ref{eq:asymtotics}), and hence
cause a $t^{-(1+n)}$ decay, with $n \ge 0$, see Appendix~\ref{connectedGnn}
for examples.)  

In (\ref{eq:alphavertexcorrel-const}), we gathered in the
first term all ``disconnected'' terms in which $\alpha_d(t)$ and $\alpha_d
(0)$ are contracted only with {\em each other}\/ (and not with any $\tilde
\alpha$'s from $\Phi_+$'s); that this contribution goes like $( i
t)^{-1}$ (as first pointed out by Fabrizio and Gogolin \cite{FG}) follows from
(\ref{eq:alphaalphadref4}) and (\ref{eq:vertexcorrel-const}). The remaining
``connected'' terms are all those in which $\alpha_d (t)$ and $\alpha_d (0)$
are contracted with some $\tilde \alpha$'s arising from the $\Phi_+$'s; of
these terms, the prefactors of those going like $\sim t^{-1}$ contribute to
the constant $C_\lambda$ in (\ref{alphavertex-result}), though most contain
more than one $t$-to-0 contractions and hence decay faster.  In
Appendix~\ref{app:connected} this is illustrated explicitly for several such
connected terms.
In Appendix~\ref{app:checkV=-1}
we also check the result $D_{\alpha_d V_\lambda}(t) \sim t^{-1}$ 
explicitly for the case $\lambda = - 1$, 
for which $D_{\alpha_d V_{-1}}(t)$
can be related to the correlator $D_\Psi (t)  \equiv
\langle \Psi_+(t) \Psi_+^\dag (0) \rangle' \sim (2v it)^{-1}$, which
we calculate exactly there.

From the result  $D_{\alpha_d V_\lambda}(t) \sim t^{-1}$
we conclude from  (\ref{eq:LL-RRcorrelation2a}) that also 
$D_{LL/RR} \sim t^{-1}$. 

The $D_{LR} (t) = D_{RL}^\ast (-t) $ 
 contributions to (\ref{DLLLRRR}) for $D_B (t)$ can
be evaluated analogously:
\begin{eqnarray}
  \label{eq:LR-RLcorrelation1}
D_{LR} (t) & \equiv & 
a^{-1} \langle G_B | e^{i H_+ t} \left( 
F_{L} e^{ - {i \over 2}  \Phi_+ } \right) 
 e^{-i H_+ t} 
\left( F_{R}^\dagger e^{- {i\over 2}  \Phi_+} \right)
| G_B \rangle \qquad \qquad \phantom{.}
\\
  \label{eq:LR-RLcorrelation2}
 &    = &  a^{-1} \langle G^\prime_B | e^{i H'_+ t} 
 \left(- F_+ i  \sqrt 2
\alpha_d e^{i \theta_B}  e^{- {i \over 2} \Phi_+ } \right)
 e^{-i \bar H'_+ t}  e^{ -{i \over 2 } \Phi_+ } | G^\prime_B \rangle \\
  \label{eq:LR-RLcorrelation3}
 &  = &   - \sqrt{ 2 / a}  \, 
 e^{i \theta_B} \langle e^{i H'_+ t}  
 \Psi_+  e^{{ i \over 2 } \Phi_+  }
 e^{-i H'_+ t}  e^{- { i \over 2 } \Phi_+  } i \alpha_d  \rangle' 
\, \sim ( i t )^{-1} . \qquad  \phantom{.}
  \label{eq:LR-RLcorrelation4}
  \label{eq:LR-RLcorrelation5}
\end{eqnarray}
For (\ref{eq:LR-RLcorrelation2}), we first commuted $F_R^\dagger$ to the front
(changing $H_+$ to $\bar H_+$), where it combines with $F_L$ 
to give $F_L F_R^\dagger = - F_+ $, and then
made the unitary transformation $U_+$, which
changes $F_+ $ to 
$ F_+ i \sqrt 2 \alpha_d e^{i \theta_B}$, see
(\ref{eq:U+definition}); 
for (\ref{eq:LR-RLcorrelation3}) we factorized $F_+ e^{-{i \over 2}
\Phi_+}$ as $\sqrt a \Psi_+ e^{{i \over 2}  \Phi_+} $,
and commuted $i \alpha_d$ to the right using (\ref{eq:aHa}).
The resulting correlator (\ref{eq:LR-RLcorrelation5})
decays as $t^{-1}$ for the same reasons as $ D_{\alpha_d V_{\lambda}}$
of (\ref{eq:alphaVlambda}); the leading term,
$\langle \Psi_+ (t) [\Phi(t) - \Phi_+ (0)] \alpha_d(0) \rangle
\sim t^{-1}$  is calculated  explicitly
in Appendix~\ref{app:leadingPsiPhialpha}. 
%

Putting everything together, we conclude that  for $g = \half$, 
$D_B(t) \sim t^{-1}$ and hence 
$  D_{phys} (t) = D_F(t) D_B (t) \sim t^{-2} $.
Thus, the sought-after exponent in (\ref{eq:Dtt}) is $\nu =
2$, hence (\ref{eq:dosDt}) implies $ \rho_{dos} (\omega) \sim \omega$.

\subsubsection{Discussion of the controversy regarding $\rho_{dos} (\omega)$.}
\label{discusscontrov}
\label{sec:OF}

Our result $\nu = 2$ agrees with 
that of Fabrizio and Gogolin (FG) \cite{FG} and Furusaki \cite{Furusaki},
who found  $\nu = {1 \over g}$ \cite{OF} for  general $g$,
 but not with that of Oreg and Finkel'stein (OF)
\cite{OF}, who found $\nu = {1 \over 2g}$, using
an exact  mapping to a Coulomb gas problem. 
In this mapping, the correlation function $D_B(t)$ of (\ref{DLLLRRR}) 
is represented as 
$D_B(t) = Z_e (t) - Z_o (t)$, where $Z_e$ and $Z_o$ 
are two Coulomb gas partition functions  whose asymptotic $t \to \infty$
behavior, $Z_{e,o} (t) \sim C_{e,o} + D_{e,o} / t^{\mu_{e,o}}$,
needs to be determined
($C_{e,o}$ , $D_{e,o}$ and $\mu_{e,o}>0$ are constants).
Using a seemingly rather natural mean-field approximation,
OF concluded that $C_e > C_o$, 
implying  that  $Z_e (t) - Z_o (t)$ 
asymptotically approaches a {\em constant}\/
$ \sim (C_e - C_o) \neq 0$].
This disagrees, however, with our exact  result that for  $g=1/2$, 
$D_B(t) $ asymptotically {\em decays}\/ as $1/t$
[see (\ref{eq:LL-RRcorrelation2a}), (\ref{alphavertex-result})].
Now,  Fabrizio and Gogolin (FG) were the first to point out
explicitly,
in a Comment \cite{FG} on OF's work, that this
decay   stems from a correlator of
Klein factors  namely 
 \begin{equation}
   \label{eq:OF-FGdispute}
\langle e^{i H_+ t} F_{L/R} e^{- i H_+ t}  F_{L/R}^\dag  \rangle
=  \langle e^{i H'_+ t}   e^{-i \bar H'_+ t}  \rangle' 
= 2 \langle e^{i H'_+ t} \alpha_d  e^{-i H'_+ t}  \alpha_d  \rangle' 
 = 2 D_{\alpha_d}(t)  \; 
 \end{equation}
[by (\ref{eq:FLRHFLR}),  (\ref{eq:aHa})]. 
This correlator occurs in
(\ref{eq:alphavertexcorrel-const}) 
(in which FG somewhat cavalierly ignored the connected terms),
and  is at the heart of the dispute ---
FG concluded that for general $g$ 
it decays as $t^{-1/2g}$ (their argument is
explained in Appendix~\ref{sec:NNcor}), in agreement
with our $g = 1/2$ result (\ref{eq:alphaalphadref4}), 
$D_{\alpha_d} \sim (it)^{-1}$. 
Hence FG concluded  that ``the neglect of Klein factors is the
likely origin'' of the missing $1/t^{\mu_{e,o}}$ in OF's result. 

In our opinion, however,
this conclusion of FG is somewhat premature and the matter is more subtle:
OF {\em did}\/ treat Klein factors correctly -- they
lead to the minus sign in $Z_e-Z_o$, as emphasized by
OF in their Reply \cite{OF97} to FG's Comment, and as 
we confirm explicitly in Appendix~\ref{app:OF}. On the other hand,
our exact $g=1/2$ result that $D_B(t) = Z_e (t) - Z_o (t) \sim 1/t$ 
 unambiguously implies that $C_e = C_o$, i.e.\ that for $g=1/2$ the 
leading constants in fact  {\em cancel} precisely. 
We  must therefore conclude that  a ``sign problem'',
i.e.\ the cancellation of two large quantities,
occurs in  OF's Coulomb gas,  and that 
{\em the mean field 
approximation which they used is not sufficiently 
accurate to correctly deal with this sign problem.}\/
Since the latter arose {\em because of Klein factors}\/, 
 Fabrizio and Gogolin evidently {\em were} correct in 
emphasizing their importance. 

In replying to FG's Comment, OF rejected 
their result for $D_{\alpha_d}$, 
arguing that 
the effect of density fluctuations,
which, as OF correctly point out,
 are {\em not} suppressed by a backscattering impurity,
were not properly taken into account by FG. 
We disagree with this criticism and believe that FG's calculation is sound. 
%
The technical details of FG's calculation and OF's objection to it are
described in Appendix~\ref{sec:NNcor}.  \vspace*{4mm}

{\em Acknowledgements:} J.v.D.\ would like to thank A. Ludwig for
teaching him the intricacies of field-theoretic bosonization; G.
Zar\'and for a stimulating collaboration during which rigorous
finite-size refermionization was first developed and applied to the
2-channel Kondo problem; M. Fabrizio for allerting him to the 
importance of the dynamics of Klein factors in impurity
problems; C. Dobler for proofreading most of this review; K.
Sch\"onhammer, Y. Gefen and V. Meden for comments on the manuscript;
and particularly Y. Oreg for an extended,  very open-minded and
stimulating discussion on the density-of-states controversy, 
 which revealed a weakness in a previous version
of our calculation and very considerably deepened our understanding of the
matter.  During this discussion, both benefited from the
hospitality of the ICTP in Trieste. 
This work was supported by ``SFB 195'' of the Deutsche
Forschungsgemeinschaft.
\vspace*{12mm}

\newpage

\appendix

\setcounter{section}{0}

\noindent
{\Large  \noindent {\bf Appendices}} \vspace*{-6mm} }

\renewcommand{\thesection}{\Alph{section}}
\renewcommand{\thesubsection}{\thesection.\arabic{subsection}}
\renewcommand{\thesubsubsection}{\thesection.\arabic{subsection}.\alph{subsubsection}}

\section{Relation between field-theoretical and constructive
 bosonization}

\setcounter{equation}{0}
\renewcommand{\theequation}{\Alph{section}\arabic{equation}}

\label{fieldconstruct}

\noindent
{\em We make explicit the connection between the
constructive and field-theoretical approaches to bosonization,
by showing how the operators used in the latter can be
constructed in terms of the former. To be specific, we shall
transcribe into the $L/R$-language of Section~\ref{LRmovers} the 
excellent treatment of Shankar \cite{Shankar}
(which we summarize here without giving details, since these are 
well explained in~Ref.~\cite{Shankar}), denoting
his notation by the subscript $\SH$.}

\subsection{Definition of boson fields}

Shankar starts from a set of boson operators $\phi_\SH (p)$ satisfying
$[\phi_\SH (p), \phi_\SH^\dagger (p')] = 2 \pi \delta (p-p')$,
where $p \in (-\infty, \infty)$ is a 1-D continuous momentum index.
His Hamiltonian is 
\be
\label{SHAM}
   H \; \equiv \;   
   \int_{- \infty}^\infty \!  {\textstyle {dp \over 2 \pi}}
     |p|   \phi_\SH^\dagger (t,p) \phi_\SH (t,p) 
\; = \; 
     \half \int_{- \infty}^\infty dx 
     \left[   \left( \partial_x  \phi_\SH (t, x) \right)^2  
       + \Pi_\SH^2 (t, x) \right] \; ,
\ee
where the 1-D boson field $\phi_\SH (t, x)$ is defined for 
$x \in (- \infty, \infty)$ as
\be
\label{SHbose}
     \phi_\SH (t,x) \equiv
      \int_{- \infty}^\infty \!  
      {\textstyle {dp \over 2 \pi \sqrt{2 |p|}}}
      \left[  \phi_\SH (p) e^{i p x} +  \phi_\SH^\dagger (p) e^{-i px} \right]
      e^{-a |p|/2}
\ee
($a>0$ being an infinitessimal regularization parameter), and 
$\Pi_\SH (t,x) \equiv \partial_t  \phi_\SH (t,x)$ is the 
canonically conjugate field. One can check that
$[ \phi_\SH (t,x), \Pi_\SH (t,x') ] = i \delta (x - x') $
in the limit $a \to 0$. From these fields, the combinations 
\be
\label{ShPM}
         \phi_{\pm \SH} (t,x) \equiv \lim_{x_0 \to - \infty}
         \half  \left[  \phi_\SH (t,x) \mp 
           \int_{x_0}^x \! dx' \, \Pi_\SH (t,x') \right]
\ee
are constructed, whose commutations relations can be checked to be
\bea
\label{ShcomPP}
         [\phi_{\pm \SH} (t,x) , \phi_{\pm \SH} (t,x') ]  &=& 
         \pm {\textstyle {i \over 4}} \, \epsilon(x-x') \; ,
        \quad \mbox{where}\quad  \epsilon (x) \equiv
        \left\{ \begin{array}{cc}
        \pm 1 & \; \mbox{for} \quad x \gol 0 \\
        0  & \; \mbox{for} \quad x =  0
        \end{array} \right. ; \qquad \phantom{.}
\\
\label{ShcomPM}
         [\phi_{+\SH} (t,x) , \phi_{- \SH} (t,x') ]  &=& 
         \pm {\textstyle {i \over 4}} \; .
\eea
(The absence in Shankar's Eq.~(3.12) of the factor $\pm i$ occuring
in \Eq{ShcomPP}   is a typo.) 
The fields  $\phi_{\pm \SH} (t,x) $  can be checked
to depend only on $(t \mp x)$
and hence are called $R$- and $L$-moving. 

\subsection{Bosonization postulate}

The  bosonization formula for $R$- and $L$-moving fermion 
fields is postulated to be 
\be
\label{Shbosformula}
    \psi_{\pm \SH} (t,x) = (2 \pi a)^{-1/2} 
    e^{\pm i \sqrt{4 \pi} \phi_{\pm \SH} (t, x)} \; ,
\ee
and its correctness is verified by checking explicitly that the 
correlation functions and anti-commutators of $\psi_{\pm \SH (x)}$
are correctly reproduced. 

Note that  \Eq{Shbosformula} 
lacks Klein factors $F_{R/L}$ that lower the number of
$R$- or $L$-moving electrons  by one.
Therefore (\ref{Shbosformula}) 
does not have the status of an operator identity in Fock space, but 
has meaning only within expectation values 
containing precisely one
 $\psi_{+ \SH}^\dagger$ (or  $\psi_{-\SH}^\dagger$) 
 for every $\psi_{+ \SH}$ (or $\psi_{- \SH}$), i.e.\ in which the product
of all Klein factors would equal one anyway. Note furthermore
that in the absence of 
(anti-commuting) Klein factors, 
 which in the constructive approach guarantee
that $\{ \tilde \psi_{R} (x), \tilde \psi^\dagger_{L} 
( x') \} = 0$, ``special tricks'' are required to 
ensure that this relation is correctly reproduced. In 
the above construction, the trick is that
$ \phi_{+ \SH}$ and $ \phi_{- \SH}$ do {\em not} commute,
so that   $\{  \psi_{+\SH} (x),  \psi^\dagger_{- \SH} 
( x') \} = 0$ follows from \Eqs{ShcomPM} and (\ref{lemma1b}).

\subsection{Relation between our notation and that of Shankar}
\label{Shankarsnotation}

Let us now transcribe the above field-theoretical
approach into our notation, as used in
Section~\ref{LRmovers}.  Because the lack of Klein factors 
in the former, such a transcription can never be completely
one-to-one. 
Our aim is therefore merely to find the relation
between Shankar's $\phi_{\pm \SH} (x)$  and our $\tilde \phi_{R/L} (x)$
fields.
There is some freedom in choice of signs, etc.,
which we shall  exploit to ensure that 
Shankar's  bosonization formula  \Eq{Shbosformula}
for $\psi_{\pm \SH}$  is consistent
with our \Eq{nubosonization} for $\tilde \psi_{R/L}$. 
To this end, we identify 
Shankar's  $\phi_\SH (p)$-operators (defined for positive and negative $p$)
with our two species of boson operators $b_{q, L/R}$
 (defined only for positive $q$), as follows:
\be
\label{SHwebos}
     \phi_\SH (p) := L^{1/2} \left( \theta(-p)  b_{|p|,L} -
       \theta (p) b_{p,R}   \right) 
\ee
The above Hamiltonian   (\ref{SHAM}) then 
equals our  $H_{kin}$ of (\ref{eq:HL001}), 
in the limit $L\to \infty$.
The $\phi_{\pm \SH} (x)$ of \Eq{ShPM} 
can be expressed in terms of the  $\tilde \phi_{R/L} (x)$ fields 
  defined in \Eq{LRbosonizewe}
 by using  \Eq{SHwebos} to 
translate $  \phi_\SH (p) $'s into $b_{|p|L/R}$'s: 
\bea
\label{SH-we-bos}
        \phi_{\pm \SH} (t,x) & \!\! := \!\! 
& \mp {\textstyle {1 \over 2 \sqrt \pi}} \, 
         \left( \tilde \phi_{R/L} (t, x) - \tilde \phi_0 (t) \right) \; ,
\\
\label{phi00}
\tilde \phi_0 (t) \!\! & = & \!\! \half \left[ \tilde \phi_L (t, x_0) + 
\tilde \phi_R (t, x_0) \right] \; ,
\\
\psi_{\pm \SH} (t,x) \!\! &=& \!\!
 (2 \pi a)^{-1/2} e^{- i (\tilde \phi_{R/L} (t , x) -
  \tilde \phi_0(t))} \; := (2 \pi)^{-1/2}
\tilde \psi_{R/L} (t,x) ] \; ,
\\
\nonumber 
: \!  \psi^\dagger_{\pm \SH} (t,x) \psi_{\pm \SH} (t,x) \! : \!\!
&=&\!\!
 {\textstyle
  {1 \over \sqrt \pi}} \partial_{x} \tilde \phi_{\pm \SH} (t , x) := 
\mp {\textstyle {1 \over 2 \pi}} \partial_{ x} \tilde \phi_{R/L} (t , x) =
{\textstyle {1 \over 2 \pi}}
: \!  \psi^\dagger_{R/L} (t,x) \psi_{R/L} (t,x) \! : 
\eea 
 \Eqs{SH-we-bos} and (\ref{phi00}) (in the limit $x_0 \to -
\infty$) allow one to readily reproduce the commutators
of \Eqs{ShcomPP} and (\ref{ShcomPM}), and in doing so to pin-point the reason
why the latter is non-zero: it is the presence of the ``zero-mode term''
$\tilde \phi_0 (t)$ in \Eq{SH-we-bos} 
[which corresponds to the terms that Shankar
calls ``oscillating pieces at spatial infinity'' after his Eq.~(3.11)].  Thus
the ability of \Eq{Shbosformula} to represent two different species of {\em
  anti-commuting}\/ fermion fields $\psi_{\pm \SH}$ purely in terms of bosonic
operators and without using Klein factors comes at a 
\mbox{price: in 
\Eq{SH-we-bos}}
one effectively adds to the $\tilde \phi_{R/L}$, 
which are built purely from operators
of the {\em same}\/ ${R/L}$ species ($b_{|p|, {R/L}}$, 
$b^\dagger_{|p|, {R/L}}$), a
zero mode term $\tilde \phi_0$, 
which contains a {\em mixture}\/ of operators of {\em
  opposite}\/ species (both $b_{|p|, {R/L}}$, $b^\dagger_{|p|, {R/L}}$ 
{\em and}\/
$b_{|p|, L/R}$, $b^\dagger_{|p|, L/R}$).  This price is rather high from a
conceptual point of view, because as soon as $\psi_{\pm \SH}$ contains such
mixtures, it no longer makes sense to try to construct, as in \Eq{defbq}, the
$b_{q R/L}$ in terms of the ``original'' fermion operators $c_{k R/L}$ in
terms of which the fermion fields were ``originally'' defined: each $\psi_{\mp
  \SH}$ would then depend on both $c_{k L}$ and $c_{k R}$, which clearly does
not make sense.  Thus, in the above way of formulating the field-theoretic
approach the $\phi_\SH (p)$ are introduced for purely formal
reasons, and the very concrete relation between boson and fermion operators
$b_{q R/L}$ and $c_{k R/L}$ that is the hallmark of the constructive approach
to bosonization is lost.  Nevertheless, the field-theoretical approach is
perfectly well-defined and produces correct answers if used with sufficient
care (though this is sometimes easier said than done, and several
consequential mistakes have been made in the past
(Ref.~\cite{KotliarSi96} discusses an example). 

\vspace*{5mm}
\section{Completeness of boson representation}

\setcounter{equation}{0}
\renewcommand{\theequation}{\Alph{section}\arabic{equation}}

 \label{app:partition}

\noindent
{\em
We prove, following Haldane \cite{Haldane81},
the following theorem   for a single species of 1-D fermions
 (i.e.\ $M=1$, hence 
the fixed index $\eta $ is  not shown explicitly below)\vspace*{2mm}:

\noindent
{\bf Theorem:}\/
The Fock space ($F_c$)
spanned by arbitrary combinations
of the fermion operators $c_{k}$ and $c_{k }^\dagger$'s
acting on the ``vacuum state''
$|N = 0 \rangle_0$,  is identical to 
the Fock space ($F_b$) spanned 
by arbitrary combinations of the bosonic operators
$b^\dagger_{q }$'s  acting on the
set of all $N$-particle ground states $\{ |N \rangle_0,
N \in \ZZ\}$\vspace*{2mm}. 
\vspace*{2mm}}

This is equivalent to proving \Eq{NFN0}:
$F_{c} = F_{b}$ implies that any $| N \rangle \in F_c$
can be written as $| N \rangle = 
\sum_{N' \in \ZZ} f_{N'} (b^\dagger_q) |N'\rangle_0$,
where $f_{N'} (b^\dagger_q)$ is some function, labelled by
$N'$, of $b^\dagger_q$'s; 
but since $[b_{q }^\dagger, \widehat N] = 0$, only the $N'=N$
term can be non-zero. Since the generalization to several species
is trivial (all operators with $\eta \neq \eta'$ commute),
\Eq{NFN0} immediatly follows. 

{\em Proof:}\/
It is evident that $F_b$ is a subspace of $F_c$, since
the $b^\dagger$'s are functions of the $c^\dagger$ and $c$'s. 
To prove that in fact $F_b = F_c$, 
we are thus confronted by the  state-counting problem of showing 
that  every state in $F_c$ also occurs in $F_b$.
This can be done elegantly by calculating 
the corresponding grand-canonical partition functions:
since both are sums over positive definite quantities,
one would find  $F_c > F_b$ if  $F_c$ had more  states than 
$F_b$; conversely, if one found   $Z_c = Z_b$, this would imply
$F_c = F_b$. Since this argument is independent of the 
form of the Hamiltonian, we are free to choose $H$ such
that the calculation of partition functions becomes tractable.
To this end, we choose the linear dispersion $\varepsilon_k = 
 k$ of \Eq{H0}, with 
$\delta_{b} = 1$, so that 
\be
\label{H0app}
        H_{0} \equiv \tpol  \sum_{k }
        (n_k - \half)  
        {\,}^\ast_\ast  c^\dagger_{k } c_{k}  {\,}^\ast_\ast \; .
\ee
The calculation of $Z_c$ is an elementary text-book excercise:
Summing over all $n_k \in \ZZ$, with the corresponding
fermion state $c^\dagger_{k } |N = 0 \rangle$
either empty or occupied, yields 
\bea
\label{Zc} 
      Z_c &=& \mbox{Tr} \mbox{[} e^{-\beta (H_0- \mu 
        \widehat N)} \mbox{]} 
\\
      &=& 
      \prod_{n_k = 1}^{\infty}
      (1 + e^{- \beta   2 \pi / L  (n_k - 1/2)} e ^{\beta \mu n_k}) 
      \prod_{n_k = - \infty}^{-1}
      (1 + e^{- \beta   2 \pi / L  |n_k - 1/2|} e ^{\beta \mu n_k})
\\
\label{Zcfinal}
      &=& \prod_{n = 1}^{\infty}
       (1 + w^{2n - 1}v^n )  (1 + w^{2n - 1}v^{-n} )
      \qquad \mbox{where} \quad
      w = e^{- \beta   \pi / L} \quad \mbox{and} \quad
       v = e^{\beta \mu } . \qquad \phantom{.}
\eea

Next we calculate  $Z_b$: We note that 
$F_b$ is spanned by the set of all states of the form
\be
\label{bosonexcited}  
          |N; \{ m_q \} \rangle = 
          \prod_{q > 0}^\infty 
          \frac{b_{q}^{\dagger m_q} }{(m_q !)^{1/2}}
          | N \rangle_0 
\ee
where $N \in \ZZ$,  and for each 
$q = \tpol n_q > 0$ (with $n_q \in \ZZ^+$)
the $m_q \ge 0$ are integers specifying
how many bosonic quanta of momentum $q$ have been excited. 
By \Eqs{EvecNeigenvalues} and (\ref{bqraises}), each
$|N; \{ m_q \} \rangle$ is an eigenstate of $H_0$,  
with eigenvalue $\tpol \half N^2 + \sum_{q>0} q \, m_q$. 
Therefore
\bea
\label{Fb}
           Z_b &=& \sum_{N = - \infty}^\infty 
                      \sum_{\{ m_q \}} 
            \langle N; \{ m_q \} | e^{- \beta (H_0 - \mu \widehat N)} 
            |N; \{ m_q \} \rangle
\\
\label{ZbZcproof}
           & = & \sum_{N = - \infty}^\infty 
                      \sum_{\{ m_q \} }
            e^{- \beta    2 \pi / L (N^2 /2 +  \sum_{ q > 0}
           n_{q} m_{q}) } e^{\mu \beta N} \; 
\\
\label{Zbfinal}
&=& \left( \sum_{N = - \infty}^\infty   
           w^{N^2} v^N \right) \left( 
            \sum_{M=0}^{\infty} P(M)  w^{2M} \right) 
            = {\sum_{N = - \infty}^\infty   w^{N^2} v^N \over
            \prod_{n=1}^{\infty} (1-w^{2n})} \; = Z_c \; .
\eea
In  the first of (\ref{Zbfinal}), we denoted
by $P(M)$ the number of   states  $|N; \{ m_q \} \rangle$ (for fixed $N$)
  satisfying  $\sum_{n_{q} = 1}^\infty n_q m_{q} = M$, and,
since this number is independent of $N$, factorized the sum.
For the second of (\ref{Zbfinal}), we noted that $P(M)$ is
 just the number of partitions of $M$, 
and hence (by simple combinatorics)
 also occurs in the series expansion of  the function
\be
\label{eq:A1.Dedekind}
                {1 \over \prod_{n=1}^{\infty} (1-y^n)}
         =  \prod_{n=1}^{\infty}
           \Bigl[\sum_{m = 0}^\infty (y^n)^m \Bigr] 
           = \sum_{M=0}^{\infty} P(M) y^M    \; .
\ee
 Finally, the last
of (\ref{Zbfinal})
follows by inserting Jacobi' triple product identity,\footnote{ 
Jacobi's triple product identity can be proven using both the 
series and product representations of 
the elliptic theta function $\theta(v;w)$: 
equating Gradshteyn and Ryzhik's  \cite{Gradshteyn}
Eqs. (8.192.3) and (8.181.2) for   $\theta (-iv/2 ;w)$
yields \protect(\ref{thetaelliptic}).}
\be
\label{thetaelliptic}
       \sum_{N = - \infty}^\infty w^{N^2} v^N
        = \prod_{n=1}^\infty (1 + w^{2n-1} v)(1 + w^{2n-1} v^{-1})
        (1-w^{2n}) \; ,         
\ee
and comparing with \Eq{Zcfinal} for $Z_c$.
The rather remarkable result 
 $Z_b = Z_c$ immediately implies
 \Eq{NFN0}, as argued above, which completes the proof. 

Incidentally,  \Eq{NFN0} can also proven more formally: Since \Eq{Zc}
has the form $Z_c = \sum_{N \in \ZZ} Z_{cN} v^N$, where 
  $Z_{cN}$ is the  {\em canonical}\/ $N$-particle  partition function,
and since $Z_c = Z_b$, \Eq{ZbZcproof} shows that  
\be
\label{Zcn}
        Z_{cN} =  \sum_{\{ m_q \}} 
            \langle N; \{ m_q \} | e^{- \beta H_0} 
            |N; \{ m_q \} \rangle \; .
\ee
This proves that the bosonic representation
of fermion states is complete also within any of the {\em fixed}\/-$N$
Hilbert spaces $H_N$.

\section{Useful identities}

\setcounter{equation}{0}
\renewcommand{\theequation}{\Alph{section}\arabic{equation}}

\label{identities}

\noindent
{\em We derive the operator identities needed for bosonization,
all of which are standard \cite{CM71,Emery79}.
\vspace*{2mm}}

Below, $A$ and $B$ are operators;
we define an operator-valued function $f(A)$ 
through its Taylor expansion, i.e.\ 
\vspace*{2mm} $f(A) \equiv \sum_{n=0}^\infty {1 \over n!} f^{(n)} (0) 
A^n$\vspace{2mm}. 

\noindent
{\em Theorem 1}\/ (Baker-Hausdorff):
Define $[A,B]_{n+1} \equiv [[A,B]_n,B]$ and $[A,B]_0 \equiv A$.
Then 
\begin{equation}
  \label{eq:Baker-Hausdorff}
  e^{-B} A e^{B} \; = \; 
 \sum_{n = 0}^\infty   {1 \over n !} [A,B]_n \; = \;
 A + [A,B] + {\textstyle {1 \over 2  !}} [[A,B],B] +
  \dots \; .
\end{equation}
\noindent
{\em Proof:}\/  
Consider the operator-valued function ${\cal A} (s) \equiv e^{-s B} A
\, e^{ s B}$, where $s \in \RR$. 
Since ${d^{n}  {\cal A}(s)  \over ds^{n}} = 
 e^{-s B} [A,B]_n \, e^{s B}$, as can be shown inductively,
the Taylor series  about $s=0$ gives 
\begin{equation}
  \label{eq:BH-s=0}
  A(s) = \sum_{n=0}^\infty  { s^n  \over n !}
   \left( {\textstyle d^n {\cal A} (s)  \over d s^n} \right)_{s=0}
   \; = \; 
    \sum_{n=0}^\infty  {s^n \over n !} 
    [A,B]_n  \; .
\end{equation}
Taking $s=1$ in \Eq{eq:BH-s=0} yields 
the Baker-Hausdorff theorem (\ref{eq:Baker-Hausdorff}). 
\hfill \vspace*{4mm}$\Box$ 

\noindent
{\em Theorem 2:}\/  If $ C \equiv  [A,B] $ 
satisfies $ [A,C] = [B,C] = 0 $, then 
\bea
\label{TrivialBH} \label{lemma1a}
 \mbox{(i)}   \qquad \quad \;\;  \,
  e^{-B} A e^{B} &=&  A + C \qquad \mbox{or} \qquad
\mbox{[}A, e^B\mbox{]} =  C e^B  \; ;\\
\label{lemma2}
\mbox{(ii)}  \qquad \qquad \quad
        e^A e^B &=&  e^{A+B} e^{C/2} =  e^{A+B +C/2} \; ;
\\
\label{lemma1}
\mbox{(iii)}  \qquad \,
        e^{-B} f(A) e^B &= & f(A+C) \;  ;
\\
\mbox{(iv)}  \qquad  \qquad \; \; \;
\label{lemma1b}
        e^A e^B &=& e^B e^A e^C  \; .
\eea
{\em Proof:}\/   (i) Inserting the stated condition into
the Baker-Hausdorff theorem (\ref{eq:Baker-Hausdorff})
yields  (\ref{TrivialBH}).  \\
(ii) The operator-valued function  ${\cal T}(s) \equiv e^{s A} e^{s B}$
($s \in \RR$) satisfies the differential  equation
\begin{eqnarray}
  \label{eq:TABtheorem}
  {\textstyle {d {\cal T} (s) \over ds }}
 = e^{s A} A e^{s B} + e^{s A}  e^{s B} B = 
  {\cal T} (s) (A  + s C + B)  \; 
\end{eqnarray}
[the second equality follows from (\ref{TrivialBH})],
with boundary condition  ${\cal T}(0) = 1$.
Since $[A+B,C] = 0$, its solution by inspection
is ${\cal T}(s) = e^{s (A+B)} e^{ s^2 C/2}$. 
Thus ${\cal T} (1)$ yields (\ref{lemma2}). \\
(iii) By induction, (\ref{TrivialBH}) implies
$e^{-B} A^n e^B = (A + C)^n$, which yields (\ref{lemma1})
when inserted into the Taylor 
expansion for $f(A)$.  
(iv) is a special cases of (iii), with $f(A) = e^{A}$. \hfill
\vspace*{4mm}
$\Box$

\noindent
\be
\label{lemma3}
\mbox{\em Theorem 3:}\/  \hfill  \quad
\mbox{If} \quad [A,B] = D B \; \mbox{and} \;
[A,D] = [B,D] = 0, \quad 
\mbox{then} \quad 
        f(A) B = B f(A+D) . \qquad \phantom{.}
\ee
\noindent
{\em Proof:}\/  
 Use $AB = B(A+D) $ to  show inductively that $A^n B
= B(A+D)^n$. This yields (\ref{lemma3}) when 
inserted into the Taylor expansion for $f(A)$. $\Box$ ---
Using (\ref{lemma3}), one readily finds:
\be
\label{lemma3a}
       e^A B = B  e^{A+D} \; , \qquad \qquad
        e^A B^n =  B^n e^{A + nD} = (B e^D)^n e^A  \; , \qquad \qquad
        e^A e^B = e^{B e^D}  e^A\; \vspace*{4mm} .
\ee

\noindent
{\em Theorem 4} (free bosons):
For a free boson Hamilton 
$H = \sum_j \hbar \omega_j (b_j^\dagger b_j + \half)$
with $[b_j, b^\dagger_{j'}] = \delta_{j j'}$, 
the thermal average of $e^{\hat B}$, where $\hat B
= \sum_j (\lambda_j b_j^\dagger + \tilde \lambda_j b_j )$
is linear in bosons, is 
\begin{equation}
  \label{eq:freebosonexp} 
  \label{expboson}
  \langle e^{\hat B} \rangle = e^{{1 \over 2} \langle \hat B^2
  \rangle},
\qquad \mbox{ where} \quad 
\langle \hat O \rangle = \mbox{Tr} \left(  e^{- \beta H} \hat O  \right) 
/ \mbox{Tr}  e^{- \beta H} \; .
\end{equation}
{\em Proof}\/ \cite{Emery79}
(see also (\ref{eq:provemagiceasy}) for a simpler proof):
Since the bosons are independent, the thermal averages separate
into independent parts,
$\langle e^{\hat B} \rangle = \prod_j C_j$ and 
$\langle \hat B^2 \rangle/2 = 
\sum_j  \lambda_j \tilde \lambda_j \langle  b_j^\dagger b_j + 1/2
\rangle$,  thus
 it suffices to show  that
\begin{equation}
  \label{eq:i-thmodeexp}
  C_j \equiv \langle e^{\lambda_j b_j^\dagger + \tilde \lambda_j b_j}
  \rangle
  = e^{ \lambda_j \tilde \lambda_j \langle b_j^\dagger b_j + 1/2 \rangle}
  \; .
\end{equation}
We denote the $j$-th mode partition
function by $Z_j$ and drop the index $j$ henceforth. Then 
\begin{eqnarray}
  \label{eq:DeriveZ_j}
  Z \!\! &=& \!\!  \sum_{m= 0}^\infty
 \langle m | e^{-\beta \hbar \omega (b^\dagger b + 1/2)} |m \rangle
 =  \sum_{m= 0}^\infty x^{m + 1 / 2} =
  {x^{1/2} \over 1- x}, \quad 
  \mbox{where} \; x \equiv e^{-\beta \hbar \omega }  ; \quad \phantom{.}
\\
  \label{eq:Derivebb_j1}
 \langle b^\dagger b \rangle \!\! &= & \!\!
Z^{-1}  \sum_{m= 0}^\infty 
\langle m | e^{- \beta \hbar \omega  (b^\dagger b + 1/2)}
 b^\dagger b |m  \rangle
 = Z^{-1} \sum_{m=0}^\infty x^{m + 1/2} m = (x^{-1} - 1 )^{-1}  \; ;
\\
  \label{eq:DeriveC_j1}
  C \!\! &=&\!\! Z^{-1} \sum_{m= 0}^\infty 
  \langle m | e^{-\beta \hbar \omega (b^\dagger b + 1/2)} 
     e^{\lambda \tilde  \lambda/2} 
e^{\lambda b^\dagger} e^{ \tilde \lambda b} 
 |m \rangle
\\
  \label{eq:DeriveC_j2}
 &=& \!\! Z^{-1} e^{\lambda \tilde  \lambda/2} 
 \sum_{m= 0}^\infty 
 x^{m+1/2} \sum_{r = 0}^m
   {\lambda^r \tilde \lambda^r \over (r !)^2} 
   { m ! \over (m \! - \! r)! } 
 \\
  \label{eq:DeriveC_j3}
 &= & \!\!  Z^{-1} e^{\lambda \tilde  \lambda/2} 
 \sum_{r= 0}^\infty  {(\lambda \tilde \lambda)^r \over r !} S_r \; ,
\\
   \label{eq:DeriveC_j4}
\mbox{where} \quad S_r \!\! & \equiv & \!\!
  \sum_{m= r}^\infty x^{m+1/2}   {m \choose r} 
  = x^{r + 1/2} \sum_{\bar m = 0}^\infty x^{\bar m} 
 {r + \bar m \choose r}  
 = {x^{r + 1/2} \over (1-x)^{r+1}} 
 = Z \langle b^\dagger b \rangle^r . \qquad \phantom{.}
\end{eqnarray} 
(\ref{eq:DeriveZ_j}) and (\ref{eq:Derivebb_j1}) are standard.  In
(\ref{eq:DeriveC_j1}) we normal-ordered the $e^{\lambda_j b_j^\dagger
  +\tilde \lambda_j b_j}$ of (\ref{eq:i-thmodeexp}) using
(\ref{lemma2}).  For (\ref{eq:DeriveC_j2}) we expanded the last two
exponentials of (\ref{eq:DeriveC_j1}) and evaluated the matrix
elements $\langle m | b^{\dagger r} b^{r'} | m \rangle$ using $b^r
|m\rangle = \sqrt m \, b^{r-1} | m \!-\! 1 \rangle = 
[m (m-1) \dots (m-r+1)]^{1/2}
|m \! -\! r \rangle$ for $r \ge m$, and 0 for $r < m$. For
(\ref{eq:DeriveC_j3}) we reordered the double sum, using
$\sum_{m=0}^\infty \sum_{r = 0}^m = \sum_{r = 0}^\infty \sum_{m =
  r}^\infty$.  To evaluate the sum $S_r$ defined 
in (\ref{eq:DeriveC_j4}), 
we wrote $\bar m
= m-r $ in the second equality,
 for the third evaluated 
the sum $\sum_{\bar m = 0}^\infty$
using  the identity ${r + \bar
  m \choose r } = (-1)^{\bar m} {-r -1 \choose \bar m}$ and the binomial
theorem, and for the fourth simplified
using (\ref{eq:DeriveZ_j}) and (\ref{eq:Derivebb_j1}).
 Inserting the last of (\ref{eq:DeriveC_j4}) into
(\ref{eq:DeriveC_j3})  yields
(\ref{eq:i-thmodeexp}). \hfill $\Box$

\section{More on Klein factors}

\setcounter{equation}{0}
\renewcommand{\theequation}{\Alph{section}\arabic{equation}}

\label{Kleinfactors}

\noindent
{\em We explain why Klein factors are often ignored,
discuss the pitfalls behind the notation $F_\eta = e^{-i \widehat
  \theta_\eta}$, and give an explicit construction
of $F_\eta$ in terms of $c^\dagger_{k   \eta}$ and $c_{k \eta}$ operators.}

\subsection{Why Klein factors are often ignored}
\label{WhynoKlein}

In very many papers, the Klein
factors are tacitly assumed but not written explicitly, 
or simply ignored. This is usually OK if one calculates
correlation functions of the form
\be
        G = \langle \psi_1 \psi_2 \dots \psi_n
                \psi^\dagger_{n} \psi^\dagger_{n-1} \dots \psi^\dagger_1
        \rangle
\ee
for a Hamiltonian that conserves each separate $\widehat N_\eta$, 
because  such functions are only non-zero
if they contain an {\em equal}\/  number of $\psi_\eta^\dagger$ and
$\psi_\eta$ and thus
of $F^\dagger_\eta$ and $F_\eta$, so that the latter
 can be commuted past each other and
all bosonic operators and combined to give 1.
Of course, one has to be careful to keep track of
the minus signs  that arise in this procedure, but
various authors have their own way of doing this
(some of which are discussed in the next subsection,
e.g.\ using a set of anti-commuting Majorana fermions, 
or adding so-called ``zero modes'' with appropriate
properties  to  the boson fields).
Moreover, for free bosons the bosonic fields themselves see to it
that only correlation functions  containing an
{\em equal}\/  number of $\psi_\eta \propto e^{- i \phi_\eta }$ and
$\psi_\eta^\dagger \propto e^{ i \phi_\eta }$
are non-zero, because  invariance
of the bosonic Hamiltonian (\ref{bosonH})
under $\phi_\eta (x) \to \phi_\eta (x) + const$
implies  that
$
        \langle e^{i (\lambda_1 \phi_\eta (x_1)
        + \dots + \lambda_n \phi_\eta (x_n) ) } \rangle
        = 0
$ 
unless $\sum_n  \lambda_n  = 0$ [see \Eq{F.vertexGreen} for details,
and Appendix~\ref{app:OF} for an example]. 

However, if the Hamiltonian does {\em not}\/ conserve
each separate $\widehat N_\eta$, the above arguments are
no longer applicable. Neglecting Klein factors is 
then very dangerous and has been known to lead to mistakes,
as discussed, for example, in Refs.~\cite{KotliarSi96}.

\vspace*{4mm}
\subsection{The Notation $F_\eta^\dagger = e^{i \hat \theta_\eta}$}
\label{thetaF} 

In the literature, the Klein factors 
are often written as 
$
        F_{\eta}^\dagger \equiv e^{i \widehat \theta_\eta}  
$ and 
$
        F_{\eta} \equiv e^{- i \widehat \theta_\eta} ,
$
complemented by the {\em mnemonical}\/ relation
\be
\label{meaningless}
        [ \widehat N_\eta, i \widehat \theta_{\eta'} ] \equiv
        \delta_{\eta \eta'} \; ,
\ee
which suggests that the ``phase operator'' $\widehat \theta$ 
is conjugate to $\widehat N$. (For an explicit construction
of $\hat  \theta$ and an enlightening discussion thereof,
see Eq.~(B.20) of Ref.~\cite{Schoenhammer2},
where it is denoted by $\hat k$) 
The motivation for this
notation is that these relations can be used to ``derive'' the
crucial relation  $ [\widehat N_\eta, F^\dagger_\eta] =
        \delta_{\eta \eta'}   F^\dagger_\eta$
[see \Eq{NF}] using identity (\ref{lemma1a}), which states that
$[A,e^B] = C e^B$ if $C = [A,B] $ commutes
with $A$ and $B$.  

Furthermore, to ensure that the $F_\eta$'s anti-commute\footnote{Some
authors (e.g.\ Fabrizio and Gogolin \cite{FGGGG}) 
instead ensure this by writing
$F_\eta = \widehat \chi_\eta e^{-i \widehat  \theta_\eta}$
with
$ [ \widehat \theta_\eta, \widehat \theta_{\eta'} ]
= 0 $, where the $\widehat \chi_\eta$ are a set of Majorana fermions
satisfying $\{ \widehat \chi_\eta, \widehat \chi_{\eta'} \} = 
2 \delta_{\eta \eta'}$.}
 for different $\eta$'s
[\Eqs{FFcommute} and (\ref{FdFdcommute})] 
one takes, for example [{\em again as mnemonic only,}\/
to be used with (\ref{lemma1b})]
\be
        [ \widehat \theta_\eta, \widehat \theta_{\eta'} ]
        \equiv \left\{
        \begin{array}{lll}
        i \pi & \mbox{if} & \eta > \eta' \\
        - i \pi  & \mbox{if} & \eta < \eta' \\
        0 & \mbox{if} & \eta = \eta'
        \end{array} \right. \; , \qquad \mbox{thus} \qquad 
\label{antiguar}
        \{ e^{i \widehat \theta_\eta} , e^{\pm i \widehat
        \theta_{\eta'}}  \}
        = 2 \delta_{\eta \eta'}
        e^{i (\widehat \theta_\eta \pm \widehat \theta_{\eta'})}
        \; .
\ee
However, the reader should be warned that Eq.~(\ref{meaningless}) 
merely has the status of a  {\em mnemonic}, to be used
{\em only}\/ in conjunction with
$[A,e^B] = c e^{B}$ 
in order to evaluate $ [\widehat N_\eta, F^\dagger_\eta]$,
  as described above. 
The reason for this warning is the following:
If Eq.~(\protect\ref{meaningless}) 
is viewed as an operator identity and $\widehat N_\eta$ is  treated
as an Hermitian operator, a contradiction arises 
(which unfortunately  is not always appreciated,
although this is 
discussed at length \mbox{by Carruthers and} 
Nieto in Ref.~\cite{CN68}).
To see this, take the diagonal 
expectation value of Eq.~(\protect\ref{meaningless}): on
the one hand, Eq.~(\protect\ref{meaningless}) gives $ \langle N_\eta |
[\widehat N_\eta , i \widehat \theta_\eta ] | N_\eta \rangle = \langle N_\eta
|1| N_\eta \rangle = 1 \; , $ and on the other \be
\label{paradox}
       \langle N_\eta | \widehat N_\eta \, i \widehat \theta_\eta
       - i \widehat \theta_\eta  \widehat N_\eta|N_\eta \rangle
       = (N_\eta - N_\eta) \langle N_\eta
       |  i \widehat \theta_\eta | N_\eta  \rangle = 0 .
\ee
To understand the origin of this contradiction, recall 
the following general result from quantum mechanics:
If $\hat X$ and $\hat Y$ are conjugate operators
(i.e. $[\hat X, i \hat Y] = 1$) and the spectrum of $\hat X$ 
are the discrete integers,  then  $\hat X$ is 
Hermitian only in the space of  states ``periodic in $\hat Y$'', 
i.e.\ obtained by acting on a reference state (say $|0 \rangle $)
with {\em periodic}\/ functions of $\hat Y$, i.e.\ functions
 depending  {\em only}\/ on  the exponentials $e^{\pm i \hat Y}$.
In the present case where $\hat X = \widehat N_\eta$
and $\hat Y = \widehat \theta_\eta$, 
the above contradiction thus arose since in \Eq{paradox}
in fact $ \langle N_\eta | \widehat N_\eta \widehat \theta_\eta
\neq N_\eta  \langle N_\eta | \widehat \theta_\eta$.
(A formal way of keeping track of the non-Hermiticity
of $\widehat N_\eta$ is discussed in the appendix of
Ref.~\cite{Loss91}.) 

To avoid contradictions, $\hat \theta$  should
be defined not through \Eq{meaningless}, but by writing
\begin{equation}
\label{noparadox}
        [\widehat N_\eta, e^{\pm i \widehat \theta_{\eta'}} ]=
        \pm \delta_{\eta \eta'}   e^{\pm i \widehat \theta_{\eta'}}  \; ,
\ee
[which is just \Eq{NF}], which evidently identifies 
the exponentials $e^{\pm i \widehat \theta_\eta}$
as raising and lowering operators. Since
these {\em do not}\/ have any diagonal matrix
elements, no contradiction occurs: one the one hand
$
         \langle N_\eta |
       [\widehat N_\eta , e^{\pm i \widehat \theta_\eta} ]
       | N_\eta \rangle =
(N_\eta - N_\eta) \langle N_\eta
       |  e^{\pm i \widehat \theta_\eta} | N_\eta  \rangle
= 0
$, and on the other, using \Eq{noparadox}, 
$ \langle N_\eta |
 \pm e^{\pm i \theta_\eta}  | N_\eta \rangle
= \pm \langle N_\eta |    N_\eta \pm 1 \rangle = 0
$.

Note that the 
above discussion also 
implies that all non-integer powers of  
raising or lowering operators, i.e. expressions of the form
$\left( e^{\pm i \widehat \theta_\eta} \right)^g$ with $g \not \in  \ZZ$,
are ill-defined, since they
 would be plagued by the same kind of inconsistencies as \Eq{meaningless}.  
Unfortunately, many authors unwittingly do use such objects: it
is quite common to ``absorb''
$ \widehat \theta_\eta$ (often called a ``zero mode'' in this context)
into the boson field $\Phi_\eta (s)$ of (\ref{bf5})
by writing $\tilde \Phi_\eta (x) \equiv  \widehat \theta_\eta + \Phi_\eta
(x)$, and to write the bosonization relation (\ref{bf5})  simply as
$\psi_\eta (x) = a^{-1/2} e^{-i \tilde \Phi_\eta (x) }$.
This  procedure is formally perfectly valid. However,  
it is also quite common to  subsequently use
expressions like $e^{-i g \tilde \Phi_\eta (x)}$ with  $g \not \in
\ZZ$, which arise, e.g., when making a linear
transformation of the form
$\tilde \Phi'_\eta (x) = A_{\eta \eta'} \tilde \Phi_\eta (x)$
[as one does for the Kondo problem\cite{vDZF,ZvD} or
a Tomonaga-Luttinger, see Section~\ref{scatterer}, \Eq{pmnewbosonizewe2}].
{\em Strictly speaking, this procedure is formally
meaningless, since the factors $e^{\pm i g  \widehat \theta_\eta}$
 contained in such expressions are ill-defined.}

\subsection{Fermionic representation of $F^\dagger_\eta$}
\label{explicitF}

In this section we  give  an explicit construction for
$F^\dagger_\eta$  in terms of  nothing but fermionic $c^\dagger_{k
  \eta}$ and $c_{k \eta}$'s, and 
 check explicitly that it satisfies  the defining  properties
\Eqs{Ucommutes} to (\ref{UNb}). Though in principle
such a construction is not necessary, since \Eqs{Ucommutes} to
(\ref{UNb}) fully define $F^\dagger_\eta$, its existence serves as
a useful consistency check for the formalism.

The construction is in fact almost trivial: by inverting
Eq.~(\ref{bf1}), $F^\dagger_\eta $
is immediately expressed in terms of $\psi_\eta$
and $\phi_\eta$, both of which are functions only of $c^\dagger_{k
  \eta}$'s: 
\be
\label{UUa}
        F^\dagger_\eta
        = \left( {\textstyle {L \over 2 \pi}} \right)^{ 1/2}
        e^{-i {2 \pi \over L} (\widehat N_\eta - {1 \over 2}
        \delta_b)x } e^{- i \varphi_\eta (x) }
        e^{- i \varphi^\dagger_\eta (x) }
        \psi^\dagger_\eta (x) \; .
\ee
Upon inserting \Eq{deffermions} for $\psi^\dagger_\eta (x)$
and Eqs.~(\ref{defbosonsa}) and (\ref{defbosonsb})
for $ \varphi^\dagger_\eta (x)$ and $\varphi_\eta (x)$, this
equation expresses $F^\dagger_\eta $ purely in terms
of electron operators.
This expression seems to be $x$-dependent, but is not, since
all its matrix-elements between
($x$-independent) states are $x$-independent, as we shall see below.

Eq. (\ref{UUa}) can be used to check whether $F^\dagger_\eta$ does
have all required properties (see Section~\ref{ladder}):
E.g., to check that $[F^\dagger, b] = 0$, 
note that $b_{q \eta}$ commutes with all operators in
\Eq{UUa} except
$ \varphi^\dagger_{\eta'} (x)$ and $ \psi^\dagger_{\eta'} (x) $,
but that the two contributions from
\bea
        \mbox{[} b_{q \eta}, e^{- i \varphi^\dagger_{\eta'} (x)} \mbox{]} &=&
        \delta_{\eta \eta'} \alpha_q (x)  e^{- i \varphi^\dagger_{\eta'} (x)}
\\
        \mbox{[} b_{q \eta},  \psi^\dagger_{\eta'} (x) \mbox{]} &=&
        - \delta_{\eta \eta'} \alpha_q (x)  \psi^\dagger_{\eta'} (x)
\eea
exactly cancel (an observation due to Haldane \cite{Haldane81}). 
The other commutators of \Eq{Ucommutes} can 
be similarly verified. 

Next, one has to verify that
\be
\label{checkU}
        {}_0 \langle \vec N' |
        \widehat T_\eta F^\dagger_\eta  |  \vec N \rangle_0 
        = \delta_{N'_1,  N_1} \dots
        \delta_{N'_\eta , N_\eta +1} \dots
         \delta_{N'_M,  N_M}  \, ,
\ee
where $\widehat T_\eta$ is the phase-counting operator
of \Eq{phasecounting}.
Insert \Eq{UUa} into the left-hand side of \Eq{checkU},
and use the identities
\bea
          e^{- i \varphi_\eta (x) } e^{- i \varphi^\dagger_\eta (x) }
        &=&  e^{- i \varphi^\dagger_\eta (x) } e^{- i \varphi_\eta (x) }
        e^{- [\sum_{q >0}           
          {1 \over n_q}          e^{-qa}]}
        \qquad  \mbox{[using \Eq{lemma1b} and (\ref{rhorhocom})]}
\qquad \phantom{.}
\\
         e^{- i \varphi_\eta (x) } \psi^\dagger_\eta (x)
        &=&  \psi^\dagger_\eta (x)  e^{- i \varphi_\eta (x) }
        e^{[ \sum_{q >0}  {1 \over n_q} e^{-qa}]}
        \qquad  \mbox{[using \Eq{lemma3a}]}
\eea
to commute $ e^{- i \varphi^\dagger_\eta (x) }$ to the very left
and $ e^{- i \varphi_\eta (x) } $ to the very right,
where they are equal to unity when acting
on ${}_0 \langle \vec N' |$ and
$ |  \vec N \rangle_0$ respectively.
Since the two $c$-number exponentials thus produced
cancel ($e^{(- 1 + 1) [\sum_{q >0} {1 \over n_q}   e^{-qa}]} = 1$), 
we get
\bea
         {}_0 \langle \vec N' |
         \widehat T_\eta F^\dagger_\eta  |  \vec N \rangle_0
        &=&
        {}_0 \langle \vec N' |
        \widehat T_\eta
         \left( {\textstyle {L \over 2 \pi}} \right)^{ 1/2}
        e^{-i {2 \pi \over L} (\widehat N_\eta - {1 \over 2}
        \delta_b)x }
        e^{- i \varphi_\eta (x) }
        e^{- i \varphi^\dagger_\eta (x) }
        \psi^\dagger_\eta (x)
         |  \vec N \rangle_0
\\ &=&
\label{NUN}
       {}_0 \langle \vec N' |
        \widehat T_\eta
         \left( {\textstyle {L \over 2 \pi}} \right)^{ 1/2}
       e^{-i {2 \pi \over L} (\widehat N_\eta - {1 \over 2}
        \delta_b)x }
        \left[ \left( \tpol \right)^{1/2} \sum_k e^{i k x} c^\dagger_{k
\eta} \right]
         |  \vec N \rangle_0 \, , \qquad \phantom{.}
\eea
where we have inserted \Eq{deffermions} for $\psi^\dagger_\eta (x)$.
Now the argument is just like that given in the determination
of $\widehat \lambda_\eta$ in Section~\ref{lambdasec}:
Commuting $ c^\dagger_{k \eta}$ past
$|N_1 \rangle \otimes \dots \otimes |N_{\eta -1}\rangle$
produces a factor $T_\eta$ which exactly cancels the
phase contributed by $ \widehat T_\eta$.
Since neither ${}_0 \langle \vec N' |$
nor $ |  \vec N \rangle_0$ contain any
particle-hole-excitations, \Eq{NUN} is  non-zero only if
$c^\dagger_{k \eta} $ adds the
$(N_\eta +1 )$-th particle [which has
momentum $k = \tpol (N_\eta + 1  - \half \delta_b)$]
to $| N_\eta \rangle$ to produce $| N_\eta +1 \rangle$,
and if at the same time
$N'_\eta = N_\eta + 1 $. Hence all c-number-exponentials
cancel, showing that indeed  \Eq{NUN} is equal to \Eq{checkU}.

Finally, it immediately follows from $[F^\dagger, b] =
0$ that  $F^\dagger_\eta$ creates no particle-hole excitations, i.e. that
\be
        {}_0\langle \vec N' |
        f(\{ b_{q \bar \eta} \} ) F^\dagger_\eta
        |  \vec N \rangle_0  =0
\qquad 
        \mbox{for all $ f(\{ b_{q \bar \eta} \} )$, $\vec N$ and 
          $\vec N'$}.
\ee

Thus, we conclude that all
 matrix elements (between $x$-independent states)
of $F^\dagger_\eta$ are indeed $x$-independent, as stated earlier.
Hence  the   $x$-independence of $F^\dagger_\eta$ 
in \Eq{UUa} can be made
explicit by 
either setting $x=0$, or by including a dummy integration
$ L^{-1} \int_{-L/2}^{L/2} dx $ 
(we also unnormal-ordering the exponentials):
\be
\label{UUdx}
\label{Kleinexplicit}
        F^\dagger_\eta = a^{1/2}  e^{- i \phi_\eta (0) }
        \psi^\dagger_\eta (0) 
        = { a^{1/2}\over L }
        \; \int_{-L/2}^{L/2} dx \;
        e^{-i {2 \pi \over L} (\widehat N_\eta - {1 \over 2}
        \delta_b)x } e^{- i \phi_\eta (x) }
        \psi^\dagger_\eta (x) \; .
\ee

\section{Remarkable cancellations
involving bosonization }

\setcounter{equation}{0}
\renewcommand{\theequation}{\Alph{section}\arabic{equation}}

\label{explicitholes}

\noindent
{\em To gain intuition into the
remarkable way in which  the bosonization identity works,
we compare the expansions of $ \psi (x) |0 \rangle_0$ and
$a^{-1/2} F e^{-i \phi (x)} | 0 \rangle_0 $.
\vspace*{2mm}}

Consider the state $\psi (x) | 0 \rangle_0$ 
for a single species of fermions (i.e.\ $M=1$, and we drop the
index $\eta$). As illustrated in Fig.~\ref{fig:psiaction},
we can obtain 
two equivalent representations for
this state by either Fourier-expanding $\psi$ using
(\ref{deffermions}), or bosonizing it using (\ref{bf1}):
\begin{eqnarray}
  \label{eq:psiongroundstate}
%
\lefteqn{
\hspace{-2.4cm}\left( \tpol \right)^{1/2}  \!\!
\sum_{n \in \ZZ} \! e^{-i  (n - {1 \over 2} \delta_b ) {2 \pi x /
    L}} c_{n} |0 \rangle_0
= F_{\eta} \! 
         \left( \tpol \right)^{1/2} \! 
        e^{-i  (\widehat N_\eta - {1 \over 2}
        \delta_b)  {2 \pi x / L} }
        e^{-i \varphi_\eta^\dagger (x)}
        e^{-i \varphi_\eta (x)}  |0 \rangle_0 }
\\
\label{astonishing}
      \sum_{n = 0}^\infty y^n c_{-n}  |0 \rangle_0 \!\! 
      && = 
      e^{ - \left( \sum_{n = 1}^\infty {1 \over n} y^n \rho_n \right)}
        c_0 |0 \rangle_0
\; , \quad \mbox{where} \quad 
 y \equiv e^{i {2 \pi x / L}} , \quad
 \rho_n \equiv \sum_{\bar n \in \ZZ} c^\dagger_{\bar n + n} c_{\bar n}
  , \quad \phantom{.}
\\ 
\label{illustrativeexp1}
&& \hspace{-6mm} = \left[ 1 - y \rho_1 + y^2 (-\half \rho_2 + \half \rho_1^2 )
+ y^3 ( {\textstyle {1 \over 3}} \rho_3 + \half \rho_1 \rho_2
- {\textstyle {1 \over 6}} \rho_1^3) + \dots \right] c_0 | 0 \rangle_0 \\ 
&&  \hspace{-6mm} = \sum_{n = 0}^\infty \left[ A_n  y^n c_{-n} + 
  B_n  y^{n+2} (c^\dagger_{n+1}     c_{-1}) c_0 
    +  C_n   y^{n+3} (c^\dagger_{n+1} c_{-2})  c_0 + \dots \right]
   | 0\rangle_0  . \quad \phantom{.}
\label{illustrativeexp2}
\end{eqnarray}
Eqs.~(\ref{UNb}), (\ref{defpsia}), (\ref{defbq}) and (\ref{a-vacuum})
were used to obtain 
from the right-hand side of (\ref{eq:psiongroundstate})
that of (\ref{astonishing}),
and some of the terms arising in the latter's expansion
are indicated in (\ref{illustrativeexp1}-\ref{illustrativeexp2}), where
the dots represent infinitely many further contributions.
\Eq{illustrativeexp2} implies that the only non-zero
coefficients in (\ref{illustrativeexp2}) are 
$A_n = 1$, whereas {\em all others}\/ are zero, $B_n = C_n = \dots =
0$. This is quite astonishing, since it
implies that when $e^{-i \varphi^\dagger (x)}$ 
acts on $  c_0 |0 \rangle_0$, all the many ways
in which excited states such as  $c_n^\dagger c_{-1} c_0 |0 \rangle_0$ can
arise must  somehow cancel each other, with only 
terms of the form $y^n c_{-n} |0 \rangle_0$ surviving. 
To get a feeling for how this can possibly be, we consider
the lowest few terms in (\ref{illustrativeexp1})
(by inserting the sums from $\rho_n$) that 
contribute to the 
$A_n$ and $B_n$ series explicitly (more general terms become
intractable):
\begin{eqnarray}
  \label{eq:ANseries}
\lefteqn{  \hspace*{-5mm} \sum_{n = 0}^\infty A_n  y^n c_{-n} | 0 \rangle_0 =
 \Biggl\{ 1 - y (c^\dagger_0 c_{-1}) + y^2 \! \left[ 
     - \half (c^\dagger_0 c_{-2}) + 
     \half  ( c^\dagger_{-1}  c_{-2})  ( c^\dagger_0 c_{-1}) \right]
 \Biggl. }
 \\
\nonumber
 &  &  +  \Biggl. y^3 \! \left[
   - {\textstyle{1 \over 3}} (c^\dagger_0 c_{-3}) + 
   \half (c^\dagger_{-2} c_{-3}) (c^\dagger_0 c_{-2}) 
   - {\textstyle {1 \over 6}} (c^\dagger_{-2} c_{-3})
   (c^\dagger_{-1} c_{-2}) (c^\dagger_{0} c_{-1}) \right] + \dots
      \Biggr\} c_0 |0 \rangle_0 
\\
 &=& 
 \Biggl\{ 1 + y c_{-1} + y^2 \! \left[ \half + \half \right] c_{-2} 
 + y^3 \! \left[{\textstyle  { 1 \over 3}} + 
   \half + {\textstyle {1 \over 6}} \right] c_{-3} + \dots 
      \Biggr\} |0 \rangle_0 
\label{Aseriesfinal}
\\
  \label{eq:ANseries2}
\lefteqn{
\hspace*{-5mm}
  \sum_{n = 0}^\infty B_n  y^{n+2} (c^\dagger_{n} c_{-1}) c_0 | 0 \rangle_0 =
 \Biggl\{ y^2 \! \left[ 
     - \half (c^\dagger_1 c_{-1}) + 
     \half  ( c^\dagger_{1}  c_{0})  ( c^\dagger_0 c_{-1}) \right]
 \Biggl. }
 \\
\nonumber
 &  &  + \Biggl. y^3 \! \left[
   - {\textstyle{1 \over 3}} (c^\dagger_2 c_{-1}) + 
   \half (c^\dagger_{2} c_{1}) (c^\dagger_1 c_{-1}) 
   - {\textstyle {1 \over 6}} (c^\dagger_{2} c_{1})
   (c^\dagger_{1} c_{0}) (c^\dagger_{0} c_{-1}) \right] + \dots
      \Biggr\} c_0 |0 \rangle_0 
\\
\label{Bseriesfinal}
 &=& 
 \Biggl\{ y^2 \! \left[ - \half + \half \right] (c^\dagger_1 c_{-1} )
 + y^3 \! \left[- {\textstyle  { 1 \over 3}} + 
   \half - {\textstyle {1 \over 6}} \right] (c^\dagger_2 c_{-1}) + \dots 
      \Biggr\} c_0 |0 \rangle_0 
\end{eqnarray}
This shows that  the first few terms of these series 
(illustrated in Fig.~\ref{fig:psiaction}) do give $A_n = 1$
and $B_n = 0$, and illustrates how the remarkable cancellations
of excited states occur. 
To confirm that this happens for all $n\ge 0 $, we note that
the systematics according to which (\ref{Aseriesfinal}) and
(\ref{Bseriesfinal}) arose imply that the $A_n$ and $B_n$ 
can be found by summing the following two series,
corresponding to taking $\rho_n = -1 $ or 1 in (\ref{astonishing})
(the sign difference arises because $c_0$ is commuted past the 
$c_{-n}$ in the factor
$(c_0^\dagger c_{-n})$ in the $A$ series, but not in the $B$ series):
\begin{eqnarray}
  \label{eq:Ashortcut}
   \sum_{n = 0}^\infty A_n  y^n \!\! & \equiv  & \!\! 
  e^{- \left( \sum_{n = 1}^\infty {- 1 \over n} y^n \right)} 
  = e^{- \ln (1-y) } = (1 - y)^{-1} = \sum_{n = 0}^\infty y^n  \; ,
   \quad  \mbox{implying} \; A_n = 1  ; \qquad \phantom{.}
\\
\sum_{n = -2}^\infty B_n  y^{n +2} \!\! & \equiv & \!\!
  e^{- \left( \sum_{n = 1}^\infty { 1 \over n} y^n \right)} 
  = e^{ \ln (1-y) } = 1 - y \; ,
  \quad \mbox{implying} \; B_{n\ge 0}  = 0 \; .
\end{eqnarray}
The $B_{-2}$ and $B_{-1}$ terms correspond to the
same terms as $A_0$ and $- A_1$, namely $c_0 |0 \rangle_0$
and $c_{-1} | - \rangle_0$. 

Doing such checks
explicitly for more general terms than the $A_n$ and $B_n$ series
becomes intractably complicated.  But we know
that the seemingly miraculous cancellations needed to make them
vanish will indeed occur and are not really miraculous at all,
since the remarkable properties of coherent states allowed us to 
rigorously derive the  bosonization identity
as an operator identity in Section~\ref{schoeller}.

\section{Checking anti-commutators}

\setcounter{equation}{0}
\renewcommand{\theequation}{\Alph{section}\arabic{equation}}

\label{checkanticom}

\noindent
{\em
We check explicitly (following Haldane
\cite{Haldane81})
that the bosonized versions of $\psi_\eta (x)$ 
correctly reproduce the anti-commutation relations (\ref{antic}).
\vspace*{2mm}}

Note first that the anti-commutation relations of
the Klein factors  [\Eq{FdFdcommute}]
trivially guarantee $\{ \psi_\eta, \psi_{\eta'} \} =
\{ \psi_\eta, \psi^\dagger_{\eta'} \} = 0$
for $\eta \neq \eta'$. That this works is of course
no miracle, since \Eq{FdFdcommute} is a consequence
of the anti-commutation relations of the
original $c_{\eta k}$-operators.

For $\eta = \eta'$, start from \Eq{bf1}, and use
\Eqs{lemma1b} and (\ref{NF})
to commute the exponentials into
the  order that occurs in the operators $O_1$ and $O_2$
defined below. One finds:
\bea
        \psi_\eta (x) \psi_\eta (x')
        &=&
        O_{1}(x,x') \,
        e^{i {2 \pi \over L} x} \,
        e^{[ - i \varphi_\eta (x), - i \varphi^\dagger_\eta (x') ]}
\\
        &=&
        O_{1}(x,x') \, e^{i {2 \pi \over L} x}
        \left( 1 -  y e^{- 2 \pi a/L}  \right)     \; ;
\\
        \psi_\eta (x) \psi^\dagger_\eta (x')
        &=&
        O_{2}(x,x') \,
        e^{- i {2 \pi \over L} (x - x')} \,
        e^{[- i \varphi_\eta (x), i \varphi^\dagger_\eta (x') ]}
\\
        &=&
        O_{2}(x,x') \, y
        \left( 1 -  y e^{- 2 \pi a/L}  \right)^{-1}        \; ;
\\
         \psi^\dagger_\eta (x') \psi_\eta (x)
        &=&
        O_{2}(x,x') \,
        e^{[ i \varphi_\eta (x'), - i \varphi^\dagger_\eta (x) ]}
\\
        &=&
        O_{2}(x,x') \,
        \left( 1 -  y^{-1} e^{- 2 \pi a/L} \right)^{-1}    \; .
\eea
Here we have defined
$
        y \equiv e^{- i {2 \pi \over L} (x - x')} \; ,
$
and the operators $O_1$ and $O_2$ are given by
\bea
        O_{1}(x,x')  &=&
        \tpol
        F_\eta F_\eta
        e^{-i {2 \pi \over L} (\widehat N_\eta - {1 \over 2}
        \delta_b) (x +x') }
        e^{-i \left( 
        \varphi_\eta^\dagger (x) +  \varphi_\eta^\dagger (x') \right)} 
       e^{-i \left( \varphi_\eta (x) + \varphi_\eta (x') \right) }
\\
        O_{2}(x,x')  &=&
        \tpol
        e^{-i {2 \pi \over L} (\widehat N_\eta - {1 \over 2}
        \delta_b) (x - x') }
        e^{-i \left( 
        \varphi_\eta^\dagger (x) -  \varphi_\eta^\dagger (x') \right)}
        e^{-i \left( \varphi_\eta (x) - \varphi_\eta (x') \right) } \; .
\eea
It follows immediately that
\bea
        \{ \psi_\eta (x), \psi_\eta (x') \}
        &=& O_{1}(x,x') \,
        e^{i {\pi \over L} (x+x')}
        \left[ y^{- {1 \over 2}} ( 1 - y e^{- 2 \pi a/L} )
        +  y^{ {1 \over 2}} ( 1 - y^{-1} e^{- 2 \pi a/L} ) \right]
\qquad \phantom{.}
\\ & \stackrel{a \to 0 }{\longrightarrow} & \; 0 \; ;
\\
        \{ \psi_\eta (x), \psi^\dagger_\eta (x') \}
        &=& O_{2}(x,x') \,
         y^{ {1 \over 2}}
        \left[ y^{ {1 \over 2}} ( 1 - y e^{- 2 \pi a/L} )^{-1}
        +  y^{ - {1 \over 2}} ( 1 - y^{-1} e^{- 2 \pi a/L} )^{-1}
         \right]
\qquad \phantom{.}
\\ & \stackrel{a \to 0 }{\longrightarrow} &
         O_{2}(x,x') \,
         \sum_{\bar n \in \ZZ} y^{\bar n} \; = \;
         O_{2}(x,x') \, L  \sum_{\bar n \in \ZZ}
        \delta(x-x' - \bar n L )
\\ &=&
        2 \pi \sum_{\bar n \in \ZZ}
        \delta(x-x' - \bar n L ) e^{i \pi \bar n \delta_b} .
\eea
The last line follows because $\varphi_\eta (x) = \varphi_\eta (x +L) $
so that         $O_{2}(x,x+ \bar n L )        =
{\textstyle {1 \over L}} e^{i \pi \bar n \delta_b}$.
Thus, we have reproduced the anti-commutation relations
\Eq{antic}.

\section{Point-splitting} 

\setcounter{equation}{0}
\renewcommand{\theequation}{\Alph{section}\arabic{equation}}

\label{point-splitting}

\noindent
{\em We discuss the regularization technique of
``point-splitting''. We introduce the general
concept of operator product expansions to 
explain why point-splitting an operator product
regularizes it,  then explain
why this is usually equivalent to normal-ordering, and
finally illustrate the care needed when using bosonization to 
evaluate point-split products of fermion fields.}

\subsection{Operator product expansions}
\label{OPEss}

Consider the product $\psi^\dagger_\eta (z+a) \, \psi_\eta (z)$ 
of two fermion fields, with $z = \tau+ix$ and $a>0$ a real
constant. When $a \to 0$,
the result diverges, because the product is not normal-ordered. 
To calculate the divergence, one simply has to normal order it explicitly:
\bea
\label{psipsip}
         \psi_\eta^\dagger (z+a) \psi_\eta (z)
         &=&
         \sum_{k \neq 0}
         e^{- k (z+a) }
          \tpol  \sum_{k'} e^{k' (z+a)}
          c^\dagger_{k'-k \eta} e^{- k' z} c_{k' \eta}
         +  \tpol  \sum_{k'} e^{k' a}
         c^\dagger_{k' \eta} c_{k' \eta}
\qquad \phantom{.}
\\
\label{psipsipB}  
         &\stackrel{a \to 0}{ \longrightarrow}&  \sum_{q > 0}
          \left({\textstyle  {2 \pi q \over L}}\right)^{1/2} 
         \left( e^{- q z} i b_{q \eta}
           -  e^{ q z} i b_{q \eta}^\dagger \right)
         +  \tpol \widehat N_\eta
         + \tpol \sum_{k' \le 0} e^{k' a}
\\
\label{pointsplit}
        &=&
        i \partial_{z} \phi_\eta (z)
        + \tpol \widehat N_\eta
        + \left[ {\textstyle {1 \over   a}}
        +  {\textstyle {\pi \over L}} (1 - \delta_b)   
        + \mbox{Order}({\textstyle {a \over L^2}}) \right]
\eea
Since    the first
($k \neq 0$) term in \Eq{psipsip} is normal-ordered, 
it is possible to set $a=0$ there. However, in the
second ($k =0$) term we first have to normal order, producing
the $\sum_{k'}$ in \Eq{psipsipB}, which diverges for $a \to 0$.
The bracketed terms in \Eq{pointsplit} are its
order $a^{-1}$ and $a^0$ contributions.
\Eq{pointsplit} agrees with \Eq{fermionOPE}, but,
since we included terms of order $1/L$, includes
finite-size corrections that were neglected in the latter.

\Eq{pointsplit} is an example of a so-called  
{\em operator product expansion}\/ (OPE). In general,
when the product $O_i (z) \, O_j (z')$ 
of two quantum fields is evaluated at points $z$ and $z'$ that
are very close to each other,
the result diverges
if the product is not normal-ordered, typically as some power $(z -z')^{-\nu}$.
By bringing the product $O_i (z) O_j (z')$ into normal-ordered form,
one can generally express  it as  a linear combination
of other (normal-ordered) fields in the theory, the coefficients
being functions of $z - z'$. In the limit $z \to z'$, a
Laurent-expansion of these  functions in powers of 
$(z  \! - \! z')$ yields an OPE of the general form
\be
\label{OPE}
        O_i (z) O_j (z') 
        \stackrel{z\to z'}{\longrightarrow}
        \sum_k {C_{ijk} \,  O_k (z') \over (z-z')^{\Delta_i
            + \Delta_j - \Delta_k}} \; .
\ee
The exponents $\Delta_j$ are known as the {\em scaling
        dimensions}\/ of the fields $O_j (z)$, and the 
$C_{ijk}$ are $c$-number coefficients. An OPE succinctly summarizes 
the short-distance behavior of a theory. For example, 
the leading ultraviolet behavior of correlation functions can directly
be read off from \Eq{OPE}:
$\langle  O_i (z) O_j (z') \rangle \to  C_{ij1} 
(z-z')^{-(\Delta_i + \Delta_j)} $, since $\langle 
O_k (z') \rangle = \delta_{k 1}$ is only non-zero if $O_k =
\mbox{\boldmath$1$}$, the unit operator, for which $\Delta_1 = 0$.
It follows that the fermion fields $\psi$ and $\psi^\dagger$,
for which $\langle \psi (z) \psi^\dagger(0) \rangle = z^{-1}$,
have scaling dimension $\Delta_\psi =  \Delta_{\psi^\dagger} = 1/2$.

\subsection{Point splitting versus normal ordering}

In field theory, it is popular to 
regularize  divergent products
of two fields at the same point by adopting 
the so-called {\em point-splitting}\/  prescription, denoted by $: \quad :$,
which evaluates the product at points a short distance apart, and then
subtracts the divergence:
\be
\label{defpointsplit}
        : \!  O_i (z) O_j(z) \! : \; \equiv \;
        O_i (z + a) O_j (z) -
        {}_0\langle \vec 0 | 
        O_i (z+ a) O_j (z) 
        | \vec 0 \rangle_0 \; .
\ee
Note that we chose to use here the same regularization parameter $a$ as the
one introduced for boson fields in \Eq{defpsia}; this is necessary
if one wants to reproduce
point-split products of fermion fields by using the
bosonization formula, as shown in Section~\ref{Bosepointsplit}.

From \Eqs{pointsplit} and (\ref{normorddens}), we find 
\be
\label{fermpointsplit}
        : \!  \psi_\eta^\dagger (z) \psi_\eta (z) \! : \; = \;
         i \partial_z \phi_\eta (z)
        + \tpol \widehat N_\eta  \; = \;
        {\,}^\ast_\ast
       \psi_\eta^\dagger (z) \psi_\eta (z)
       {\,}^\ast_\ast \; = \; \rho_\eta (z) \; ,
\ee
showing that the point-split and normal-ordered versions 
of the electron density agree. 
This illustrates the fact that
point-splitting  simply subtracts a constant that would not
have arisen at all had we
started  with a normal-ordered expression. 
Therefore, in most cases point-splitting and normal ordering
are equivalent ways of regularizing.
There are, however, exceptions: if  the term that diverges as 
$a \to 0$ is not a $c$-number but an operator, 
such as $O_k(z')$ in the general OPE (\ref{OPE}), 
and if the expectation value of this operator is zero, 
then the point-splitting prescription does not succeed
in subtracting this divergence.

Point-splitting is a popular regularization scheme in
field-theoretical texts, because field theorists typically ``know''
from experience the various standard OPEs of a given theory,
so that it is a simple
matter to subtract out the appropriate divergences.
However, if one is less familiar with standard OPEs than
a an experienced field-theorist, one would have to
derive them first by normal-ordering all products of fields.
But then one might as well simply adopt the prescription
that  from the beginning only normal-ordered products of fields 
at the same point  are to be used, thereby eliminating the need
to point-split. For the purposes of this tutorial, written
for ``non-field-theorists'', we always adopt the latter procedure.

\subsection{Evaluating point-split products
of Fermion Fields using Bo\-so\-ni\-za\-tion}
\label{Bosepointsplit}

As a consistency check on the bosonization rules, we now check that
point-split products of fermion fields can also be calculated
using their bosonized versions (\ref{bf1}) or (\ref{fermionvertex}).

\noindent
{\em Density:---}\/
Eq.~(\ref{pointsplit}) can be rederived as follows
[we abbreviate $\tilde {\cal N}_\eta \equiv {2 \pi \over L}
(\widehat N_\eta - {1 \over 2}  \delta_b)$]:
\bea
\label{checkpointb1}
         \psi_\eta^\dagger (z + a) \psi_\eta (z)
         &=&
        \tpol e^{ \tilde {\cal N}_\eta (z+a) }
        e^{i \varphi_\eta^\dagger (z+ a)} e^{i \varphi_\eta (z+ a)}
                   e^{ - \tilde {\cal N}_\eta z }
        e^{- i  \varphi^\dagger_\eta (z)} e^{- i \varphi_\eta (z)}
\\
\label{checkpointb2}
       &=& \tpol e^{ \tilde {\cal N}_\eta a }
        e^{i ( \varphi_\eta^\dagger (z+ a) -  \varphi_\eta^\dagger (z))}
        e^{i ( \varphi_\eta (z+ a) -  \varphi_\eta (z))}
        e^{[\varphi_\eta (z+ a), \varphi_\eta^\dagger (z) ] }
\\
\label{checkpointb3}
        &=&
        \tpol
        \left[ 1 +  i a \partial_z
          (  \varphi_\eta^\dagger (z) +  \varphi_\eta (z))
          + a  \tilde {\cal N}_\eta
          \right]
           \left( {\textstyle {L \over 2 \pi  a} + {1 \over 2}} \right)
\\
\label{checkbosonpoint}
         &=&
          i \partial_z \phi_\eta (z)
        + \tpol (\widehat N_\eta - \half \delta_b)
        +  {\textstyle {1 \over   a}}
       +  {\textstyle {\pi \over L}} 
       + \mbox{Order}({\textstyle {a \over L^2}}) \; .
\eea
To get \Eq{checkpointb2}, we 
normal-ordered \Eq{checkpointb1},
using \Eq{lemma1b}. For \Eq{checkpointb3}, we Taylor-expanded 
the normal-ordered expressions in $a$,
and evaluated the boson commutator using \Eq{comln}. Note that
the latter had to be  expanded to next-to-leading order in
$a$, giving 
$ \left( {\textstyle {L \over 2 \pi  a} + {1 \over 2}} \right)$,
 in order to correctly reproduce 
the subleading (non-diverging) terms of \Eq{pointsplit}.
Note also that in (\ref{checkpointb3})
the cancellation of the $1/a$ arising from the boson commutator 
and the linear $a$ factors arising  from expanding functions of $z+a$
{\em only}\/ occurs if we use the same short-distance regularization
parameter $a$ when point-splitting as the one used for the boson fields.

\noindent 
{\em Hamiltonian:---}\/
Next we consider the Hamiltonian $H_{0\eta}$. In analogy to
\Eq{psipsip}, we have 
\be
\label{psidxpsi}
       - \psi_\eta^\dagger (z + a)  \partial_z \psi_\eta (z)
         =  \sum_{k \neq 0}
         e^{- k (z+a) }
          \tpol  \sum_{k'} k' e^{k' (z+a)}
          c^\dagger_{k'-k \eta} e^{- k' z} c_{k' \eta}
         +  \tpol  \sum_{k'} e^{k' a} k' 
         c^\dagger_{k' \eta} c_{k' \eta} \; .
\ee
It follows immediately that 
\be
\label{H0point}
      -  \int_{-L/2}^{L/2} \! 
        {\textstyle {dx \over 2 \pi}}
         : \!  \psi_\eta^\dagger (z) \partial_z \psi_\eta  (z) \! :
        \; = \;
        \sum_{k }
        k  {\,}^\ast_\ast  c^\dagger_{k \eta} c_{k \eta}
        {\,}^\ast_\ast
        \; = \;  H_{0\eta} \; ,
\ee
since $ \int_{-L/2}^{L/2} \! 
        {\textstyle {dx \over 2 \pi}} e^{-i k x} = 0 $ for $k \neq 0$,
and the point-splitting subtraction eliminates the non-normal-ordered 
contributions in the second term. Comparison with  \Eq{Hpsipoint}
shows that the normal-ordered and point-split fermionic
versions of $H_{0\eta}$ are equal. 

Next we show that the point-split bosonic version (\ref{bosonH})
of $H_{0 \eta}$ can be obtained from \Eq{H0point} by diligent use
of the bosonization formula. 
Proceeding as in \Eq{checkpointb1}, we find 
\bea
\nonumber
\lefteqn{    -     \psi_\eta^\dagger (z + a) \partial_z \psi_\eta (z)}
\\
\label{checkpointc1}
        & & \hspace{-8mm} =
        \tpol e^{ \tilde {\cal N}_\eta (z+a) }
        e^{i \varphi_\eta^\dagger (z+ a)} e^{i \varphi_\eta (z+ a)}
                   e^{ - \tilde {\cal N}_\eta z }
        e^{- i  \varphi^\dagger_\eta (z)} e^{- i \varphi_\eta (z)} \!
        \left\{ \tilde {\cal N}_\eta + i \partial_z \phi_\eta (z) 
          + [ \partial_z \varphi^\dagger_\eta (z), \varphi_\eta (z)] \right\}
\qquad \phantom{.}
\\
\label{checkpointc2}
        & & \hspace{-8mm} = 
\tpol e^{ \tilde {\cal N}_\eta a }
        e^{i ( \varphi_\eta^\dagger (z+ a) -  \varphi_\eta^\dagger (z))}
        e^{i ( \varphi_\eta (z+ a) -  \varphi_\eta (z))}
           \left( {\textstyle {L \over 2 \pi  a} + {1 \over 2}} \right)
        \left\{ \tilde {\cal N}_\eta + i \partial_z \phi_\eta (z) -
        \left( {\textstyle {1 \over  a} - {\pi \over L}} \right) 
        \right\}
\\
\nonumber
        & & \hspace{-8mm} =
       \left[ 1 +  
          a \! \left( \tilde {\cal N}_\eta + i \partial_z \phi_\eta (z) \right)
          + \half a^2 \!
          \left(
          \tilde {\cal N}_\eta^2 + 
          2 \tilde {\cal N}_\eta \,  i \partial_z \phi_\eta (z)
          + {\,}^\ast_\ast (i \partial_z \phi_\eta (z) )^2 {\,}^\ast_\ast
          + i  \partial^2_z \phi_\eta (z)
          \right)  \right]
\\
\label{checkpointc3}
  & &     \times \,
        \left\{ (\tilde {\cal N}_\eta + i \partial_z \phi_\eta(z) ) 
        \left( {\textstyle {1 \over  a} + {\pi \over L}} \right) 
       -    \left( {\textstyle {1 \over  a^2} + {\pi^2 \over L^2}} \right)
       \right\} \; .
\eea
The commutator in \Eq{checkpointc1} arises  from  commuting the
$i \partial_z \varphi_\eta^\dagger (z)$, produced by
differentiating an exponent, to the left past 
$e^{-i \varphi_\eta (z)}$ using \Eq{lemma1a}. Using \Eq{comln} it
 is evaluated to subleading order in $1/a$, 
giving $ - \left( {\textstyle {1 \over  a} - {\pi \over L}} \right) $.
The factor $ \left( {\textstyle {L \over 2 \pi  a} + {1 \over 2}} \right)$
in \Eq{checkpointc2} comes from normal-ordering the exponentials
of \Eq{checkpointc1},
as in \Eq{checkpointb3}. Since the leading divergence is $1/a^2$,
we had to Taylor-expand all exponentials to order $a^2$.
It is now merely a matter of algebra to verify that 
\be
\label{H0bospoint}
        -  \int_{-L/2}^{L/2} \! 
        {\textstyle {dx \over 2 \pi}}
         : \!  \psi_\eta^\dagger (z) \partial_z \psi_\eta  (z) \! :   
         \, =  
         \int_{-L/2}^{L/2} {\textstyle {dx \over 2 \pi}}
        {\textstyle {1 \over 2}}
         : \! (i \partial_z \phi_\eta (z) )^2 \! :
          \, + 
         \tpol  \half \widehat N_\eta ( \widehat 
         N_\eta + 1 - \delta_b )  \, = \,          H_{0 \eta}  ,
\ee
which is the point-split version of the boson Hamiltonian (\ref{bosonH}).
To this end, note that the $\int dx$ integral gives zero
for all expressions linear in $\partial_z \phi$ and
$\partial_z^2 \phi$, because $\phi$ is periodic in $x$. Furthermore,
we used the fact, easily verified, that 
$ 
      : \! (i \partial_z \phi_\eta (z) )^2 \! : = 
      {\,}^\ast_\ast (i \partial_z \phi_\eta (z) )^2 {\,}^\ast_\ast 
$ (which sometimes  is also written as 
$\half : \! \Pi^2 (z) + (i \partial_z \phi(z))^2 \!:$, compare
footnote~\ref{Pi}).

The above way of deriving the bosonic form (\ref{H0bospoint}) 
of $H_{0 \eta}$ is often used
in the field-theoretical approach to bosonization. Note
that it is considerably more arduous than the derivation
given in Section~\ref{ham}. We
included it here for two reasons:
Firstly, to illustrate how careful one needs to
be if one wants to correctly produce $1/L$ terms
using  point splitting methods; and secondly, because
when performed in reverse order
it forms the rigorous basis for refermionizing a bosonic
Hamiltonian, as in Eqs.~(\ref{oldHphi}-\ref{eq:H0cbk}).

\section{Free Green's functions}

\setcounter{equation}{0}
\renewcommand{\theequation}{\Alph{section}\arabic{equation}}

\label{freefermbos}

\noindent
{\em We calculate the  free fermion and boson
 Green's functions
$\langle {\cal T} \psi_\eta \psi^\dagger_\eta \rangle$
and 
$\langle {\cal T} \phi_\eta \phi_\eta \rangle$,
for $L \neq \infty$ at $T=0$, and also for $L\to \infty$ at $T \neq 0$.
\vspace*{2mm}}

We shall calculate the desired time-ordered Green's functions
in the imaginary-time Heisenberg picture,
in which they depend 
on the imaginary-time variable $\tau \in (- \beta, \beta]$. 
Real time-ordered Green's functions
can obtained from these by simply analytically continuing $\tau \to i t$
(or, for  $v_F \hbar \neq 1$,
 $\tau \to i v_F \hbar t$ and $\beta \to v_F \hbar \beta$).

The linear dispersion of the Hamiltonian of 
Eqs.(\ref{H0}) and (\ref{Hboson}) implies the following
thermal expectation values and imaginary-time development
of the fermion and boson operators:
\bea
\label{fermocc}
        \langle c^\dagger_{k \eta} c_{k' \eta'} \rangle 
        & = & {  \delta_{\eta \eta'} \delta_{k k'}  \over
          e^{\beta k} + 1 } \; ,
\qquad \qquad 
        c_{k \eta} (\tau) = e^{- k \tau}    c_{k \eta} \; , 
\qquad \qquad 
        c^\dagger_{k \eta} (\tau) = e^{ k \tau}    c^\dagger_{k \eta} \; , 
\\
\label{bosonocc}
        \langle b^\dagger_{q \eta} b_{q' \eta'} \rangle 
        & = & {  \delta_{\eta \eta'} \delta_{q q'}  \over
          e^{\beta q} - 1 } \; ,
\qquad \qquad 
        b_{q \eta} (\tau) = e^{- q \tau}    b_{q \eta} \; ,
\qquad \qquad 
        \; b^\dagger_{q \eta} (\tau) = e^{q \tau}    b^\dagger_{q \eta} \; .
\eea

\subsection{The limit $T=0$ for $L \neq \infty$}
\label{app:freefermbosT=0}
For  $T=0$, the factors in (\ref{fermocc}-\ref{bosonocc}) reduce to
$(e^{\beta k} + 1)^{-1} = \theta (- k)$ and
$(e^{\beta q} - 1)^{-1} = - \theta (-q )$,
hence correlation functions are easy to evaluate.

\subsubsection{Fermion correlation function}
\label{sec:fermT=0Lneqinf}
For fermions fields, 
$ - G^{>}_{\eta \eta'} (\tau,x) 
 \equiv  \langle \psi_\eta (\tau,x) \psi_{\eta'}^\dagger (0,0) \rangle$
and
$  G^{<}_{\eta \eta'} (\tau,x)  \equiv 
\langle  \psi_{\eta'}^\dagger (0,0) \psi_\eta (\tau,x) \rangle$
(defined only for $\tau \gol 0$, respectively) 
are two distinct functions, that must be calculated independently.
The time-ordered Green's function $ G_{\eta \eta'} (\tau,x)$
is a convenient combination
of both, which can be evaluated as follows
(with $\sigma \equiv \mbox{sgn}(\tau)$ and 
$y \equiv e^{-{2 \pi \over L}(\sigma \tau + \sigma i x + a)} $):
\bea
\label{GFF1}
      - G_{\eta \eta'} (\tau,x) & \equiv &
\theta (\tau)  G^{<}_{\eta \eta'} (\tau,x) + \theta(-\tau) 
 G^{<}_{\eta \eta'} (\tau,x)  \\
\label{GFF1T=02}
&=& \delta_{\eta \eta'} 
\Bigl[ \theta( \tau) \tpol \sum_{k > 0} 
e^{-k (\tau + ix + \sigma a)}
- \theta (- \tau)  \tpol \sum_{k<0 } e^{- k (\tau + ix + \sigma a)}
\Bigr]
\\
&=& \delta_{\eta \eta'}
\tpol \sigma y^{- \sigma \delta_b/2} \sum_{n = 1}^\infty
   y^n  = \delta_{\eta \eta'}
\tpol \sigma {y^{-( \sigma \delta_b + 1)/2}
     \over y^{- 1/2} - y^{1/2}}  
  \\
&=& 
\label{GF:app1}
{\delta_{\eta \eta'} e^{{\pi \over L} (\delta_b + \sigma) (\tau + ix) } 
      \over
     {\textstyle {L \over \pi} \sinh [{\pi \over L}
(\tau + ix + \sigma a)]}} \; 
\stackrel{L \to \infty}{\longrightarrow}
{\delta_{\eta \eta'}  \over \tau + ix + \sigma a} \; . 
\eea
For (\ref{GFF1T=02}) we inserted the
definition (\ref{deffermions}) of $\psi_\eta$, simplified
using (\ref{fermocc}) and inserted a factor $e^{-k \sigma a}$ 
to ensure con\-ver\-gence\footnote{\label{f:convergence}
It is not {\em a priori}\/ clear that the regularization
 parameter $a$ to be used in the cut-off factor $e^{-k \sigma a}$ 
 in Eqs.~(\ref{GFF1T=02}) or (\ref{fermint}) for the 
fermion correlator 
 must be precisely the same $a$ as the one 
occuring in the  cut-off factor $e^{-q a/2}$ introduced
for the  boson fields in \Eq{defpsia}. 
In Section~\ref{bosfer} we checked that 
 it must indeed be the same, 
else the results (\ref{GF:app1}) or  (\ref{appfinalF})
for the fermion correlator would not be consistent with
those obtained by evaluating these correlators via bosonization
[see (\ref{eq:GreensT=0}) and (\ref{bbGT=0}) or  
(\ref{GF2a}) and (\ref{bbG})].}
 when $\tau \to  0$. 
In the $k$ sums, we took $ k = \tpol (n_k - \delta_b/2)$,
see (\ref{momentumquant}).

\subsubsection{Boson correlation function}

For boson fields, $- \,{\cal G}^>_{\eta \eta'} (\tau,x)
\equiv  \langle \phi_\eta (\tau, x) \phi_{\eta'} (0,0) \rangle$
and 
$- \, {\cal G}^<_{\eta \eta'} (\tau,x)
\equiv  \langle \phi_{\eta'} (0,0) \phi_{\eta}  (\tau, x) \rangle$
(defined only for $\tau \gol 0$, respectively) 
are in fact {\em not}\/ independent, since 
${\cal G}^<_{\eta \eta'} (\tau,x) = {\cal G}^>_{\eta' \eta} (-
\tau,- x)$. Hence, the time-ordered Green's function
can be evaluated as follows: 
\bea
\label{bosoncorr}
\lefteqn{
   - {\cal G}_{\eta \eta'} (\tau,x) \equiv 
        \theta (\tau) {\cal G}^>_{\eta \eta'} (\tau,x)
       \; + \; 
       \theta (- \tau) {\cal G}^<_{\eta \eta'} (\tau,x)
       \; = \; {\cal G}^>_{\eta \eta'} (\sigma \tau,\sigma x)}
\\
\label{bosoncorrT=0}
 &=& \delta_{\eta \eta'}
 \sum_{q > 0}^\infty {1 \over n_q} e^{- q ( \sigma \tau + \sigma i x + a)} 
= \delta_{\eta \eta'}
\sum_{n_q = 1}^\infty {1 \over n_q} y^{n_q}
= - \delta_{\eta \eta'}
\ln ( 1 -y) \\
\label{bosoncorrT=0final}
&=& - \delta_{\eta \eta'}
\ln \left(1 - e^{-{2 \pi \over L}(\sigma \tau + \sigma i x + a)}\right)
\; \stackrel{L \to \infty}{\longrightarrow}
- \delta_{\eta \eta'} 
\ln \left[ \tpol ( \sigma \tau + \sigma i x + a) \right]  . 
\qquad \phantom{.}
\eea
For (\ref{bosoncorrT=0}) we inserted the
definition  (\ref{defphi}) of $\phi_\eta$ and simplified
using (\ref{bosonocc}). \\

\subsection{The limit $L \to \infty$ for $T \neq 0$}
\label{L-to-infty-T=0}

We next consider the continuum limit $L\to \infty$ for $T \neq 0$, 
in which the $1/L$ terms in \Eq{Hboson} can be neglected,
and discrete sums can be treated as integrals,
$\tpol \sum_{n_k} \to \int \! dk$. 
(In particular, if also $T\to 0$, the order of limits considered
here is   $L/\beta \to \infty$.)
Calculating the free fermion and
boson Green's functions is again
straightforward, though for $T\neq 0$ the integrals we shall
encounter are non-trivial (readers that do not  enjoy 
doing contour integrals can look them up in \cite{Bateman}). 

\subsubsection{Fermion correlation function}
\label{corfermion}

Starting from (\ref{GFF1}), simplifying
using (\ref{fermocc}) and again inserting
the convergence${}^{\protect\ref{f:convergence}}$ factor $e^{-k \sigma a}$, 
we proceed as follows
(with   $\sigma \equiv \mbox{sgn} (\tau) $ and 
$\bar y \equiv e^{-2 \pi i (\sigma \tau + \sigma ix + a)/\beta}$):
\bea
      - G_{\eta \eta'} (\tau,x) 
\label{fermint}
        &=& \delta_{\eta \eta'} \:
         \int_{-\infty}^\infty \! \! dk \,
        {e^{- k (\tau + i x + \sigma a)} \over
        \sigma(  1 + e^{-\sigma \beta k})}
\\
     &=&  \delta_{\eta \eta'} \:
     (2 \pi i/ \beta) 
     \left[ - \theta (x) 
         \sum_{\bar n = - \infty}^{0} \bar y^{ (\bar n - 1/2) \sigma} 
            \; +  \; \theta (-x)  \sum_{\bar n = 0 }^\infty
           \bar y^{ (\bar n + 1/2)\sigma } \right]
\\
\label{appfinalF}
     &=&  {\delta_{\eta \eta'}  \over
     {\textstyle {\beta \over \pi} 
       \sin [{\pi \over \beta}(\tau + i x + \sigma a)]}}
   \qquad \stackrel{T \to 0}{\longrightarrow}  \qquad 
     {\delta_{\eta \eta'}  \over \tau + i x + \sigma a}
\eea
The integral was done using  contour methods, by closing 
it along a semi-circle 
in the lower (upper) half of the complex $k$ plane for $x > 0$
$(x<0)$. The poles  at $k = 2 \pi i (\bar n + 1/2)/ \beta$
have residues $\bar y^{(\bar n +1/2) \sigma } /\beta$;
summing their contributions readily yields the $1/\sin$ behavior. \\

\subsubsection{Boson correlation function}

Starting from (\ref{bosoncorr}) and simplifying
using (\ref{bosonocc}), we obtain
\bea
   - {\cal G}_{\eta \eta'} (\tau,x) 
\label{trueconv}
 &=&  \delta_{\eta \eta'} 
         \int_{{2 \pi \over L}}^\infty \! \! dq \,
         {e^{-q a}\over q} \left[
         {e^{- q (\sigma \tau + \sigma i x)} \over
        (  1 - e^{- \beta q})}
        +  {e^{ q ( \sigma \tau + \sigma i x)} \over
        ( e^{ \beta q} - 1) } \right] \; .
\eea
Since  the original discrete sum $\sum_{q>0}$ does not include
$q=0$, we took the lower integration limit in \Eq{trueconv} at 
$q = 2 \pi / L$ to regularize the infrared-divergence at $q=0$. 
When $L\to \infty$, the integral diverges
as $\ln(L/a)$ (as is particularly obvious for $\tau = x = 0$ and  $T= 0$). 
To be able to perform the integral by contour methods
and nevertheless correctly keep track of this divergent constant, 
we shall proceed  as follows: 
We take the lower integration limit in \Eq{trueconv} at $q=0$
and regularize the divergence using the principle-value prescription,
which  gives a finite ($L$-independent) function of $(\tau + ix)/\beta$.
To this we add a (diverging) $L$-dependent constant $C$, whose value 
we shall find at the end by requiring that the final result agree with
$      -  {\cal G}^{(T=0, L \neq \infty)}_{\eta \eta'} (0,0) $. 
The integral (\ref{trueconv}) can then be evaluated as follows:
\bea
\label{easyconv}
\lefteqn{        - {\cal G}_{\eta \eta'} (\tau,x)
        =   \delta_{\eta \eta'} \Bigl[
         \int_{-\infty}^\infty \hspace{-4.5ex}P \hspace{1.5ex}  dq \;
         {e^{- q (\sigma \tau + i \sigma x + a )} \over
        q (  1 - e^{- \beta q})} \; + \; C \Bigr] }
\\
\label{halfresidues}
      &=& 
       \delta_{\eta \eta'}  
      \Bigl[ - \theta (\sigma x) 
       \Bigl(\half  \ln \bar y
         + \sum_{\bar n = - \infty}^{-1} 
          {\bar y^{  \bar n} \over \bar n} \Bigr)
            + \theta (- \sigma x)  \Bigl( 
           \half \ln \bar y
         + \sum_{\bar n = 1 }^\infty
            {\bar y^{  \bar n} \over \bar n} \Bigr) + C \Bigr]
            \qquad \phantom{.}
\\
      &=& 
        \delta_{\eta \eta'} 
     \Bigl[ -
        \theta (\sigma x) 
      \Bigl( \ln \bar y^{1/2} +   \ln (1 - \bar y ^{-1}) \Bigr)
        - \theta (-\sigma x) 
      \Bigl( \ln \bar y^{-1/2} +   \ln (1 - \bar y ) \Bigr)
   +  C \Bigr] \qquad \phantom{.}
\label{ln-1-y}
\\
    &=&    \delta_{\eta \eta'}  \Bigl[-
        \ln \left[ \mbox{sgn}(\sigma x) 
        \left( \bar y^{1/2} - \bar y^{- 1/2} \right) \right] + C
        \Bigr]
\label{last-C}
\\
\label{Gresult}
     &=&  
     -  \; \delta_{\eta \eta'} \: 
     \ln \left(  {\textstyle {2 \beta \over L}
         \sin [{\pi \over \beta}
         (\sigma \tau  + \sigma ix + a )]} \right) \; . 
\eea
Since in  (\ref{trueconv}) the convergence factor $e^{-qa}$
is needed only in the first term (and there only when $\tau = 0$),
we replaced it in the second term by  $e^{q a}$ 
(which causes errors of at most $a / \beta \simeq 0$),
since both terms can then be combined into the
single $\int^\infty_{-\infty} dq$  integral of (\ref{easyconv}).
The $\int \! d q$ integral of (\ref{easyconv})
can   be done by contour methods, closing the contour along a semi-circle 
in the lower (upper) half of the complex $q$ plane for $\sigma x > 0$
$(\sigma x<0)$. The poles  of order one at $q = 2 \pi i \bar n / \beta$
(with $\bar n \neq 0$)
have residues $\bar y^{ \bar n} / (2 \pi i \bar n)$,
where again $\bar y = e^{-i 2 \pi
 (\sigma \tau + \sigma ix +  a)/\beta}$. The pole 
of order two at $q=0$ has residue $- (\sigma \tau + i \sigma x + a)/ \beta
= (\ln \bar y) / (2 \pi i) $, which is
multiplied by ${1 \over 2}$ in (\ref{halfresidues})
due to the principle-value prescription. 
The sums $\sum_{\bar n}$ directly yield 
the $\ln (1 - \bar y^{\mp 1})$ terms of (\ref{ln-1-y}).
We find the constant $C$
by requiring that (\ref{last-C}) 
be  compatable  with (\ref{bosoncorrT=0final})
for $\tau = x = 0$, namely 
$      -  {\cal G}^{(T=0, L \neq \infty)}_{\eta \eta'} (0,0) 
       =  \delta_{\eta \eta'} \: \ln (L/2 \pi a) \; ;
$ 
this readily yields $C =\ln [ - i L \, \mbox{sgn} (\sigma x) / \beta]$,
and hence also (\ref{Gresult}).

That $ {\cal G}_{\eta \eta'} (\tau,x)$
 diverges as $L \to \infty$ is in itself no cause for concern, since
it turns out that only those combinations of Green's functions 
are needed in which the divergence is subtracted out. 
For the  combination  most often encountered, namely
\bea
    - \left[ {\cal G}_{\eta \eta'} (\tau,x)  - 
      {\cal G}_{\eta \eta'}  (0 ,0) \right]
    &=&
  \delta_{\eta \eta'} \:
     \ln\left( { \pi a / \beta \over 
         \sin [{\pi \over \beta}
         (\sigma \tau  + \sigma ix + a )]} \right) 
\\
\label{t0bos}
    & \stackrel{T \to 0}{\longrightarrow} & \;
        \delta_{\eta \eta'} \:
        \ln\left( { a \over  \sigma \tau + \sigma i x + a } \right) \; ,
\eea
it would not have been necessary to 
worry about the divergent term $- \ln(2 \beta / L)$ term 
in (\ref{Gresult}) at all.
However, there are 
cases for which it {\em is}\/ needed explicitly,
e.g.\ when obtaining \Eq{VVVexp} from \Eq{EEEepx}, where it
yields the important $L$-dependent prefactor of Eq.~(\ref{vertexproduct}).

As a final consistency check, note that using \Eq{t0bos}, we can
readily reproduce \Eq{phidphiaa}:
\be
\label{bcheck}
  \partial_{x'} \langle \phi_\eta ( 0^+,x) \, 
                        \phi_\eta (0,x') \rangle \; = \; 
                        {2 a i \over a^2 + (x-x')^2 }
                         \; \stackrel{a \to 0 }{\longrightarrow}  \;
                         2 \pi i \:  \delta (x-x') \; .
\ee

\section{Finite-size diagonalization of backscattering Hamiltonian $H'_+$}

\setcounter{equation}{0}
\renewcommand{\theequation}{\Alph{section}\arabic{equation}}

\label{app:fsdiagonalization}

\noindent
{\em
We diagonalize the refermionized  
impurity backscattering term 
in a  $g=\half$ Tomonaga-Luttinger liquid
of finite size $L$, exploiting its similarity
to the 2-channel Kondo problem. 
\vspace*{2mm}}

The refermionized $H'_+$ of
\Eq{eq:h+prime},  Section~\ref{fsrefermionization},
 which describes backscattering off an impurity
in a Tomonaga-Luttinger liquid with coupling constant $g = \half$, has the form
\begin{equation}
  \label{eq:h+prime:app}
  H'_+ \equiv U_+ H_+ U_+^{-1} =
  \Delta_L {\textstyle {P\over 8}} + 
 \sum_\bk  \left[\varepsilon_\bk 
         {}_\ast^\ast  c^\dagger _{\bar k} c_{\bar k} {}_\ast^\ast
         +   \sqrt{\Delta_L \Gamma} \left( c_\bk^\dagger  + 
    c_\bk \right) \left( i \sqrt{2} \, \alpha_d \right) \right] \; ,
\end{equation}
with $ \{ \alpha_d, \alpha_d \} = 1$, $\alpha_d^\dagger = \alpha_d$
and $\{ c_\bk, c_{\bk '}^\dagger \} = \delta_{\bk \bk'}$.
As pointed out by Furusaki \cite{Furusaki}, 
its form is related to that arising  after bosonizing and refermionizing
the 2-channel Kondo model,
whose solution in  Ref.~\cite{ZvD} inspired that 
 presented below. 

Perhaps the cleanest and  most instructive  way to diagonalize
$H'_+$ is to do so for finite $L$ with\footnote{The case
$\bk = \tpol n_\bk $ (i.e.\ the $P=1$ subspace)
can be treated analogously,
but requires a bit more care due to the presence of
a $\bk = 0$ state that does not arise for $P=0$,
see \cite{vDZF,ZvD}.}
 $\bk = \tpol (n_\bk - \half)$
(i.e.\ in the $P=0$ subspace  of Section~\ref{fsrefermionization}) 
and take the continuum limit $L \to \infty$
at the end. Determining the unitary transformation
from the  $c_\bk$'s and $\alpha_d$ to the
``eigenoperators'' $\tilde \alpha_\varepsilon$ 
that diagonalize $H'_+$  (and inverting this transformation)
is easier for finite $L$ than 
in the continuum limit, since the discrete state $\alpha_d$  
can easier be kept track of if all states are discrete than
if the $\bk$'s form a continuum. Moreover, one sees explicitly
how each exact eigenvalue $\varepsilon$ develops from 
its unperturbed value $\varepsilon_\bk$ as the scattering
interaction is turned on (the energy shift being
of order $\Delta_L$). This is useful and instructive,
but not possible if $L\to \infty$ is taken from the outset,
since then the spectrum is dense and shifts of order $\Delta_L$
are negligible. 

We begin by making a further transformation to a new set of fermions
$\alpha_\bk$ and $\beta_\bk$,
\begin{eqnarray}
  \label{eq:alphabeta}
\left(\!\!\begin{array}{l} 
      \alpha_{\bk} \\  \beta_\bk
      \end{array}\!\!
\right) 
=  {1\over \sqrt{2}}
   \left(\!\! \begin{array}{rr}
                     1 & 1 \\ - i & i
                     \end{array}\!\! \right)
 \left(\!\! \begin{array}{c} 
      c_\bk \\ c ^\dagger_{-\bk}
      \end{array} \!\! \right), 
\qquad  
\left(\!\! \begin{array}{l} 
       c_\bk \\  c^\dagger_{-\bk}
      \end{array}
\!\! \right) 
&\equiv&  {1\over \sqrt{2}}
   \left(\!\! \begin{array}{rr}
                     1 & i \\ 1 & - i
                     \end{array} \!\! \right)
 \left(\!\! \begin{array}{c} 
     \alpha_{\bk} \\  \beta_\bk
           \end{array} \!\! \right),   
\end{eqnarray}
which have the following properties (the index $n$ takes the values $\bk$ 
and $d$, with $-d \equiv d$):
\begin{eqnarray}
\label{abcommute}
 \alpha_{n}^\dagger = \alpha_{-n}, \quad
\beta^\dagger_\bk = \beta_{-\bk}, \quad
\{ \alpha_n, \alpha_{- n'} \} = \delta_{n n'} , \quad 
\{ \beta_\bk, \beta_{- \bk'} \} = \delta_{ \bk  \bk'} , \quad
\{\alpha_n, \beta_{\bk'} \} = 0 \, .
\end{eqnarray}
These have the advantage that the $\beta_\bk$ decouple completely,
since \Eq{eq:h+prime:app} becomes:
\begin{equation}
  \label{eq:alpabetaold}
   H'_+ 
  =  \Delta_L {\textstyle {P\over 8}}  +  \sum_{\bk >0}  
 \varepsilon_\bk 
    \left(  \alpha^\dagger_{\bar k} \alpha_{\bar k} + 
     \beta^\dagger_{\bar k} \beta_{\bar k} \right) 
         + i \, 2  \sqrt{\Delta_L \Gamma}  \sum_{\bk >0} 
         \left( \alpha^\dagger_\bk + \alpha_\bk \right)
          \alpha_d   ; .
\end{equation}
We seek a set of orthonormal fermions, $\{ \tilde \alpha_\varepsilon,
\tilde\alpha_{\varepsilon'}^\dagger \} = 
\delta_{\varepsilon \varepsilon'}$, that diagonalize
$H'_+$, i.e.\ for which 
\begin{equation}
  \label{eq:H+diagonalfinal}
   H'_+   \equiv  \sum_{\varepsilon >0} \,\varepsilon 
        \, \tilde \alpha^\dagger_{\varepsilon} \tilde \alpha_{\varepsilon}
                  \; + \; 
          \sum_{\bk>0} \, \varepsilon_\bk   \,
          \beta^\dagger_{\bar k} \beta_{\bar k} \; + \; E_G', 
\end{equation}
where the constant $E_G'$ represents a shift in ground state
energy due to the interaction.
Now, when diagonalizing, it is convenient to use as  independent operators
not the set\footnote{
Working only with $\alpha_\bk$'s having $\bk > 0$ 
 would have required instead of Eq.~(\protect\ref{eq:linearAnsatz})
the more cumbersome
 Bogoljubov-like Ansatz $\tilde \alpha_\varepsilon \equiv 
A^\dagger_{\varepsilon d}
  \alpha_d + \sum_{\bk > 0} (A^\dagger_{\varepsilon \bk} \alpha_\bk +
  \bar A^\dagger_{\varepsilon \bk} \alpha^\dagger_\bk) $.} 
 $\{
\alpha_d; \, \alpha_\bk, \alpha_\bk^\dagger,  \forall \bar k > 0\} $ 
but  the set
$\{\alpha_d; \, \alpha_\bk, \forall \bar k \gol 0\} \equiv \{ \alpha_n \}$ 
(which by the first of
\Eqs{abcommute} is equivalent to the former),
since then the scattering term in \Eq{eq:alpabeta}
looks simpler:
\begin{equation}
  \label{eq:alpabeta}
   H'_+ 
  =  \Delta_L {\textstyle {P\over 8}}  +  \sum_\bk  \left[\half \varepsilon_\bk 
    \left(  \alpha _{- \bar k} \alpha_{\bar k}  - \theta(-
  \bk) \right) 
         + i \, 2 s_B \sqrt{\Delta_L \Gamma} \alpha_\bk
          \, \alpha_d \right]  + 
        \sum_{\bk>0} \, \varepsilon_\bk   \,
          \beta^\dagger_{\bar k} \beta_{\bar k} \; .
\end{equation}
Analogously, we define 
\begin{equation}
  \label{eq:tildealphaminus}
   \tilde\alpha_{-\varepsilon}\equiv  \tilde\alpha_\varepsilon^\dagger \; ,
\qquad \mbox{with} \qquad
\{ \tilde \alpha_\varepsilon, \tilde \alpha_{- \varepsilon'} \} \equiv
         \delta_{\varepsilon \varepsilon'}   \, 
\end{equation}
($ \tilde \alpha_0 $ will turn out to be the Majorana fermion that $\alpha_d$
develops into when the interaction is turned on), 
and use not the set
$\{ \tilde \alpha_0 ; \,
\tilde \alpha_\varepsilon, \tilde \alpha_{\varepsilon}^\dagger,
\forall \varepsilon > 0 \}$, but instead the set
$\{\tilde \alpha_0 ;  \,
\tilde \alpha_\varepsilon, \forall \varepsilon \gol 0 \} \equiv 
\{ \tilde \alpha_\varepsilon \}$ (and  below
$\sum_\varepsilon$  sums over {\em all} 
these $\varepsilon$). 
Then the  desired diagonal form of $H'_+$ is:
\begin{equation}
  \label{eq:H+diagonal}
   H'_+   \equiv  \sum_{\varepsilon } \half \,\varepsilon 
        \left( \tilde \alpha _{- \varepsilon} \tilde \alpha_{\varepsilon}
          - \theta(-\varepsilon) \right)
         \; + \; E'_G \; + \; 
         \sum_{\bk>0} \, \varepsilon_\bk   \,
          \beta^\dagger_{\bar k} \beta_{\bar k} \; .
    \end{equation}
Since $H'_+$ is quadratic, the $\alpha$'s and $\tilde \alpha$'s are linearly
related, hence we make the Ansatz ($A$ is a matrix, with
$A^\dagger_{\varepsilon n} \equiv A^\ast_{n \varepsilon}$)
\begin{equation}
  \label{eq:linearAnsatz}
  \tilde \alpha_\varepsilon = \sum_{n= d, \bk} A^\dagger_{\varepsilon n} \,
  \alpha_n \; , \qquad \mbox{with}  \quad 
  (A^\dagger_{\varepsilon n})^\ast =
  A^\dagger_{- \varepsilon  - n} = A^\ast_{-n -\varepsilon},  \quad 
  \mbox{to ensure} \quad 
  \tilde\alpha_\varepsilon^\dagger \equiv \tilde\alpha_{-\varepsilon} \; .
\end{equation}
Inserting the first of \Eq{eq:linearAnsatz}
into \Eq{eq:tildealphaminus} shows that 
\begin{equation}
  \label{eq:invertlinear}
  \sum_{n} A^\dagger_{\varepsilon n} A_{n \varepsilon' } = 
\delta_{\varepsilon  \varepsilon'} \; \qquad
\left(\mbox{i.e.} \; A_{\varepsilon n }^\dagger 
= (A^{-1})_{\varepsilon n} \right) \; ,
\qquad \mbox{thus} \qquad
\alpha_{n} = \sum_{\varepsilon} A_{n \varepsilon} \,
  \tilde \alpha_\varepsilon 
\end{equation}
is the inverse transformation of \Eq{eq:linearAnsatz}. 
To determine the coefficients $A^\dagger_{\varepsilon n}$, insert
 Ansatz (\ref{eq:linearAnsatz}) into
the Heisenberg equation
$\varepsilon \tilde \alpha_\varepsilon = [ \tilde \alpha_\varepsilon ,H'_+]$
[which follows from  (\ref{eq:H+diagonal})]. This  yields 
\begin{eqnarray}
  \label{eq:solveAs}
  \varepsilon A^\dagger_{\varepsilon d} &=& i \, 2 \sqrt{\Delta_L \Gamma} 
\sum_\bk A^\dagger_{\varepsilon \bk} \; , \qquad
  A^\dagger_{\varepsilon \bk } = 
-{ i  2 \sqrt{\Delta_L \Gamma} A^\dagger_{\varepsilon d} 
\over \varepsilon - \varepsilon_\bk} \, , \qquad \mbox{implying}
\\
  \label{eq:eigenvalues}
  {\varepsilon \over 4 \Gamma} &=&  S_1 (\varepsilon) \; , \qquad \mbox{where} \qquad
  S_1(\varepsilon) \equiv \Delta_L \! \! \sum_{\bk = - \infty}^\infty 
 {1 \over \varepsilon -\varepsilon_\bk}
  = - \pi \tan (\pi \varepsilon / \Delta_L)
\end{eqnarray}
(the latter equality is a standard identity for 
$\varepsilon_\bk  =  \Delta_L (n_\bk - \half)$, $n_\bk \in \ZZ$).
The first of \Eq{eq:eigenvalues} is an eigenvalue equation determining
the allowed $\varepsilon$'s as functions of $\Gamma$. Analyzing
it (e.g. graphically, cf. [\cite{ZvD}]) shows that (apart from one 
$\varepsilon=0$ solution) each $\varepsilon_k$ is shifted
to a corresponding $\varepsilon(\bk) \equiv
\varepsilon_\bk + \mbox{sgn}(\varepsilon) \delta_\bk \Delta_L$, where the shift
$\delta_\bk \simeq \half$ if $|\varepsilon_\bk| \ll \Gamma$,
and $\delta_\bk \simeq 0$ if $|\varepsilon_\bk| \gg \Gamma$.
This identifies $\Gamma$ as the cross-over scale below or above
which the spectrum is strongly or weakly shifted,  respectively.

$A^\dagger_{\varepsilon d}$ can be determined as follows from the first
of \Eq{eq:invertlinear}, with $\varepsilon = \varepsilon'$:
\begin{eqnarray}
  \label{eq:Aed}
  1 \!\! &=& \!\! \sum_n A^\dagger_{\varepsilon n} A_{n \varepsilon} =
|A^\dagger_{\varepsilon d}|^2 \left[ 1 + 4 \Gamma S_2 (\varepsilon) \right] \; ,
\qquad \mbox{where}
\\
\label{eq:S2}
S_2(\varepsilon) \!\! &\equiv& \!\! \Delta_L \sum_\bk 
{1 \over (\varepsilon - \varepsilon_\bk)^2}
=
- {\partial S_1(\varepsilon) \over \partial \varepsilon}  = 
{\pi^2 \over \Delta_L} \left[1 + \tan^2 (\pi \varepsilon / \Delta_L) \right]
= {1 \over \Delta_L} \left[ \pi^2 + {\varepsilon^2 \over 16 \Gamma^2}
\right] \! .
 \qquad \phantom{.}
\end{eqnarray}
The second and third equalities follow from the
first and second for $S_1(\varepsilon)$ in \Eq{eq:eigenvalues},
the fourth from the first of \Eq{eq:eigenvalues}.\footnote{
As consistency check, we verify that the first
of Eq.~(\protect\ref{eq:invertlinear}), divided by
$A^\dagger_{\varepsilon d} A_{d \varepsilon'}$, also 
holds for $\varepsilon \neq \varepsilon'$
[the last equality follows from the first of
Eq.~(\protect\ref{eq:eigenvalues})]:
\begin{eqnarray}
  \label{eq:S3} \nonumber
 [A^\dagger_{\varepsilon d} A_{d \varepsilon'}]^{-1} 
{\textstyle {\sum_n}} A^\dagger_{\varepsilon n} A_{n \varepsilon'} &=&
 1 + 4 \Gamma \Delta_L {\textstyle {\sum_\bk}}
{\textstyle {1 \over (\varepsilon - \varepsilon_\bk) (\varepsilon' -
  \varepsilon_\bk)}}  = 
 1 + 4 \Gamma 
\Delta_L {{\textstyle {\sum_\bk}}
{\textstyle {1 \over \varepsilon' - \varepsilon} }\left(
{\textstyle {1 \over \varepsilon - \varepsilon_\bk} - {1 \over \varepsilon' -
  \varepsilon_\bk}} \right) = 0 \; .}
\end{eqnarray}
}
Combining \Eqs{eq:S2}, (\ref{eq:Aed}) and the second of (\ref{eq:solveAs}), 
gives 
\begin{eqnarray}
  \label{finalAen:app}
A_{d \varepsilon} \!\! &=&\!\! 
 (A^\dagger_{\varepsilon d })^\ast = 
- i \,  \mbox{sgn} (\varepsilon)  \left[{ 4 \Delta_L \Gamma \over
4 \Delta_L \Gamma + \varepsilon^2 + (4 \pi \Gamma)^2 } \right]^{1/ 2} ,
\quad (\mbox{with sgn} (\varepsilon =0 )  \equiv i) \; , 
 \qquad \phantom{.}
\\
\label{finalAen:app-2}
   A_{\bk \varepsilon} \!\! &=& \!\!  (A^\dagger_{\varepsilon \bk })^\ast = 
{ i  2  \sqrt{\Delta_L \Gamma} A_{d \varepsilon} 
\over \varepsilon - \varepsilon_\bk} .
\end{eqnarray}
The phases in \Eqs{finalAen:app} and (\ref{finalAen:app-2})
where chosen such that 
$A^\ast_{n \varepsilon } =   A_{- n - \varepsilon }$
[as required by \Eq{eq:linearAnsatz}] and that $A_{\bk \varepsilon}$
is real [the latter somewhat arbitrary choice ensures
consistency with Ref.~\cite{ZvD},
with $(\alpha_\bk)_{here} = {1 \over \sqrt 2} (\gamma_{\bk +} 
+ i \gamma_{\bk -})_{there}$ and
$(\alpha_d)_{here} =  (\gamma_{d-})_{there}$]. 

With \Eq{finalAen:app}, the desired unitary transformation 
that maps the refermionized $H'_+$ of \Eq{eq:h+prime:app} into the 
diagonal form (\ref{eq:H+diagonalfinal})  is complete.\footnote{
It is straightforward and instructive to verify directly that inserting
the last of \Eq{eq:invertlinear} for $\tilde \alpha_n$ into
the original form (\ref{eq:h+prime:app}) for $H'_+$ indeed
yields the diagonal form  (\ref{eq:H+diagonal}).}
That it indeed diagonalizes $H'_+$ can
be checked explicitly 
by inserting the last of \Eqs{eq:invertlinear} for $\alpha_n$
 into \Eq{eq:alpabeta}.
After some rearrangement and use of \Eqs{eq:eigenvalues}
and (\ref{eq:S2}) to do the $\bk$ sums,
 one readily recovers \Eq{eq:H+diagonal}, and
in the process finds that the ground state energy shift is 
$E'_G = \sum_{\bk > 0} \varepsilon_\bk - \sum_{\varepsilon > 0}
\varepsilon$ (for details,  see Ref.~\cite{ZvD}).

To calculate ($T=0$)  expectation values  with
respect to the ground state $|G'_B \rangle$ of
$H'_+$, denoted by $\langle \; \rangle'$,
of expressions involving  the original operators $c_\bk$ and $\alpha_d$, 
one uses the 
inverse transformations [obtained from 
 the last of \Eq{eq:invertlinear} and the second
of \Eq{eq:alphabeta}]:
\begin{eqnarray}
  \label{eq:oldcbetaalpha}
  c_\bk \!\! &=& \!\! {\textstyle {1 \over \sqrt 2}}
( \alpha_\bk + i \beta_\bk ) = 
 {\textstyle {1 \over \sqrt 2}} \left( i \beta _\bk +
\sum_\varepsilon A_{\bk \varepsilon} \tilde \alpha_\varepsilon \right)  \; , 
\qquad 
\alpha_d = \sum_\varepsilon A_{d \varepsilon}  \tilde \alpha_\varepsilon \; ,
\\
  \label{eq:expectab}
\langle \beta_\bk \beta_{- \bk'} \rangle' \!\! &=& \!\! 
\langle \beta_\bk \beta^\dagger_{ \bk'} \rangle' = 
\delta_{\bk \bk'} \theta(\varepsilon_{\bk'}) \, , \quad  
  \langle \tilde\alpha_\varepsilon \tilde\alpha_{- \varepsilon'} \rangle' =
\langle \tilde\alpha_\varepsilon \tilde \alpha^\dagger_{ \varepsilon'} \rangle'
  = \delta_{\varepsilon \varepsilon' } \theta(\varepsilon') \, , \quad
[\theta(\varepsilon' = 0) \equiv \half]  .
 \qquad \phantom{.}
\end{eqnarray}

\section{Asymptotic analysis of various correlators}
\setcounter{equation}{0}
\renewcommand{\theequation}{\Alph{section}\arabic{equation}}
\label{asymptotic}

\noindent
{\em We evaluate explicitly the asymptotic $t \to \infty$ 
behavior of a number of correlators occuring in the
refermionized theory of scattering off a Luttinger liquid
of Sections~\ref{scatterer} and \ref{dos}. Since they
can all be expressed in terms of the fermionic operators
$\beta$ and $\tilde \alpha$, this is possible using
Wick's theorem. The corresponding Feynman diagrams are 
shown in Fig.~\ref{fig:feynman}. 
Throughout this Appendix, we use the shorthand $c \equiv 4 \pi \Gamma$.
\vspace*{2mm}}

\begin{figure}[htbp]
  \begin{center}
        \leavevmode 
    \epsfig{
width=0.8\linewidth,%
file=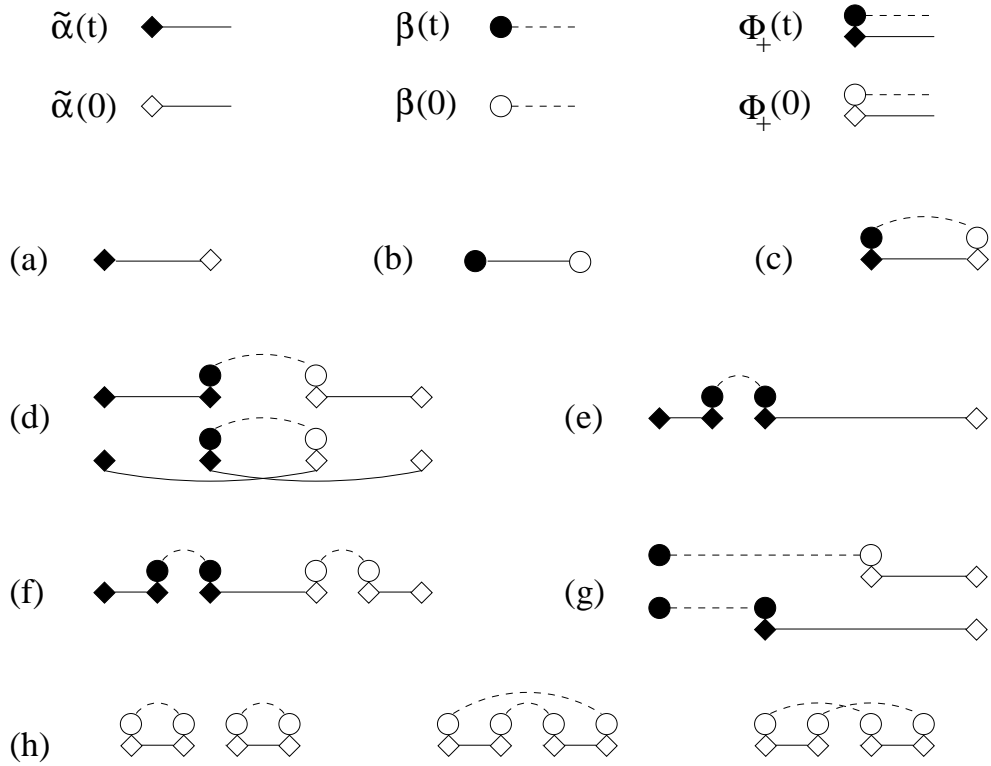}
\vspace*{5mm}
\caption{\label{fig:feynman} Feynman diagrams 
indicating the contractions that give the asymptotically
leading contributions to the following correlators: 
(a) $D_{\alpha_d}$ of (\protect\ref{eq:alphalphadref1}),
the subleading term of $D_\Psi$ of (\protect\ref{eq:psipsiref1}),
and also $D_{\Psi \alpha_d}$ of (\protect\ref{eq:psialphadref1});
(b) $D_\beta$ of (\protect\ref{eq:betacorrelator}) and the leading
term of $D_\Psi$ of (\protect\ref{eq:psipsiref1}); 
(c) $D_{\Phi_+}$ of (\protect\ref{eq:phi(t)phi(0)})
and $D_{N_+}$ of (\protect\ref{eq:DNNcor-1});
(d) $D_{11}$ of (\protect\ref{eq:G11});
(e) $D_{20}$ of (\protect\ref{eq:G20-1});
(f) $D_{22}$ of (\protect\ref{eq:G22-1}); 
(g) $D_{\Psi \Phi_+ \alpha_d}$ of (\protect\ref{eq:psiPhialphadref1});
and (h)  $ \langle \hat B^4 \rangle $ of (\ref{eq:non-pair-contractions}).}
\end{center}
\end{figure}

\subsection{The ``total current'' correlator 
$\langle \widehat N_+ (t) \widehat N_+ (0) \rangle' $}
\label{sec:NNcor}
\label{connectedGnn-NN}

\noindent
{\em We describe the dispute between Fabrizio \& Gogolin 
and  Oreg \& Finkelstein,
mentioned in Section~\ref{sec:OF}, regarding the 
calculation of the correlator $D_{\alpha_d} (t)$
in terms of the ``total current'' correlator
$\langle \widehat N_+ (t) \widehat N_+ (0) \rangle' $.
We believe that OF's critique of FG is unfounded;
to illustrate our view, we confirm FG's calculation
explicitly for $g=1/2$.\vspace*{2mm}}\/

To evaluate $D_{\alpha_d}(t)$ for general $g$, 
FG \cite{FG} exploited the fact that 
$\alpha_d = 
e^{i \pi \widehat {\cal N}}/\sqrt 2$
depends on the ``total current'' operator $2 \widehat {\cal N} 
= (\widehat N_L - \widehat N_R)$, which they call $J$.
They  concluded that $D_{\alpha_d}(t) \sim t^{-1/2g}$ (as explained below)
by citing the result 
$D_{N_+} (t) \equiv 
\langle \widehat  N_+ (t) \widehat  N_+ (0) \rangle'
\sim - (\ln t)/(2g\pi^2)$ from 
previous papers \cite{FGGGG,GogolinProkofev}.
In Ref.~\cite{FGGGG} FG had arrived at the latter result by assuming, following
the RG results of Kane and Fisher \cite{KF},  that 
{\em as far as current fluctuations are concerned}\/
(which is of course all that matters for $D_{N_+}$),
the effect of a backscattering impurity can
be mimicked by using ``open boundary conditions'',
$\Psi_{phys} (x=0) = \Psi_{phys} (x=L) = 0$,
since both suppress current fluctuations. 
[Recall that current and density fluctuations
are governed by  $\Phi_+$, $\widehat  N_+$
and $\Phi_-$, $\widehat  N_-$, respectively, since 
 $\tilde \rho_L (0) \mp  
\tilde \rho_R (0) = \sqrt{2} g^{\mp/2} \partial_x \Phi_\pm (x)|_{x=0} 
+ 2 \tpol \widehat N_\pm$, see (\ref{eq:phizero})].
In their Reply \cite{OF97}, OF objected to this assumption
of \cite{FGGGG} (without commenting  on \cite{GogolinProkofev}),
pointing out that 
``cutting the wire'' (i.e.\ open boundary conditions)
is {\em not}\/ fully equivalent to a backscattering impurity, 
since the latter affects only  current but {\em not the density
fluctuations}\/ 
at the impurity site. As a general statement, 
this assertion is certainly correct:
the density at $x=0$ is clearly unaffected by backscattering,
since $[H_B, \Phi_-] = 0$.
[Free density fluctuations, in fact, are responsible for the 
decay of $D_F (t) \sim t^{-1/2g}$ of (\ref{DFeval}).] 
Nevertheless,  in our opinion OF's critique 
is misguided, simply because {\em  $D_{\alpha_d}$
depends solely on current fluctuations}\/
(i.e.\ solely on $\Phi_+$ and $\widehat N_+$ --- though
in the field-theoretical
bosonziation formalism employed by FG and OF,
this fact is perhaps not as obvious as here);
but for the calculation of {\em current}\/  fluctuations
it is irrelevant whether {\em density}\/ fluctuations are present or not,
since the two types of fluctuations are completely decoupled
($[ \Phi_+, \Phi_- ] = 0$ and $[\widehat N_+, \Phi_- ] = 0)$. 
Therefore,  FG's strategy in Ref.~\cite{FGGGG} 
for finding $\langle \widehat  N_+ (t) \widehat  N_+ (0) \rangle'$
is sound.

To illustrate our view and confirm FG's results,  
we now calculated $D_{N_+}$ explicitly for $g=1/2$. 
Using (\ref{oldck}) and (\ref{alpha-tildealpha}),
 the number operator $\widehat N_+$ of (\ref{N_k})
(with $P=0$) can be written as 
\begin{equation}
  \label{eq:Nrefermionize}
 \widehat N_+ = \sum_{\bk >  0} i(\alpha_\bk^\dag \beta_\bk - 
\beta_\bk^\dag \alpha_\bk ) = \sum_{\bk} i(\alpha_{-\bk} \beta_{\bk}^\dag )=
\sum_{\bk \varepsilon} i A_{-\bk, \varepsilon}
\tilde \alpha_\varepsilon \beta_\bk \; .  
\end{equation}
Its correlator $D_{N_+} (t)$ [Fig.~\ref{fig:feynman}(c)]
can thus be evaluated as follows: 
\begin{eqnarray}
  \label{eq:DNNcor-1}
D_{N_+}(t)  &\equiv& \langle \widehat N_+ (t) \widehat N_+ (0) \rangle'  =
  \sum_{\bk \bk' \varepsilon \varepsilon'}
 A_{- \bk, \varepsilon} A^\ast_{- \bk',  \varepsilon'}
 \langle \tilde \alpha_\varepsilon(t)  \beta_\bk (t) 
\tilde  \beta_{\bk'}^\dag (0) \alpha_{\varepsilon'}^\dag  (0) \rangle'
\\
\label{eq:DNNcor-2}
&=& \sum_{\varepsilon, \bk \ge 0}  \theta (\varepsilon)
|A_{- \bk, \varepsilon}|^2
e^{-i(\varepsilon + \varepsilon_\bk)t} 
\label{eq:DNNcor-3int}
\; \stackrel{L \to \infty}{\longrightarrow} \; 
{c^2 \over \pi^2} 
P \!\!\!\!\!\! \int_{\Delta_L}^\infty \!\! d \varepsilon \; 
P \!\!\!\!\!\! \int_0^\infty \!\! d \varepsilon_\bk 
{ e^{-i(\varepsilon + \varepsilon_\bk)t} \over
(\varepsilon^2 + c^2) (\varepsilon + \varepsilon_\bk)^2}
\qquad \phantom{.}
\\
\label{eq:DNNcor-3}
& = & 
 \left\{ \begin{array}{l}
 - {1 \over \pi^2} \ln (\Delta_L/c ) \quad (t = 0 )
 \vspace{2mm} ;
\qquad \phantom{.}
\\
 - {1 \over  \pi^2} \ln (r \Delta_L i t) \quad ( c t \gg 1, \Delta_L t \ll 1)  .
\qquad \phantom{.}
\end{array} \right. 
  \end{eqnarray}
When taking the limit $L \to \infty$ using 
(\ref{eq:continuumlimit}), we cut off the low-energy
divergence in the double integral of (\ref{eq:DNNcor-3int})
by $\Delta_L = v 2 \pi / L$, 
the level spacing for finite $L$. 
 For $t=0$, the logarithmic 
divergence of $\langle \widehat N_+^2 \rangle'$ with
system size when  $L \to \infty$ reflects
the fact that due to backscattering $\widehat N_+$ is {\em not conserved}.
The divergence is sufficiently slow, however, 
that factors of order $\widehat N_+ / L$
{\em can}\/ be safely neglected when taking the continuum
limit (as done in Section~\ref{dos}).  
The simplest way to find the  asymptotic  $c t \gg 1$ result is
to first show that  $\partial_{t^2} D_{N_+} (t) \sim (it)^{-2}$
using (\ref{eq:asymtotics}),  then integrating 
twice w.r.t.\ $t$. This yields $-\ln (r \Delta_L i t)$,
where $r$ is a constant of order unity,
(and not, for example, $- \ln (r c i t)$),  
since it is $\Delta_L$ (not $c$) which cuts off the  low-energy divergence of 
$(\varepsilon + \varepsilon_\bk)^{-2}$ 
as long as $\Delta_L t \ll 1$, just as for the case $t=0$. 
The numerical value of $r$ depends on the precise 
way in which this infra-red cut-off is introduced.

These $g=1/2$ results confirm those   for general $g$ 
of Gogolin and Prokof'ev \cite{GogolinProkofev},
who found $\langle \widehat N_+^2 \rangle' \sim \ln (L c)$,
and those of Fabrizio and Gogolin \cite{FGGGG} 
for  $D_{N_+} (t) $ mentioned above.
FG used $D_{N_+}$ in  their Comment \cite{FG} on Oreg and Finkel'stein's work 
  to calculate the correlator
$D_{\alpha_d}$ for {\em general}\/ $g$, by essentially proceeding  as follows:
\begin{eqnarray}
  \label{eq:alphad-NN}
  D_{\alpha_d} (t) \equiv \langle \alpha_d (t) \alpha_d (0) \rangle' 
 = {\textstyle {1 \over 2}}
 \langle e^{i \pi \widehat N_+(t)} e^{-i \pi \widehat N_+(0)}
\rangle'  \; \mbox{``=''} \;  {\textstyle {1 \over 2}}
e^{\pi^2 [D_{N_+}(t)  - D_{N_+}(0)]}
\sim  {\textstyle {1 \over 2}} (r c i t)^{- 1 / 2g} \; .
\quad \phantom{.}
\end{eqnarray}
For $g = 1/2$ this yields $(ict)^{-1}$ behavior, in agreement 
with our  result 
$ D_{\alpha_d} (t) \sim (\pi c i t)^{-1}$ 
of (\ref{eq:alphaalphadref4}), 
but the numerical value of the prefactor (which FG did not specify)
is $1/2r$ instead of $1/\pi$, i.e.\ it depends 
on the way the infrared cut-off in (\ref{eq:DNNcor-3int})
was performed. Actually, since 
the identity (\ref{magicid}), which would 
make the ``='' in (\ref{eq:alphad-NN}) a true equality,
holds only for {\em free}\/ bosonic operators
(cf.\ the end of Appendix~\ref{app:checkV=-1}),
the prefactor in  (\ref{eq:alphad-NN})
will be further renormalized by additional
contributions, not contained in $e^{\pi^2[D_{N_+}(t) - D_{N_+} (0)]}$;
these will depend on $r$ but not on $t$ and 
must evidently conspire to change the prefactor from $1/2r$ to $1/ \pi$.

It is instructive to identify the nature of these additional
contributions. For a free bosonic operator $\hat B$,
the relation (\ref{Vexp}) can be 
proven as follows:
\begin{eqnarray}
  \label{eq:provemagiceasy}
 \langle e^{\hat B_0} \rangle = \sum_{n=0}^\infty
{1 \over 2n!} { \langle \hat B_0^{2n} \rangle}
= \sum_{n=0}^\infty {1 \over 2n!}
{(2n-1)! \over 2^{n-1} (n-1)!} \langle \hat B_0^2 \rangle^n
= \sum_{n=0}^\infty {1 \over 2^n n!}
 \langle \hat B_0^2 \rangle^n
=  e^{\langle \hat B_0^2 \rangle/2} .
\end{eqnarray}
For the first equality  we used  $\langle \hat B_0^{2n+1} \rangle = 0$,
and for the second evoked Wick's theorem to
reduce $\langle \hat B_0^{2n} \rangle$ to a sum 
of $(2n-1)(2n-3) \dots $ identical terms,
each equal to  $\langle \hat B_0^2 \rangle^n$.
Relation (\ref{magicid}) can be similarly proven,
though the combinatorics is more involved.

Now, the bosonic operators $\hat N_+$ and $\Phi_+$ are not
free, but both of the general form 
$\hat B = \sum_{\varepsilon \bk} B_{\varepsilon \bk}
\tilde \alpha_\varepsilon \beta_\bk$, with non-trivial
coefficients $ B_{\varepsilon \bk}$. When evaluating
$\langle\hat B^{2n} \rangle$ using Wick's theorem for
free fermions, one thus obtains two  types of contributions:
firstly, those 
containing only ``pairwise'' contractions, in which
{\em both}\/ operators from one pair $(\tilde\alpha \beta)$
are contracted with {\em both}\/ operators from 
another pair; this yields  
$(2n-1)(2n-3)$ times  $\langle \hat B^2 \rangle^n = 
[\sum_{\varepsilon, \bk \ge 0} | B_{\varepsilon \bk}|^2]^n $,
just as for free bosons.
Secondly, there are 
``non-pairwise'' contractions, in which $(\tilde\alpha \beta)$
is contracted with the $\tilde \alpha$ of one pair
and the $\beta$ of another, which
has no analogue for free bosons. For example, in 
\begin{eqnarray}
  \label{eq:non-pair-contractions}
  \langle \hat B^4 \rangle = 
  3 \biggl\{ \biggl[ 
\sum_{\varepsilon, \bk \ge 0} | B_{\varepsilon \bk}|^2 \biggr]^2
+
\sum_{\varepsilon, \bk, \varepsilon', \bk' \ge 0 }
\biggl[ 
 B_{\varepsilon \bk} B_{- \varepsilon \bk'} B^\ast_{-\varepsilon' \bk'}
 B^\ast_{\varepsilon' \bk}
+
 B_{\varepsilon \bk} B_{- \varepsilon \bk'} B^\ast_{-\varepsilon' \bk}
 B^\ast_{\varepsilon' \bk'} 
\biggr] \biggr\}
\end{eqnarray}
the first term arises from pairwise contractions,
the second and third from non-pairwise contractions
[see Fig.~\ref{fig:feynman}(h)].
It follows that in general 
\begin{eqnarray}
\label{non-magic}
 \langle e^{\hat B} \rangle =  
e^{\langle \hat B^2 \rangle/2} + \dots \; , \quad \qquad
 \langle e^{\hat B} e^{\hat B'} \rangle =  
e^{\langle \hat B \hat B' + (\hat B^2 + \hat B^{\prime 2})/2 \rangle}
 + \dots \; , 
\end{eqnarray}
where on the right-hand sides the exponentials and dots
arise from pairwise and non-pairwise contractions, respectively.
The latter will change the numerical value of the  pre\-factor
(\ref{eq:alphad-NN}), but not the leading asymptotic
behavior $t^{-1}$. (It is  straightforward but cumbersome
to check this, by evaluating, for example,
the non-pairwise contracted terms for 
$\langle \widehat N_+^2 (t) \widehat N_+^2 (0) \rangle$.)

\subsection{Checking that $D_{\alpha_d V_{-1}}(t) \sim (it)^{-1}$}
\label{app:checkV=-1}

\noindent
{\em We check 
 the result $D_{\alpha_d V_\lambda} (t) \sim (it)^{-1}$
 for  $\lambda = -1$,
by  relating $D_{\alpha_d V_{-1}}$
to  $D_\Psi \equiv \langle \Psi_+(t) \Psi_+^\dag(0) \rangle'$
and calculating the latter explicitly.\vspace*{2mm}}
\begin{eqnarray}
\nonumber 
 D_{\alpha_d V_{-1}} (t) 
&=&  { 1 \over a}  \langle e^{ i H_+' t}
\alpha_d e^{ - i  \Phi_+ } e^{ - i H_+' t}
 e^{ i  \Phi_+} \alpha_d \rangle'
\\
&=& {1 \over 2 a} \langle e^{ i H_+' t}
F_+ e^{-i \Phi_+  } e^{ - i H_+' t}
 e^{ i  \Phi_+ } F_+^\dagger \rangle'
=   {\textstyle {1 \over 2}}  D_\Psi (t) 
\; . 
\label{V1=psi}
\end{eqnarray}
We used  (\ref{eq:aHa}) and (\ref{eq:FHF}) to trade $\alpha_d H_+'
\alpha_d$ for ${ 1 \over 2} F_+ H_+' F_+^\dagger$, 
then used (\ref{eq:refermionize})
to identify \linebreak
$a^{-1/2} F_+ e^{-i \Phi_+}$ as $\Psi_+$. The correlator
$D_\Psi$ [Fig.~\ref{fig:feynman}(a,b)] can be  evaluated as follows:
\begin{eqnarray}
  \label{eq:psipsiref1}
\lefteqn{
D_\Psi (t)  \equiv  \langle \Psi_+ (t) 
\Psi_+^\dagger (0) \rangle'  \, = \,
\tpol  \sum_{\bk \bk'} 
\half \langle \left( \alpha_\bk (t)  + i \beta_{\bar k} (t) 
\right) \left( 
\vphantom{ \alpha^\dagger_\bk }
\alpha_{\bar k'}^\dagger (0) - i \beta_{\bar k'}^\dagger (0) \right)
\rangle' } \\  
\label{eq:psipsiref2}
&=& \!\! {\Delta_L \over 2 v} \Biggl[  D_\beta(t) + 
\sum_{\varepsilon \ge 0} 
e^{- \varepsilon (i t + a/v)} \theta (\varepsilon)  
 \sum_{\bk \bk'} A_{\bk,  \varepsilon} A^\ast_{\bk',  \varepsilon}
\Biggr] \\
  \label{eq:psipsiref3}
  \label{eq:psipsiref4}
& = & \! \! {1 \over 2 v} \left[ D_\beta (t) + 
\!  \int_0^\infty \! d \varepsilon \,
{ e^{- \varepsilon (i t + a/v)}
\varepsilon^2 \over \varepsilon^2 + c^2}
\right]
\, = 
 \left\{ \begin{array}{l}
  {1 \over a} ( 1 - \pi a c/4) 
\quad (t = 0 , \;  c a \ll 1)
 \vspace{2mm} ;
\qquad \phantom{.}
\\
 {1 \over  2 i v t} \left[ 1 + {\cal O} \left( {1 \over c^2 t^2  }
\right) \right]
\quad (\Gamma t \gg 1)  .
\end{array} \right. 
  \label{eq:psipsiref5} 
  \label{eq:psipsiref6}
\end{eqnarray}
To obtain \Eq{eq:psipsiref1}, (\ref{eq:psipsiref2}) and (\ref{eq:psipsiref3})
 we used, respectively,
the first of \Eqs{eq:refermionize} and 
(\ref{oldck}) for $\Psi_+$,
 (\ref{eq:betacorrelator}) for $D_\beta$,  
(\ref{alpha-tildealpha}) and (\ref{finalAen}) for $\alpha_{\bar k} (t)$
[with $c \equiv 4 \pi \Gamma$], (\ref{eq:eigenvaluesmain})
to do the $\sum_\bk$ sums in (\ref{eq:psipsiref2}),
then (\ref{eq:continuumlimit}) to take the
continuum limit, and (\ref{eq:asymtotics}) for the asymptotic integral. 
 
Eqs.~(\ref{V1=psi}) and (\ref{eq:psipsiref6}) together evidently
confirm that $D_{\alpha_d V_{-1}} (t) \sim (it)^{-1}$. 
Moreover, they also yield the leading prefactor of $ D_{\alpha_d V_{-1}}$
in (\ref{alphavertex-result}), 
namely $C_{-1} = 1 /4$. This implies, 
via  (\ref{eq:alphavertexcorrel-const}) and (\ref{eq:alphaalphadref4})
that $a^{-1}   \langle  e^{i \lambda \Phi_+(t) } 
e^{-i \lambda \Phi_+(0)} \rangle'$ is proportional to $\Gamma$,
which is consistent with what we would get 
from $a^{- 1} e^{-  \langle \Phi_+(0,0)^2 \rangle' } = 
(e^\gamma 4 \pi \Gamma /v )$
[by (\ref{scatteredbosoncor})].
That the numerical prefactors obtained from 
$a^{- 1} e^{-  \langle \Phi_+(0,0)^2 \rangle' }$ 
and  $a^{-1}   \langle  e^{i \lambda \Phi_+(t) } 
e^{-i \lambda \Phi_+(0)} \rangle'$
differ is of course no surprise,
since according to the 
discussion preceding Eq.~(\ref{non-magic})
these two quantities are  identically equal
only for free boson fields.

Incidentally, it is straightforward to check that
the result $D_{\alpha_d V_1} \sim (it)^{-1}$ can also be 
derived by writing $\alpha_d e^{-i \Phi_+} = 
e^{-i (\pi \widehat N + \Phi_+)}$ 
and using (\ref{non-magic}), with $\hat B = 
\pi \widehat N + \Phi_+$.

\subsection{Leading connected contributions to 
$\langle \alpha_d (t) \Phi_+^n(t) \Phi_+^{n'} (0)
\alpha_d(0) \rangle'$}

\label{app:connected}
\label{connectedGnn}

\noindent
{\em The result  $D_{\alpha_d V_\lambda} \sim (it)^{-1}$
of (\ref{alphavertex-result}) rests on  the fact
that each ``connected'' contribution
to the correlator 
$D_{nn'}(t) \equiv {i^{n - n'} \over n! n'!}
\langle \alpha_d (t) \Phi_+^n(t) \Phi_+^{n'} (0)
\alpha_d(0) \rangle'$  occuring in (\ref{eq:alphavertexcorrel-expand})
asymptotically decays at least as fast as $1/t$
(most decay much faster), since it 
contains at least one contraction between two operators at 
times $t$ and  0, which yields at least one factor of $1/t$.
Here we illustrate this by considering 
the leading connected contributions to $D_{11}$,        $D_{20}$
 and $D_{22}$ explicitly\vspace*{2mm}.}

\label{f:I1I2}
Apart from  (\ref{eq:asymtotics}),
the following integrals will be found useful in the
asymptotic evaluation of $D_{nn'}$:
\begin{eqnarray}
\label{I_1}
  I_1 &=& 
P \!\!\!\!\!\! \int_0^\infty \!\! d \varepsilon
{\varepsilon \over (\varepsilon + {\varepsilon'} ) (\varepsilon^2 + c^2) }
= {c \over {\varepsilon'}^2 + c^2}
\left[ 
{\pi \over 2 } + {{\varepsilon'} \over c } \ln \left( { | {\varepsilon'} | \over c} \right) 
\right] \; , 
\\
\label{I_2}
  I_2  &=&
P \!\!\!\!\!\! \int_0^\infty \!\! d \varepsilon
{c \over (\varepsilon + {\varepsilon'} ) (\varepsilon^2 + c^2)}
= {c \over {\varepsilon'}^2 + c^2}
\left[
- \ln \left( { | {\varepsilon'} | \over c} \right) + {\pi {\varepsilon'} \over 2 c} 
 \right] \; ,
\\
\label{I_3}
I_3 &=& P \!\!\!\!\!\! \int_0^\infty \!\! d \varepsilon
{\varepsilon^n [\ln(\varepsilon / \bar c)]^{\bar n} e^{- \varepsilon (i t +a)} 
 \over (\varepsilon^2 + c^2)^m} \, \sim \, 
 {  n! [- \ln | \bar c t|]^{\bar n} \over c^{2 m} (i t)^{n + 1} }  \;
 , 
 \end{eqnarray} 
where $I_3$ assumes $ n,\bar n, m \ge 0$ and integer, and $t/a , t c \gg 1$.

The leading connected contributions to $D_{11}$ 
 [Fig.~\ref{fig:feynman}(d)] are evaluated as follows:
\begin{eqnarray}
  \label{eq:G11}
D_{11}(t) & \equiv& 
 \langle \alpha_d (t) \Phi_+ (t) 
\Phi_+ (0) \alpha_d  (0) \rangle' 
\\ &\sim&
\sum_{\bar k \bar k' \varepsilon \varepsilon'
\bar \varepsilon \bar \varepsilon'} 
A_{d \varepsilon}
\Phi_{\bar k , \bar \varepsilon} \Phi^\ast_{\bar k' , \bar \varepsilon'}
A^\ast_{d \varepsilon'}
\langle \Bigl( \tilde \alpha_{\varepsilon} 
\beta_{\bar k} \tilde \alpha_{\bar \varepsilon}\Bigr)  (t)
\Bigl( \tilde \alpha_{\bar \varepsilon'}^\dagger  \beta_{\bar k'}^\dagger 
\tilde \alpha^\dagger_{\varepsilon'}\Bigr) (0) \rangle
\\
&= &
\label{G11lead+sub}
\sum_{\bar k, \varepsilon, \varepsilon' > 0} 
 e^{-i \bar k t }
A_{d \varepsilon}
\left[  \Phi_{\bar k, - \varepsilon} \Phi^\ast_{\bar k, - \varepsilon'}
-  e^{-i (\varepsilon + \varepsilon')t }
 \Phi_{\bar k, \varepsilon'} \Phi^\ast_{\bar k ,\varepsilon} \right]
A^\ast_{d \varepsilon'} \; . 
\end{eqnarray}
Among the contractions that produced the first or second terms
of (\ref{G11lead+sub}),
there were one or three $t$-to-0
contractions (i.e.\ connecting operators at $t$ and 0), respectively, 
yielding one or three oscillatory factors, respectively.  Thus
the first term is the one that decays slower when $t \to \infty$; to 
determine its asymptotics, we first
do the sums on $\varepsilon$ and $\varepsilon'$
for a slightly more general expression (which occurs in the
leading terms of all $D_{nn'}$):
\begin{eqnarray}
  \label{eq:Fkk}
\lefteqn{ 
F_{\bar k, \bar k'} \equiv 
\sum_{\varepsilon, \varepsilon' > 0}
A_{d \varepsilon}
\Phi_{\bar k, - \varepsilon} \Phi^\ast_{\bar k', - \varepsilon'}
A^\ast_{d \varepsilon'}  }
\\ &=&
\nonumber
P \!\!\!\!\!\! \int_0^\infty \!\! d \varepsilon
\, d  \varepsilon' 
{ \Delta_L \, c \, \varepsilon \, \varepsilon' 
\over
  \pi [\varepsilon^2 + c^2 ]   [{\varepsilon'}^2 + c^2 ]}
\left[   { 1 \over \varepsilon_{\bar k} - \varepsilon } 
- {\pi \varepsilon \over c} \delta( \varepsilon_{\bar k} - \varepsilon )
\right]
\left[ { 1 \over  \varepsilon_{\bar k'}  - \varepsilon' } 
- {\pi \varepsilon' \over c} \delta(  \varepsilon_{\bar k'}  - \varepsilon' )\right]
\\ & \sim &
{\pi \Delta_L c^3   \over
4 \, [ \varepsilon_{\bar k}^2 + c^2] [ \varepsilon_{\bar k'}^{2} + c^2]}
\left[ 1 + {\cal O}\left(  \varepsilon_{\bar k}  \over c \right) +
{\cal O}\left(  \varepsilon_{\bar k'}  \over c \right) \right] \; . 
   \end{eqnarray}
After using $I_1$ of (\ref{I_1}) twice to
perform the  double integral, 
we retained only the term with the lowest powers of $ \varepsilon_{\bar k} /c$
 and $ \varepsilon_{\bar k'} /c$,
 since [by (\ref{eq:asymtotics})] it is the one giving 
the leading asymptotic behavior
for $D_{11}$:
\begin{eqnarray}
  \label{eq:G11final}
D_{11}(t) \sim \sum_{\bar k } F_{\bar k ,  \bar k}  e^{-i  
\varepsilon_{\bar k}  t}  
\sim
 \int_0^\infty \!\! d  \varepsilon_{\bar k}   \, 
e^{-i \varepsilon_{\bar k} t}  
{\pi c^3  \over
 4  \, [ \varepsilon_{\bar k}^2 + c^2 ]^2 }
\sim {\pi  \over  4 c i t }  \; . 
\end{eqnarray}
Thus leading term in 
$D_{11} (t)$ decays just as fast as the disconnected terms
proportional to $D_{00} = D_{\alpha_d}(t) \sim 1/(cit)$, and hence its prefactor
contributes to the prefactor $C_\lambda$ in (\ref{alphavertex-result})
for $D_{\alpha_d V_\lambda}$.
For the second term of (\ref{G11lead+sub}),
we do the $\varepsilon_{\bar k}$ before the $\varepsilon, \varepsilon'$
integrals (using  (\ref{eq:asymtotics}) for all three integrals),
obtaining
\begin{equation}
  \label{eq:G11-subleading}
P \!\!\!\!\!\! \int_0^\infty \!\! d \varepsilon \, d \varepsilon' \, d 
\varepsilon_{\bar k}
\, {c \, e^{-i (\varepsilon_{\bar k} + \varepsilon + \varepsilon')t}
\over
 \pi \,
 [\varepsilon^2 + c^2 ]   [{\varepsilon'}^2 + c^2   ]}
 \left ( {\varepsilon'  \over \varepsilon_{\bar k} + \varepsilon'} \right) 
 \left ( {\varepsilon  \over \varepsilon_{\bar k} + \varepsilon} \right) 
\; \sim \; {1 \over \pi (i c t)^3}  \, .
\end{equation}
This  illustrates that the more contractions there
are between times  $t$ and 0, the more powers of $1/(c t)$ 
are produced. 

Next we consider $D_{20}$  [Fig.~\ref{fig:feynman}(e)]
, whose leading connected term differs from
that of $D_{11}$ only in the oscillatory factor
(the $1/2!$ is cancelled by a combinatorical factor $2!$):
\begin{eqnarray}
  \label{eq:G20-1}
D_{20}(t) & \equiv & 
{1 \over 2!}
  \langle \alpha_d (t) \Phi_+^2 (t) \alpha_d  (0) \rangle' 
\sim 
{2! \over 2!} \sum_{\bar k, \varepsilon, \varepsilon' > 0} 
A_{d \varepsilon}
\Phi_{\bar k, - \varepsilon} \Phi^\ast_{\bar k, - \varepsilon'}
A^\ast_{d \varepsilon'} \,  e^{-i \varepsilon' t }
\\
\label{eq:G20-2}
& \sim &
P \!\!\!\!\!\! \int_0^\infty \!\! 
\, d  \varepsilon'  \, d {\varepsilon_\bk} \, d  \varepsilon
{c \, \Delta_L  \, \varepsilon \, \varepsilon' \,  e^{-i \varepsilon' t }
\over  \pi \,
 [\varepsilon^2 + c^2 ]   [{\varepsilon'}^2 + c^2 ]
[{\varepsilon_\bk} - \varepsilon ][{\varepsilon_\bk} - \varepsilon']}
\\ &\sim &
\label{eq:G20-3}
P \!\!\!\!\!\! \int_0^\infty \!\! d \varepsilon' \, 
{c^2 \, \varepsilon' \, \ln|\varepsilon'/c|  e^{-i \varepsilon' t} 
\over 2 \, [{\varepsilon'}^2 + c^2]^2 }
\, \sim \, - {\ln | c t | \over 2 (c i t)^2} \; . 
\end{eqnarray}
To obtain (\ref{eq:G20-3}) we did, in that order,
the $\varepsilon$,  ${\varepsilon_\bk}$ and $\varepsilon'$
integrals, using $I_1$, $I_2$ and $I_3$ of 
(\ref{I_1}), (\ref{I_2}) and (\ref{I_3}), 
respectively, keeping at each step only the 
asymptotically leading term. Evidently, $D_{20}$ decays
faster than $D_{00} = D_{\alpha_d}$ by a factor $\ln |c t|/(c i t)$.

Finally, we consider the leading contribution to 
$D_{22}$  [Fig.~\ref{fig:feynman}(f)], namely
\begin{eqnarray}
  \label{eq:G22-1}
D_{22}(t) & \equiv & 
{1 \over (2!)^2}
  \langle \alpha_d (t) \Phi_+^2 (t) \Phi_+^2(0)  \alpha_d  (0) \rangle' 
\sim 
{(2!)^2  \over (2!)^2} 
\sum_{\bar k ,\bk' , \varepsilon > 0} 
F_{\bk \bk'}
\Phi_{- \bar k, \varepsilon} \Phi^\ast_{- \bar k' ,  \varepsilon}
 \,  e^{-i \varepsilon t }  
\qquad \phantom{.}
\\ &\sim & 
\label{eq:G22-2}
P \!\!\!\!\!\! \int_0^\infty \!\! d \varepsilon \, 
{\pi \, c^3 \, \varepsilon^2 \, (\ln|\varepsilon/c|)^2  e^{-i \varepsilon t} 
\over  4 \, [\varepsilon^2 + c^2]^3 }
\, \sim \,  {\pi (\ln | c t |)^2  \over 4 \, (c i t)^3} \; . 
\end{eqnarray}
The $F_{\bk \bk'}$ in  (\ref{eq:G22-1}) arises in the
same way as in (\ref{G11lead+sub}); the ${\varepsilon_\bk}$, 
$\varepsilon_{\bk'}$ 
integrals and the $\varepsilon$ integral can be done
with $I_2$ and $I_3$ of (\ref{I_2}) and (\ref{I_3}), respectively.

These examples illustrate that the integrals that have to
be done rapidly become very complicated  when $n,n'$ increase,
so that  it would be a daunting
task to give a general formula for the leading
asymptotic behavior of $D_{nn'}$. However, it 
also is evident that the leading
term will always decay as least as fast as $\sim (it)^{-1}$,
simply because it always contains at least one $t$-to-0 contraction.\footnote{
The only exception to this rule occurs for 
$D_{N+}$ of (\ref{eq:DNNcor-1}), for which
two $t$-to-0 contractions yield $\ln t$; the reason
why $D_{N_+}$ is special is that it contains
a factor $|A_{-\bk, \varepsilon}|^2 \sim (\varepsilon + \bk)^{-2}$,
which produces an infrared divergence leading to
$\ln t$. In contrast, however, the coefficients occuring
when $\langle \alpha_d(t) \dots ... \alpha_d (0) \rangle$
is involved 
are less infrared divergent (since $A_{d, \varepsilon} \to $ const
for $\varepsilon \to 0$).}

\subsection{Leading contributions to $D_{LR}(t)$}
\label{app:leadingPsiPhialpha} 

\noindent
{\em
We asymptotically evaluate  the leading terms arising 
when $D_{LR}(t)$ of (\ref{eq:LR-RLcorrelation2})
is expanded in powers of $\Phi_+$, namely 
$\langle \Psi_+(t) \alpha_d (0) \rangle' \sim (it)^{-2}$
and $\langle \Psi_+(t) [\Phi_+(0) - \Phi_+(t) ]\alpha_d (0) \rangle'
 \sim (it)^{-1}$.\vspace*{2mm}}

The calculation of the first of these  [Fig.~\ref{fig:feynman}(a)]
is  analogous
to that of $D_\Psi (t)$ of (\ref{eq:psipsiref1}): 
\begin{eqnarray}
  \label{eq:psialphadref1}
\lefteqn{
D_{\Psi \alpha_d} (t)  
\equiv \sqrt{2/a} \, \langle \Psi_+ (t ) i \alpha_d (0)  \rangle'
 = 
\sqrt {2 \pi / a L} \sum_{\bk } 
\langle \left[ \alpha_\bk (t) + i \beta_{\bar k} (t) \right]
i \alpha_d (0)
\rangle'} \\  
\label{eq:psialphadref2}
&=& i  \sqrt {\Delta_L  / v a} 
\sum_{\varepsilon } e^{- \varepsilon (i t +a/v)}
  \theta (\varepsilon)  
 \sum_{\bk} A_{\bk,  \varepsilon} A^\ast_{d, \varepsilon}
 \langle \tilde \alpha_\varepsilon
\tilde \alpha_\varepsilon^\dag  \rangle'
 \\
  \label{eq:psialphadref4}
&=&  \! \! -  \sqrt{{c \over \pi a v}} 
  \int_0^\infty \! d \varepsilon \,  
{ e^{- \varepsilon (i t +a/v)} \, 
\varepsilon  
\over  \varepsilon^2 + c^2}
  \label{eq:psialphadref5}
= \left\{ \begin{array}{l}
   \! \sqrt{{c \over \pi a v}} \ln\left( e^\gamma c a /v  \right) 
\qquad (t = 0 , \;  c a \ll 1) \vspace{1mm}
; \qquad \phantom{.}
\\
\label{psialphaasym}
  \!  \sqrt{{c  \over \pi a v}} 
{ 1 \over (c t)^2} \left[ 1 + 
{\cal O} \left( { 1 \over c t}  \right) \right] 
\quad ( c t \gg 1 ) .
\end{array} \right. 
\end{eqnarray}
Here $\gamma = 0.577 \dots$ is Euler's constant, and the last line's 
asymptotic $ct \gg 1$ result 
follows from (\ref{eq:asymtotics}). 
Note that the non-zero result  for  $D_{\Psi \alpha} (t=0)  $ 
implies by (\ref{eq:h+prime}) 
that $\langle H_B' \rangle' \neq 0$, as 
expected. 

Next we consider the correlator $D_{\Psi \Phi_+ \alpha_d} (t)  $, 
which is linear in $\Phi_+$
 [Fig.~\ref{fig:feynman}(g)];
it  is non-zero, since
the $\beta$ 
in $\Phi_+$ can be contracted  with that in $\Psi_+$:
\begin{eqnarray}
  \label{eq:psiPhialphadref1}
\lefteqn{
D_{\Psi \Phi_+ \alpha_d} (t)  
\equiv 
 \sqrt{2/a} \, \langle \Psi_+ (t ) 
\left[ \Phi_+(0) - \Phi_+ (t) \right]  \alpha_d (0)  \rangle'}
 \\
&\sim &
\label{psiPhialpha-2}
\sqrt {\Delta_L  / v a} \sum_{\bk \bk' \varepsilon \varepsilon' } 
i \, \Phi_{-\bk', \varepsilon'}   A_{d, \varepsilon}^\ast
 \langle  \beta_{\bar k} (t)
 \left[
(\beta^\dag_{\bk'} \tilde \alpha_{\varepsilon'} )(0)
- (\beta_{\bk'}^\dag \tilde \alpha_{\varepsilon'} )(t) \right]
 \alpha^\dag_\varepsilon (0) \rangle'
\\
\label{psiPhialpha-3}
&\sim&  \! \! -  \sqrt{{c \over \pi a v}} 
P \!\!\!\!\!\! \int_0^\infty \!\! d \varepsilon \, d \varepsilon_\bk \,
\left( e^{- i\varepsilon_\bk  t } - e^{- i\varepsilon  t } \right)
\,  {\varepsilon  
\over  \varepsilon^2 + c^2}
\left[   { e^{- | \varepsilon - \varepsilon_{\bar k} |a/2v}
 \over \varepsilon - \varepsilon_{\bar k} } 
\, + \, {\pi \varepsilon \over c} \delta( \varepsilon - \varepsilon_{\bar k} )
\right]
\qquad \phantom{.}
\\
\label{psiPhialpha-4}
&\sim&  \! \! -  \sqrt{{c \over \pi a v}} \left\{
P \!\!\!\!\!\! \int_0^\infty \!\!  d \varepsilon_\bk \,
 {e^{- i\varepsilon_\bk  t } \, c \over 
\varepsilon_\bk^2 + c^2} \left[{\pi \over 2 }-
{ \varepsilon_\bk \over c} \ln \left| {\varepsilon_\bk \over c} \right| 
\right]
\, - \, 
P \!\!\!\!\!\! \int_0^\infty \!\! d \varepsilon \, 
{e^{- i\varepsilon  t } \, \varepsilon \, \ln | \varepsilon a / 2v | 
 \over  \varepsilon^2 + c^2} 
\right\}
\qquad \phantom{.}
\\ 
\label{psiPhialpha-5} & \sim &
   \! - \sqrt{{ \pi \over c a v}} \, 
{ 1 \over (2 i t)} \left[ 1 + {2 \ln ( 2 c v t^2 /a ) \over
\pi c i t}  \right] 
\qquad ( c t \gg 1 ) .
\end{eqnarray}
The first term of  (\ref{psiPhialpha-4}) was obtained
from the $e^{-i \varepsilon_k t}$ term
 of (\ref{psiPhialpha-3}) by doing the 
$\varepsilon $
integral using $I_1$ of (\ref{I_1}), the second 
term of  (\ref{psiPhialpha-4}) was obtained
from the $e^{-i \varepsilon t}$ term
 of (\ref{psiPhialpha-3}) by doing the 
$\varepsilon_\bk $
integral using $\int_0^\infty \! d \varepsilon 
e^{-a \varepsilon}/(\varepsilon - c) = - \ln |a c|$ for $a c \ll 1$.
Eq.~(\ref{psiPhialpha-5}) follows from (\ref{psiPhialpha-4}) 
by using (\ref{eq:asymtotics}) for the leading term and 
$I_3$ of (\ref{I_3}) for the logarithmic terms. 
-- Remarkably, the   $(it)^{-1}$ decay of 
$D_{\Psi \Phi_+ \alpha_d}  $
is  slower than the $(it)^{-2}$ of $D_{\Psi \alpha_d}$.
The reason is that  the coefficients $C_\bk$ in its
$\sum_\bk C_\bk \langle \beta_\bk (t) \beta_\bk^\dag (0) \rangle'$
contraction contain less powers of $\varepsilon_\bk$ than the 
powers of $\varepsilon$ contained in the coefficient
$C_\varepsilon$  arising 
in the contraction $\sum_\varepsilon C_\varepsilon
\langle \tilde \alpha_\varepsilon(t)
\tilde \alpha_\varepsilon^\dag (0) \rangle'$ in 
(\ref{eq:psialphadref2}) for $D_{\Psi \alpha_d}$.
[A similar observation applies for the $\beta $ and
$\tilde \alpha$ contributions to $D_\Psi$ of (\ref{eq:psipsiref1}).]

\section{Coulomb gas representation for $D_B$}
\setcounter{equation}{0}
\renewcommand{\theequation}{\Alph{section}\arabic{equation}}
\label{app:OF}

\noindent
{\em 
We rederive Oreg and Finkel'stein's \cite{OF}
exact mapping of the  correlator $D_B$ of (\ref{DLLLRRR})
onto a difference of Coulomb gas partition
functions, $D_B = Z_e - Z_o $,
and confirm that their 
treatment of fermionic anti-commutation relations was
correct, contrary to recent suggestions in the literature \cite{FG}.
\vspace*{2mm}}

In this Appendix we  write $F_{\pm} \equiv F_{L/R} \equiv 
F_\nu$,  where $\nu = (L,R) = (+,-)$
(in contrast to our refermionization notation
$F_+ \equiv F^\dag_R F_L$ of Section~\ref{fsdiagonalization}). 
The backscattering term  $H_B$ of (\ref{H_B}), 
with $\theta_B = 0$ for simplicity,
and the  $T=0$, imaginary-time version
of the correlator $D_B$ of (\ref{DLLLRRR}),
with $\tau = it$ and $\tau \in [0, \infty]$,  then read 
\begin{eqnarray}
  \label{eq:newH_B}
      H_B &=&  {v \lambda_B \over 2 \pi a}
\sum_{\nu = \pm} F_\nu^\dagger F_{-\nu} e^{i \nu c \Phi_+ } \, , 
\qquad \mbox{with} \quad c \equiv \sqrt{2 g} \; , 
\\
      D_B (\tau) &=& {1 \over a}  \sum_{\nu_0, \nu_\tau  = \pm}
      \langle G_B| e^{ (H_{0+} +H_B) \tau} 
        F_{-\nu_\tau} e^{{i \over 2} \nu_\tau c \Phi_+}
       e^{ -(H_{0+} +H_B) \tau}
        F_{\nu_0}^\dag e^{{i \over 2} \nu_0 c \Phi_+}
         | G_B \rangle      
\\
     \label{eq:expandDB-1}
&=&      
{\sum_{\nu_0, \nu_\tau  = \pm}
     \langle 0_+ | {\cal T} \left\{
      e^{- \int_0^\infty d \tau' \, H_B(\tau')}
      F_{-\nu_\tau} (\tau)  e^{{i \over 2} \nu_\tau c \Phi_+ (\tau)}
   F_{\nu_0}^\dag (0) e^{{i \over 2} \nu_0 c \Phi_+}(0) \right\} 
         | 0_+ \rangle       
\over 
a \, \langle 0_+ | {\cal T} 
      e^{- \int_0^\infty d \tau' \, H_B(\tau')} |0_+\rangle }\; . 
         \qquad \phantom{.}
\end{eqnarray}
In (\ref{eq:expandDB-1}) we wrote $D_B$ in
the $T=0$, imaginary-time interaction representation \cite{Negele}, 
in which  $\Phi_+ (\tau) = e^{H_0 \tau} \Phi_+ e^{- H_0 \tau}$
and $F_\nu (\tau) =  F_\nu$ (from (\ref{Kleintime}), with
$1/L$ terms neglected), and $|0_+\rangle$ is the ground
state of $H_{0+}$. Expanding $D_B$ in powers of
$H_B$ and keeping only connected terms (since disconnected ones
are cancelled by the denominator), we readily obtain
\begin{eqnarray}
     D_B (\tau) 
     \label{eq:expandDB-2}
        \!  &=& \!\! 
     {1 \over a } \sum_{n = 0}^\infty 
     \left( {v \lambda_B \over 2 \pi a}\right)^n
       \!\! \int_0^\infty \!\!\! d \tau_1 
       \!\!     \int_0^{\tau_1} \!\!\! d \tau_2 \dots  \!\! 
    \int_0^{\tau_{n-1}} \!\!\! d \tau_n \!\!
       \sum_{  \nu_1 \dots \nu_n, \nu_\tau \nu_0  = \pm}
    \!\!    D_{ \nu_1 \dots \nu_n, \nu_\tau \nu_0}^{
            \tau_1 \dots \tau_n, \tau} 
      S_{ \nu_1 \dots \nu_n, \nu_\tau \nu_0}^{
            \tau_1 \dots \tau_n, \tau} 
\qquad \phantom{.}
\\
     \label{eq:expandDB-3}
       D_{ \nu_1 \dots \nu_n, \nu_\tau \nu_0}^{
            \tau_1 \dots \tau_n, \tau} 
\! & \equiv & \! 
\langle 0_+ | {\cal T} \biggl\{ e^{i \nu_1 c \Phi_+ ( \tau_1) } \dots
       e^{i \nu_n c \Phi_+ ( \tau_n) } 
\,     e^{{i \over 2} \nu_\tau c \Phi_+ (\tau)}
       e^{{i \over 2} \nu_0 c \Phi_+ (0)} \biggr\} 
     | 0_+ \rangle 
\\
\nonumber
& = & \!    
      \left( {2 \pi a \over L} \right)^{{1 \over 2} Q^2}
 \exp \, c^2 \biggl\{ {\textstyle {1 \over 4}} \nu_\tau \nu_0 
\ln\left( | \tau|/a + 1\right) +
     \sum_{i=1}^n  {\textstyle  {1 \over 2}} \nu_0 \nu_i 
\ln\left( | \tau_j|/a + 1\right) 
 \biggr.
\\
     \label{eq:expandDB-4}
& & \qquad \biggl. + \sum_{i=1}^n  {\textstyle  {1 \over 2}} \nu_\tau \nu_i  
\ln\left( | \tau - \tau_j|/a + 1\right)
+ \sum_{i < j}^n
 \nu_i \nu_j \ln\left( | \tau_i - \tau_j|/a + 1\right) \biggr\}
\\
     \label{eq:expandDB-5}
  Q & \equiv & \sum_{j = 1}^n \nu_j +  \half \nu_\tau + \half \nu_0 
\\
     \label{eq:expandDB-7}
      S_{ \nu_1 \dots \nu_n, \nu_\tau \nu_0}^{
            \tau_1 \dots \tau_n, \tau} 
     \! & \equiv & \! 
      (-1)^n \langle 0_+| {\cal T} \left\{
        \left(F^\dag_{\nu_1} F_{- \nu_1} \right)(\tau_1)  \dots
        \left(F^\dag_{\nu_n} F_{- \nu_n} \right)(\tau_n) \, 
        F_{-\nu_\tau }( \tau)  F_{\nu_0}^\dag (0) 
\right\} |0_+ \rangle 
\qquad \phantom{.}
\\
     \label{eq:expandDB-8}
     & = & \!  (-1)^{N_\tau} \delta_{Q,0} 
\end{eqnarray}
In (\ref{eq:expandDB-2}) we exploited the fact
that all boson fields  commute with all  Klein factors to
factorize each term into
two factors, $D$ and $S$, that depend only on 
$\Phi_+$'s and $F$'s, respectively.
$D$ of (\ref{eq:expandDB-3})
is a time-ordered expectation value of exponentials of 
free boson fields, which gives (\ref{eq:expandDB-4}) when
evaluated using 
(\ref{EEEepx}) [analogous to our derivation of (\ref{vertexproduct})].
The argument of the exponential  in 
(\ref{eq:expandDB-4})
can be interpreted as the potential energy
of a ``1-D Coulomb gas of charged particles'',
interacting with a logarithmic inter-particle
potential  $c^2 \nu_1 \nu_2 \ln (| \tau_1 - \tau_2|/a + 1)$,
which has two charges $\half \nu_0$ and  $\half \nu_\tau$
 placed at positions 0 and $\tau$, 
and $n$ further charges $\nu_1, \dots, \nu_n$  at
positions  $\tau_j \in (0,\infty)$, where all $\nu \in \pm$. 
The total charge of the configuration
is $Q$ of (\ref{eq:expandDB-5}), and 
since the $Q$-dependent prefactor in (\ref{eq:expandDB-8}) is
non-zero in the limit $a/L \to 0$ only if $Q=0$,
only  ``neutral'' configurations of the Coulomb gas 
contribute.

This also follows from the Klein factor correlator 
 $S$ of (\ref{eq:expandDB-7}),
 which  by inspection is  non-zero only if it contains
as many $F^\dag_\pm$ as $F_\pm$ operators, which implies $Q=0$
(illustrating our comments of Appendix~\ref{WhynoKlein}).
For all neutral configurations, $S=1$ or $-1$
if $N_\tau$ is even or odd, respectively,
where $N_\tau$ is the number of charges $\nu_i$ 
occuring between $0$ and $\tau$, i.e.\ for which  $\tau_i \in (0,\tau)$.
 To see this,
one simply has to rearrange  the $F$'s under
the time-ordering symbol in (\ref{eq:expandDB-7}) until they all ``disappear''
via $F^\dagger_\nu F_\nu = 1$, and count the number
of minus signs produced by anticommuting
an $F_+$ or $F_+^\dag$
past an $F_-$ or $F_-^\dag$,
as illustrated by the following 
simple examples: 
\begin{eqnarray}
\nonumber 
N_\tau = 0: \qquad \; S_{+,--}^{\tau_1 > \tau} 
& = & 
 (-1)^1 \langle \left[F_+^\dag(\tau_1) F_-  (\tau_1)\right]  
  F_+(\tau) \, F^\dag_-(0) \rangle = 1
\\
\nonumber 
N_\tau = 1: \qquad \; S_{+,--}^{\tau > \tau_1} &=&
 (-1)^1 \langle    F_+(\tau) 
\left[F_+^\dag(\tau_1) F_-  (\tau_1)\right]  F^\dag_-(0) \rangle = - 1
\\
\nonumber 
N_\tau = 0: \quad
S_{+-,-+}^{\tau_1 > \tau_2 > \tau} &=&
 (-1)^2 \langle \left[F_+^\dag(\tau_1) F_-  (\tau_1)\right]  
\left[F_-^\dag(\tau_2) F_+  (\tau_2)\right]  
  F_+(\tau) \, F^\dag_+(0) \rangle = 1
\qquad \phantom{.}
\\
\nonumber 
N_\tau = 1: \quad
S_{+-,-+}^{\tau_1 > \tau > \tau_2} &=&
 (-1)^2 \langle \left[F_+^\dag(\tau_1) F_-  (\tau_1)\right]  
  F_+(\tau) 
\left[F_-^\dag(\tau_2) F_+  (\tau_2)\right]  
 F^\dag_+(0) \rangle = -1
\qquad \phantom{.}
\end{eqnarray}

It follows that $D_B (\tau) = Z_e (\tau) - Z_o (\tau)$,
where $Z_e$ and $Z_o$ contain only configurations
with $N_\tau = 1$ and $-1$, respectively. 
Thus  $Z_e$ and $Z_o$ can be interpreted as the grand-canconical
partition functions of a neutral Coulomb gas with
fugacity ${v \lambda_B \over 2 \pi a}$,
two charges $\pm 1/2$ at positions 0 and $\tau$,
and either an even or an odd number of
charges $\pm 1$ between them, respectively (and arbitrarily many charges
$\pm$ beyond $\tau$). 

This completes the exact mapping
of $D_B$ into a Coulomb gas representation
derived by Oreg and Finkel'stein \cite{OF}.
Evidently the minus sign in $Z_e - Z_o$ arose
from the anti-commutativity of fermion operators,
as emphasized by Oreg and Finkel'stein. Note
that their derivation
of it using field-theoretic bosonization 
\cite{OF,OFphysrevB} is rather more involved 
than ours using constructive bosonization, which illustrates the benefits of
using Klein factors.
Their conclusions \cite{OF} about the
asymptotic behavior of $Z_e - Z_o$ for $\tau \to \infty$,
and our criticism thereof, are discussed in Section~\ref{sec:OF}.




\begin{thebibliography}{99}
%
{\small 
%
\bibitem{Lud94a}  
A. W. W. Ludwig,
Methods of Conformal Field Theory in Condensed Matter Physics:
An Introduction to Nonabelian Bosonization,
in: Low-Dimensional Quantum Field Theories for Condensed Matter Physicists, 
S. Lundqvist, G. Morandi, Y. Lu (eds.) 
World Scientific 1995

\vspace*{-3.3mm} \bibitem{vondelftthesis}
Jan von Delft, Ph.D. Thesis, Cornell University 1995

\vspace*{-3.3mm} \bibitem{Emery79}
V. J. Emery, 
in: Highly Conducting One-dimensional Solids,  
J. T. Devreese, R. P. Evrard, V. E. Van Doren (eds.)
Plenum 1979 p.\ 247

\vspace*{-3.3mm} \bibitem{Haldane81}
F.  D. M. Haldane,
J. Phys.\ C. {\bf 14}  (1981) 2585



\vspace*{-3.3mm} \bibitem{Tomonaga}
S. Tomonaga,
Prog. Theor. Phys.\ (Kyoto) {\bf 5} (1950)  544

\vspace*{-3.3mm} \bibitem{MattisLieb}
D. C. Mattis,  E. H. Lieb, 
J. Math.\ Phys.\ {\bf 6} (1965) 304

\vspace*{-3.3mm} \bibitem{Luttinger}
J. M. Luttinger, 
J. Math.\ Phys.\ {\bf 4} (1963)  1154

\vspace*{-3.3mm} \bibitem{SchotteSchotte}
K. D. Schotte,  U. Schotte,
Phys.\ Rev.\ {\bf 182} (1969) 479

\vspace*{-3.3mm} \bibitem{Mattis74}
D. C. Mattis,
J. Math.\ Phys.\ {\bf 15} (1974)   609

\vspace*{-3.3mm} \bibitem{LutherPeschel}
A. Luther, I. Peschel, 
Phys.\ Rev.\ B {\bf 9} (1974) 2911



\vspace*{-3.3mm} \bibitem{Heidenreich75}
R. Heidenreich, B. Schroer, R. Seiler, D. Uhlenbrock,
Phys.\ Lett.\ {\bf 54A} (1975) 119

\vspace*{-3.3mm} \bibitem{Coleman78} 
S. Coleman,
Phys.\ Rev.\ D {\bf 11} (1975) 2088

\vspace*{-3.3mm} \bibitem{Mandelstam75}
S. Mandelstam,
Phys.\ Rev.\ D {\bf 11} (1975) 3026

\vspace*{-3.3mm} \bibitem{Haldane79}
F. D. M. Haldane,
J. Phys.\ C {\bf 12} (1979) 4791


\vspace*{-3.3mm} \bibitem{Shankar}
R. Shankar, Lectures given at the BCSPIN School, Katmandu, 1991,
in:  Condensed Matter and Particle Physics,
Y. Lu, J. Pati, Q. Shafi (eds.)  World Scientific 1993

\vspace*{-3.3mm} \bibitem{KF}
C. L. Kane,  M. P. A. Fisher, 
Phys.\ Rev.\ Lett.\ {\bf 68} (1992)  1220;
Phys.\ Rev.\ B {\bf 46}  (1992) 15233

\vspace*{-3.3mm} \bibitem{vDZF}
J. von Delft, G. Zar\'and, M. Fabrizio,
Phys.\ Rev.\ Lett. {\bf 81} (1998) 196 

\vspace*{-3.3mm} \bibitem{ZvD}
G. Zar\'and, J. von Delft, 
to be submitted to Phys.\ Rev.\ B

\vspace*{-3.3mm} \bibitem{Schoenhammer1}
K. Sch\"onhammer, V. Meden, 
Fermion-boson transmutation and
comparision of statistical ensembles in one dimension,
Am.\ J. Physics {\bf 64} (1996) 1168-1176 

\vspace*{-3.3mm} \bibitem{Schoenhammer2}
K. Sch\"onhammer, Interaction fermions in one dimension:
The Tomonaga-Luttinger model, cond-mat/9710330

\vspace*{-3.3mm} \bibitem{OF}
Y. Oreg, A. M. Finkel'stein, Phys.\ Rev.\ Lett.\ {\bf 76} (1996)  4230

\vspace*{-3.3mm} \bibitem{Furusaki}
A. Furusaki, 
Phys.\ Rev.\ B {\bf 56} (1997)  9352 

\vspace*{-3.3mm} \bibitem{FG}
M. Fabrizio, A. O. Gogolin,
Phys.\ Rev.\ Lett.\ {\bf 78}  (1997) 4527

\vspace*{-3.3mm} \bibitem{OF97} 
Y. Oreg, A. M. Finkel'stein, Phys.\ Rev.\ Lett.\ {\bf 78} (1997)  4528

\vspace*{-3.3mm} \bibitem{OFphysrevB}
Y. Oreg, A. M. Finkel'stein, Phys.\ Rev.\ B {\bf 53}  (1996) 10928

\vspace*{-3.3mm} \bibitem{Gelfand}
I. M. Gel'fand,  G. E. Shilov,
Generalized Functions, Vol.~1,
Academic Press, New York 1964 p. 331

\vspace*{-3.3mm} \bibitem{KotliarSi96}
G. Kotliar, Q. Si,
Phys.\ Rev.\ B {\bf  53} (1996)  12373

\vspace*{-3.3mm} \bibitem{Heidenreich}
R. Heidenreich,  R. Seiler, D. A. Uhlenbrock,
J. Stat. Phys.\ {\bf 22} (1980)  27

\vspace*{-3.3mm} \bibitem{Neuberger}
H. Neuberger, Tel Aviv University Thesis 1975

\vspace*{-3.3mm} \bibitem{Kleinhistory}
We thank
X. Si for pointing out to us that
in the field theory literature on bosonization, $F_\eta$ is referred to
as either the Klein factor or Klein transformation [e.g.\
Ha, Phys.\ Rev.\ D {\bf 29}  (1984) 1744;
 Halpern, Phys.\ Rev.\ D {\bf 12} (1975)  1684]; 
and that its historical origin is discussed by Klaiber
in the conference proceedings
Lectures in theoretical physics,  A. Barut, W. Britten (eds.)
1968




\vspace*{-3.3mm} \bibitem{Negele}
J. W. Negele, H. Orland,
Quantum Many-Particle Systems,
Addison-Wesley 1988

\vspace*{-3.3mm} \bibitem{Affleck}
I. Affleck, A. W. W. Ludwig,
Nucl. Phys.\ B {\bf 360}  (1991) 641


\vspace*{-3.3mm} \bibitem{Haldane80}
F.  D. M. Haldane,
Phys.\ Rev.\ Lett.\ {\bf 47} (1981)   1840 

\vspace*{-3.3mm} \bibitem{Matveev}
K. A. Matveev, Phys.\ Rev.\ B {\bf 51}  (1995)  1743;
see also A. Furusaki, K. A. Matveev,
Phys.\ Rev.\ B {\bf 52}  (1995)  16676

\vspace*{-3.3mm} \bibitem{FGGGG}
M. Fabrizio, A. O. Gogolin,
Phys.\ Rev.\ B {\bf 51} (1995)  17827;
N. V. Prokof'ev, Phys.\ Rev.\ B {\bf 49}  (1994) 2148


\vspace*{-3.3mm} \bibitem{KEG} 
A. Komnik, R. Egger,  A. O. Gogolin, 
Phys.\ Rev.\ B {\bf 56} (1997)  1153

\vspace*{-3.3mm} \bibitem{Fabrizio:priv}
M. Fabrizio, private communication.

\vspace*{-3.3mm} \bibitem{Gradshteyn}
I. S. Gradshteyn, I. M. Ryzhik,
Table of Integrals, Series and Products,
Academic Press 1980


\vspace*{-3.3mm} \bibitem{CM71}
F. Constantinescu, E. Magyari, 
``Problems in Quantum Mechanics'',
Pergamon Press 1971 p.\ 7 and 17 

\vspace*{-3.3mm} \bibitem{CN68}
P. Carruthers, M. M. Nieto, Rev.\ Mod. Phys.\ {\bf 40}
  (1968) 411; see particular section V. 




\vspace*{-3.3mm} \bibitem{Loss91}
D. Loss, K. Mullen,
Phys.\ Rev.\ A, {\bf 43}  (1991) 2129

\vspace*{-3.3mm} \bibitem{Bateman}
Bateman Manuscript Project, 
Tables of Integral Transforms, Vol.~1, 
McGraw Hill 1954, p.~120, Eqs.~(14) and~(15)

\bibitem{GogolinProkofev}
A. O. Gogolin and N. V. Prokof'ev,
Phys.\ Rev.\ B {\bf 50} (1994) 4921

}

\end{thebibliography}
\end{document}